\documentclass[]{jfm}
 \usepackage{amssymb}
 \usepackage{amsmath}
\usepackage{natbib}
\usepackage{psfrag}
\usepackage{graphicx}
\usepackage{rotating}
\usepackage{color}

\def\der#1#2{ {{\partial #1 } \over {\partial #2}} }

\def\tU{\widetilde{U}}
\def\tS{\widetilde{S}}

\begin{document}           % End of preamble and beginning of text.

\title
[
Transitional and turbulent flows in rectangular ducts 
]
{
DNS of transitional and turbulent flows in rectangular ducts 
}
\author[ P.~ Orlandi  and S.~ Pirozzoli\\]
%                AUTHORS
{
P.\ns  O\ls R\ls L\ls A\ls N\ls D\ls I ,
and
S.\ns  P\ls I\ls R\ls O\ls Z\ls Z\ls O\ls L\ls I
}
\affiliation
{
Dipartimento di Ingegneria Meccanica e Aerospaziale\\
Universit\`a La Sapienza, Via Eudossiana 16, I-00184, Roma
}
\date{\today}
\maketitle

\begin{abstract}
We carry out Direct Numerical Simulation (DNS) of flows in closed rectangular ducts with several aspect ratios. The Navier-Stokes equations  are discretized through a second-order finite difference scheme, with non-uniform grids in two directions. The duct cross-sectional area is maintained constant as well as the flow rate, which allows to investigate which is the appropriate length scale in the Reynolds number for a good scaling in the laminar and in the fully turbulent regimes. We find that the Reynolds number based on the half length of the short side leads to a critical Reynolds number which is independent on the aspect ratio ($A_R$), for ducts with $A_R>1$. The mean and rms wall-normal velocity profiles are found to scale with the local value of the friction velocity. At high friction Reynolds numbers, the Reynolds number dependence is similar to  that in turbulent plane channels, hence flows in rectangular ducts allow to investigate the Reynolds number dependency through a reduced number of simulations. At low Re the profiles of the statistics differ from those in the two-dimensional channel due to the interaction of  flow structures  of different size. The projection of the velocity vector and of the Reynolds stress tensor along the eigenvectors of the strain-rate tensor yields reduced Reynolds stress anisotropy, and simple turbulent kinetic energy budgets.
\end{abstract}

\section{Introduction}

Many efforts have been directed to understanding laminar, 
transitional and turbulent flows near walls. The turbulent
channel has been largely considered in Direct Numerical Simulations
(DNS), where two homogeneous directions allow
to get satisfactory statistics profiles with a limited number of fields. 
This flow  can not be exactly reproduced in laboratory experiments
where the effects of the lateral walls can not be eliminated.
Several studies, for instance the most recent by \cite{vinuesa_14}, and
by \cite{vinuesa_16}) were devoted to investigate the differences between
ideal two-dimensional turbulent channels and
rectangular ducts with high and low aspect ratio. The
simulations and the experiments were performed at intermediate Reynolds number 
($Re_\tau \approx 500$).  
The transitional regime for the square duct was considered numerically
by \cite{uhlmann_07} and experimentally by \cite{owolabi_16}.
Numerically it is easy to relate the friction 
$Re_\tau=u_\tau L/\nu$, with $L$ the length of the side of the square
duct and $u_\tau$ the mean friction velocity, to the bulk Reynolds number $Re_L= U_b L/\nu$.
They found that the turbulent regime is observed
above $Re_L=1077$, and up to $Re_L=2000$ there is a linear
relationship between the two Reynolds numbers. In the laminar
regime the relationship was given by \cite{tatsumi_90}
$Re_\tau=\sqrt{a Re_L}$ with $a=3.3935$. Experimentally it is
rather difficult to have an exact value of the global
friction velocity since the wall shear stress varies along
the walls. Therefore \cite{owolabi_16} defined the critical Reynolds
number as that, at which, a sharp decrease of the mean
streamwise velocity $U_1$ at the center of the duct is measured. 
They obtained values in good agreement with those in the DNS
of  \cite{uhlmann_07}. The value $Re_L=1077$  does not differ too much from
that in circular pipes ($Re_C=1125$ by \cite{Faisst_04})                
and in a plane channel ($Re_C=1000$ by \cite{Carlson_82}).
\cite{orlandi2015} through DNS of Poiseuille and Couette flows
observed a jump on the total turbulent kinetic energy respectively
at $Re \approx 1800$ for Poiseuille and $Re=1000$ for the Couette 
flows.  These Reynolds numbers are defined as $Re=U_M H/\nu$ with
$H$ half channel width and $U_M$ the maximum of the laminar
parabolic Poiseille profile and the wall velocity for Couette.
It is important to recall that the initial amplitude disturbances in the numerical
or the inlet conditions in the laboratory experiments can affect
the value of the critical Reynolds. 
The sensitivity to the disturbances was carefully 
investigated by \cite{Fitzgerald_04} reporting the 
results of sophisticated experiments in circular pipes~\citep{Hof_03},
showing that reduction of the amplitude of the 
disturbance may lead to an increase of  the transition Reynolds number 
up to $Re = 18 000$, much greater than the value  obtained by 
\cite{REY895}. \cite{Orlandi_08} performed numerical experiments
to further analyse the influence of the initial disturbances
in circular pipes.
From the observation that there  are not large
differences in the critical Reynolds number 
between flows with well localised secondary flows
and flows without it (circular pipe and plane channel)
it is worth analysing in rectangular ducts which
is the appropriate length scale giving a fixed critical
Reynolds number. 

In non circular ducts several length scales can be defined.
One largely used is the hydraulic diameter $D_h=4P/A$, with
$P$ the perimeter and $A$ the cross-section area of the duct ($Re_D=U_b D_h/\nu$). 
In the present simulations the reference length is assumed to be the radius of 
an equivalent pipe, $r_p$ ($A=\pi r_p^2$), hence the relevant computational
Reynolds number is $Re=U_b r_p/\nu$. A further length scale appropriate
for rectangular ducts may be half the length of the short
side, $L_3$. Let $L_2$ is the length of the long side, 
then $L_2=L_3 A_R$, with $A_R$ the aspect ratio.
As a first check of the differences in the profiles of the friction
factor $C_f=2u^2_\tau/U^2_b$ versus the three Reynolds numbers
above reported can be obtained by using equation (3-48) at Pg.113
of \citet{white_74}. The analytical linear profiles
together with the present simulations allow to see the 
different trends of $C_f$ versus the Reynolds numbers.

The secondary motions, widely
analysed in several DNS papers, starting from \cite{gavrilakis_92}
at low Re, and ending with \cite{pirozzoli_18} at 
much higher Re, is rather weak with
respect to the main motion, hence it is likely
that they do not alter substantially the statistical profiles 
with respect to canonical wall-bounded flows. In particular, this should
be the case at high values of the Reynolds numbers,
at which the strongest vorticity
becomes localised in a smaller and smaller region~\citep{pirozzoli_18}.
Secondary motions have been deeply investigated by \cite{Joung_07},
and a clear picture of its effect can be observed in their figure 6
reporting undulations of $\tau_w/\overline {\tau_w}$ 
near the corner, with differences among the profiles 
at different Reynolds numbers. The
reduction of friction approaching the corner
should  also appear  on the shape and size of the 
streamwise vortical structures, and
therefore on the distribution of the turbulent kinetic energy.  
The decrease of the wall shear stress should be different along the short
and the long side of the rectangular  duct.  The present simulations are focused
to investigate the variations with the Reynolds number and
with the aspect ratio of several statistics in particular to
demonstrate whether the wall scaling with the averaged or the local
friction velocity hold.
In the corners the local $Re_\tau$ decreases, and therefore,
in the same flow it is possible to investigate whether
the Reynolds number dependence of the statistics in wall units
shown  by \cite{orlandi2015} for the plane channel
are also recovered.

In plane channel, DNS are often performed by
pseudospectral methods, similar or equal to that described in \cite{kim_87}.
These results can be considered as reference solutions to validate
those obtained by other numerical methods. 
However, \cite{bernardini_13} demonstrated that by using the same resolution
the streamwise spectra by second order schemes were as good as those by
pseudospectral methods in a convecting reference frame. 
However, in the steady reference frame
the profiles  of the statistics did not show any appreciable difference 
with those by pseudospectral methods.
The improvement achieved in the convective reference
frame was detected in the streamwise velocity spectra,
at high wavenumbers with low energy 
content. Based on these observations, in the present study 
a second-order staggered finite-difference scheme is used,
with the further advantage of using non-uniform grids
in two space directions,
%The other option was to use numerical methods
%similar to those used by \cite{vinuesa_14}, which from our
%point of view are more cumbersome. 
adapted from a code previously developed 
for the simulations of Poiseuille and Couette
flows.
The same procedure
of low-storage Runge-Kutta time integration of the nonlinear terms,
and implicit treatment of the viscous terms was also used.
The fundamental difference from the method developed for flows 
with two homogeneous directions (spanwise and streamwise)
resides in the solution of the elliptical equation. 
In the case of two homogeneous direction, the use of two Fast Fourier Transformations (FFT) and
a tridiagonal solver \citep[see][Chapter 9]{orlandi_12},
allows to solve Poisson equation within round-off errors.
In order to get a fast  solution at high Reynolds numbers a large number of processors
can be used through the MPI (Message Parallel Interface) directives,
by subdividing the computational domain into pencil-shaped sub-domains.
In presence of two direction of grid non-uniformity, the Poisson equation can be solved
either through a multigrid method~\citep{Joung_07}, or
direct solvers based on the cyclic reduction algorithm as FISHPACK, developed at NCAR by \citet{adams_75}.
The convergence of the multigrid is linked to the
coordinate stretching, that set the eigenvalues of the associated matrices, hence
to avoid possible slow convergence, the FISHPACK subroutine is a good choice, 
clearly less efficient than FFT-based direct solvers.
The main disadvantage of the FISHPACK consists in a limitation of the number of processors, 
since the computational domain can only be divided into slabs along the streamwise direction.
Another possible alternative is the use of compressible flow solvers adapted for
low-Mach-number flows~\citep{pirozzoli_18}, which were shown to 
yield nearly identical solutions as incompressible solvers.

\section{Flows set-up }

A large number of flow cases have been simulated, with resolution 
depending on the Reynolds number. At low and intermediate
$Re$ the computational mesh used is $161 \times 161 \times 161$, 
up to  $513 \times 385 \times 257$,
respectively in the streamwise $x_1$, lateral $x_2$ and vertical $x_3$
directions. The smaller number of points in $x_3$ than those
in $x_2$ is due to the
reduction of $L_3$ with the increase of the aspect
ratio $A_R$. In the present simulations the
flow rate is maintained constant by adding at each time step a
mean pressure gradient $\Pi$ balancing the friction losses due to the
walls. In the evaluation of the flow rate the cross-sectional
area appears which has been maintained constant. 
At each $Re$, several $A_R$ were considered, namely $1,2,4,5,6,7$.
Simulations with $A_R=1.5$ were also performed
in a range of $Re$ close to the critical one.
For the governing equations the reader may refer to 
\citet[][Chapter 9]{orlandi_12}, with no-slip boundary
conditions imposed at the duct walls. The distribution of
the initial streamwise velocity is irrelevant, in fact
it has been observed that in few time units the distribution adjusts
to the shape of the duct. However it is mandatory  that
$U_b={\int \int \int u_1 dx_1dx_2dx_3}/({L_1L_2L_3})=1$.
Previous DNS~\citep{uhlmann_07} showed that for square
ducts and for $L_1/(0.5 L_3)> 10$, the wall shear stress does not change.
In the present simulations for all $A_R$, we use $L_1/r_p=16$,
hence at $A_R=1$ we have $L_1/(0.5 L_3)=18.05$. For 
high $A_R$, $L_3$ decreases and since, as later on shown, $L_3$
is the relevant length scale, the duct is long enough
to resolve the energy-containing longitudinal structures. 

Depending on $A_R$ and on the value of $Re$ the 
simulations evolve for a different time. At 
high Reynolds number the transient time to reach
the instant at which $\Pi$ oscillates around 
the averaged $\overline {\Pi}$ is short. However, the simulations
must evolve for a sufficient time in order to have distributions 
of the statistics in the $x_2-x_3$ planes respecting as much as possible the
geometrical symmetries. At low $Re$, and, in 
particular, under laminar conditions a long initial transient
is necessary to damp the initial disturbances through viscous diffusion.
The averaged wall shear stress
$\overline {\tau_w}$, calculated through 
$\overline {\tau_w}= \overline{\Pi} {L_2 L_3}/{(L_2+L_3)}$, allows to define
the averaged friction velocity 
$\overline {u_\tau}=\sqrt{\overline {\tau_w}}$, and the friction coefficient, $C_f=2\overline {\tau_w}/U_b^2$.

\begin{figure}
\centering
\hskip -1.8cm
\includegraphics[width=5.6cm,clip]{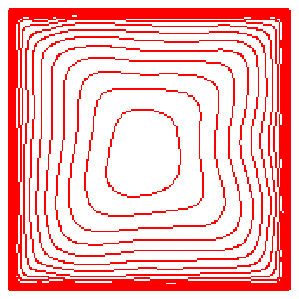}
\hskip -0.6cm
\includegraphics[width=5.6cm,clip]{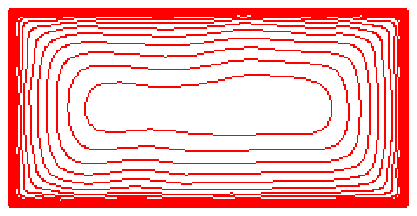}
\hskip -0.6cm
\includegraphics[width=5.6cm,clip]{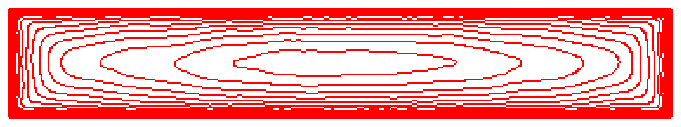}
\vskip -0.5cm
\hskip 0.5cm a) \hskip 2.7cm b) \hskip 2.7cm c)
\caption{Contours of mean streamwise velocity $<U_1(x_2,x_3)>$ for DNS of 
ducts at $Re=7750$, for a) $A_R=1$ b) $A_R=2$, c) $A_R=6$,
in increments $\Delta=0.05 U_b$.
}
\label{fig1}
\end{figure}

The shapes of the duct sections are depicted in figure~\ref{fig1}
through contour plots of the mean streamwise velocity component 
at $Re=7750$.  The plots are shown only for $A_R=1, 2, 6$ 
to appreciate how $L_3$ 
reduces and $L_2$ increases.
The mean quantities, here indicated by capital letters
are evaluated by averaging in the streamwise direction $x_1$
of the duct, and in time. The averages in time
were estimated by storing a sufficient number of fields saved
every $1.0$ time units. It may be observed
that this is in general insufficient to reproduce the expected geometrical
symmetries. Hence, further averaging is carried out
by quarters of the cross section. It is important
to recall that in turbulent plane channels 
the profiles in half channel are given. 
%In turbulent circular pipes
%\cite{orlandi_18} found that a secondary motion does exists
%that does not affect the radial profiles of the statistics.

The table listing some of  the global results, 
for the cases simulated 
and the grid used follows.

\section{ Results      }

\subsection{Square ducts}

\subsubsection{Friction factor}

%\begin{table}[htb]
\begin{table}
\centering
\begin{tabular*}{1.\textwidth}
{@{\extracolsep{\fill}}ccccccccccc}
\hline
Flow case & $Re $   & $Re_\tau$ & $N_f$ & $N_x$ & $N_y$ & $N_z$ &
$\overline{\tau_2}/\overline{\Pi}$ & $10 \overline{\Pi}$ & $Re_{D}$ & $Re_3$\\
\hline
$ A1_{2K}  $ & 2500 & 174  & 400  & 161 & 161 & 129 & $0.50$ & $0.341$ & 2500 & 2216\\
$ A2_{2K}  $ & 2500 & 174  & 400  & 161 & 161 & 129 & $0.69$ & $0.364$ & 2357 & 1567\\
$ A5_{2K}  $ & 2500 & 178  & 400  & 161 & 161 & 129 & $0.87$ & $0.478$ & 1863 & 991\\
$ A7_{2K}  $ & 2500 & 147  & 200  & 161 & 161 & 129 & $0.92$ & $0.376$ & 1750 & 837\\
$ A1_{5K}  $ & 5000 & 323  & 215  & 257 & 257 & 193 & $0.50$ & $0.295$ & 5000 & 4431\\
$ A2_{5K}  $ & 5000 & 328  & 215  & 257 & 257 & 193 & $0.69$ & $0.323$ & 4714 &3133\\
$ A5_{5K}  $ & 5000 & 340  & 182  & 257 & 257 & 193 & $0.87$ & $0.440$ & 3727 &1982\\
$ A7_{5K}  $ & 5000 & 342  & 152  & 257 & 257 & 193 & $0.91$ & $0.501$ & 3307 &1675\\
$ A1_{15K} $ & 15000 & 859  & 140  & 513 & 385 & 257 & $0.50$ & $0.232$ & 15000 &13293\\
$ A2_{15K} $ & 15000 & 872  & 140  & 513 & 385 & 257 & $0.69$ & $0.254$ & 14140 &9400\\
$ A5_{15K} $ & 15000 & 915  & 120  & 513 & 385 & 257 & $0.87$ & $0.353$ & 11180 &5945\\
$ A7_{15K} $ & 15000 & 927  & 120  & 513 & 385 & 257 & $0.91$ & $0.409$ & 9922 &5024\\
\hline
\end{tabular*}
\caption{
List of parameters for the turbulent cases, the first
number after $A$ indicates $A_R$, the subscript indicates the
$Re$ number given in the first column. 
$N_ff$ is the number of fields used
to evaluate the statistics.
$N_x$, $N_y$, $N_z$ are the number of grid points in the axial,
in the wider lateral direction, and in the shorter vertical direction.
The area is fixed and equal to $\pi$. The total friction is indicated
by $\Pi$ and the contribution of the longer side is $\tau_2$.
$Re_{D}$ is the Reynolds number based on the hydraulic diameter, and $Re_3$ is the Reynolds
number scaled with the shorter side half-length ($L_3/2$).
}
\label{tab1}
\end{table}

\begin{figure}
\centering
\vskip -0.0cm
\hskip -1.8cm
\psfrag{ylab} {$C_f       $}
\psfrag{xlab}{ $ $}
\includegraphics[width=8.0cm]{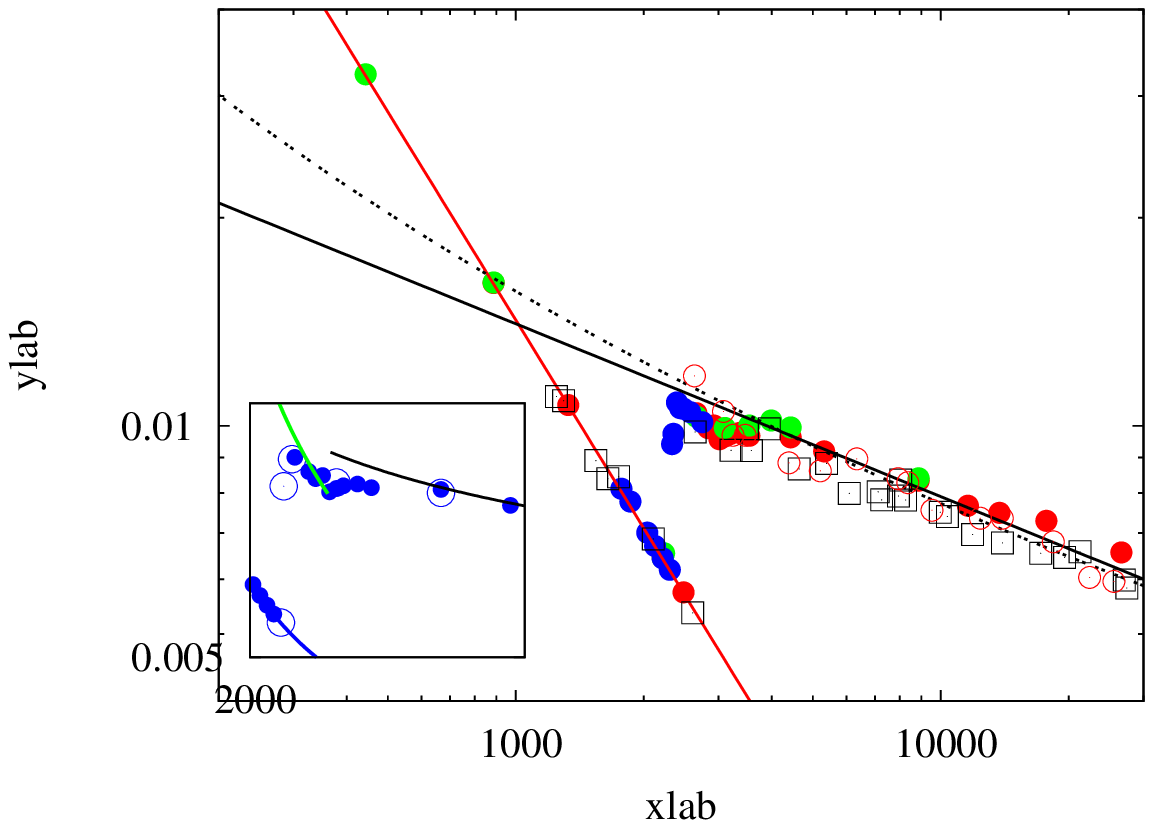}
\psfrag{ylab} {$Re_\tau    $}
\psfrag{xlab}{ $ $}
\includegraphics[width=8.0cm]{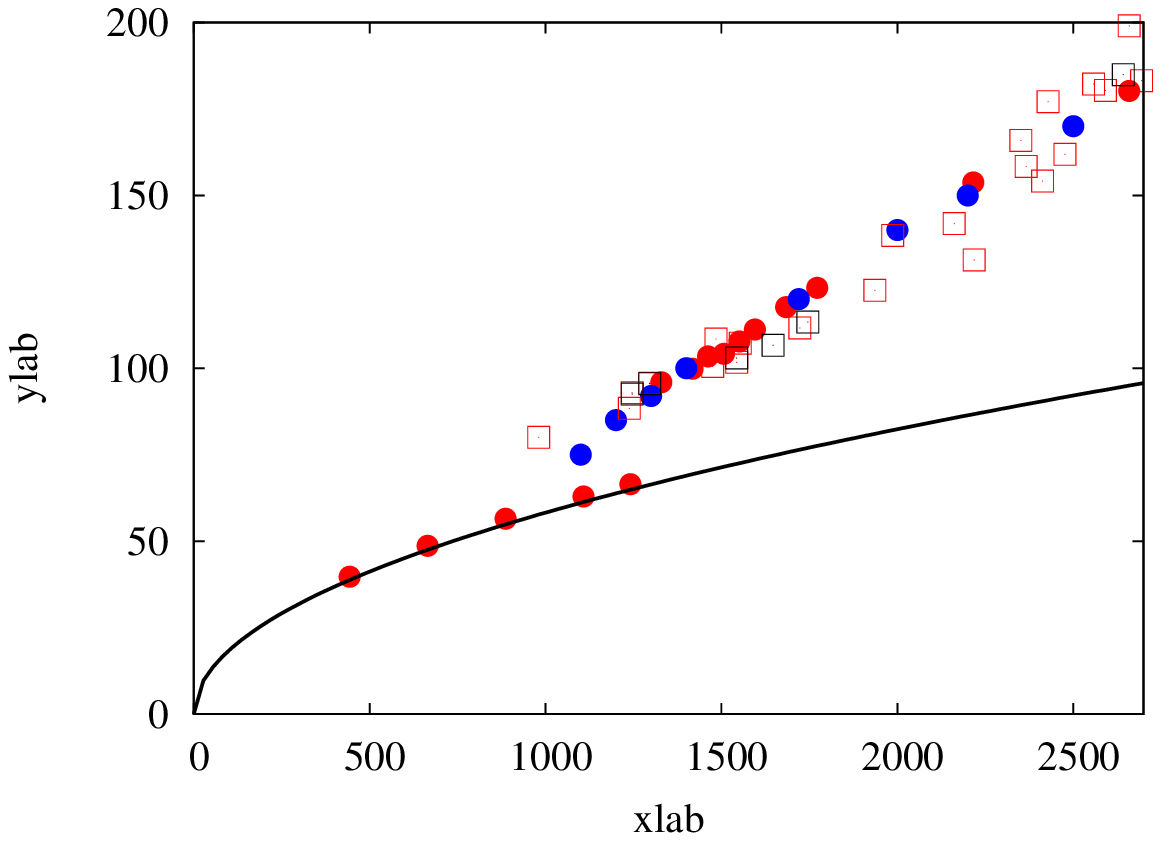}
\vskip -.5cm
\hskip 5.0cm  a)   \hskip 7.5cm  b)
\vskip -.2cm
\hskip -1.8cm
\psfrag{ylab} {$U_1/U_b       $}
\psfrag{xlab}{ $Re$}
\includegraphics[width=8.0cm]{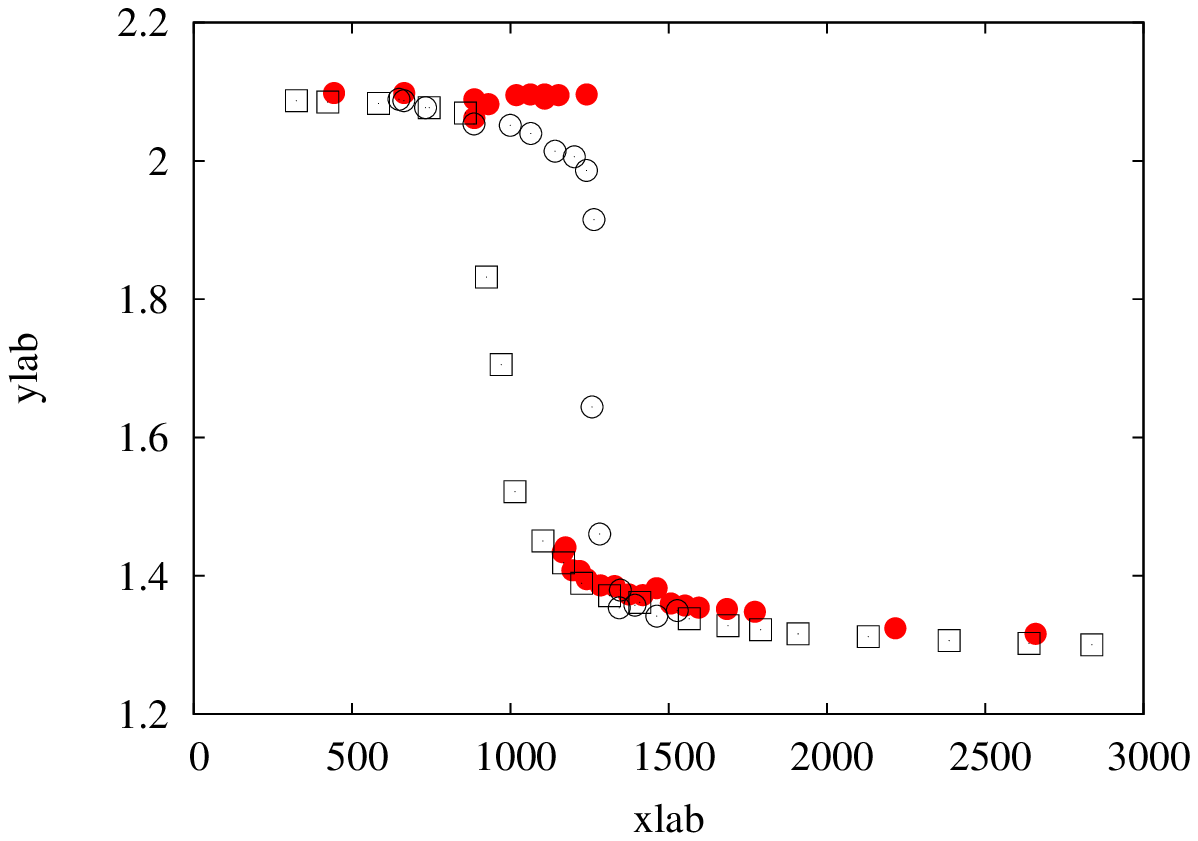}
\psfrag{ylab} {$u_{rms}/U_b       $}
\psfrag{xlab}{ $Re$}
\includegraphics[width=8.0cm]{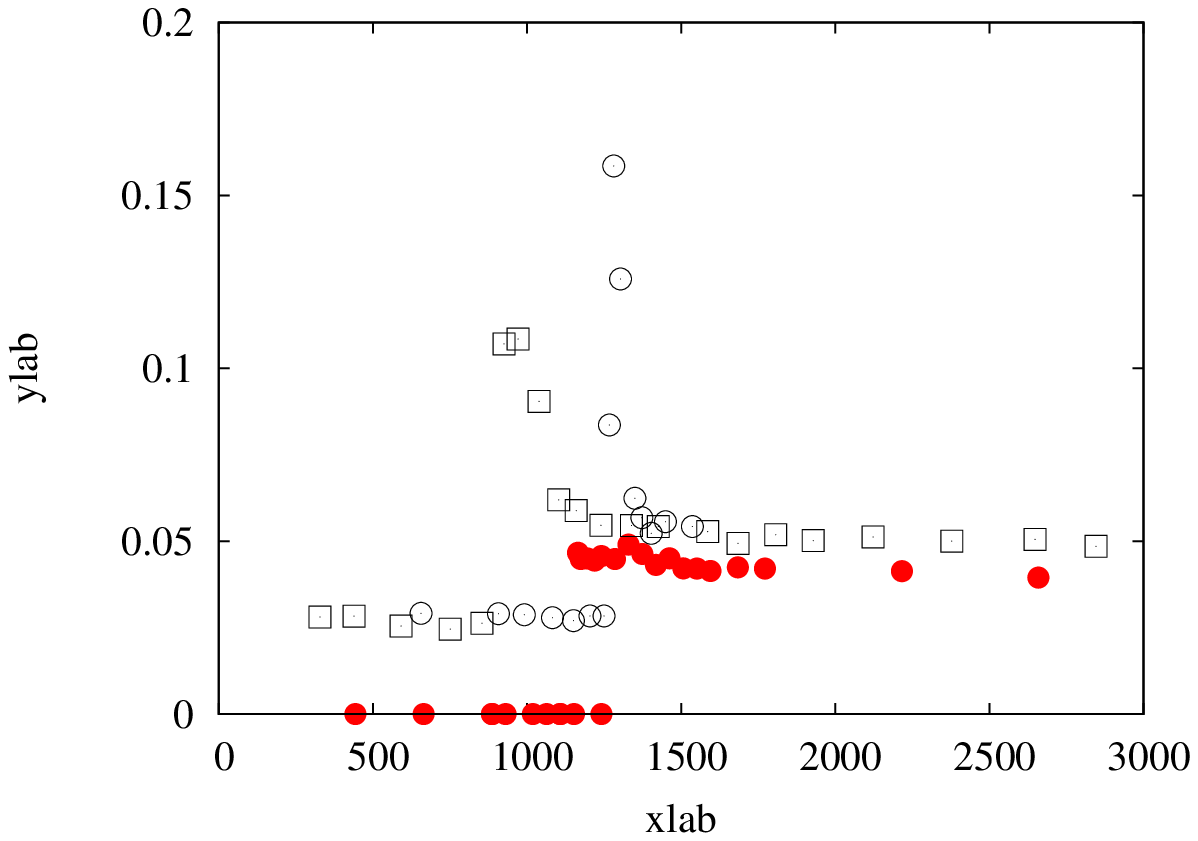}
\vskip -.5cm
\hskip 5.0cm  c)   \hskip 7.5cm  d)
\vskip 0.0cm
\caption{
Square ducts: a) friction coefficient $C_f$ versus bulk Reynolds number
$Re_b=U_b H/\nu$ ($H=2L_2$) solid circle present, red different
resolution in $x_2$ and $x_3$, green equal resolution both with
$L_x=4\pi$, blue $L_x=6\pi$ and $160$ non-uniform grid intervals
in $x_2$ and $x_3$, Prandtl (black), Blasius (blue) 
and laminar (red) theoretical curves,
open squares data from \cite{jones_76} red \cite{schiller_23} 
black \cite{hartnett_62};
b) $Re_\tau$ versus $Re_b$ symbols as in a)
and blue  solid triangle \cite{uhlmann_07} DNS,
solid black line $Re_\tau=\sqrt{3.3935 Re}$ ;
c) $U_1/U_b$, d) $u_{rms}/U_b$ at the center of the duct,
red solid circle present, \cite{owolabi_16} black open square with
inlet tripping, circle no tripping.
}
\label{fig2}
\end{figure}

Wall-bounded flows are characterised by variation of
the friction coefficient
with the bulk  Reynolds number. 
In canonical flows as circular pipes and two-dimensional plane channels,
the relevant length scale is well defined, 
the pipe diameter in the first case, and the half channel height in the latter.
For ducts with complex shape the 
hydraulic radius $D_h$ is frequently used.
%Since the circular pipe can be considered
%as a reference flow the area of
%the duct, for any aspect ratio, is set equal to that of a pipe $A_p$.
The validation of  the present results is first carried out for square
ducts, widely investigated under laminar, 
transitional and fully turbulent conditions 
at high Reynolds numbers. Several laboratory data were reported
by \cite{jones_76}, among which we extract the values by \cite{schiller_23}  and
\cite{hartnett_62} of the friction factor, $f=4 C_f$.
More recently, \cite{owolabi_16} investigated in greater detail the transitional
regime in square-duct flow through laboratory experiments, 
with and without inlet tripping. They measured the mean
and the $rms$ streamwise velocity at the duct center and at a distance
$0.3 h$ (with $h$ the side of the duct) from the wall.
We consider here only the data at the duct center, where
the mean velocity $U_1$ is high in laminar flows, and 
sharply decreases past transition.
Correspondingly, its variance abruptly increases starting from zero.
Figure ~\ref{fig2}c shows that the present data reproduce
well the reduction of $U_1$, and in figure~\ref{fig2}d the
increase of $u_{rms}$.  
Figure ~\ref{fig2}b
shows the different relationships between $Re_\tau$ and $Re_b$ in the laminar
and the turbulent regimes. The laminar results agree well with the 
theoretical relation $Re_\tau=\sqrt{3.3935 Re}$~\citep{tatsumi_90}, and fully
turbulent DNS data~\citep{uhlmann_07}.
This figure further
shows that DNS reproduces the laboratory
data. Data scatter is observed around the
critical Reynolds number, which may be due to
the different initial conditions in DNS, or to inlet
disturbances in the laboratory. To better investigate 
this influence, our data have been compared with the
mean velocity (figure~\ref{fig2}c) and 
$rms$ velocity (figure~\ref{fig2}d),
measurements by~\citet{owolabi_16}.
Those authors inserted a trip at the
inlet of the square duct observing that in these circumstances 
transition occurs at $Re\approx 940$ (figure~\ref{fig2}c). 
Without the tripping device,
transition occurs at $Re=1350$ (figure~\ref{fig2}c). The present results
agree with the data obtained without trip.
The DNS data show constant velocity in the laminar conditions, 
and sharp transition to a value that decreases
slowly in the turbulent regime. The value reached at the highest
$Re$ considered ($Re_b=2670$) is $U_1/U_b=1.281$. The same trend 
for the laboratory and numerical experiments is shown for 
$u_{rms}/U_b$ in figure~\ref{fig2}d. In this case 
the values of the maxima in the DNS are
smaller than those in the experiments. This
quantity should be zero in the laminar regime, hence it is not clear why 
\cite{owolabi_16} report a non-zero value both with
and without the tripping device.

In figure~\ref{fig2}a, an undulation of $C_f$ is observed
just after transition. To emphasise
this behavior the data are shown in linear scale in the figure inset.
showing an initial decay according a line (in green)
with a value twice that in the laminar regime.
At $Re_b \approx 3000$ a small growth of $C_f$ leads to the 
point where $C_f$ starts decaying according to 
Blasius friction law (black line).
The complex behavior of $C_f$ in the transitional range
was not reported in the DNS of \cite{uhlmann_07}, nor
in the measurements of \cite{owolabi_16}. 
However, those authors stated that in this
range of Reynolds numbers the secondary
motion is characterised by four eddies, rather than the
conventional pattern of eight eddies.
To investigate whether this change is also found here, 
in figure~\ref{fig3} we show the profiles of the wall shear stress,
$\tau_w=S_n/Re$, and of the wall-normal velocity gradient, $S_n=d U_1/d n|_w$,
as a function of the distance from the corner ($r$). 
We point out that $S_n$ is here averaged over
both walls and in each quadrant.

\subsubsection{Wall friction profiles}

\begin{figure}
\centering
\vskip -0.0cm
\hskip -1.8cm
\psfrag{ylab} {$\tau_w    $}
\psfrag{xlab}{ $r$}
\includegraphics[width=8.0cm]{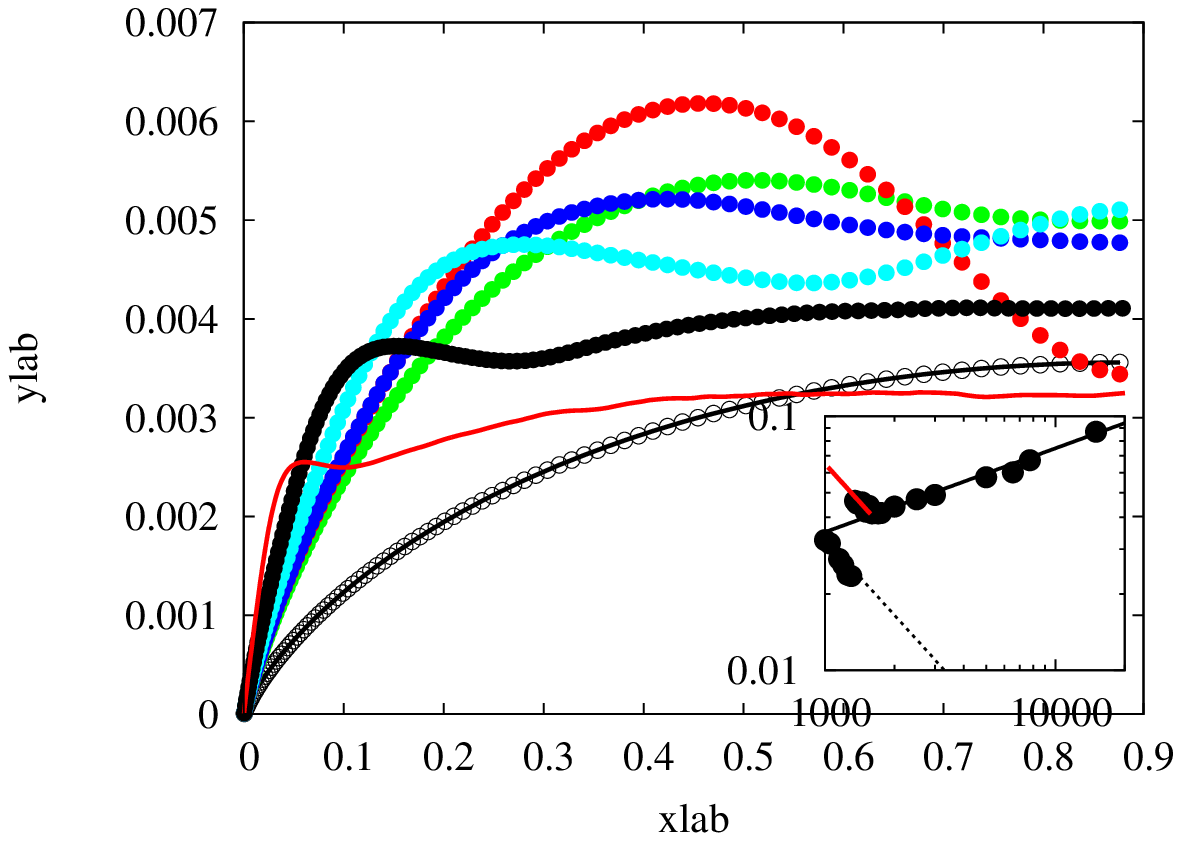}
\psfrag{ylab} {$S_n$}
\psfrag{xlab}{ $r$}
\includegraphics[width=8.0cm]{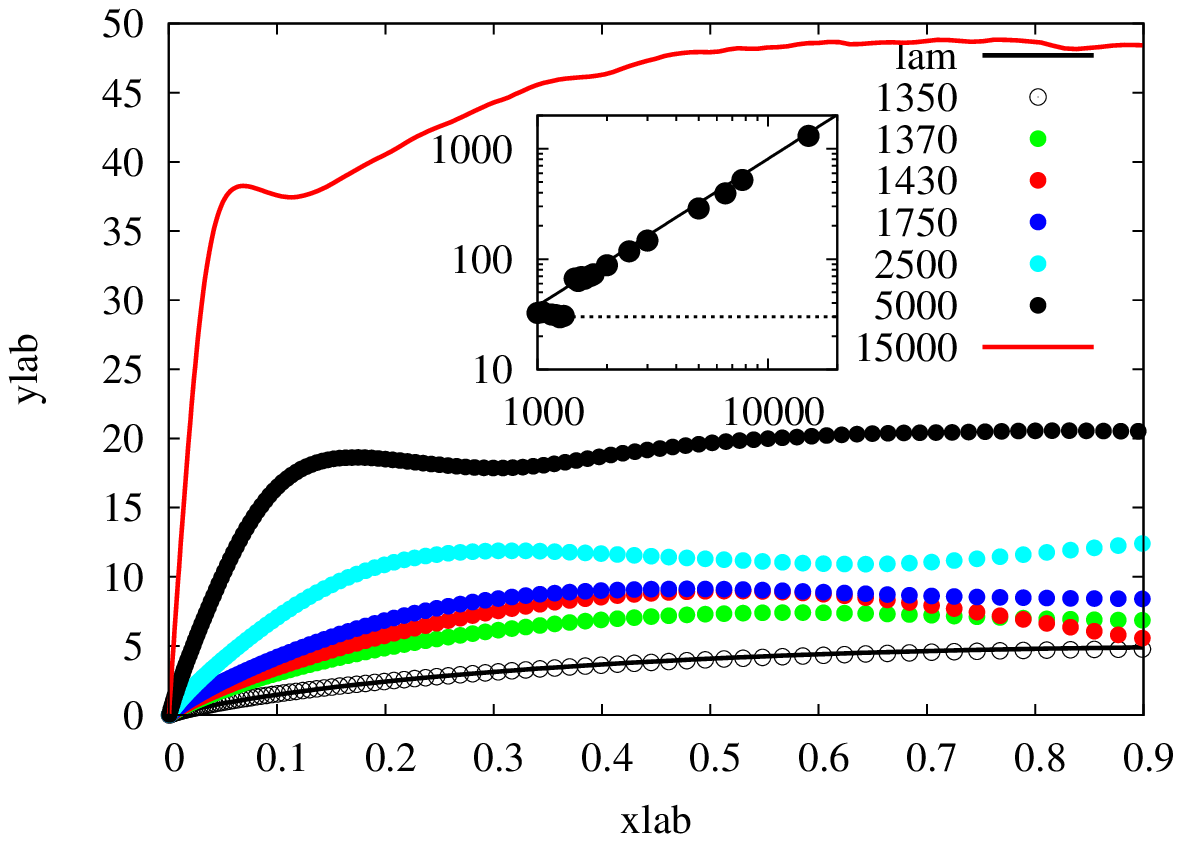}
\vskip -.5cm
\hskip 5.0cm  a)   \hskip 7.5cm  b)
\caption{Square ducts: profiles of 
a) wall shear stress, $\tau_w$ and b) wall-normal velocity gradient, $S_n$, 
at the wall, versus the distance from the corner ($r$), at different $Re$,as listed in
the legend of panel b).
}
\label{fig3}
\end{figure}

To understand the discussed behavior of $C_f=\int \tau_w dr$, it
is worth analysing the  drastic changes of the profiles of $\tau_w$
going from  the laminar through the transitional to the fully
turbulent regimes. In figure~\ref{fig3}a        
$\tau_w$ is zero at $r=0$ (the corner) and the 
strength of the secondary motion dictates the
different trends near the corners. As expected, the velocity gradients 
in the laminar regime do not
depend on the Reynolds numbers. In particular, in figure~\ref{fig3}b,
at $Re=1300$, $S_n$ 
is in perfect agreement with the expression given at
page 113 of \citet{white_74}. 
On the other hand, in the transitional
and turbulent regime, $S_n$ grows with different trends depending
on the Reynolds number.  The slope near the
corner is plotted in the inset of figure~\ref{fig3}b
showing a growth proportional to $Re^{4/3}$. 
Accordingly, the gradient of $\tau_w$ grows 
as $Re^{1/3}$ in the fully turbulent regime, as shown
in the inset of figure~\ref{fig3}a. 
In the transitional regime the 
variation of $d\tau_w/dr|_{r=0}$ with $Re$ 
equals that in the laminar regime, as given           
by the red line in the inset of figure~\ref{fig3}a.
The transitional regime ends at $Re=1600$, and from $Re=1750$
the values of $d\tau_w/dr|_{r=0}$ are aligned with those at high $Re$ numbers.
%The profiles of $\tau_w$ 
%in figure~\ref{fig3}a are obtained by multiplying the
%profiles of $S_n$ in figure~\ref{fig3}b for its $Re$ number.
Strong shape variation
in the various flow regimes are apparent. In the laminar regime, 
$\tau_w$ decreases linearly following as predicted
by equation (3-47) of \citet[p. 113]{white_74}.
In the transitional regime, and in particular in
the range of $Re$ with  $C_f$ decreasing  as in the laminar
regime  (see the insert of figure~\ref{fig2}a),
a maximum of $\tau_w$ occurs at distance $r=0.5$ from the
corner (red dots in in figure~\ref{fig3}a). 
As discussed later on, this behaviour 
does not occur at both the $x_3$ and $x_2$ walls, but rather 
on either one, depending on the Reynolds number.
Increasing $Re$, two peaks appear
well depicted by the blue and
black dots in figure~\ref{fig3}a. One of the peaks
moves closer to the corner at higher $Re$, whereas the other 
nearly remains at the center of the side,
with small oscillations in a region which becomes wider at higher $Re$.
\citep[figure 10]{pirozzoli_18} presented distributions of
$\tau_w$ in DNS evolving for times much longer than the present ones.
At sufficiently high $Re$ the shape of the profiles is however equivalent to that
shown in figure~\ref{fig3}a.

\begin{figure}
\centering
\vskip -0.0cm
\hskip -1.8cm
\includegraphics[width=3.5cm]{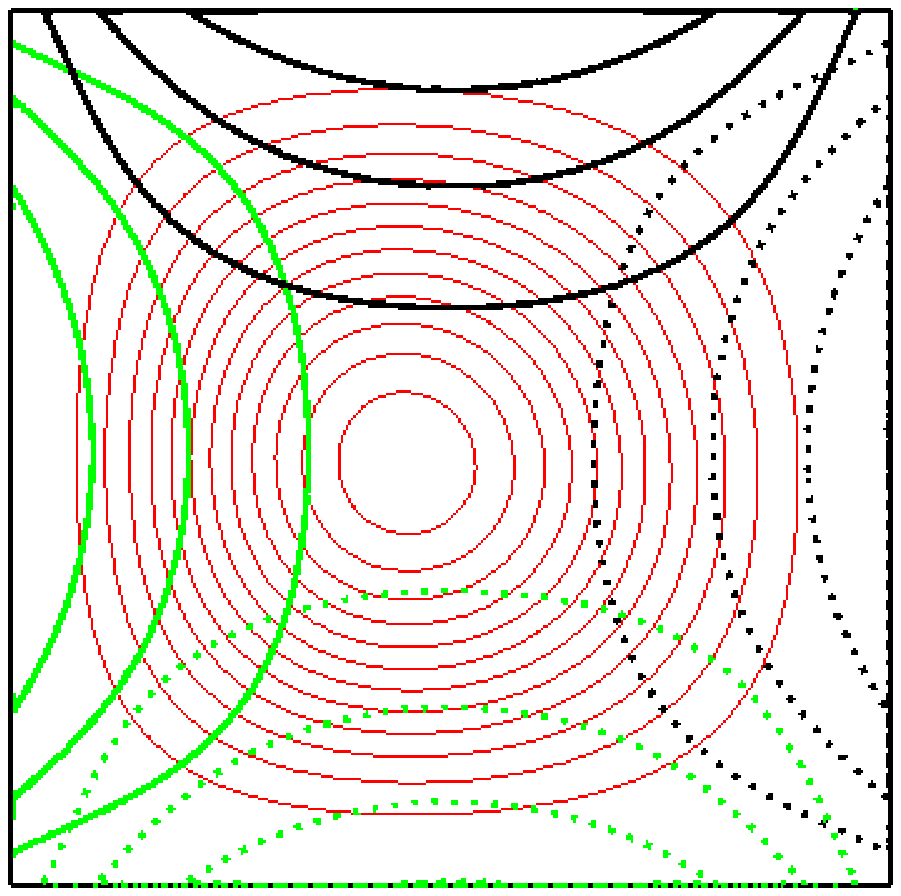}
\hskip +0.2cm
\includegraphics[width=3.5cm]{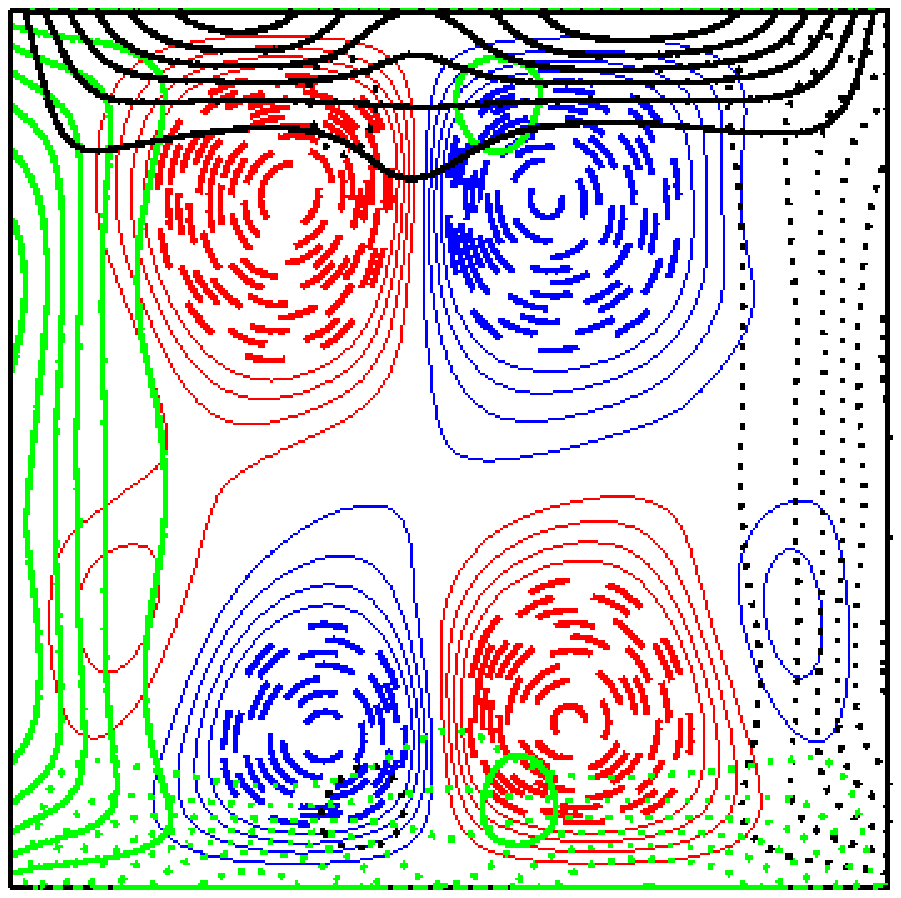}
\hskip +0.2cm
\includegraphics[width=3.5cm]{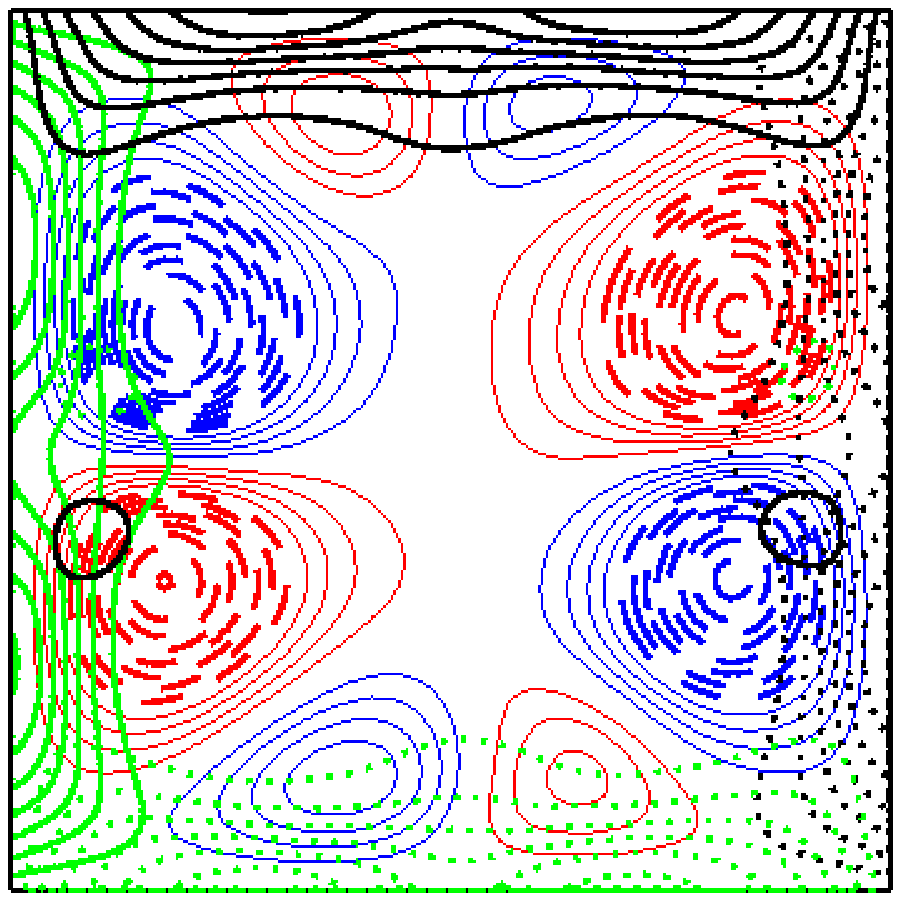}
\hskip +0.2cm
\includegraphics[width=3.5cm]{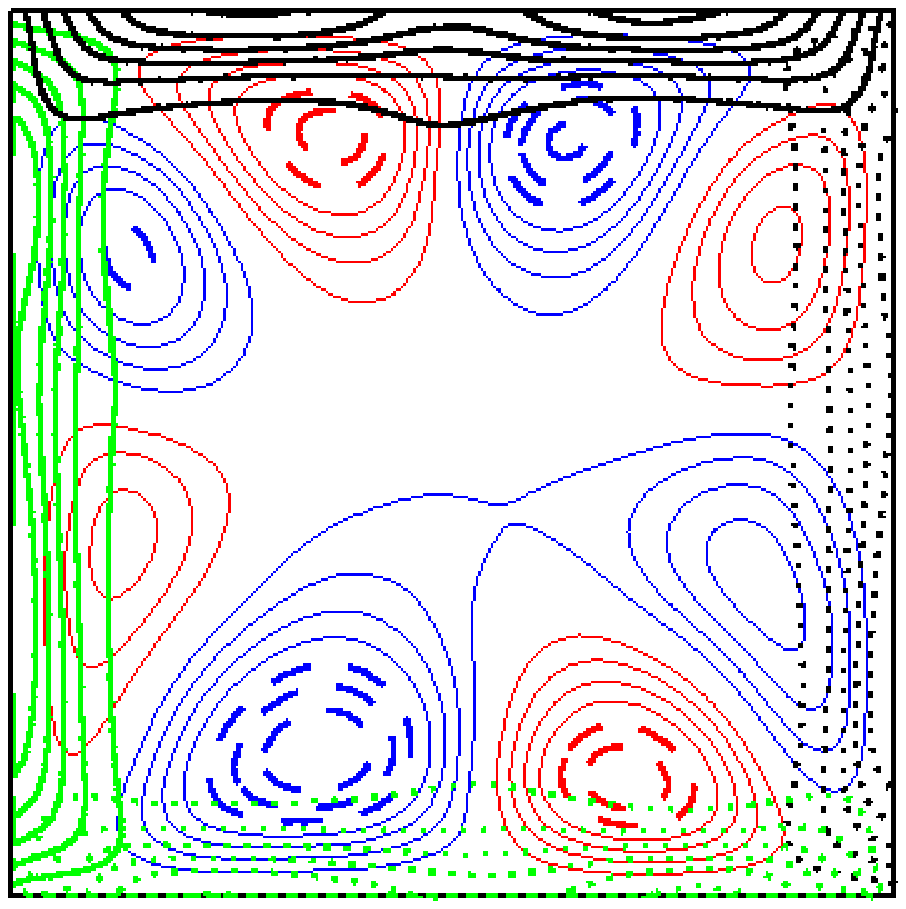}
\vskip -.2cm
\hskip 1.0cm  a)   \hskip 3.5cm  b) \hskip 3.5cm  c)   \hskip 3.5cm  d)
\vskip 0.0cm
\hskip -1.8cm
\psfrag{ylab} {$\tau_w    $}
\psfrag{xlab}{ $s$}
\includegraphics[width=4.0cm]{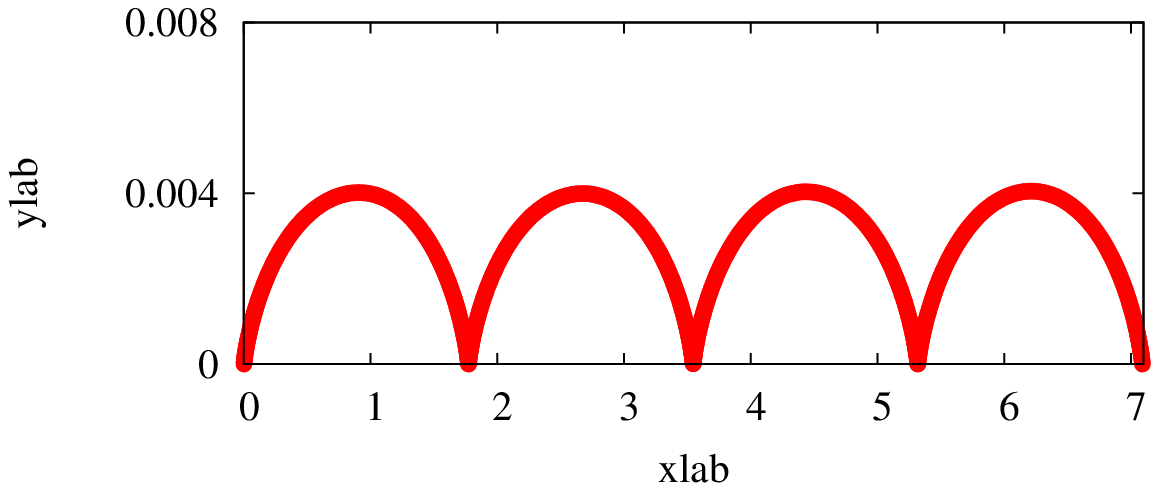}
\psfrag{ylab} {$  $}
\includegraphics[width=4.0cm]{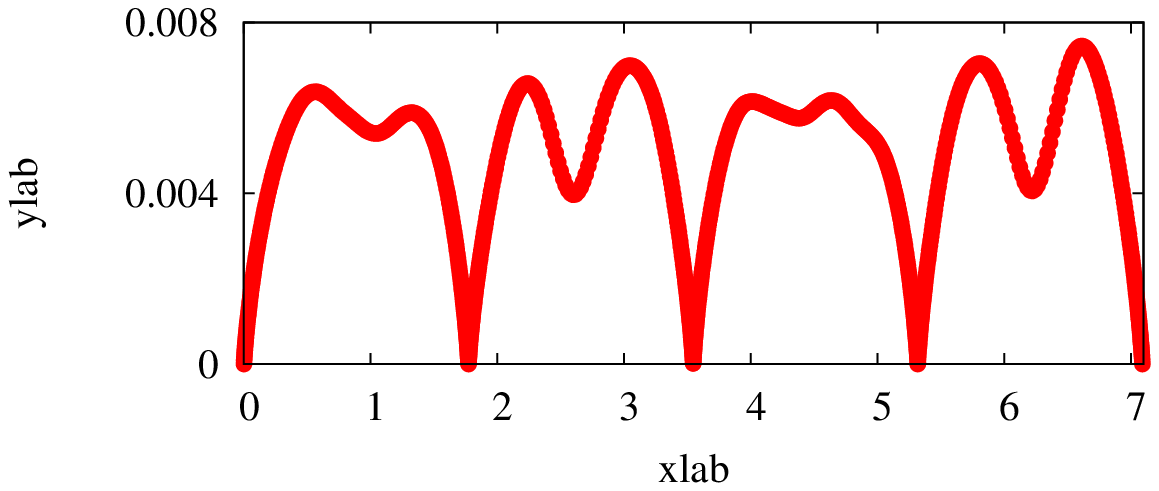}
\hskip -0.5cm
\psfrag{ylab} {$  $}
\includegraphics[width=4.0cm]{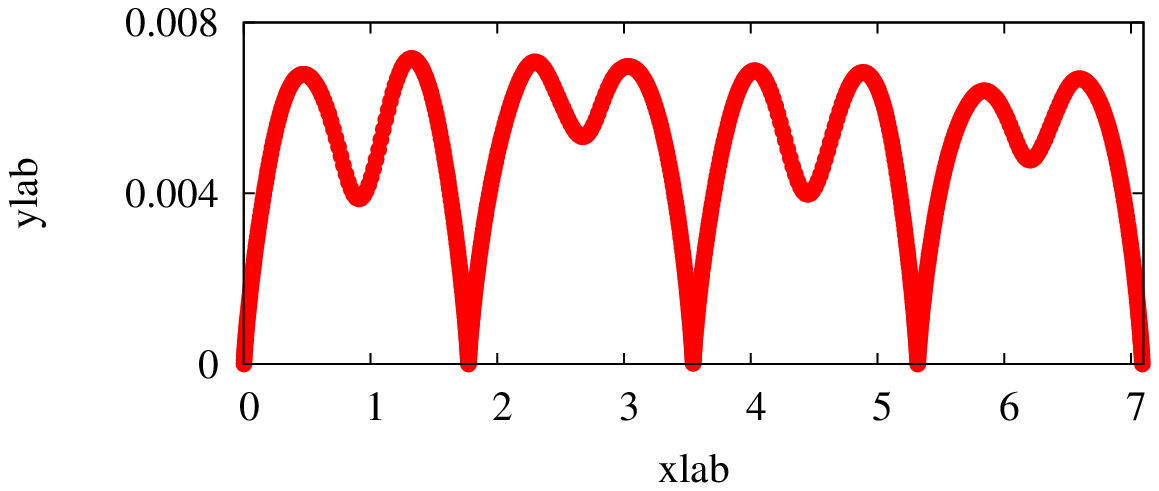}
\hskip -0.5cm
\psfrag{ylab} {$  $}
\includegraphics[width=4.0cm]{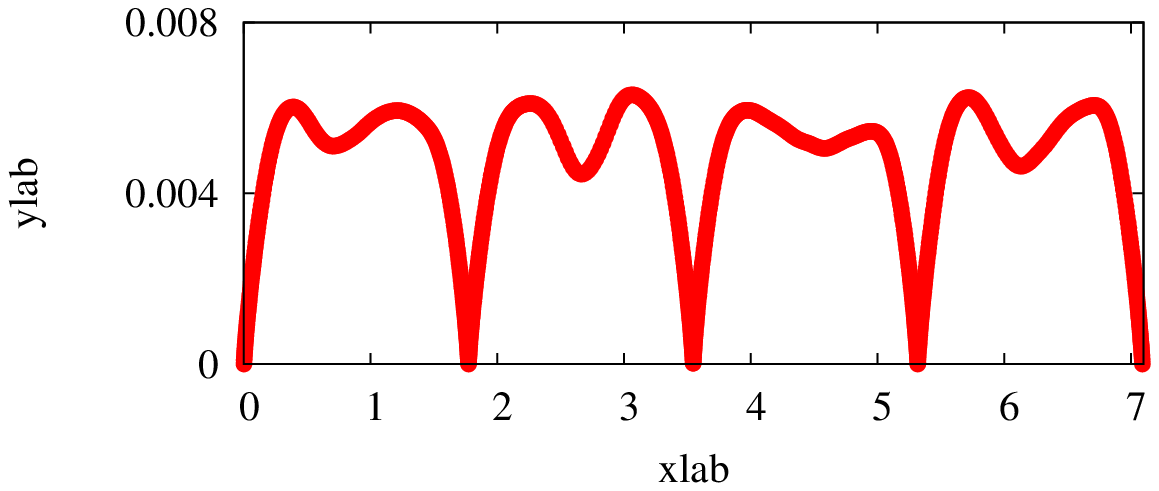}
\vskip +1.0cm
\hskip -1.8cm
\includegraphics[width=3.5cm]{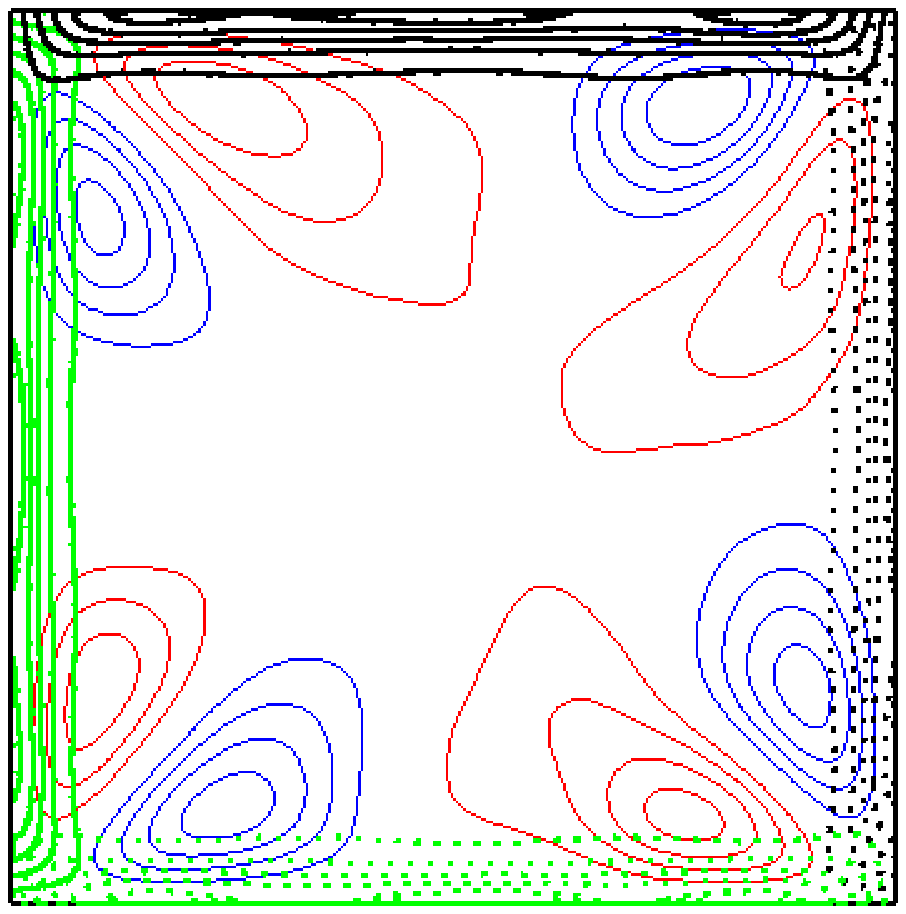}
\hskip +0.2cm
\includegraphics[width=3.5cm]{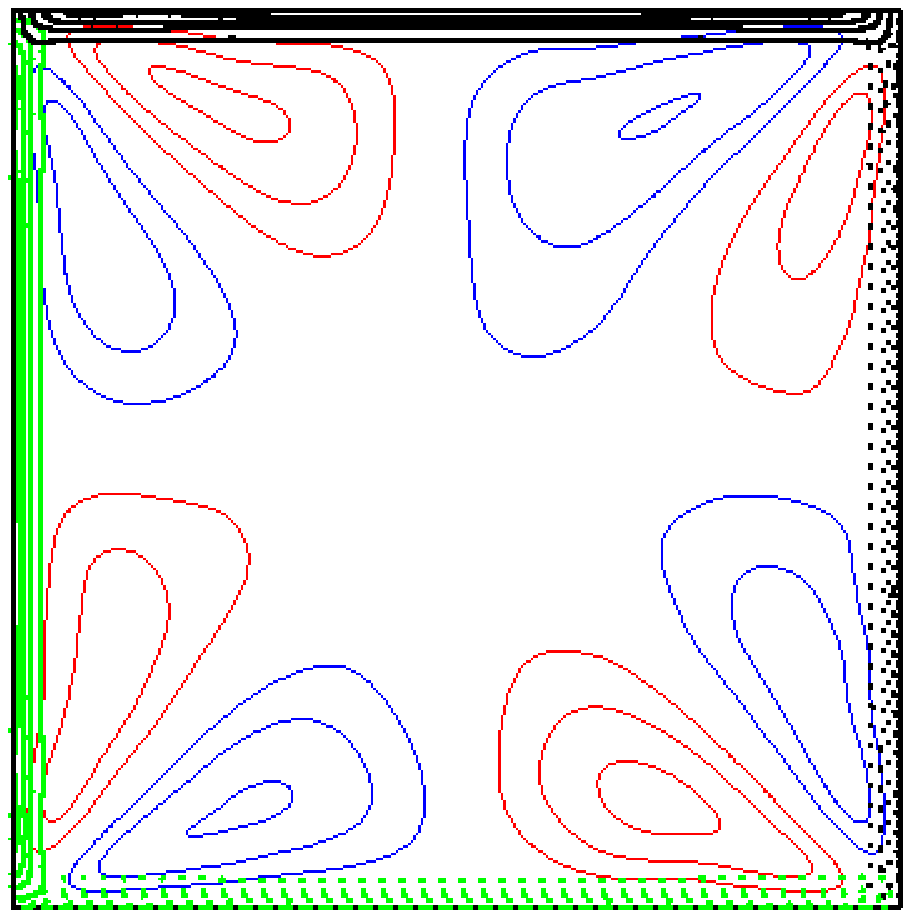}
\hskip +0.2cm
\includegraphics[width=3.5cm]{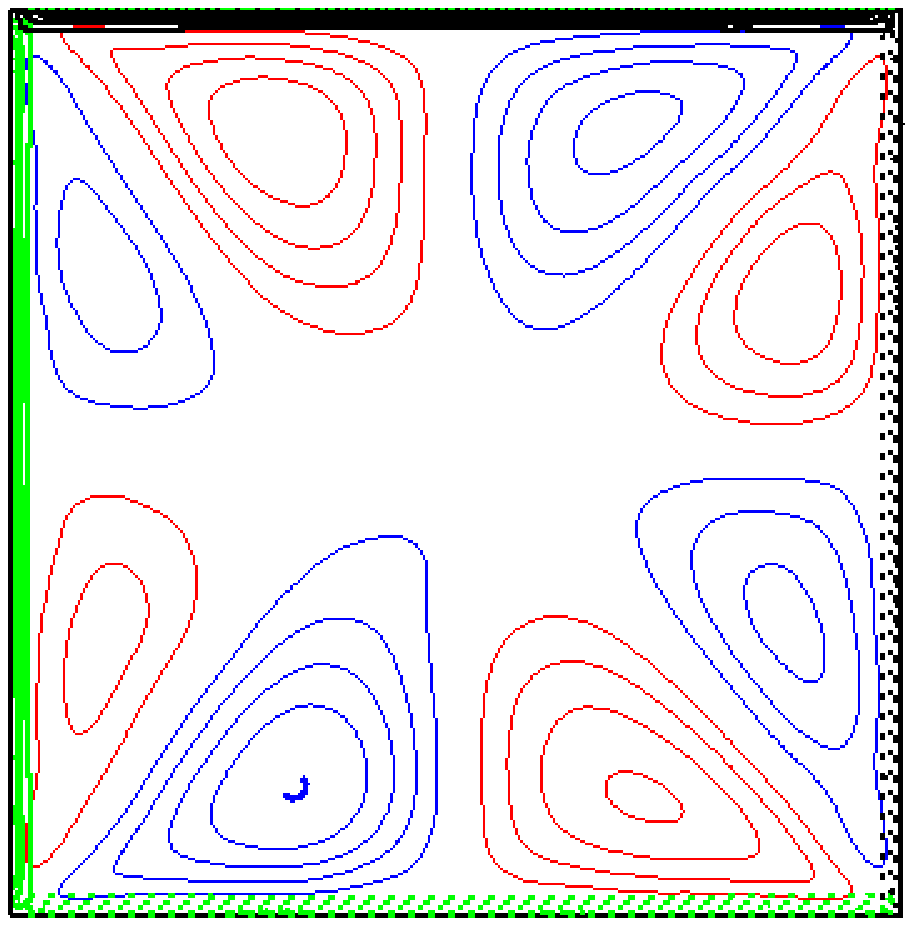}
\hskip +0.2cm
\includegraphics[width=3.5cm]{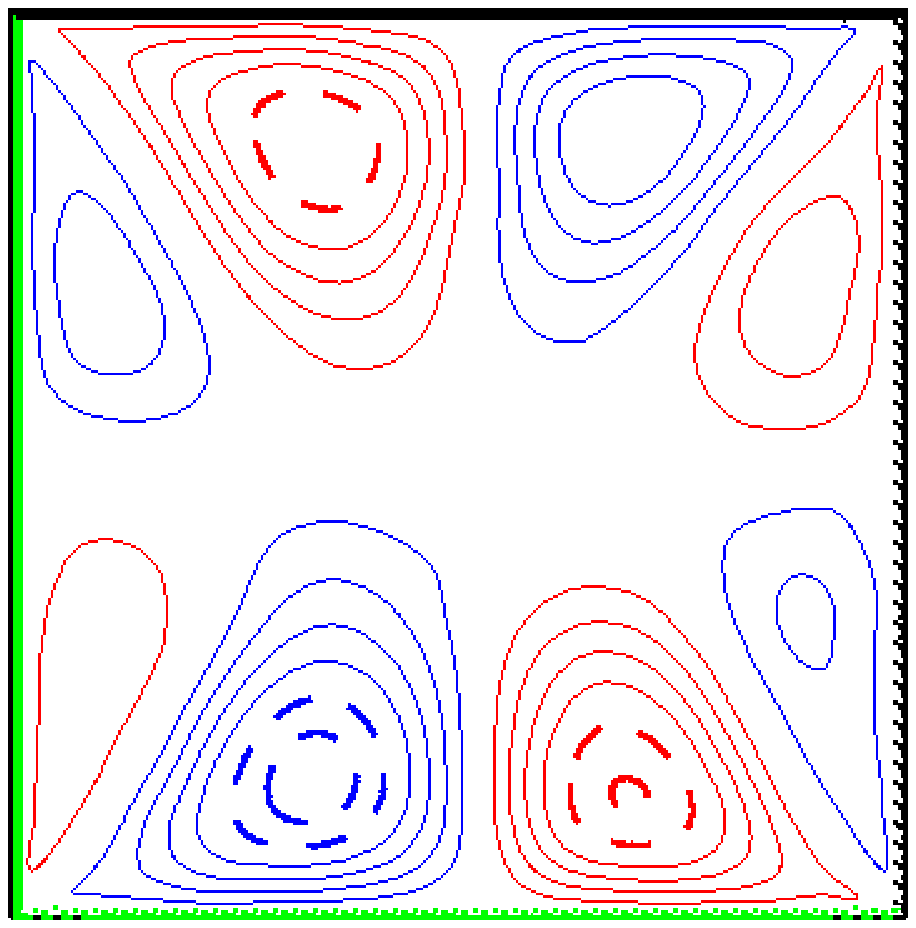}
\vskip -.2cm
\hskip 1.0cm  e)   \hskip 3.5cm  f) \hskip 3.5cm  g)   \hskip 3.5cm  h)
\vskip 0.0cm
\hskip -1.8cm
\psfrag{ylab} {$\tau_w    $}
\psfrag{xlab}{ $s$}
\includegraphics[width=4.0cm]{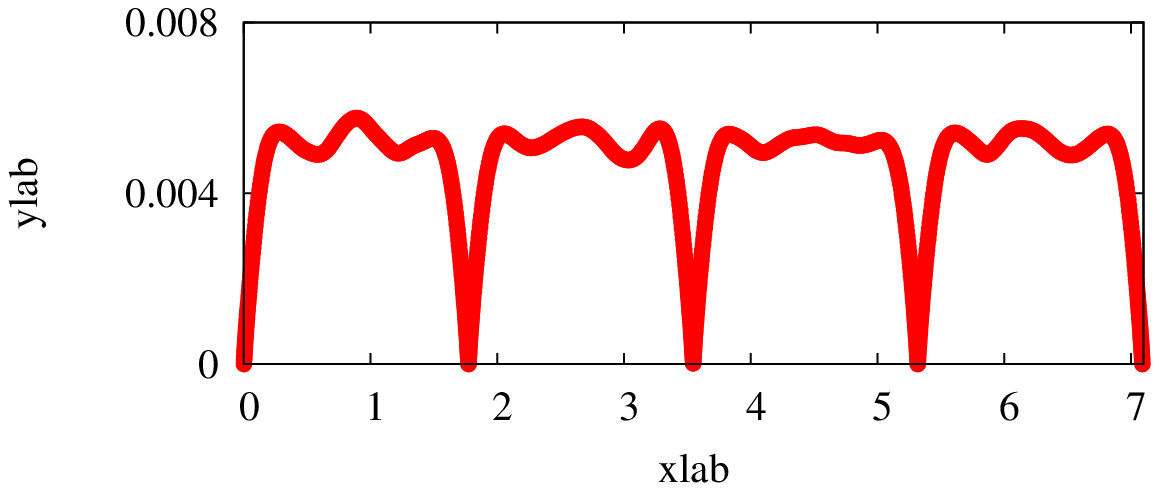}
\psfrag{ylab} {$  $}
\includegraphics[width=4.0cm]{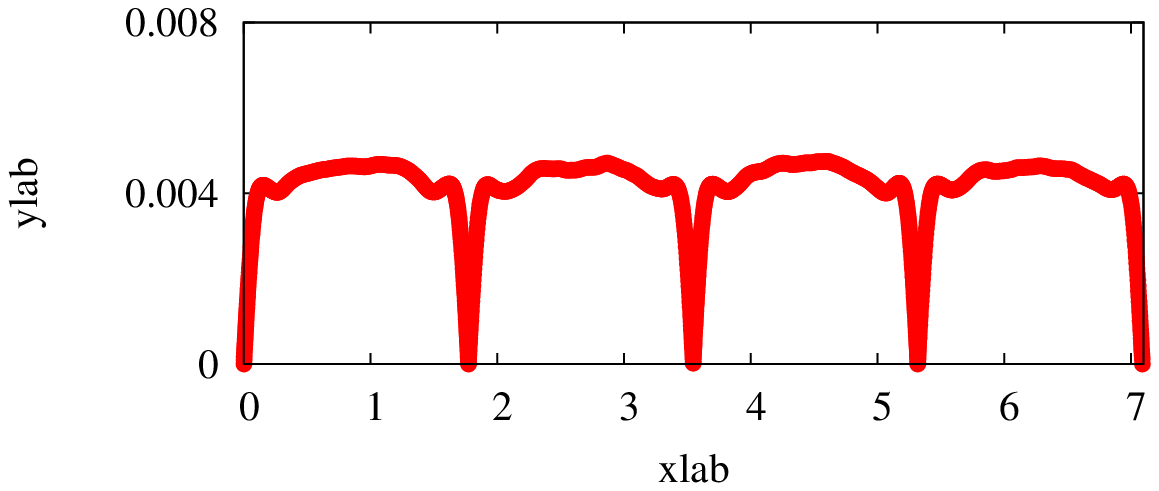}
\hskip -0.5cm
\psfrag{ylab} {$  $}
\includegraphics[width=4.0cm]{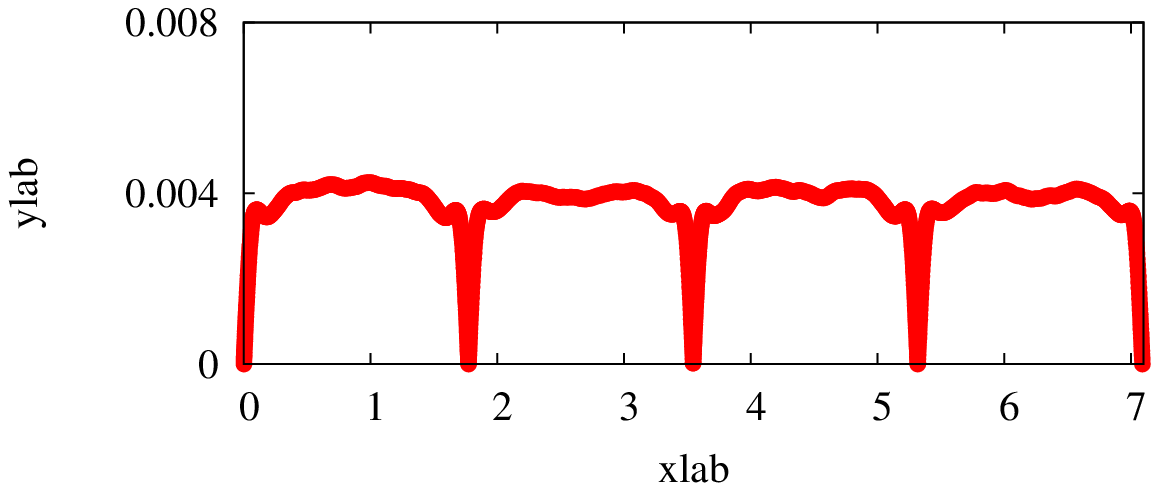}
\hskip -0.5cm
\psfrag{ylab} {$  $}
\includegraphics[width=4.0cm]{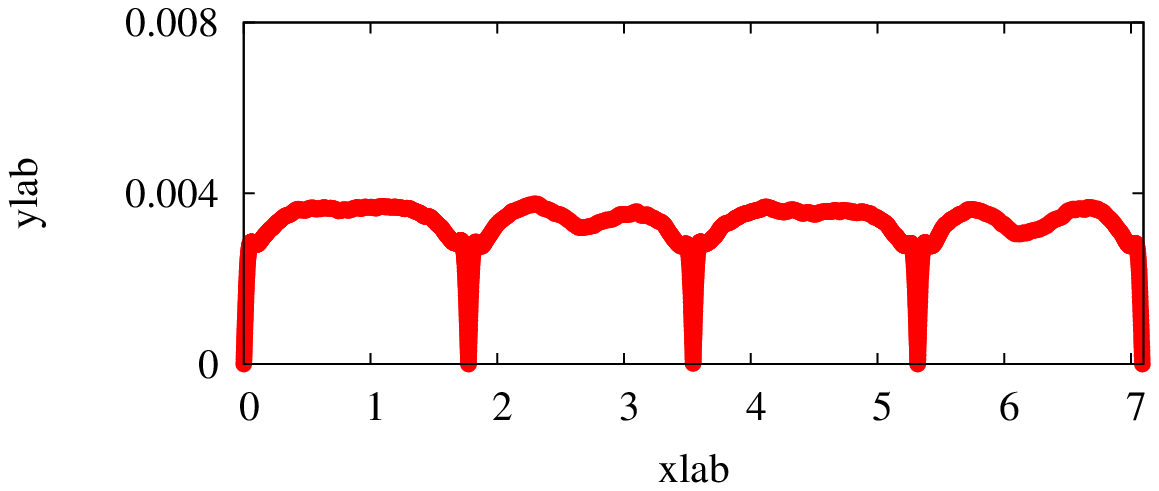}
\vskip 0.0cm
\caption{Square duct: iso-lines of $\omega_2/Re$ and $\omega_3/Re$ 
(with spacing $\Delta=0.001$ green positive dotted, black negative solid),
superimposed to iso-lines of the secondary stream-function $\psi$ 
(with spacing $\Delta=0.0005$ thin up to $.002$ from $0.025$ thick lines 
red positive blue negative), at different $Re$.
In panel a) $\Delta \psi=10^{-6}$.
The small figures under each panel show $\tau_w$ along the perimeter
of the duct.  The Reynolds numbers are
a) $1350, b)\ 1370, c) \ 1430, d)\ 1750, e)\ 2500, 
f)\ 5000, g)\ 7750, h)\ 15000$.
}
\label{fig4}
\end{figure}                                            

\subsubsection{Secondary motions     }

To better understand the differences noted above, and see
whether the statistical quantities reproduce the expected
symmetries it is worth looking              
at the contours of the stream-function of the mean secondary 
motion superimposed to the mean vorticity components
($<\omega_2>={\partial U_1}/{\partial x_3}$,
$<-\omega_3>={\partial U_1}/{\partial x_2}$),
divided by the Reynolds number, over the entire duct cross-section. 
It is important
to keep in mind that the $<\omega_i>$ contribute
to turbulent kinetic energy production, to be  discussed later on. 
When averaged on all the duct walls, 
viscous strains return the wall shear stress distributions shown
in figure~\ref{fig3}a. The profiles of $\tau_w$
along the whole duct perimeter are shown in figure~\ref{fig4},
under the corresponding streamfunction and vorticity contours.
%The contours of the viscous strains, $\frac{\partial U_1}{\partial x_3}/Re$
%solid and $\frac{\partial U_1}{\partial x_2}/Re$ dashed lines,
%in the entire section show, at the different Reynolds
%numbers, in which region are concentrated. 
In the laminar regime (figure~\ref{fig4}a), characterised
by the absence of 
secondary motions, the $U_1$ contours do not change with $Re$. The mean
strain decreases moving  from the walls  towards the central
region. Immediately after the critical Reynolds number, 
$Re_C\approx 1350$, the secondary
motion consists on four recirculating regions,
that, at high $Re$, leads to the well documented eight symmetric regions.
In figure~\ref{fig4} the increments of the stream-function
contours have been maintained fix for all the Reynolds numbers,
hence the comparison among the different regimes
leads to the conclusion that the strength of the secondary motion
decreases in the transitional and in the fully turbulent regimes  
by increasing $Re$, up to $Re=5000$. At higher $Re$ the
strength does not change very much. To emphasise that 
the flow structures change 
in a sharp range of Reynolds number near $Re_C$,
visualizations are shown at $Re=1370$ in figure~\ref{fig4}b,
and at $Re=1430$ in figure~\ref{fig4}c.
The corresponding values of $C_f$ in the inset of figure~\ref{fig2}a
are given by the open circles, corroborating the value
of $Re_C$ and that at $Re=1370$ the value of $C_f$ is
close to the green line, corresponding to the presence of four recirculating regions. 
These figures demonstrate that there is an equal probability
to have secondary structures in one or in the other side,
depending on the growth of disturbances  
either near one or the other side, during the initial transient. 
Animations allow to see the
different time history, and see where disturbances 
form and grow. A slight increase of Reynolds
number ($Re=1430$) leads to a secondary motion with
four pairs of large-scale structures, two strong and
two weak (figure~\ref{fig4}c). It has
been observed by the time history of the
mean pressure gradient $\Pi$ and of the 
total turbulent kinetic energy $K=<u^{\prime 2}_i>/2$,
that the convergence to a steady state requires
simulations lasting for very long time. At $Re=1750$ 
figure~\ref{fig4}d shows that the intensity
of the four  couples tends to be the same, 
and that the magnitude of $\tau_w$ along
the perimeter is slightly reduced with respect to that at $Re=1430$.
At this $Re$, the inset of figure~\ref{fig2}a shows
the start of tendency towards the Blasius law corresponding to a
fully turbulent regime, which is characterised by a secondary motion
with four pairs of recirculating regions.
The size of the secondary structures is comparable to
half of the wall length, hence they can transport
$U_1$ towards the wall at $r=L_2/4$ and far from the wall
at $r=L_2/2$. Further increase of the Reynolds number
($Re=2500$) yields (figure~\ref{fig4}e)
reduction of the strength
of the secondary structures, which in addition
become confined to the duct corners. Two maxima appear in the profiles
of $\tau_w$, and their amplitude decreases as was
observed in figure~\ref{fig3}a.
The thickness of the vorticity layers in figure~\ref{fig4}f,
at $Re=5000$,
reduces, implying that the near-wall turbulence
is not largely affected by secondary motion.
This behavior continues by increasing the Reynolds number, as
shown in figure~\ref{fig4}g and
figure~\ref{fig4}h. From the last two figures it emerges that
it is quite difficult to have perfect statistical
convergence with four
secondary flow structures of equal  strength.  
Even without this accomplishment
the figures with  the distribution of the $\tau_w$ along
the perimeter show the changes with the Reynolds number.

\begin{figure}
\centering
\vskip -0.0cm
\hskip -1.8cm
\psfrag{ylab} {$ \Pi Re   $}
\psfrag{xlab}{ $t$}
\includegraphics[width=8.0cm]{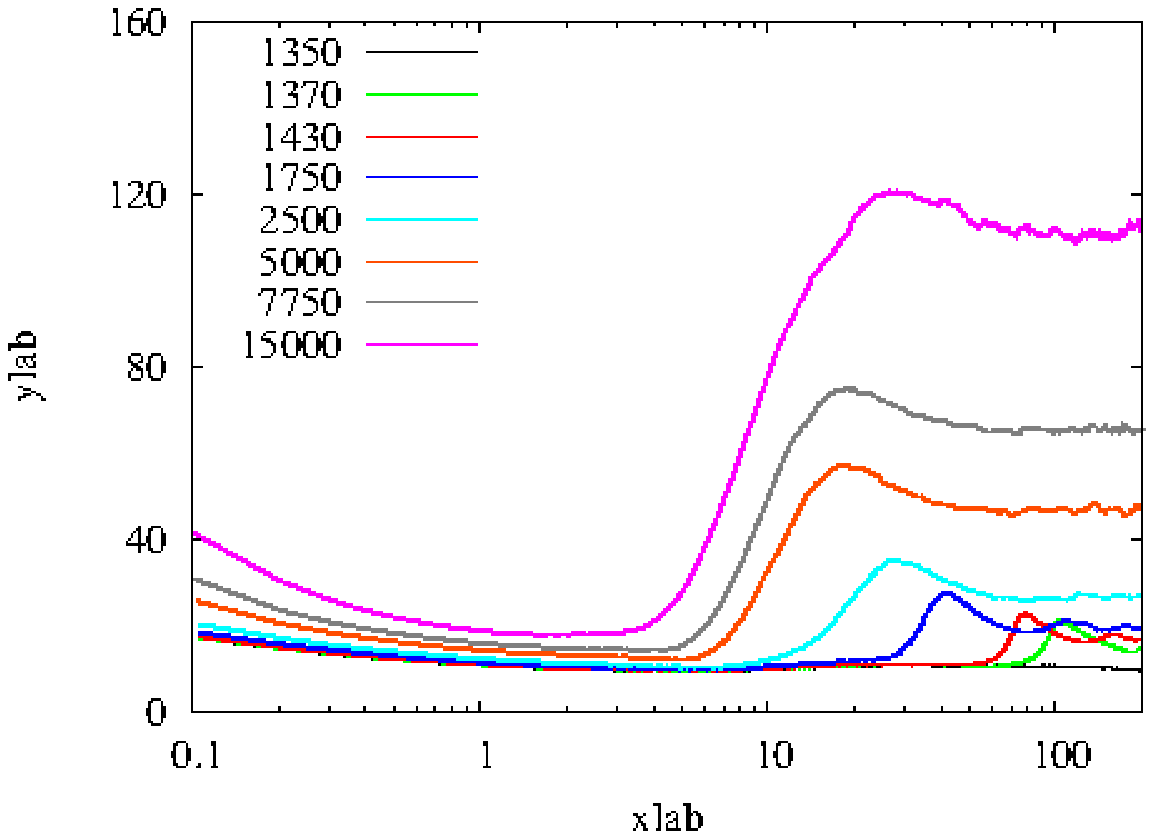}
\psfrag{ylab} {$\Pi$}
\psfrag{xlab}{ $t$}
\includegraphics[width=8.0cm]{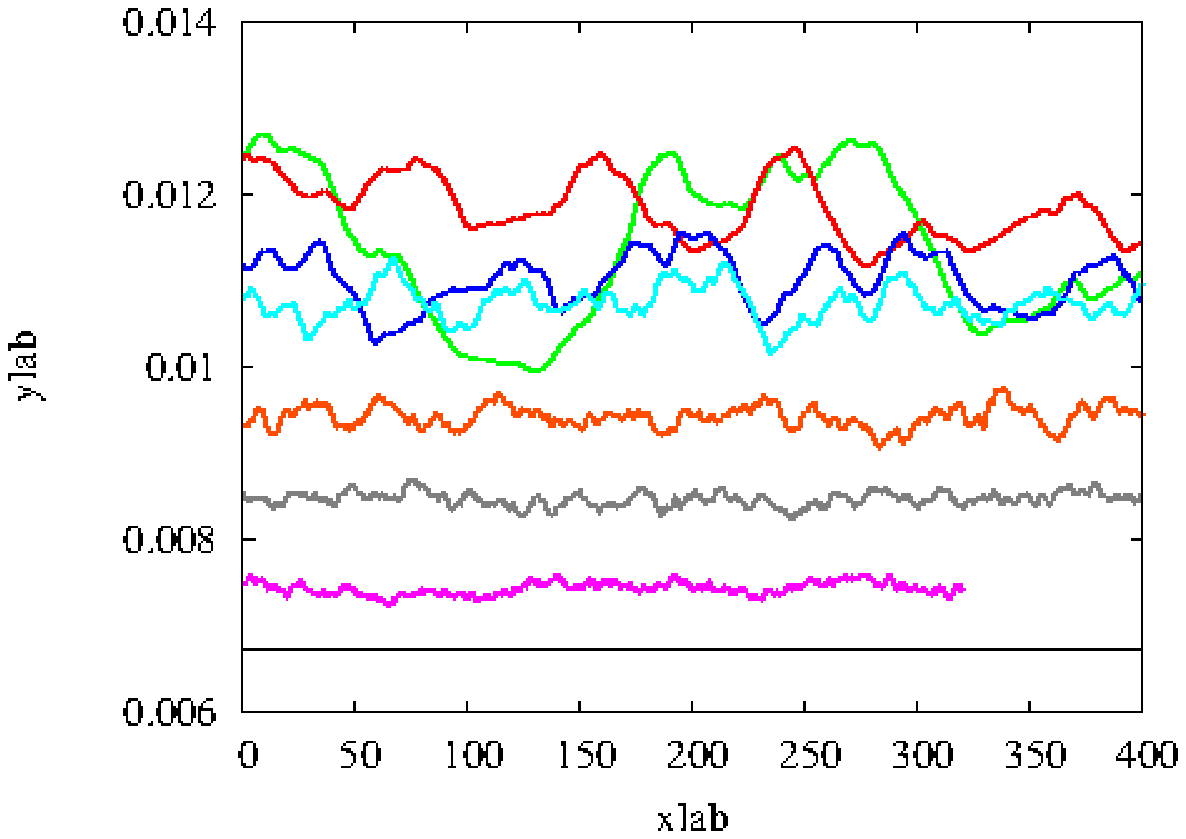}
\vskip -.5cm
\hskip 5.0cm  a)   \hskip 7.5cm  b)
\caption{ 
Square duct: temporal evolution of the mean pressure gradient
in a) divided by $Re$ and with the time in log
scale to emphasise the transition to turbulence,
in b) the time history is evaluated in the 
last $400$ time units to emphasise the frequency 
and the amplitude of of the variations at each
$Re$ given in the legend of a); the simulation
at $Re=15000$ was stopped at $t=373$ due to the
small variations of $\Pi$.
}
\label{fig5}
\end{figure}

\subsubsection{Mean pressure gradient}

The most complex flows physics in square ducts
occurs in our opinion in the
transitional regime, in which secondary motions 
help promoting mixing and heat transfer. 
In the DNS the flow dynamics 
can be studied by animations of several
quantities of interest. However a first impression
of the complexity for the cases  depicted in figure~\ref{fig4}
may be drawn from the time evolution of $\Pi$ in the early
stages (figure~\ref{fig5}a), and in the last $400$ time units 
(figure~\ref{fig5}b). 
As previously mentioned, the initial conditions  
are different from the laminar distribution,
and random disturbances are added, hence 
high friction occurs at the walls.
The sharp streamwise velocity gradients
gradients decrease in a short time due to viscous effects, 
and organized flow structures form earlier at higher $Re$.
Transition is frequently characterised by exponential growth, 
which is also the case of square ducts, as can be appreciated
in figure~\ref{fig5}a. In this figure $\Pi$ is
multiplied by the Reynolds number, hence it is a measure
of the averaged wall-normal streamwise velocity gradient.
Despite the mentioned differences, during this transient 
transitional and turbulent flows are characterised by an initial decrease
followed by exponential growth, terminated by an absolute maximum.
A brief decay then leads to the instant when the pressure gradient 
begins to oscillate, at which we start the evaluation  
of averaged properties. The time evolution in
figure~\ref{fig5}b shows that for $Re\ge 2500$
small-amplitude, high-frequency oscillations occur,
typical of fully turbulent flows, and
steady state is reached in short time.
On the other hand, large-scale, low-frequency oscillations occur
at Reynolds numbers close to the critical
one, implying that a long time evolution is necessary to
achieve statistical convergence.

\begin{figure}
\centering
\vskip -0.0cm
\hskip -1.8cm
\psfrag{ylab} {$ \Delta \Pi  $}
\psfrag{xlab}{ $t$}
\includegraphics[width=12.0cm]{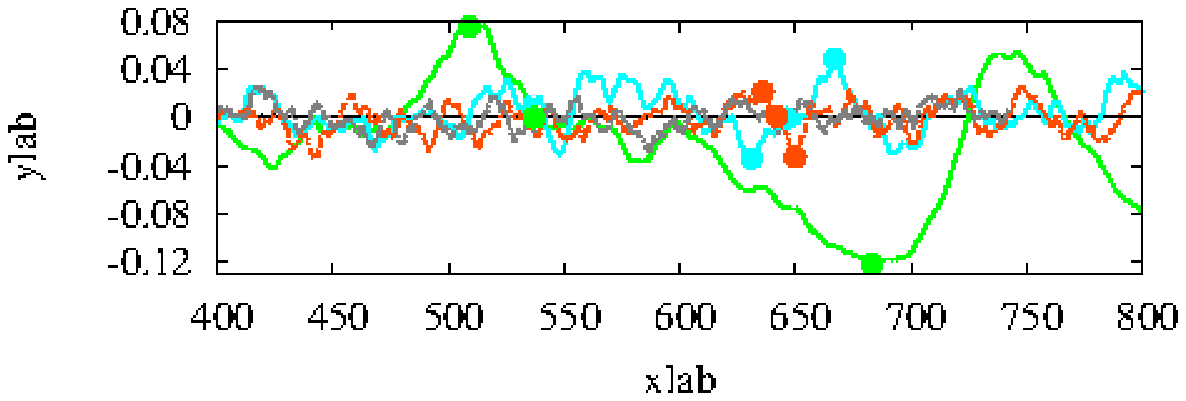}
\vskip  0.5cm
\hskip -1.8cm
\psfrag{ylab} {$ \tau      $}
\psfrag{xlab}{ $ $}
\includegraphics[width=3.7cm]{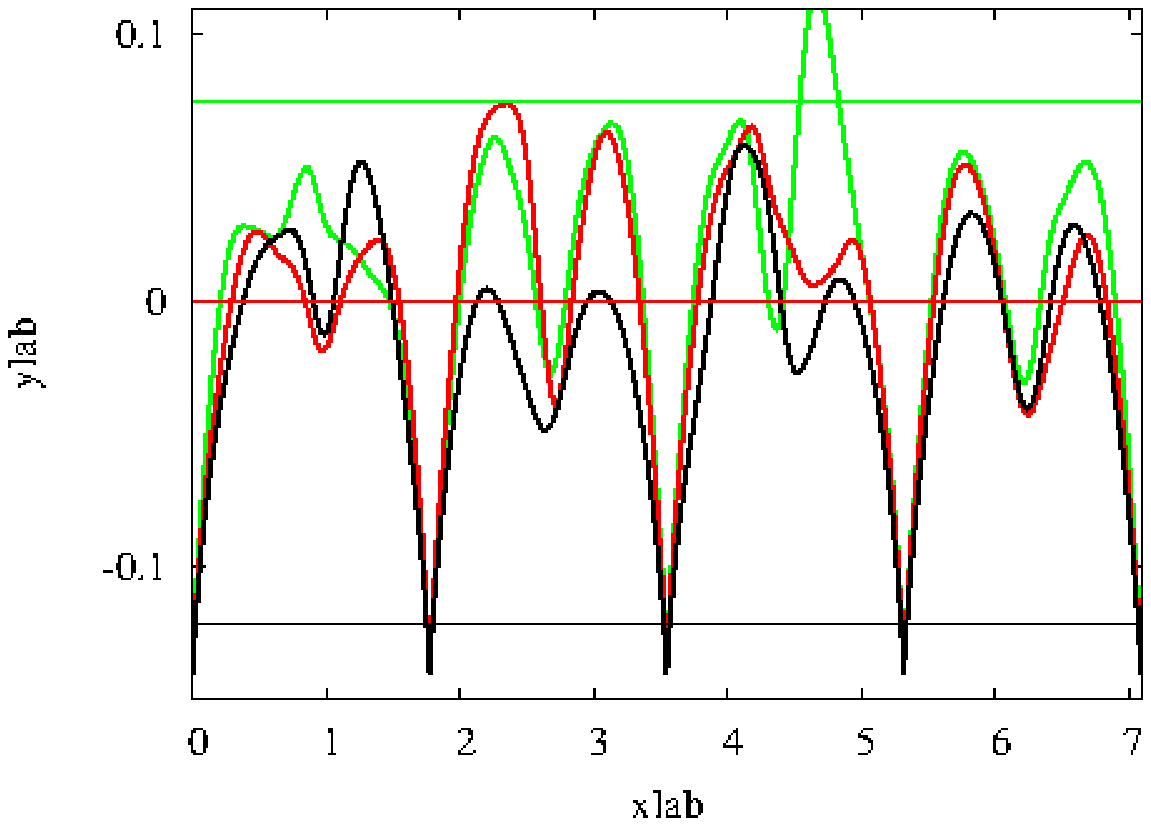}
\hskip -0.0cm
\includegraphics[width=2.35cm]{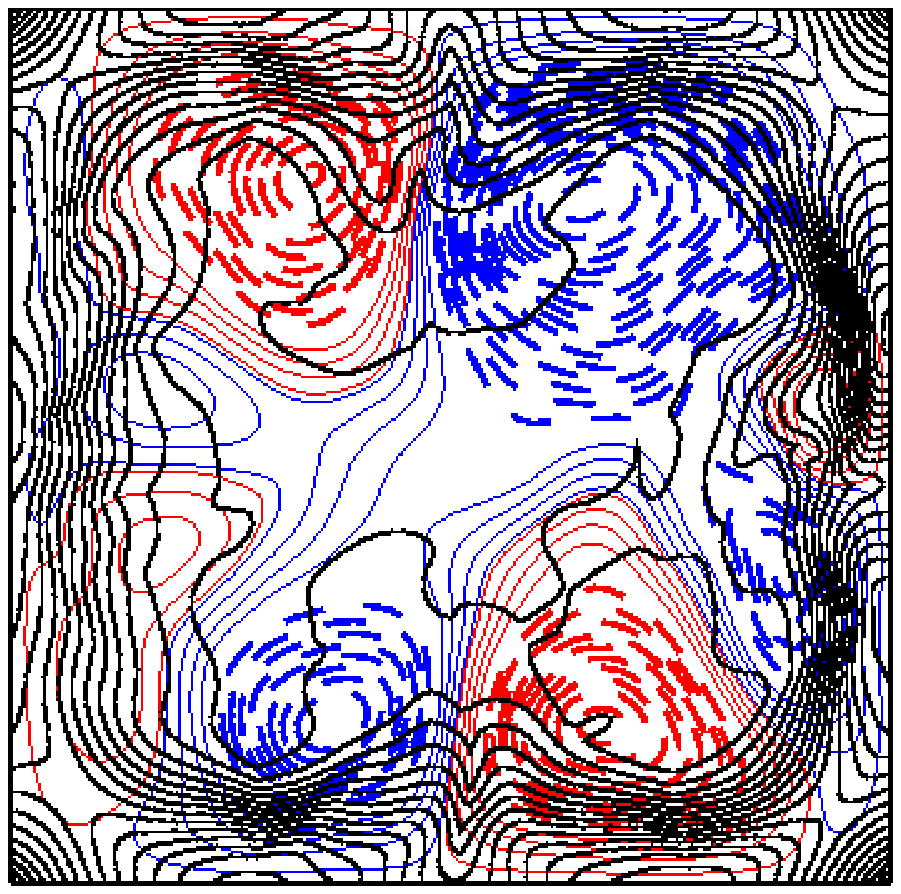}
\hskip 0.2cm
\includegraphics[width=2.35cm]{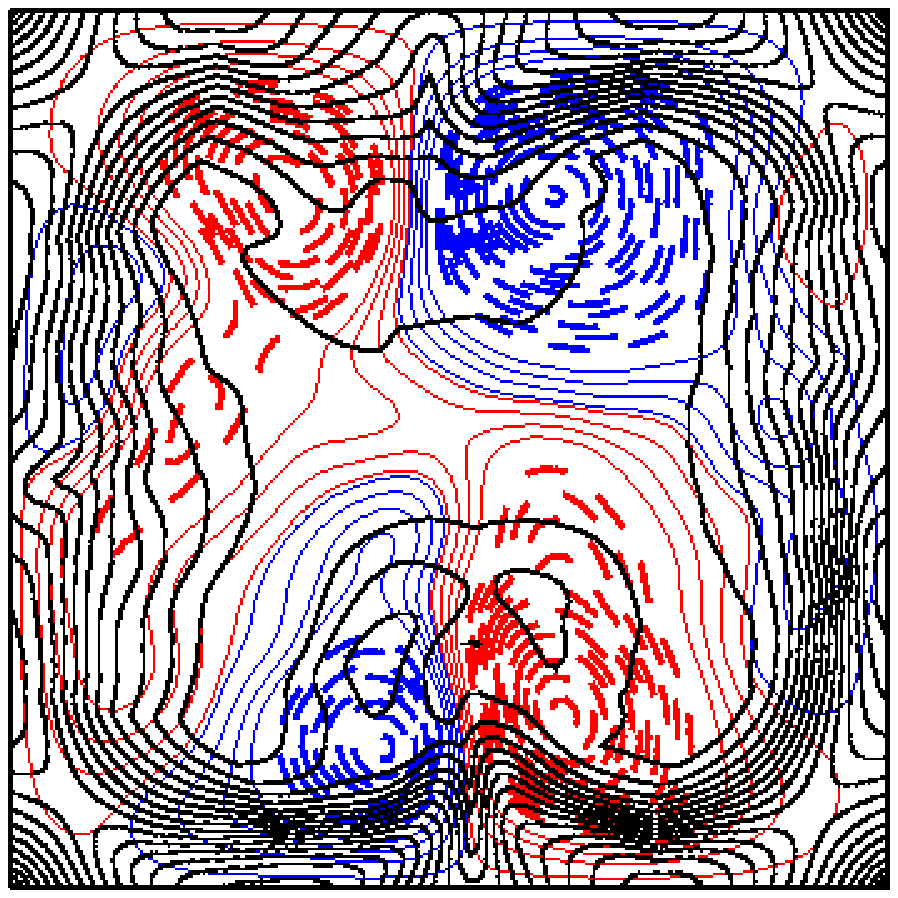}
\hskip 0.2cm
\includegraphics[width=2.35cm]{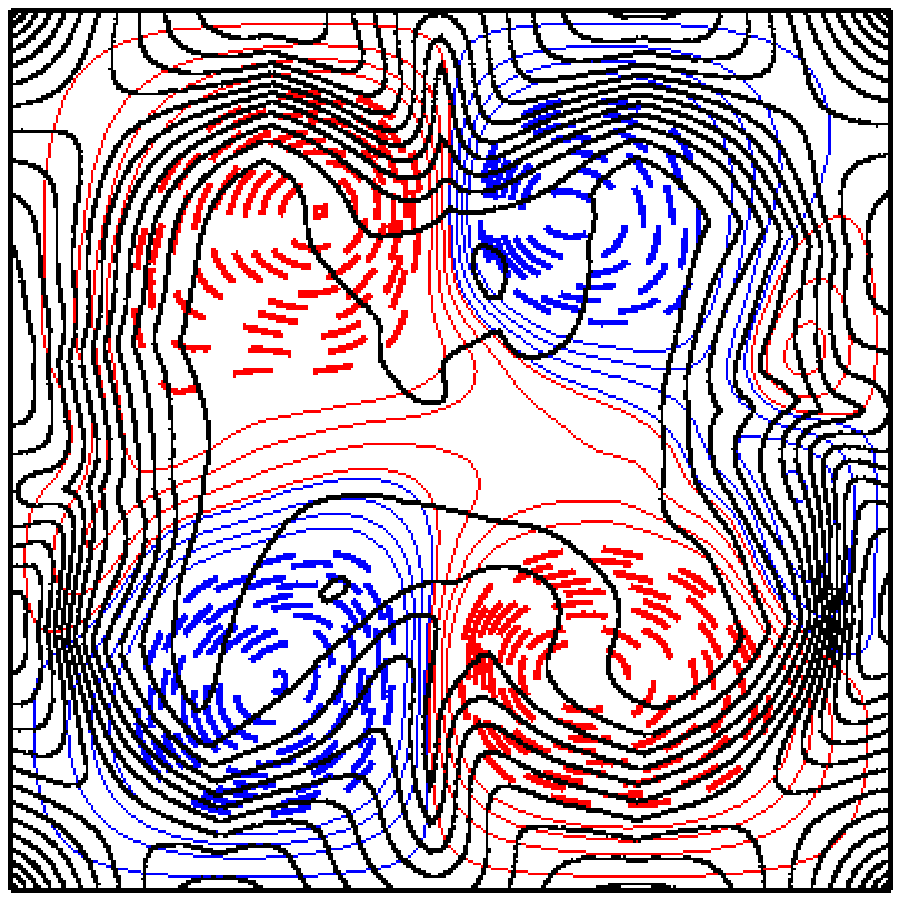}
\vskip  0.5cm
\hskip -1.8cm
\psfrag{ylab} {$ \tau      $}
\psfrag{xlab}{ $ $}
\includegraphics[width=3.7cm]{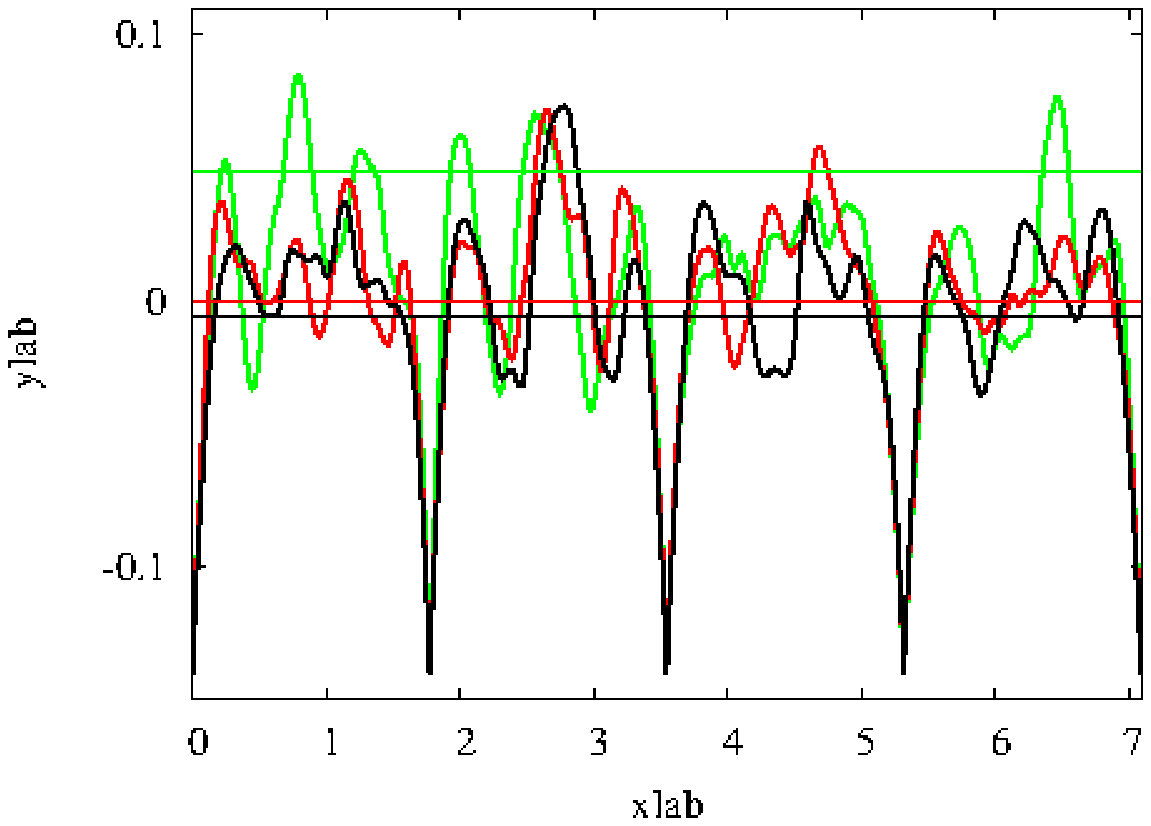}
\hskip -0.0cm
\includegraphics[width=2.35cm]{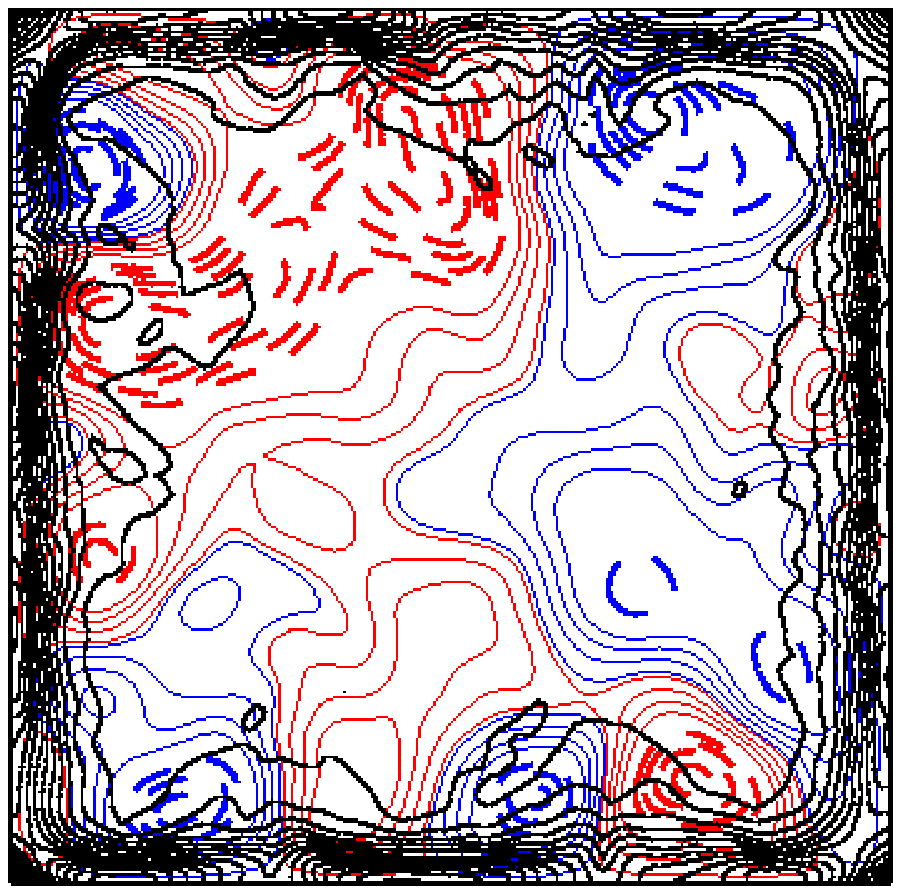}
\hskip 0.2cm
\includegraphics[width=2.35cm]{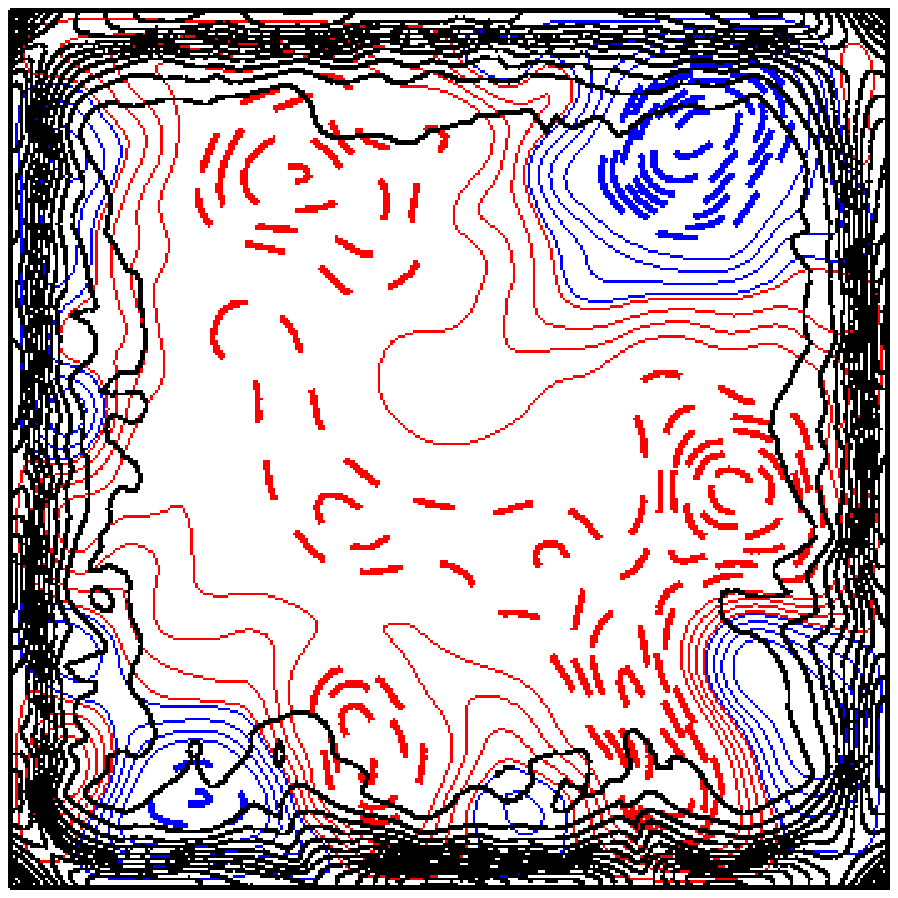}
\hskip 0.2cm
\includegraphics[width=2.35cm]{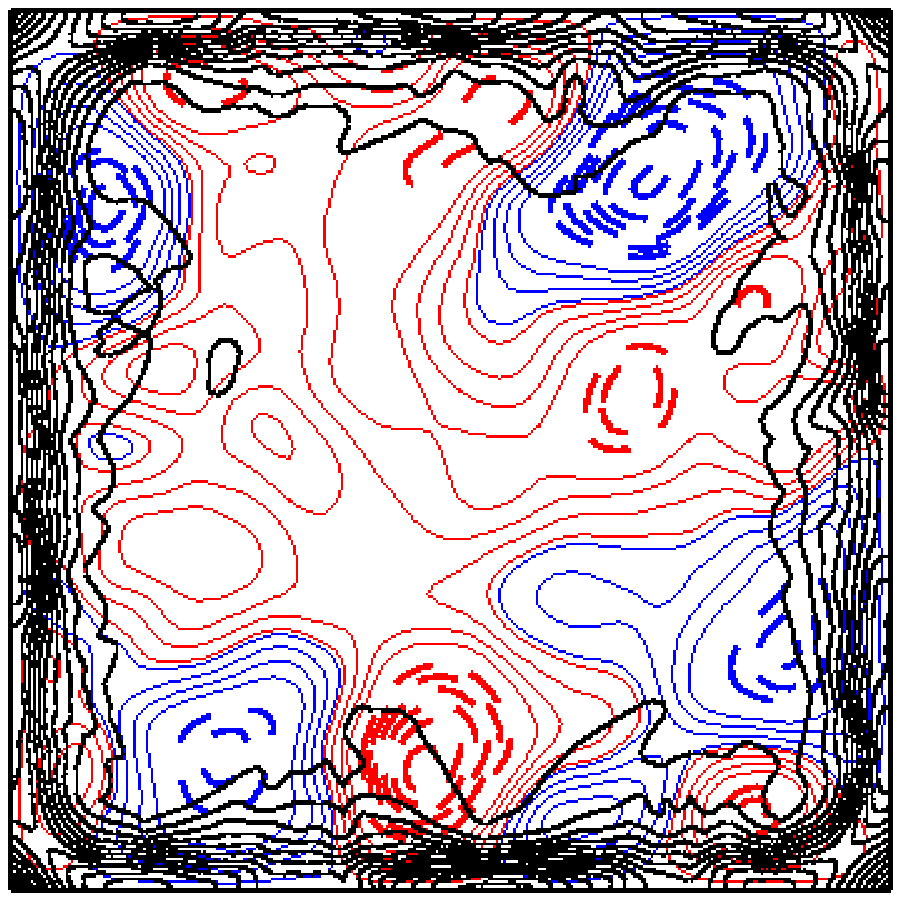}
\vskip  0.5cm
\hskip -1.8cm
\psfrag{ylab} {$ \tau      $}
\psfrag{xlab}{ $t $}
\includegraphics[width=3.7cm]{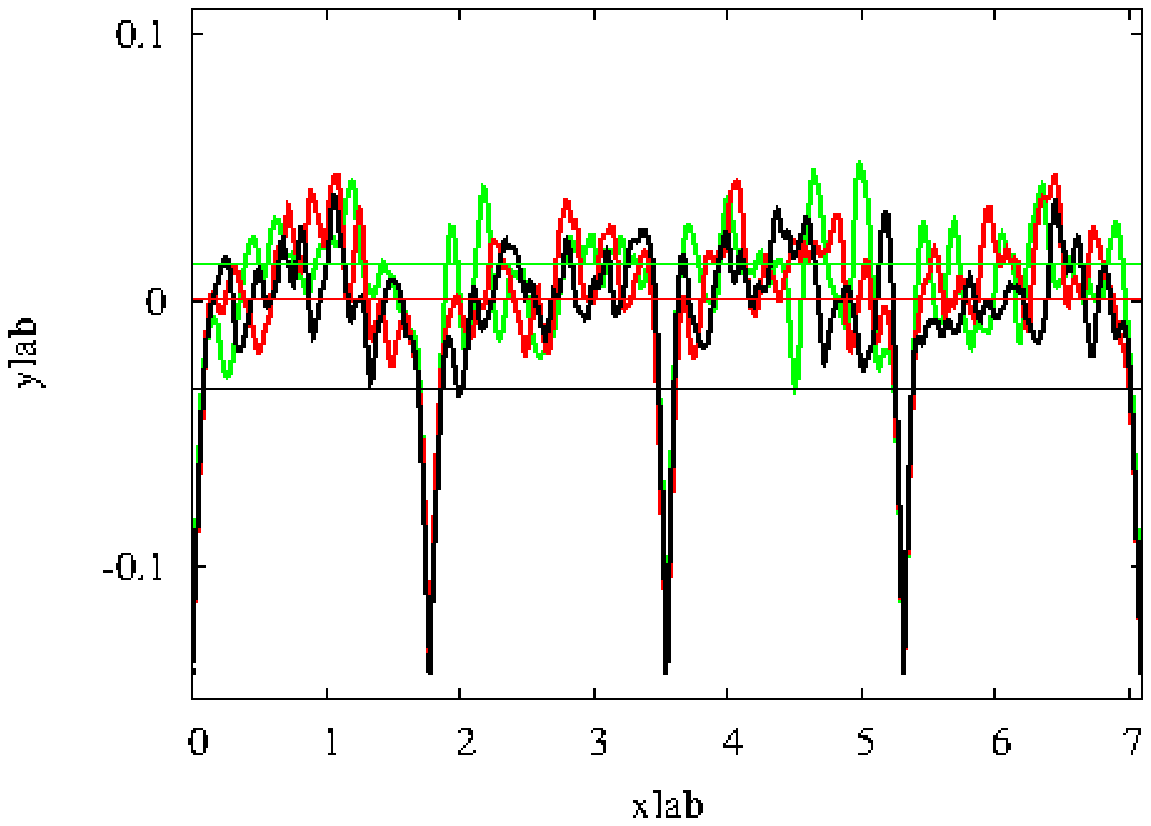}
\hskip -0.0cm
\includegraphics[width=2.35cm]{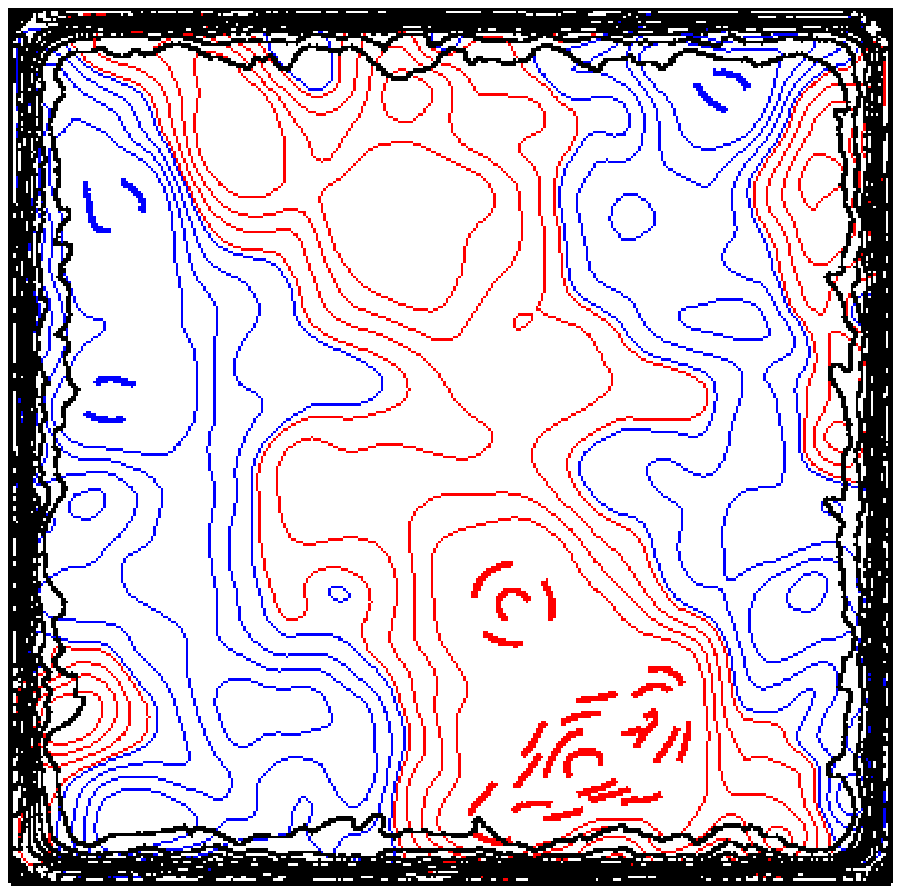}
\hskip 0.2cm
\includegraphics[width=2.35cm]{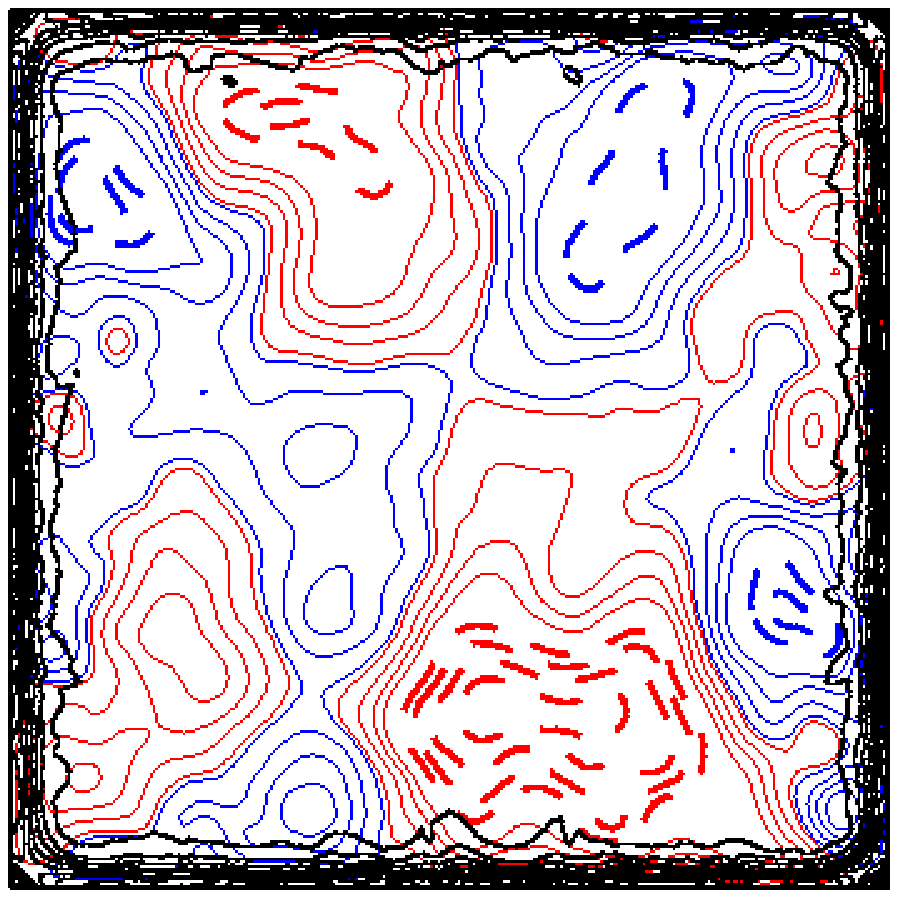}
\hskip 0.2cm
\includegraphics[width=2.35cm]{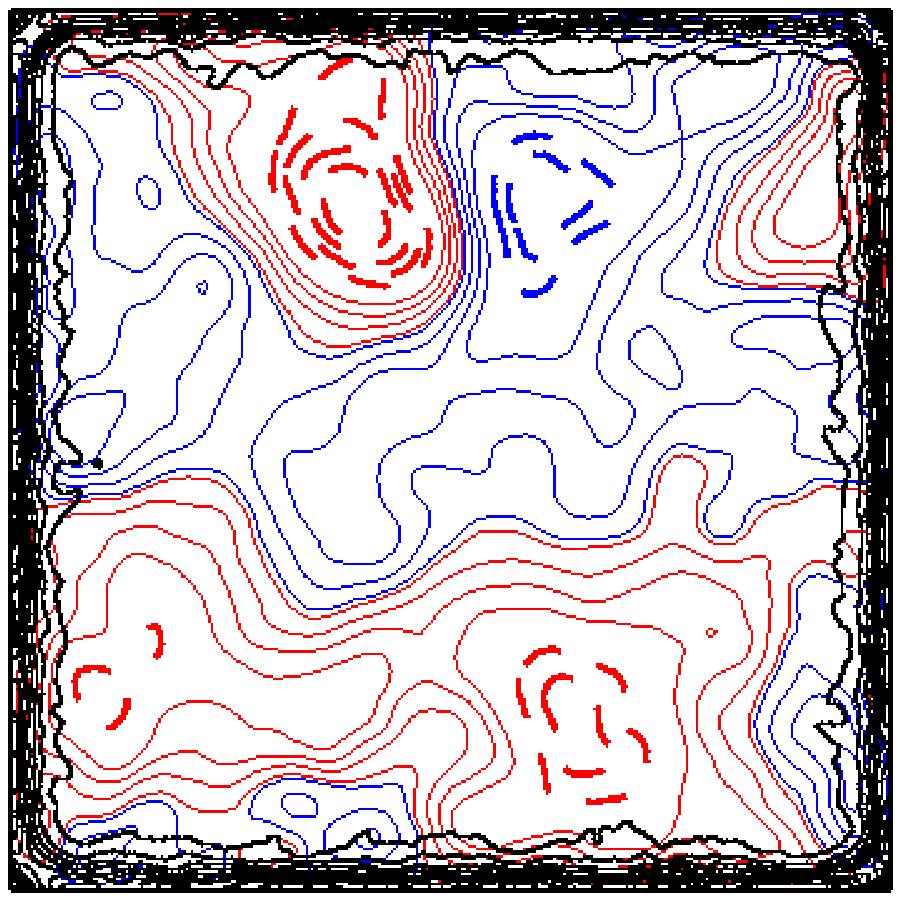}
\caption{ 
Square duct: top, temporal evolution of $\Delta \Pi=\Pi/\overline{\Pi}-1$
in the last $400$ time units for $Re=1370$,
$Re=2500$, $Re=5000$ and $Re=7750$, the colours being
the same as those in figure \ref{fig5}a, the bullets
indicate three time instants at which $\Delta \Pi$ is maximum, minimum, and average;
bottom, instantaneous friction distribution
along the four sides, starting from the left bottom corner, at
$Re=1370$, $Re=2500$, $Re=5000$;
green lines correspond to maximum $\Delta \Pi$,
red to $\Delta \Pi=0$, and black to minimum $\Delta \Pi$;
at the same instants we also show on the side
the contours of the streamwise averaged streamfunction
(red positive, blue negative, same increments as in figure \ref{fig4}) 
superimposed to contours
of $|\frac{\partial U}{\partial x_2}|+|\frac{\partial U}{\partial x_3}|$,
in increments $\Delta=0.0005$.
}
\label{fig5bis}
\end{figure}

To understand in greater detail the cause of the $\Pi$ oscillations
and their connection with the flow structures, in the top
panel of figure~\ref{fig5bis} the time evolution of $\Delta \Pi$ 
(defined in the caption) over $400$ time units is shown,
for four values of $Re$. At $Re=1370$ the scaled oscillations are
very large, and the reasons is understood by looking
at the instantaneous flow visualizations, similar to those
in figure~\ref{fig4}, corresponding 
to two time instants at which $\overline{\Pi}$ is maximum,
minimum, and average (marked with green solid points in the top panel).
The first impression is that the formation of large-scale 
secondary motions is the cause for the large amplitude
oscillations of $\Pi$. The profiles of $\tau=\tau_w-\overline{\tau_w}$,
along the perimeter shows that the peaks are located near
the center of the four large eddies. A  $Re=2500$ the secondary
eddies are quite large and cannot move inside the duct,
hence the variation in strength produce oscillations of $\tau$ in 
figure \ref{fig5bis} at the same spatial position. On the other hand, at 
$Re=5000$ the size of the secondary eddies reduce, hence they
can move in the duct, as may be deduced by comparing the streamfunction
contours at different instants.
This unsteadiness leads to small-amplitude oscillations of $\tau$,
which are quantified by the location and magnitude of the peaks 
of the green and black lines at $Re=5000$, which are not fixed as at $Re=2500$.
At $Re=7750$ the unsteadiness of the eddies increases, as clearly
shown by the $\psi$ contours and by the $\tau$ profiles in figure \ref{fig5bis}.

\subsubsection{Mean and $rms$ velocity profiles }

\begin{figure}
\centering
\vskip -0.0cm
\hskip -1.8cm
\psfrag{ylab} {$ U_1^+    $}
\psfrag{xlab}{ $ $}
\includegraphics[width=4.3cm]{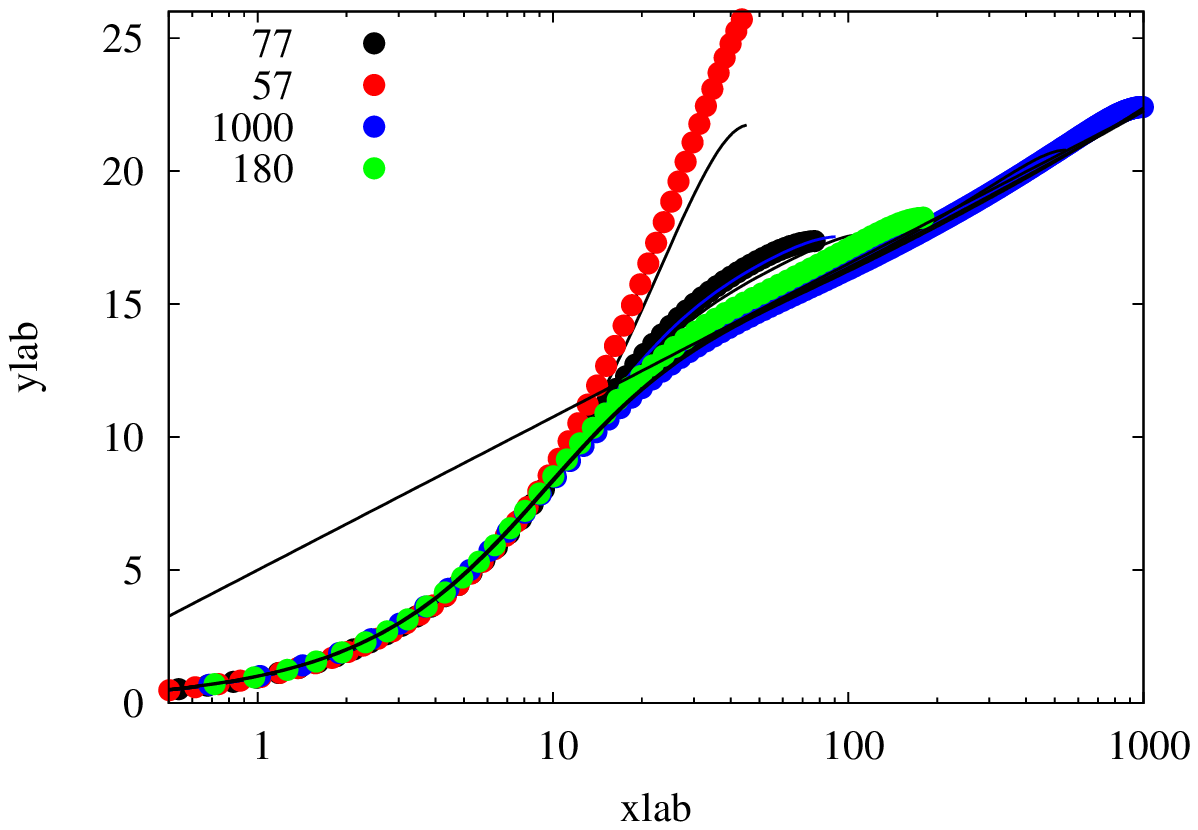}
\hskip -0.3cm
\psfrag{ylab} {\hskip -0.5cm $ u_1^{\prime +}    $}
\psfrag{xlab}{ $ $}
\includegraphics[width=4.3cm]{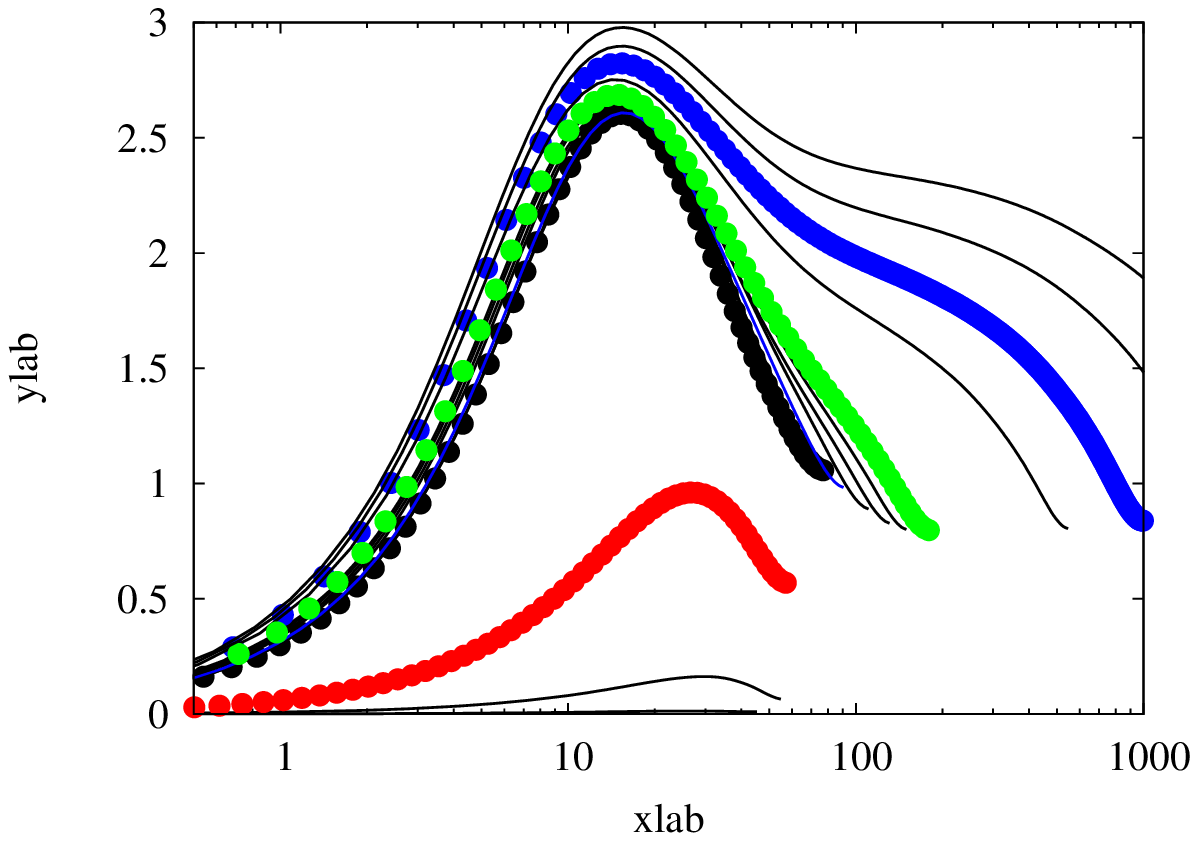}
\hskip -0.3cm
\psfrag{ylab} {\hskip -0.5cm $ u_2^{\prime +}    $}
\psfrag{xlab}{ $ $}
\includegraphics[width=4.3cm]{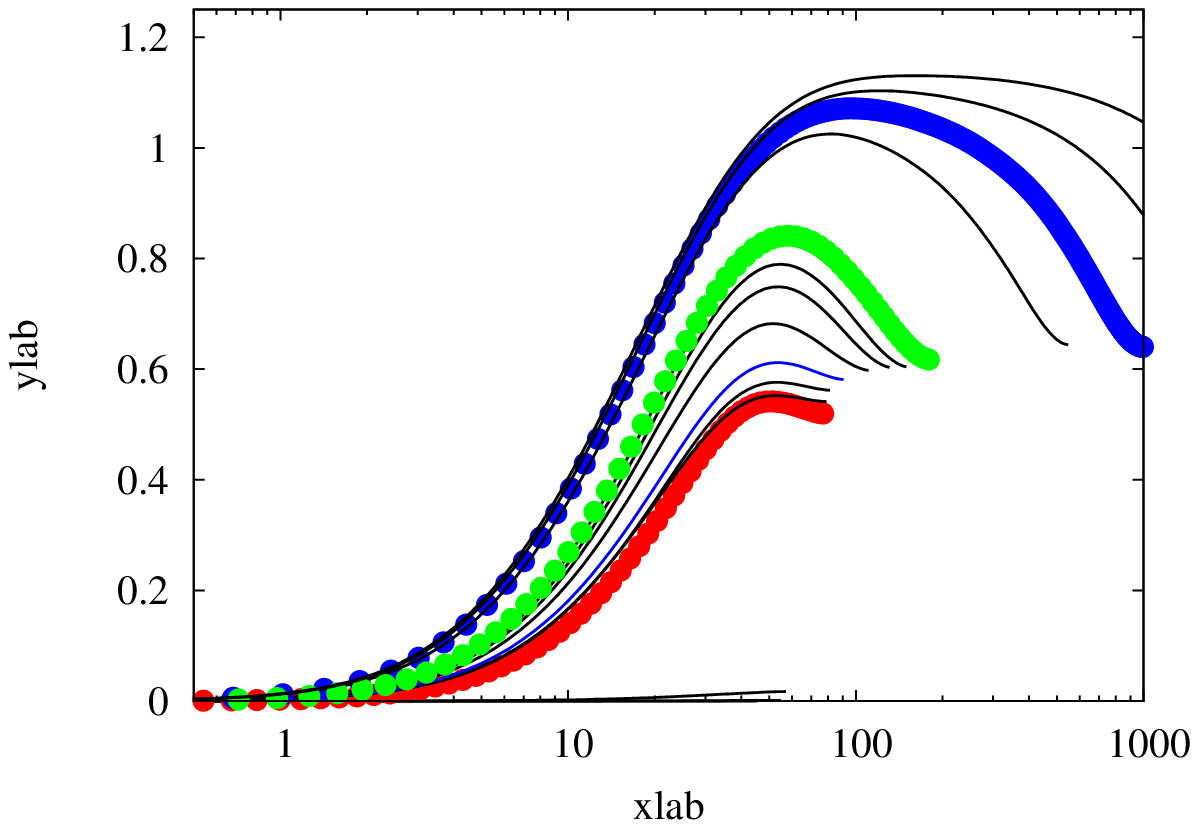}
\hskip -0.3cm
\psfrag{ylab} {\hskip -0.5cm $ u_3^{\prime +}    $}
\psfrag{xlab}{ $ $}
\includegraphics[width=4.3cm]{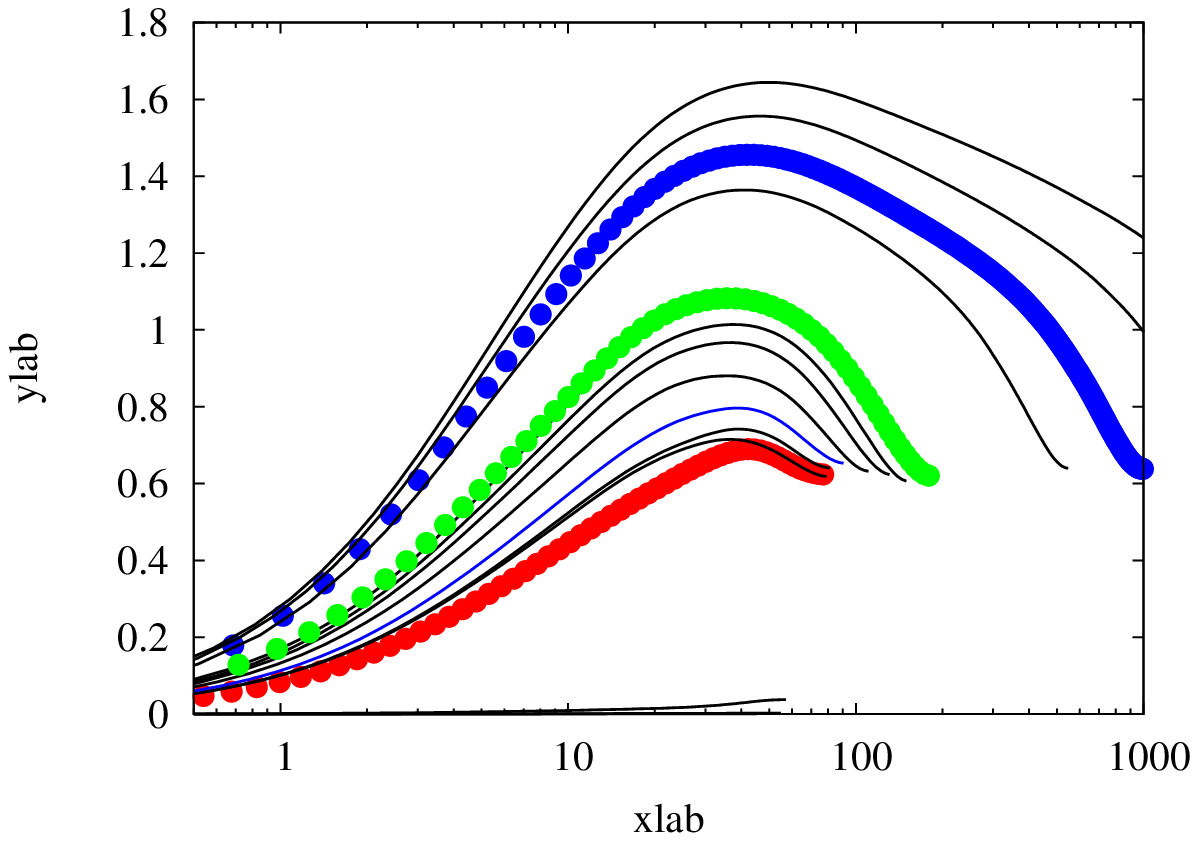}
\vskip -.5cm
\hskip 1.0cm  a)   \hskip 3.5cm  b) \hskip 3.5cm  c)   \hskip 3.5cm  d)
\vskip -0.0cm
\hskip -1.8cm
\psfrag{ylab} {$ U_1^+    $}
\psfrag{xlab}{ $y^+ $}
\includegraphics[width=4.3cm]{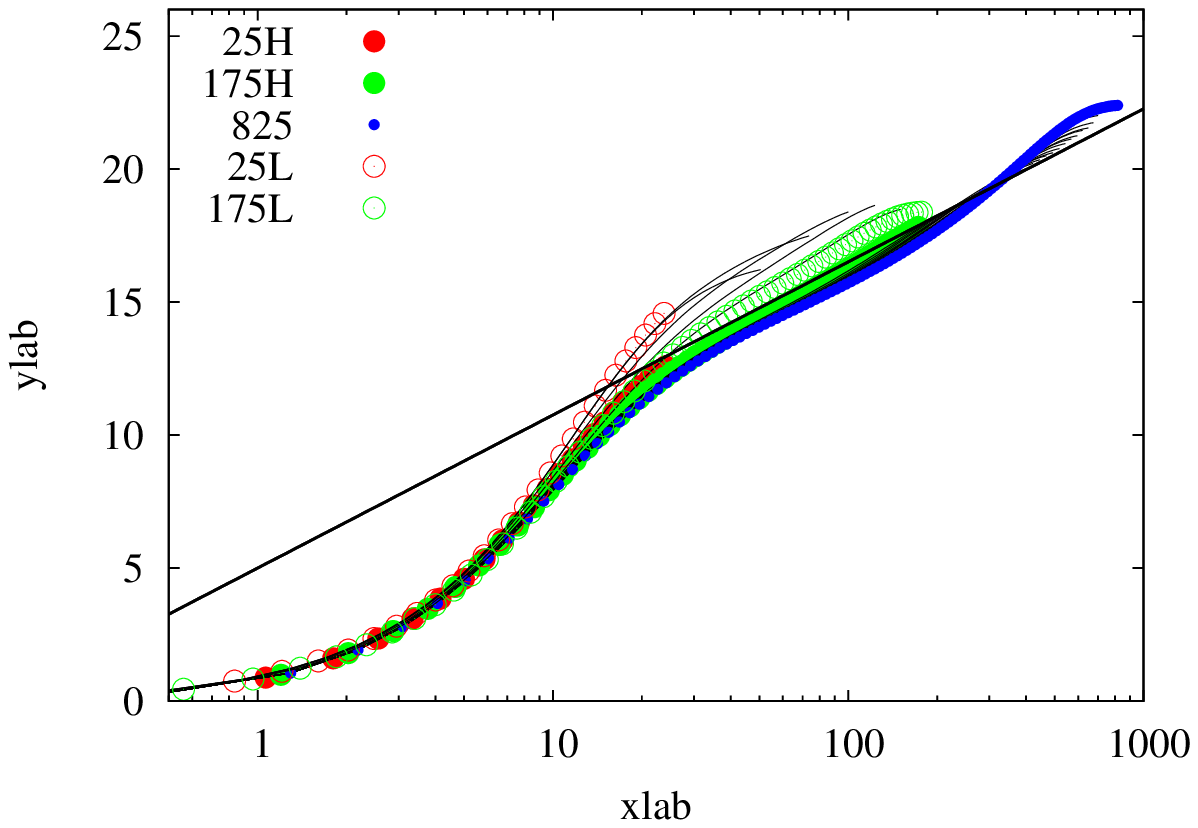}
\hskip -0.3cm
\psfrag{ylab} {\hskip -0.5cm $ u_1^{\prime +}    $}
\psfrag{xlab}{ $y^+ $}
\includegraphics[width=4.3cm]{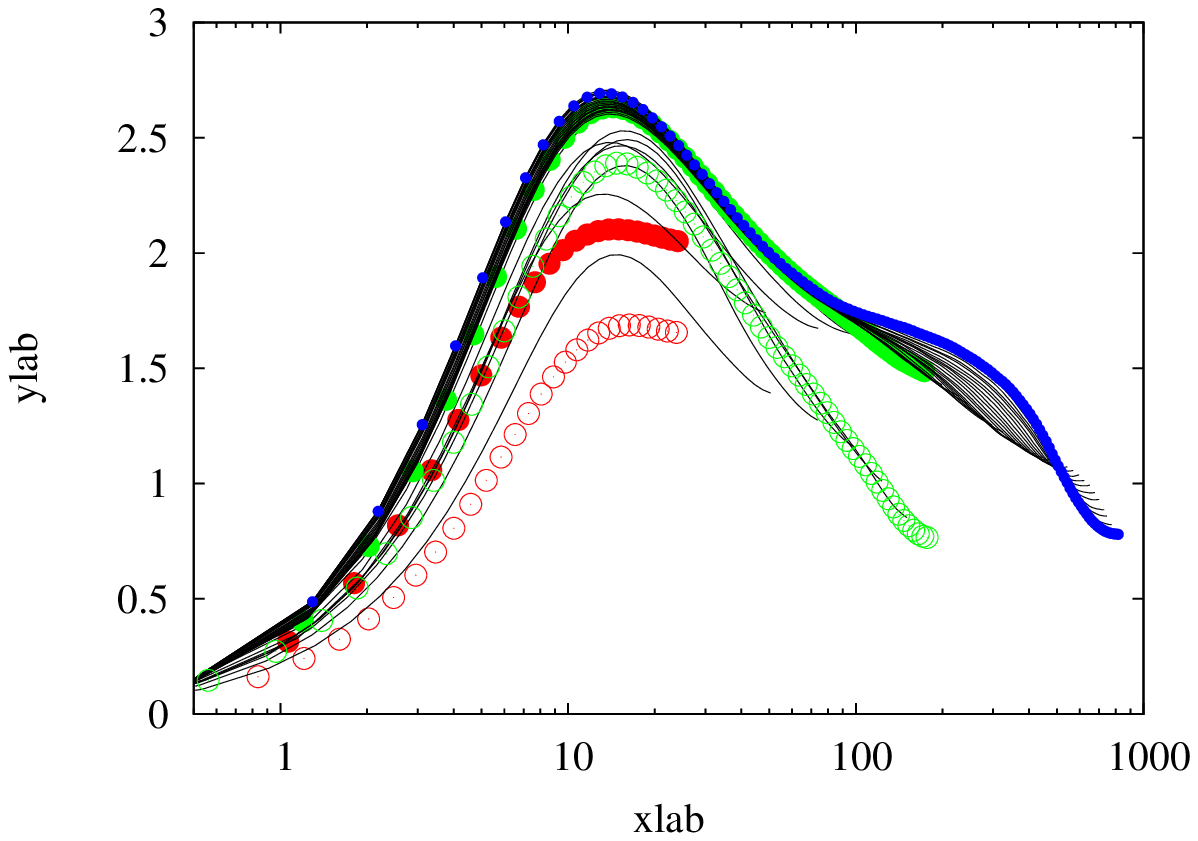}
\hskip -0.3cm
\psfrag{ylab} {\hskip -0.5cm $ u_2^{\prime +}    $}
\psfrag{xlab}{ $y^+ $}
\includegraphics[width=4.3cm]{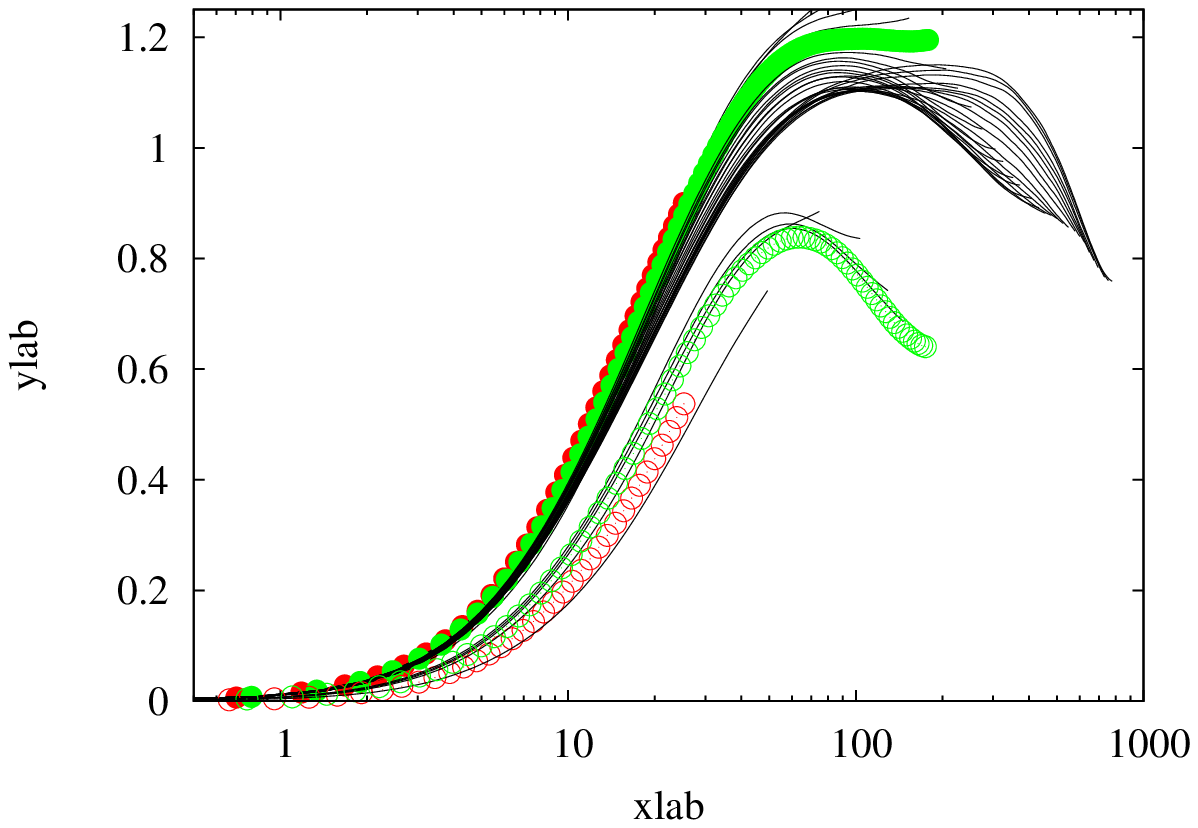}
\hskip -0.3cm
\psfrag{ylab} {\hskip -0.5cm $ u_3^{\prime +}    $}
\psfrag{xlab}{ $ y^+$}
\includegraphics[width=4.3cm]{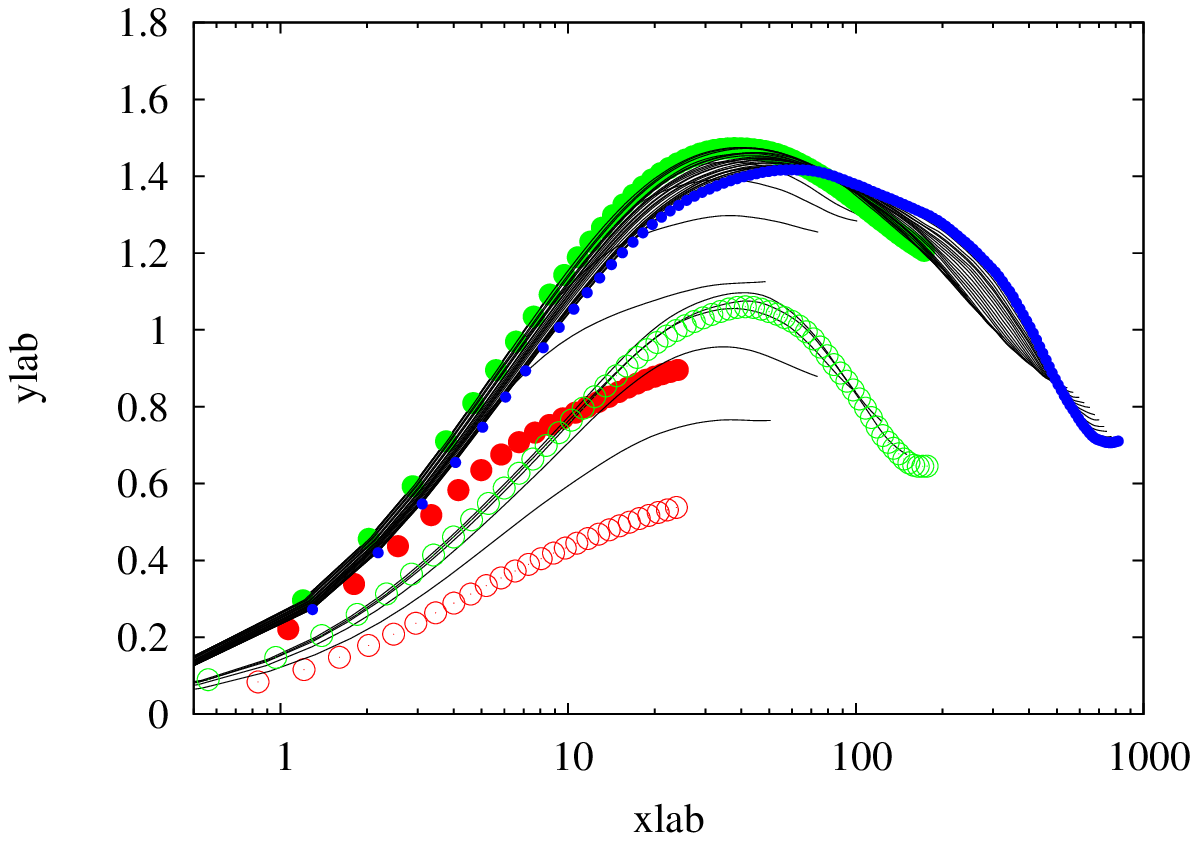}
\vskip -.5cm
\hskip 1.0cm  e)   \hskip 3.5cm  f) \hskip 3.5cm  g)   \hskip 3.5cm  h)
\caption{Square duct: profiles of mean velocity and normal turbulent
stresses, $u_i^{\prime +}=<u_i^2>^{1/2})^+$.
In the top panels we show statistics for a two-dimensional
channel, and below for the square channel, with quantities scaled
with the local friction velocity, and reported up to the corner bisector. 
Values of $Re_\tau$
for the profiles with symbols are given in the legend. Additional
profiles indicated with lines are shown to see the trends
at different Reynolds numbers. Note that to produce panels e-h,
the results at $Re=2500$ (open) and $Re=15000$ (solid) calculated at both
walls are used.
}
\label{fig6}
\end{figure}

At statistically steady  state it is interesting to analyse  the
profiles of the mean streamwise velocity $U_1$ and of
the second-order velocity statistics and to investigate 
differences with respect to a two-dimensional turbulent
channel. This comparison allows to get a general view of 
the complexity of wall-bounded flows.
Square and rectangular ducts are characterised by profiles
of the mean wall shear stress $\tau_w$ which decrease in magnitude
moving from the center of the walls towards the corners (see
figure \ref{fig3}a), hence the wall-normal profiles 
should scale with the local friction velocity, $u_\tau$.
The Reynolds number dependence in the canonical channel
has been studied in several papers~\citep[e.g.][]{bernardini_14,lee_15} 
through statistics derived from DNS. 
The data utilised by \citet{orlandi2015} are used for the profiles
shown in figure~\ref{fig6}a-d. It may be observed that for all the
components of the Reynolds stresses there is
a strong Reynolds number dependence at low $Re_\tau$.
A large jump of the peak value of $u_1^{\prime +}$ 
occurs between $Re_\tau=78$ and $Re_\tau=180$.
Considering the wall region, the inner-scaled profiles do
not show large $Re$ dependence at high Reynolds number.
In the outer region the statistics
do not scale well in wall units, leading to large differences
in the profiles of figure~\ref{fig6}. The data reported are
plotted up to  $y^+=1000$, being
the highest value reached in simulations of
ducts~\citep{pirozzoli_18}.  The results  at $Re=2500$ and $Re=15000$
are considered to evaluate the statistical profiles
at several distances from the corner. Figure \ref{fig3}a 
shows that at $Re=2500$ the $\tau_w$ profile has
a short flat region, which becomes more elongated at $Re=15000$.
The comparison between the statistics for the channel and
those for the square duct highlight 
differences and similarities. The profiles of $U_1^+$ in
channel flow (figure~\ref{fig6}a) is parabolic up to $Re_\tau\approx 77$, and at 
slightly greater $Re_\tau$ attains larger
than higher $Re_\tau$, for the same value of $y^+$. 
At this Reynolds number there is no separation between outer
and near-wall structures, and a single very large unsteady eddy is
present, causing the overshoot with respect the canonical
logarithmic velocity profile.  The same also overshoot
occurs in ducts (figure~\ref{fig6}e) at all locations, 
hence also at a position corresponding to $Re_\tau=25$, which
would be too low to have fully turbulent flow in a two-dimensional
channel. The formation of a mean velocity profile similar to that
of fully turbulent flows is therefore due to the dynamics
of the near-wall structures produced in the neighbouring regions.
This explains why the $U_1^+$ profile at $Re_\tau=25$ obtained
from the simulation at $Re=15000$ does not show the
overshoot, which instead occurs in the simulation at $Re=2500$.
Some difference may be observed in the wake region, mainly due
to evaluation of the profiles up to the diagonal line
in the square duct.
Strong interactions among the turbulent
structures may be also inferred from the $u_i^\prime$ profiles.
In two-dimensional channels, their peak value at $Re_\tau=57$ (figure \ref{fig6}b) 
is one order of magnitude smaller than at high $Re$.
On the other hand, in the square duct (figure \ref{fig6}f), the peak at $Re_\tau=25$ is of
the same order as at higher $Re$. In addition, the peak
at $Re_\tau=25$ evaluated from DNS at $Re=15000$ (red
solid symbol) is higher than at $Re=2500$ (open red circle). This difference
is also visible at $Re_\tau=175$ (solid and open green circles). 
At higher values of $Re_\tau$ there
is good collapse of the profiles, not found for the channel in figure~\ref{fig6}b, 
where differences appear between $Re_\tau=500$ (black line between
the green and the blue solid circles) and $Re_\tau=1000$.
To further appreciate Reynolds number
dependence of the outer region in planar channels, profiles (the two black solid
lines above the blue solid circles) at  $Re_\tau=2000$ and
$Re_\tau=4000$ are also shown in figure~\ref{fig6}a-d.
The effect of the interaction among turbulent structures of different size
in the ducts can be inferred by the profiles of $u_2^\prime$
(figure~\ref{fig6}g) and of $u_3^\prime$ (figure~\ref{fig6}h).
Also for these normal stresses the red and green profiles at the same $Re_\tau$
obtained from the simulations at $Re=2500$ (open circles) and $Re=15000$
(solid circles) differ in magnitude. For the square duct the maximum of 
$u_2^\prime$ is slightly higher than for the channel.
whereas for $u_3^\prime$ the peak value 
at the highest $Re_\tau$ is comparable.

\subsubsection{Turbulent stresses in the principal strain axes}

\begin{figure}
\centering
\vskip -0.0cm
\hskip -1.8cm
\includegraphics[clip,width=4.0cm,angle=90]{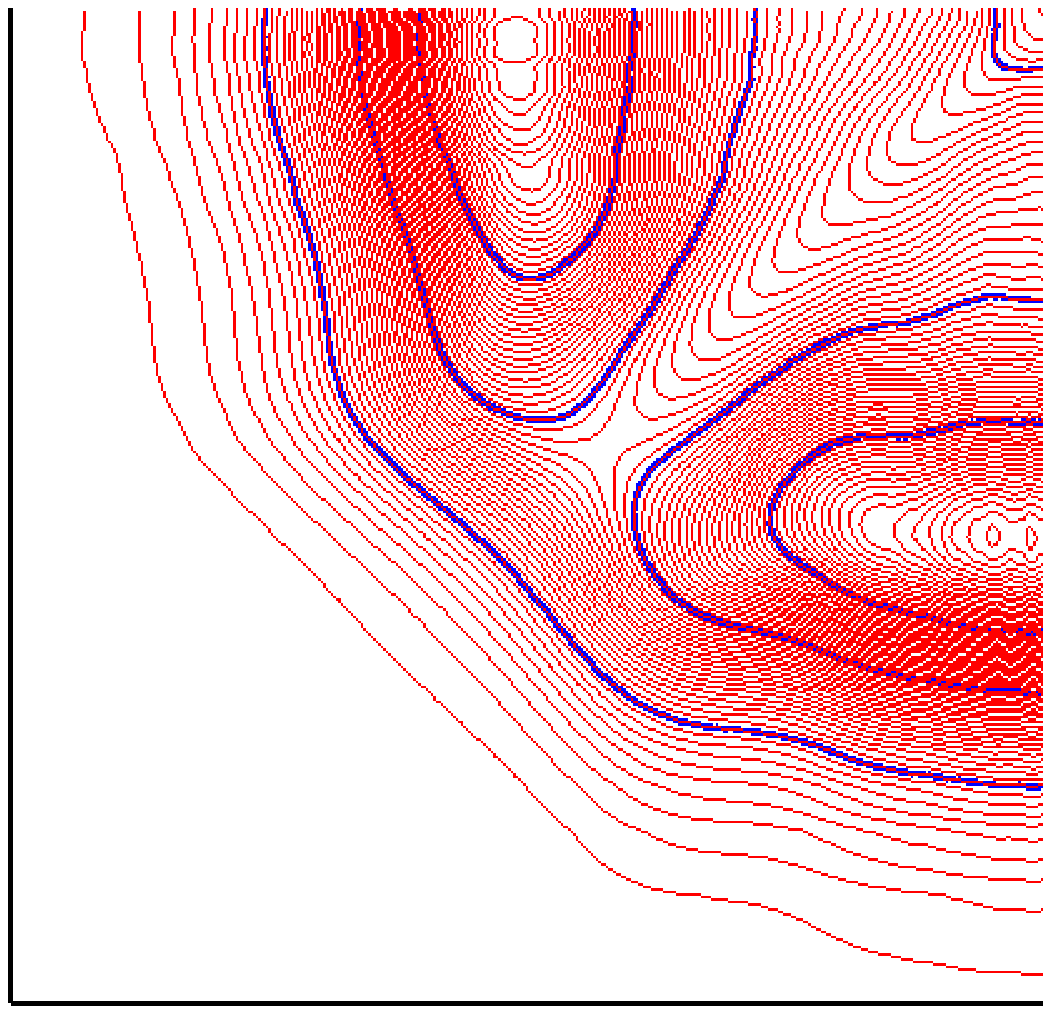}
%\hskip 0.5cm
\includegraphics[clip,width=4.0cm,angle=90]{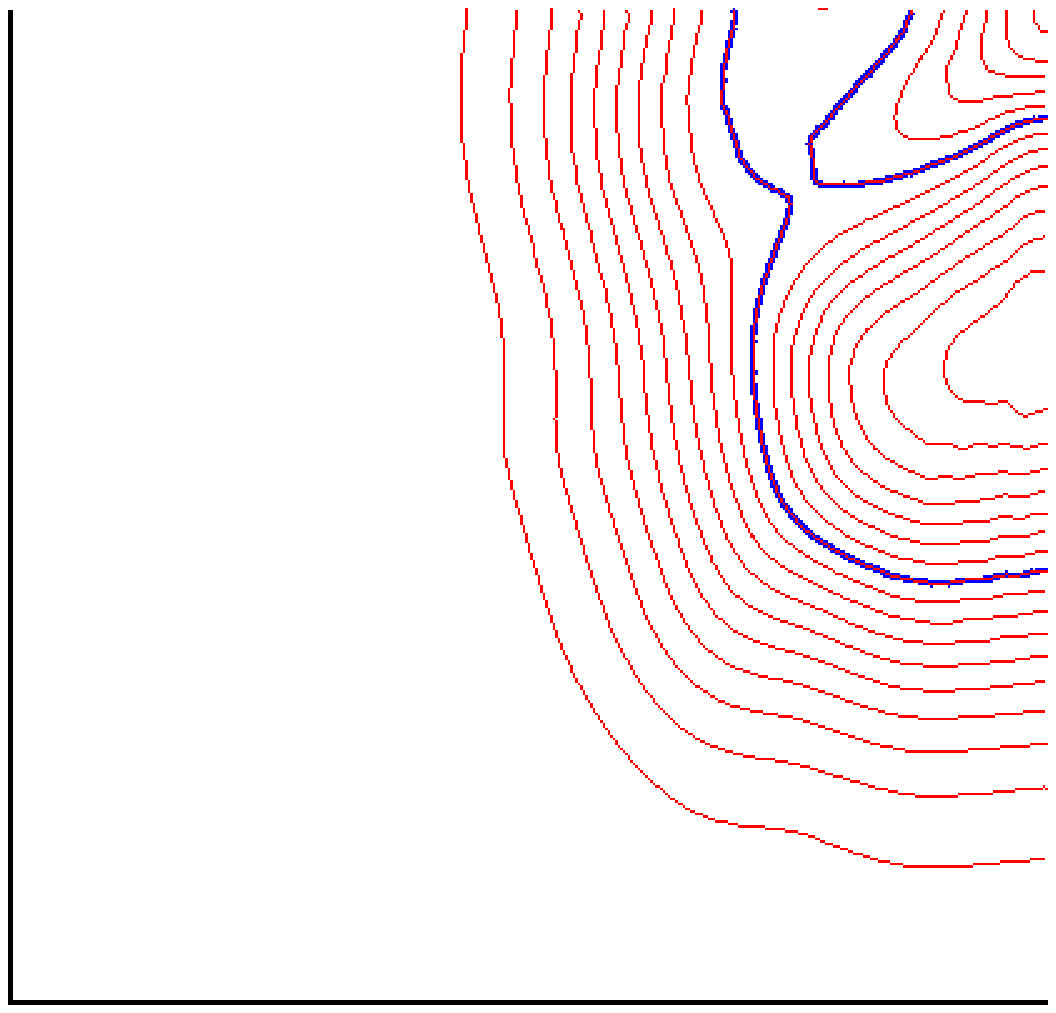}
%\hskip 0.5cm
\includegraphics[clip,width=4.0cm,angle=90]{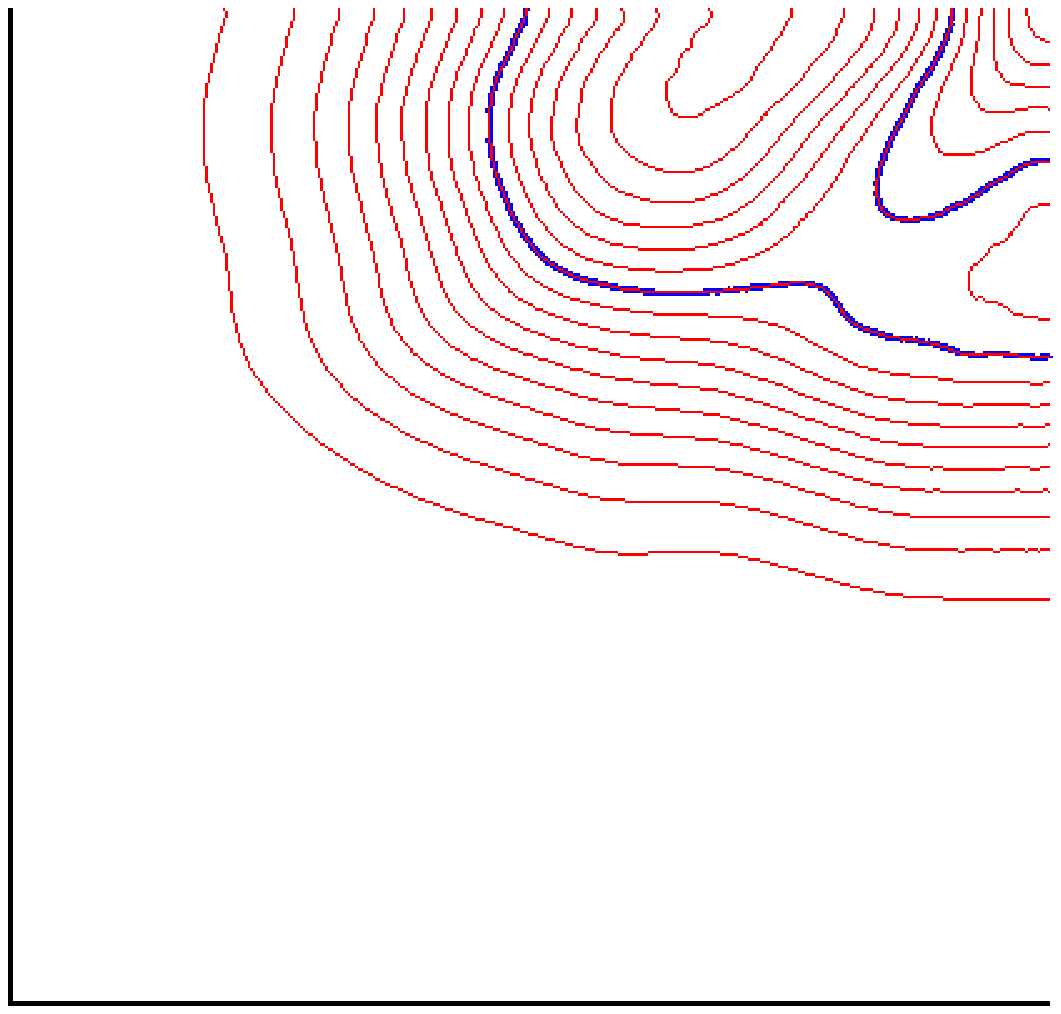}
\vskip 0.0cm
\hskip 1.0cm a) \hskip 4.2cm  b) \hskip 4.2cm  c)
\vskip 0.2cm
\hskip -1.8cm
\includegraphics[clip,width=4.0cm,angle=90]{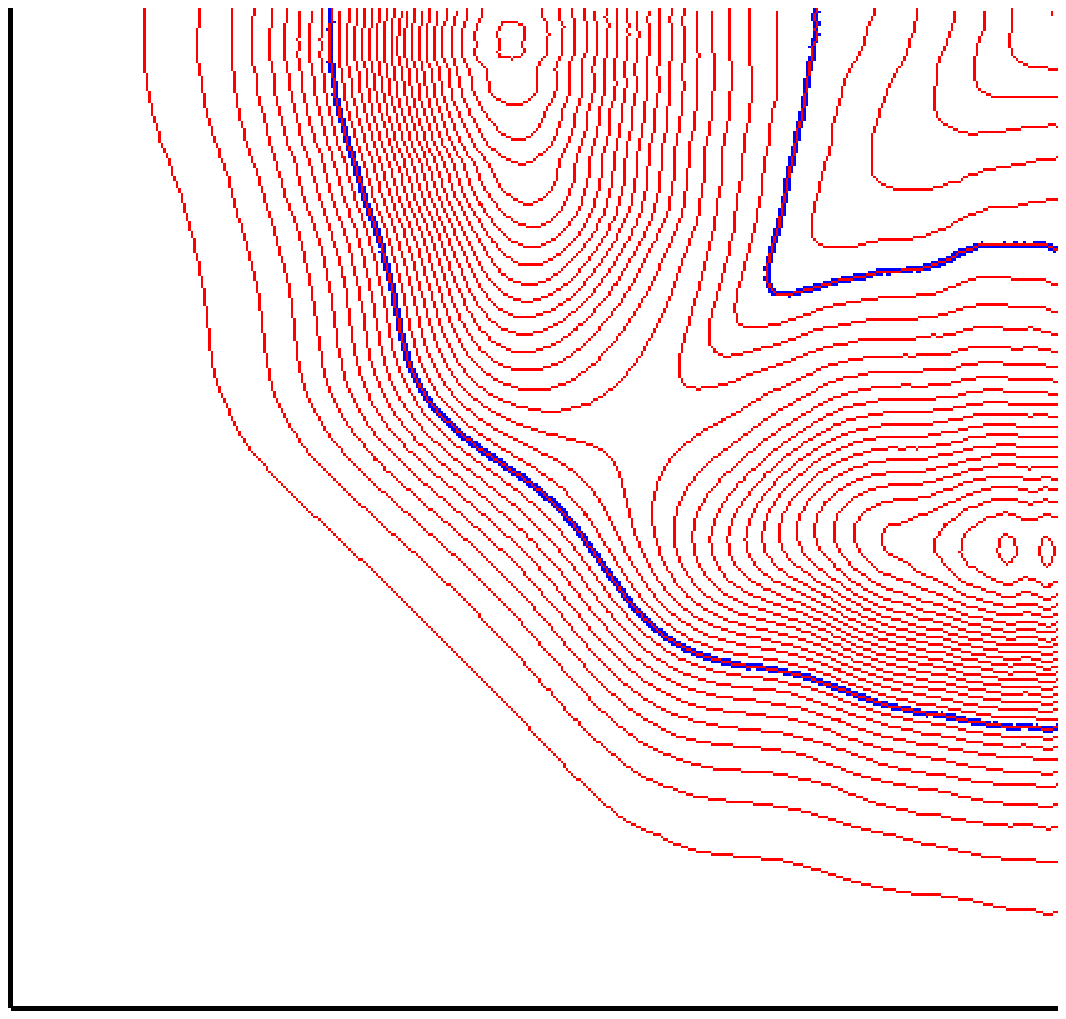}
%\hskip 0.5cm
\includegraphics[clip,width=4.0cm,angle=90]{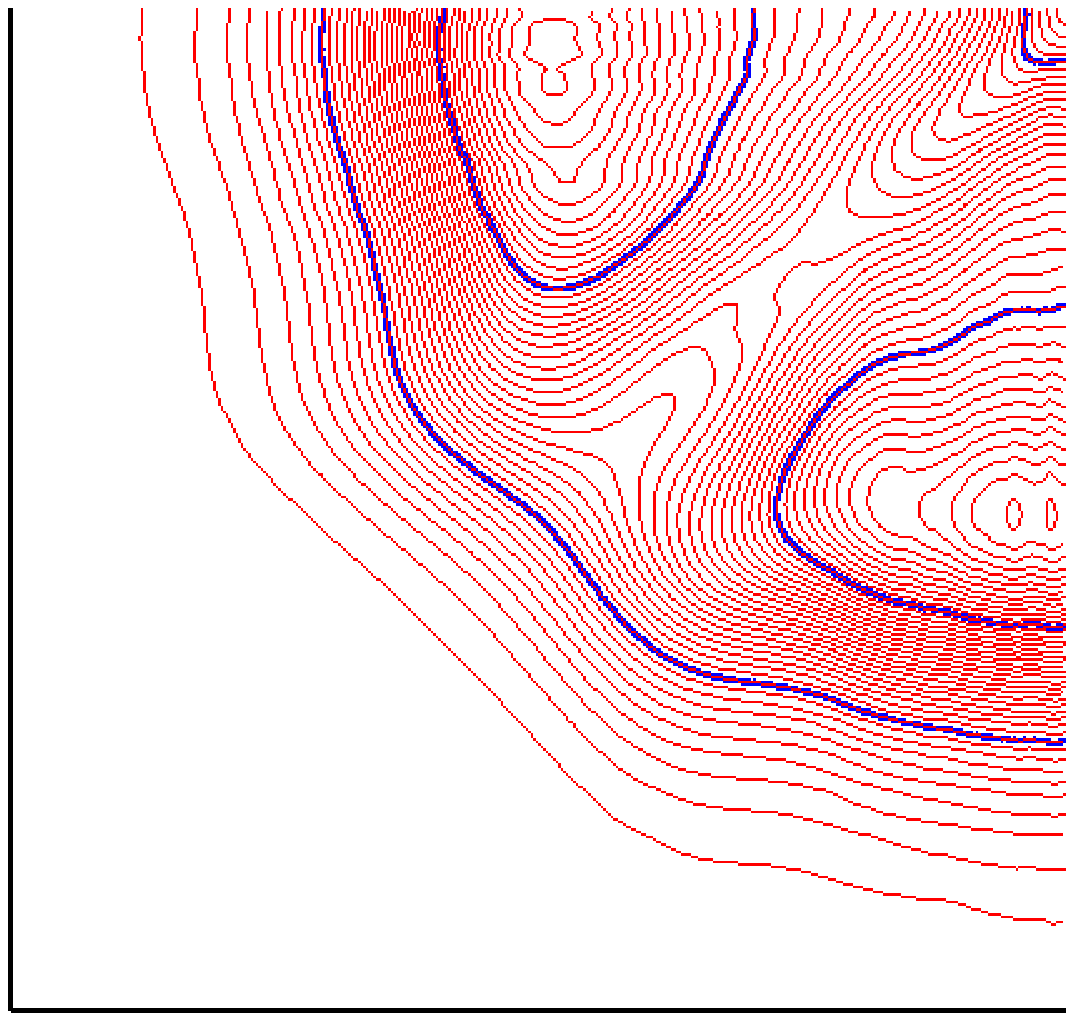}
%\hskip 0.5cm
\includegraphics[clip,width=4.0cm,angle=90]{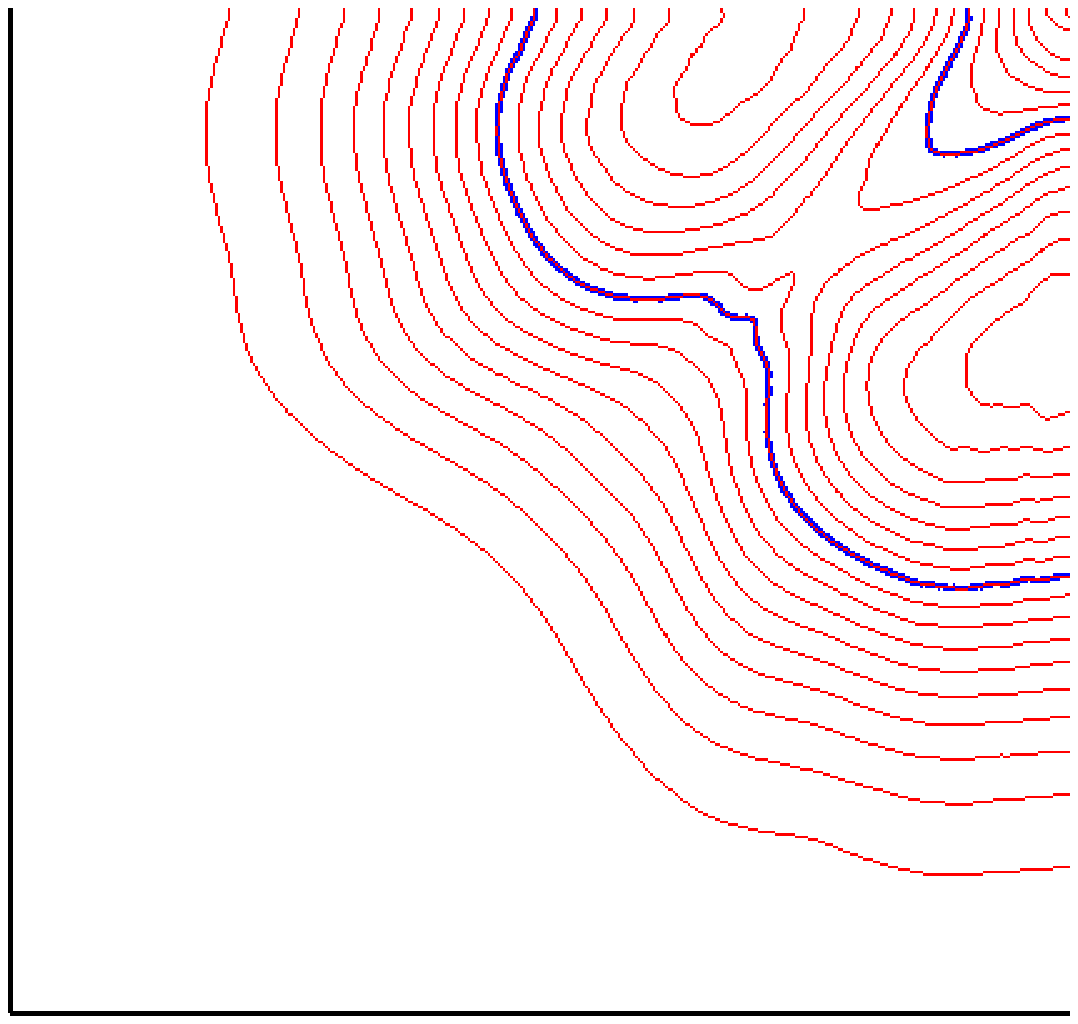}
\vskip -0.0cm
\hskip 1.0cm d) \hskip 4.2cm  e) \hskip 4.2cm  f)
\vskip -0.0cm
\caption{Square duct: contours of turbulent stresses in wall units at
$Re=7750$, in Cartesian basis (top), and in the strain eigenvector basis (bottom); 
logarithmic coordinates are used for the distances from the corner,
with $d^+ \le 420$.
Red lines are spaced by $\Delta=0.1$, and blue lines start from $1$ and are spaced by $\Delta=2$.
a)  $R_{11}$, b)  $R_{22}$, c)  $R_{33}$,
d)  $R_{\alpha \alpha}$, e)  $R_{\beta \beta}$, 
f)  $R_{\gamma \gamma}$
}
\label{fig7}
\end{figure}

The profiles of the $rms$ velocity fluctuations in channel and the ducts
show large anisotropy, which may be ascribed to differences in the respective production,
resulting from interaction of the strain rate shear tensor 
$S_{ij}=(\der{ U_i}{x_j}+\der{U_j}{x_i})/2$,
and the Reynolds stress tensor $R_{ij}=-\langle u_i u_j\rangle$. 
In particular, the production term 
in the transport equation for the stresses turbulent stresses is
$P_{ij}=-(R_{ik}\der{ U_j}{x_k}+R_{jk}\der{ U_i}{x_k})$,
hence it may be stressed that the large scales
due to the mean motion are responsible for creating 
turbulence anisotropy in wall-bounded flows. 
It may then be interesting to evaluate the eigenvalues
of $S_{ij}$ (say extensional, $S_\alpha>0$, intermediate, $S_{\beta}$, and compressional, $S_\gamma<0$), 
and project the flow statistics along the eigenvectors of $S_{ij}$.
Evolution equations for the vorticity components 
in the local strain eigenvector basis were given by~\cite{nomura1998}, 
applied to the case of homogeneous turbulence. 
\citet{orlandi2018} exploited channel flow DNS at high 
Reynolds numbers~\cite{bernardini_14,lee_15,yamamoto_18}, 
to evaluate the Reynolds stresses in the strain eigenvector basis.
In channel $S_\beta=0$, hence one of
the Reynolds stresses is unchanged, whereas the
difference between the other two is reduces. 
The turbulent kinetic production
in the strain eigenvector basis is $P_k=-(P_\alpha+P_\beta+P_\gamma)$,
with $P_\alpha=R_{\alpha \alphaį} S_\alpha>0$, and
$P_\gamma=R_{\gamma \gamma} S_\gamma<0$, and larger than $P_k$ in absolute value. 
In channels it was found that at any Reynolds number the compressive
strain generates more kinetic energy than is destroyed by extensional one.
These results may be useful to construct more reliable turbulence closures.
For duct flows it is
difficult to get satisfactory results with models
based on the linear eddy viscosity assumption~\citep{speziale_82}.
Therefore the evaluation of the Reynolds stresses in
the reference system based on the eigenvalues of $S_{ij}$
may be of interest. A comparison between the stresses in
the Cartesian reference system and those in the new reference system
are shown in figure~\ref{fig7}. In this figure a logarithmic scale 
is used for distances from the corner  
to emphasize the near-wall behavior.
Wall units are based on the mean friction velocity $\overline{u_\tau}$, 
and also used to scale the normal stresses.  The results at $Re=7750$ 
may be regarded as representative of flows at high Reynolds number. 
The red contours are separated by $\Delta=0.1$, and to
emphasise the tendency towards an isotropization in the new strain eigenvector basis,
blue contours are shown starting from unit value, and
separated by $\Delta=2$. The three top figures show large
anisotropy of the normal stresses, in fact in
figure~\ref{fig7}a there are three blue lines, whereas
the other two stresses there only have one blue contour.
In addition, the stresses 
in the wall-normal direction are not symmetric
about the corner bisector. This asymmetry should also
be reproduced in RANS models. On the other hand, 
the three bottom panels of figure~\ref{fig7} show that
turbulent stresses become symmetric with respect to the bisector
in the strain eigenvector basis.
The anisotropy level is also reduced, in fact only the compressive stress
in figure~\ref{fig7}f has two blue contours. As
found by \cite{orlandi2018} in two-dimensional channels, $R_{\gamma \gamma}$
is found to be larger than $R_{\alpha \alpha}$, leading to
greater turbulent kinetic energy
production through $P_\gamma$ than the destruction by the extensional
strain, $P_\alpha$.

\subsubsection{Turbulent kinetic energy budgets }

\begin{figure}
\centering
\vskip -0.0cm
\hskip -1.8cm
\includegraphics[clip,width=3.0cm,angle=90]{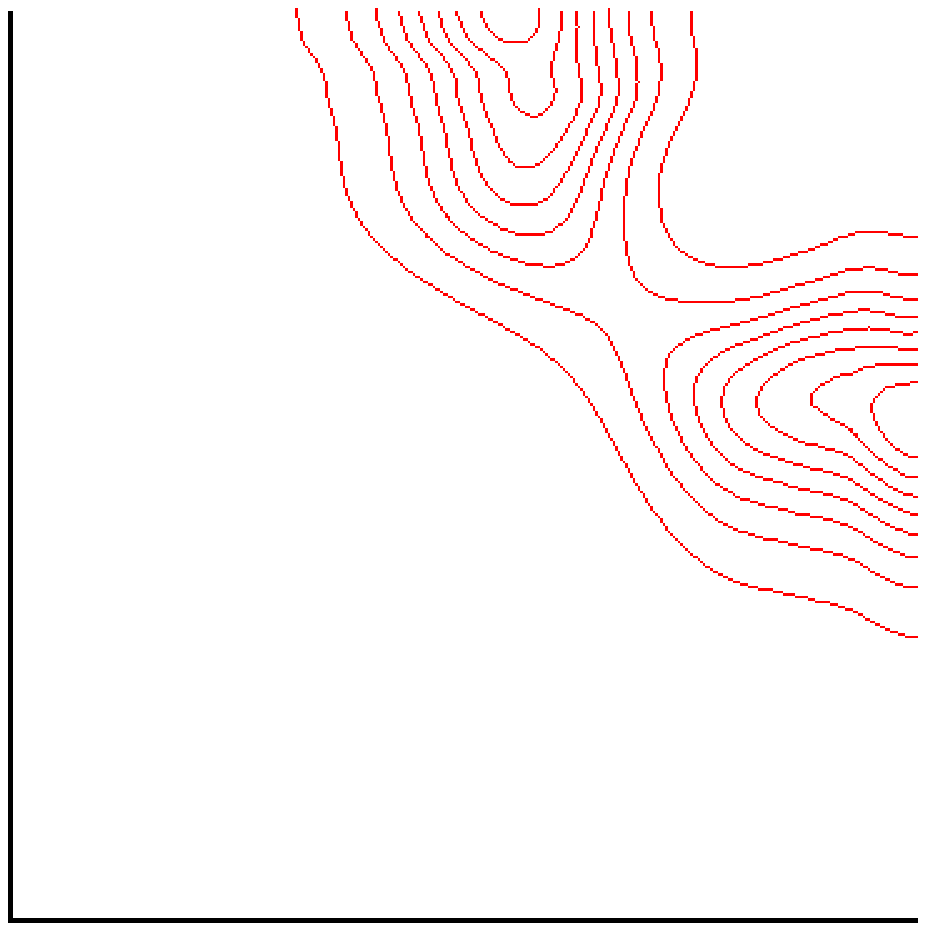}
%\hskip 0.5cm
\includegraphics[clip,width=3.0cm,angle=90]{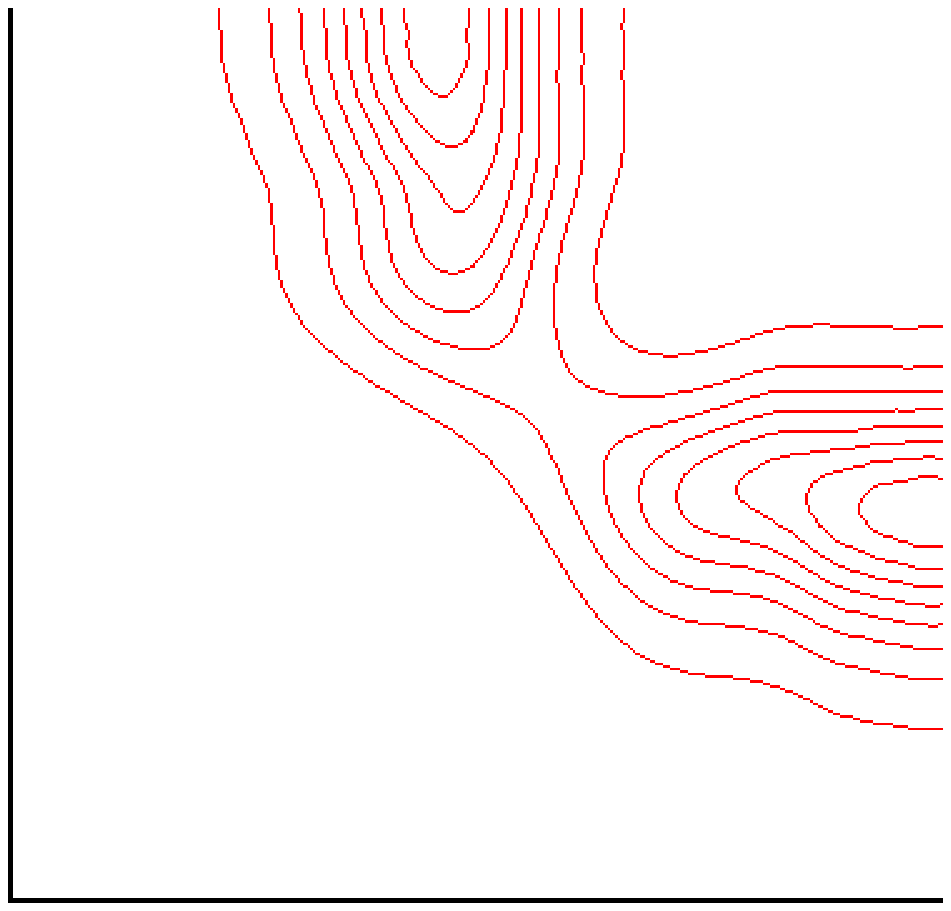}
%\hskip 0.5cm
\includegraphics[clip,width=3.0cm,angle=90]{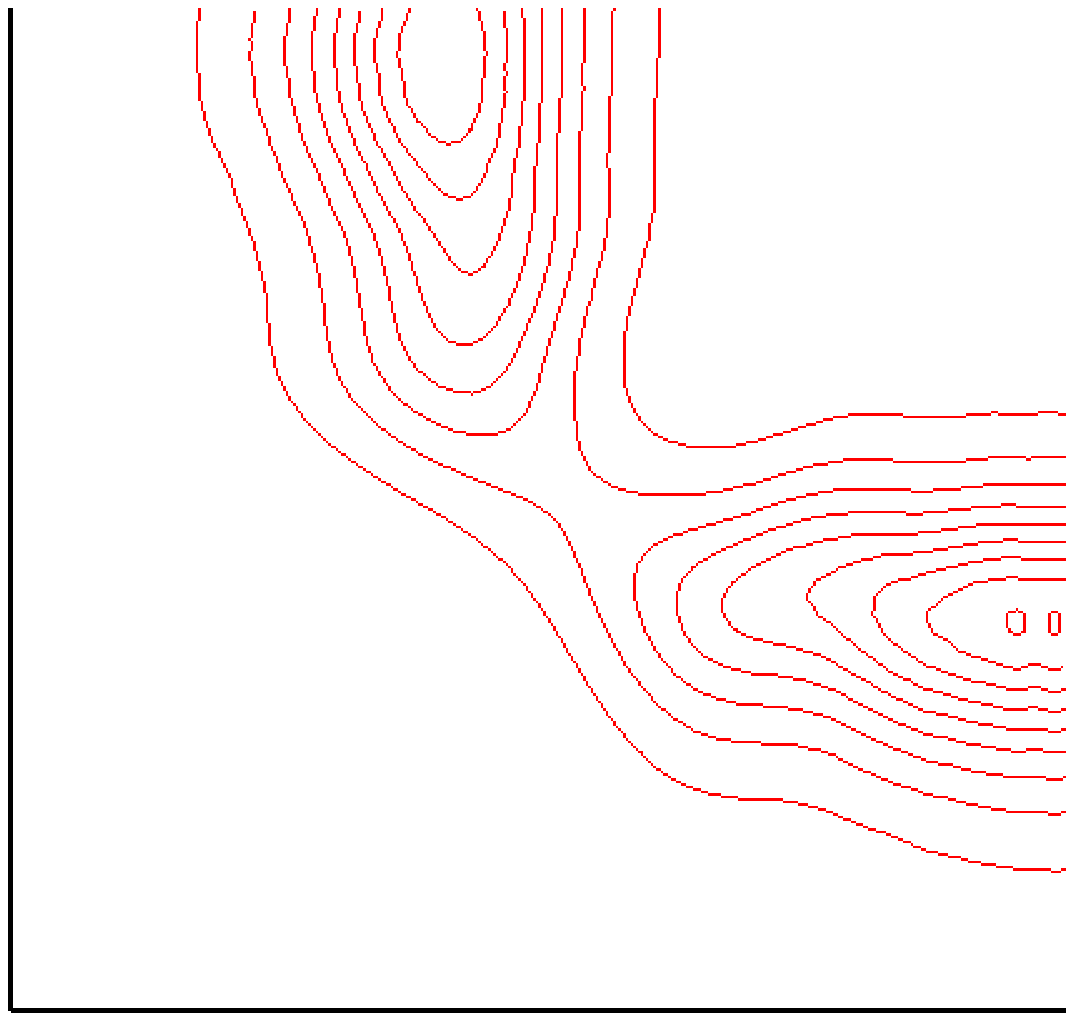}
%\hskip 0.5cm
\includegraphics[clip,width=3.0cm,angle=90]{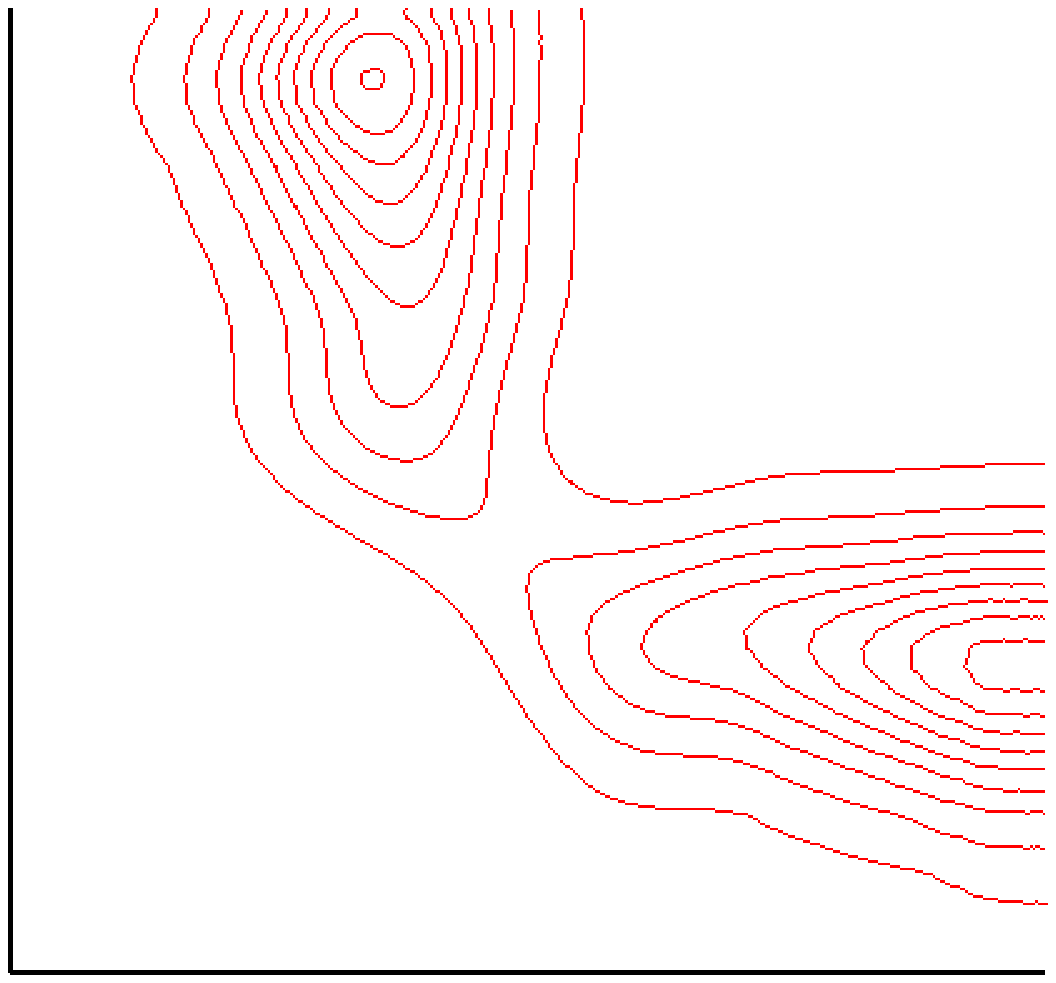}
\vskip 0.2cm
\hskip 1.0cm a) \hskip 3.2cm  b)  \hskip 3.2cm  c)  \hskip 3.2cm  d)
\vskip 0.0cm
\hskip -1.8cm
\includegraphics[clip,width=3.0cm,angle=90]{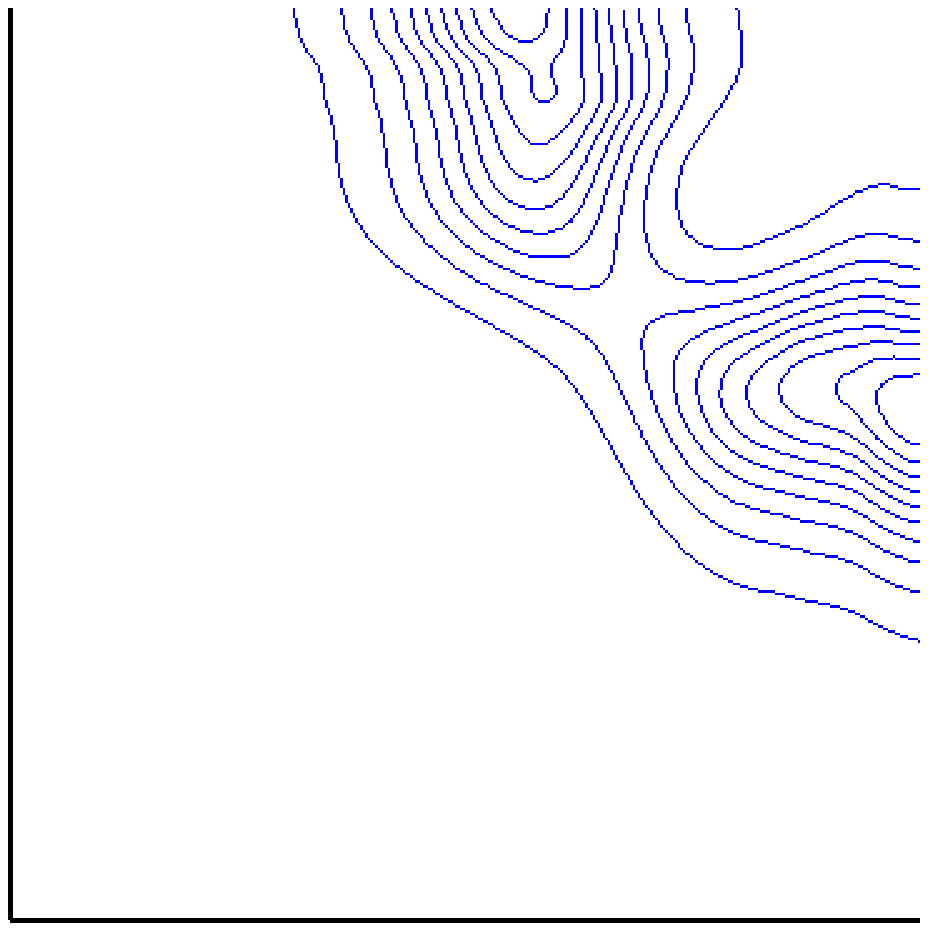}
%\hskip 0.5cm
\includegraphics[clip,width=3.0cm,angle=90]{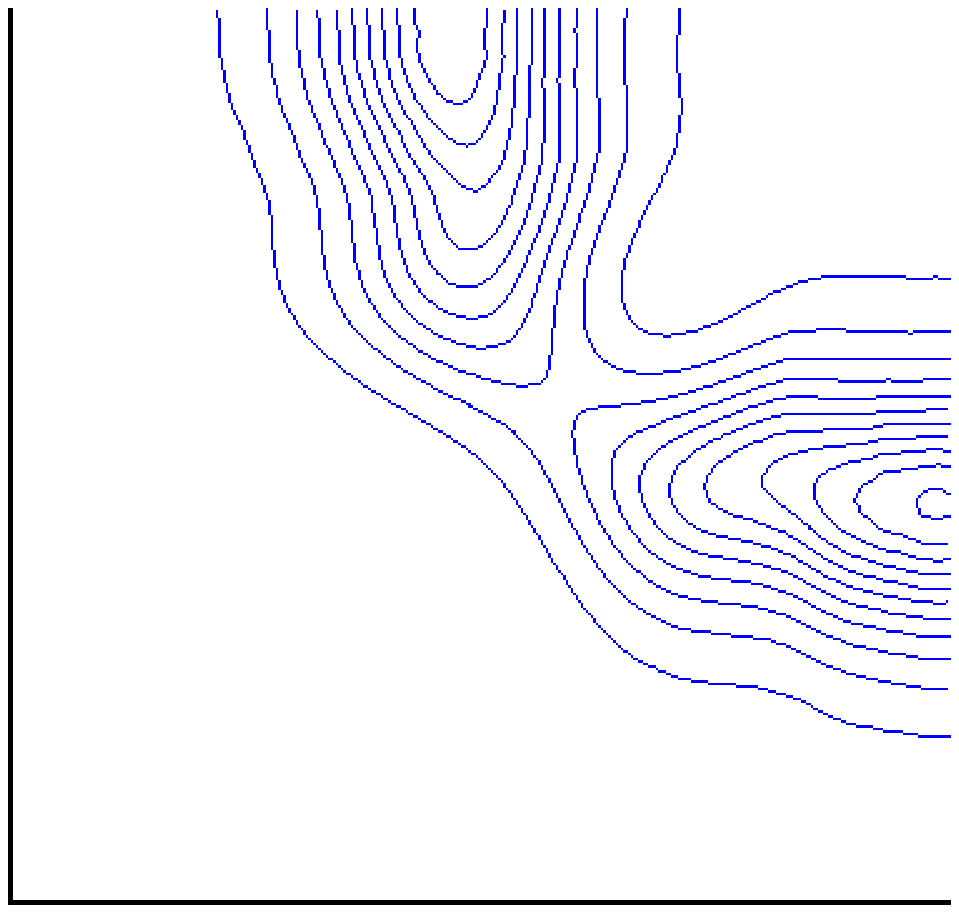}
%\hskip 0.5cm
\includegraphics[clip,width=3.0cm,angle=90]{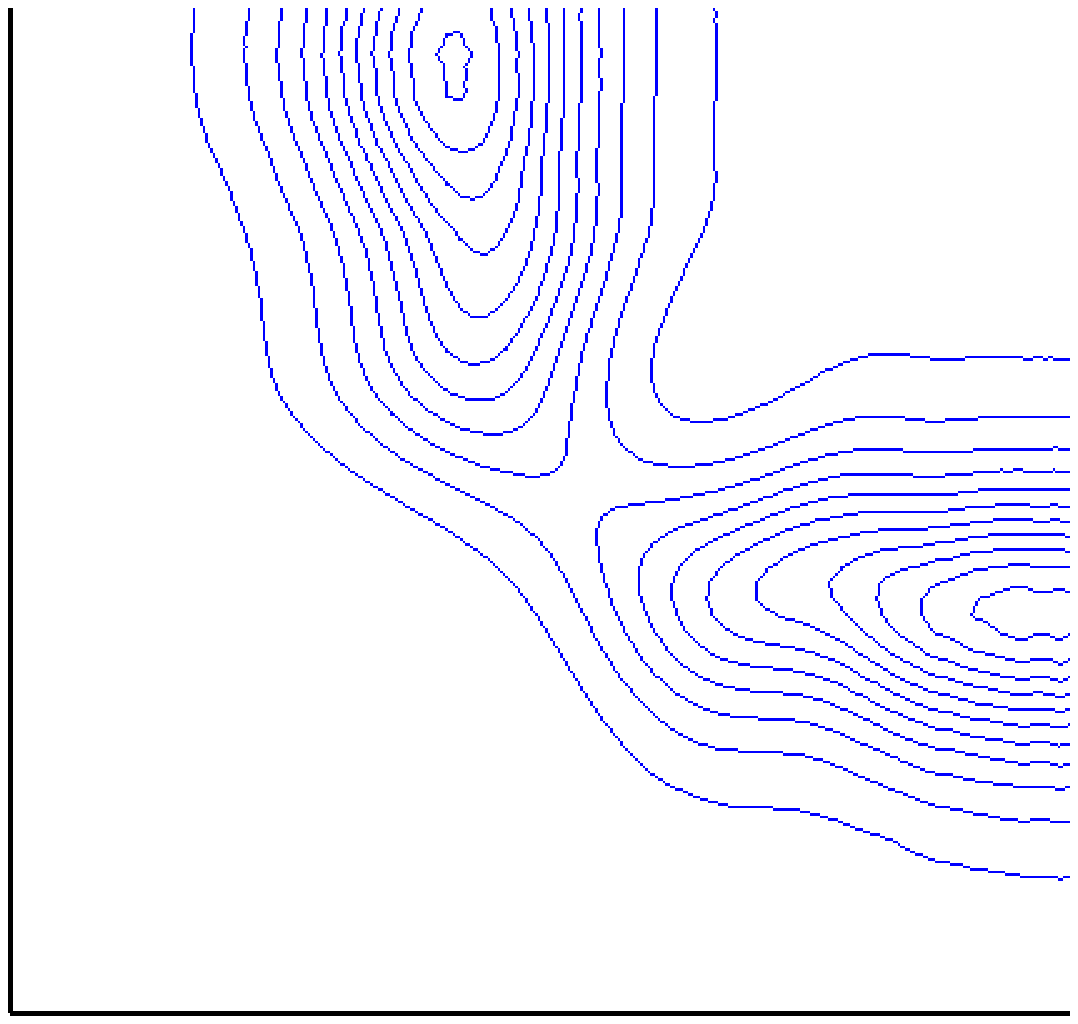}
%\hskip 0.5cm
\includegraphics[clip,width=3.0cm,angle=90]{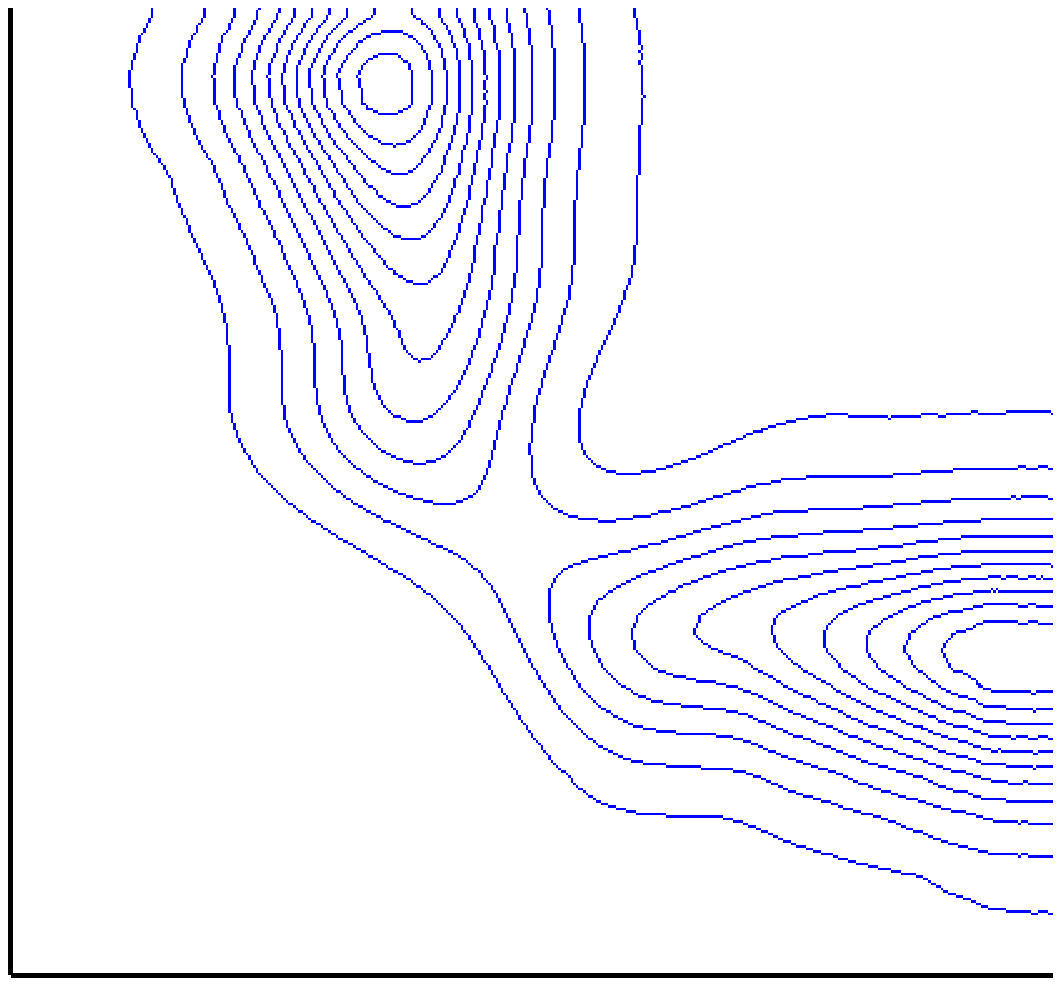}
\vskip -0.0cm
\hskip 1.0cm e) \hskip 3.2cm  f)  \hskip 3.2cm  g)  \hskip 3.2cm  h)
\vskip -0.0cm
\caption{Square duct: contours of turbulent kinetic energy 
production in the strain eigenvector basis in wall units,
at different $Re$ (a),e) $Re=2500$; b),f)  $Re=5000$; 
c),g) $Re=7750$; d),h) $Re=15000$.
Panels (a-c) show $P_{\alpha}$,
and panels (d-f) show $P_{\gamma}$, in increments 
$\Delta=0.1$, red positive, blue negative.
Logarithmic coordinates are used for the distances from the corner, with 
$d^+ \le 151$ (a,e),
$d^+ \le 284$ (b,f),
$d^+ \le 420$ (c,g),
$d^+ \le 765$ (d,h).
}
\label{fig8}
\end{figure}

The contours of the two extensional and compressional contributions to the turbulent
kinetic energy production are shown in a quadrant in figure~\ref{fig8}.
The contribution due to the intermediate principal strain,
$P_\beta=R_{\beta \beta}S_\beta$ is found to be 
symmetric with respect to the corner bisector, but
much smaller than the other, hence it is not shown. 
The symmetry of $P_\alpha$ and $P_\gamma$
with respect to the corner bisector is quite good up to $Re=7750$, 
whereas slight asymmetry near the two peaks of $P_\gamma$ and $P_\alpha$
is found at $Re=15000$.
At all $Re$ number, the larger number of contours
for $P_\gamma$ depicts the formation of a 
positive production ($-P_\gamma$) due to the compressive larger than the 
negative destruction ($-P_\alpha$) due to the extensional strain.
%It is worth recalling that the turbulent kinetic energy
%production is $P_K=-(P_\gamma+P_\alpha+P_\beta)$.
Overall, figure~\ref{fig8} 
shows satisfactory good Reynolds number independence 
of magnitude and spatial distribution of the productions terms.
Similar behavior of the profiles of $-P_\gamma$ and
$-P_\alpha$ was reported by \cite{orlandi2018} in planar
channels, hence we may assert that this behavior is typical
of turbulent flows in presence of smooth walls.

To look with greater detail into the flow physics of square ducts
it is worth evaluating the budget of the turbulent kinetic energy, $K$,
as found in several textbooks~\citep[p. 315]{pope_00},
\begin{equation}
\underbrace{
\frac{1}{2}U_k\frac{\partial  \langle u_iu_i\rangle}{\partial x_k} 
+\frac{1}{2} \frac{\partial \langle u_iu_iu_k\rangle}{\partial x_k}
+\langle u_i\frac{\partial p}{\partial x_i}\rangle 
}_{T_K}
- 
\underbrace{
\frac{1}{Re} \langle u_i\nabla^2 u_i \rangle
}_{D_K}
\underbrace{
-\langle u_iu_k\rangle \frac{\partial U_i}{\partial x_k}
}_{P_K}  
=0 ,
\label{eqbudgk}
\end{equation}

\begin{figure}
\centering
\vskip -0.0cm
\hskip -1.8cm
\includegraphics[clip,width=3.0cm,angle=90]{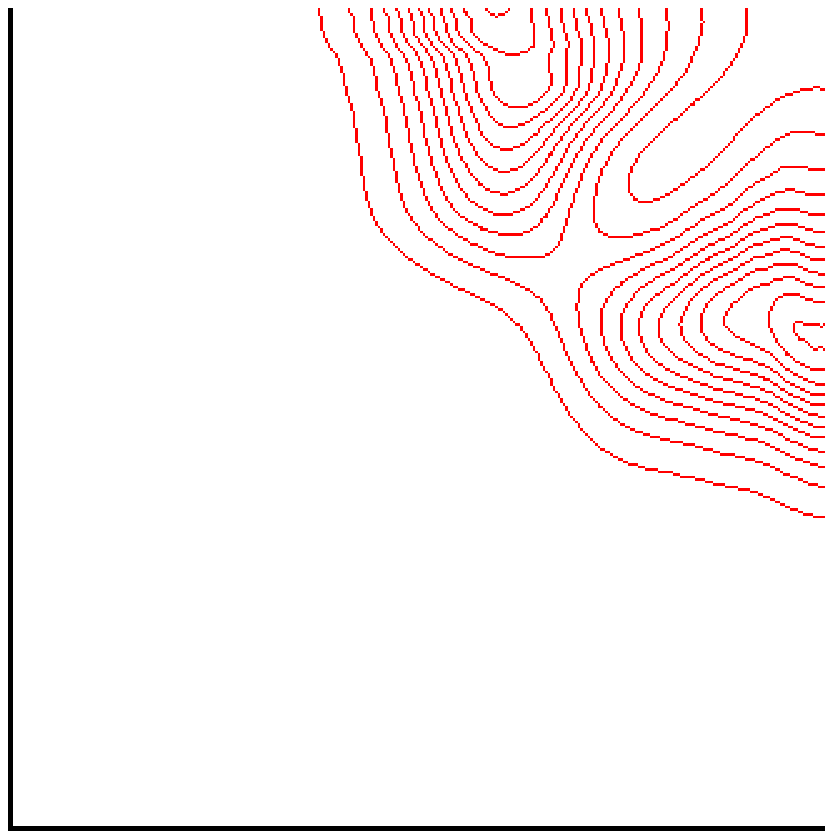}
%\hskip 0.5cm
\includegraphics[clip,width=3.0cm,angle=90]{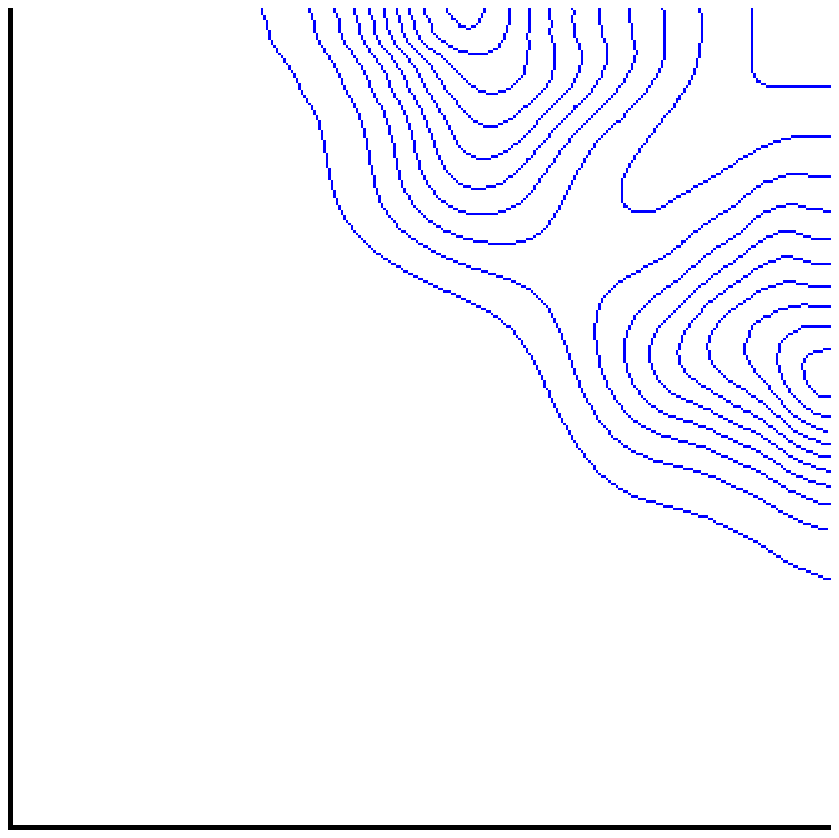}
%\hskip 0.5cm
\includegraphics[clip,width=3.0cm,angle=90]{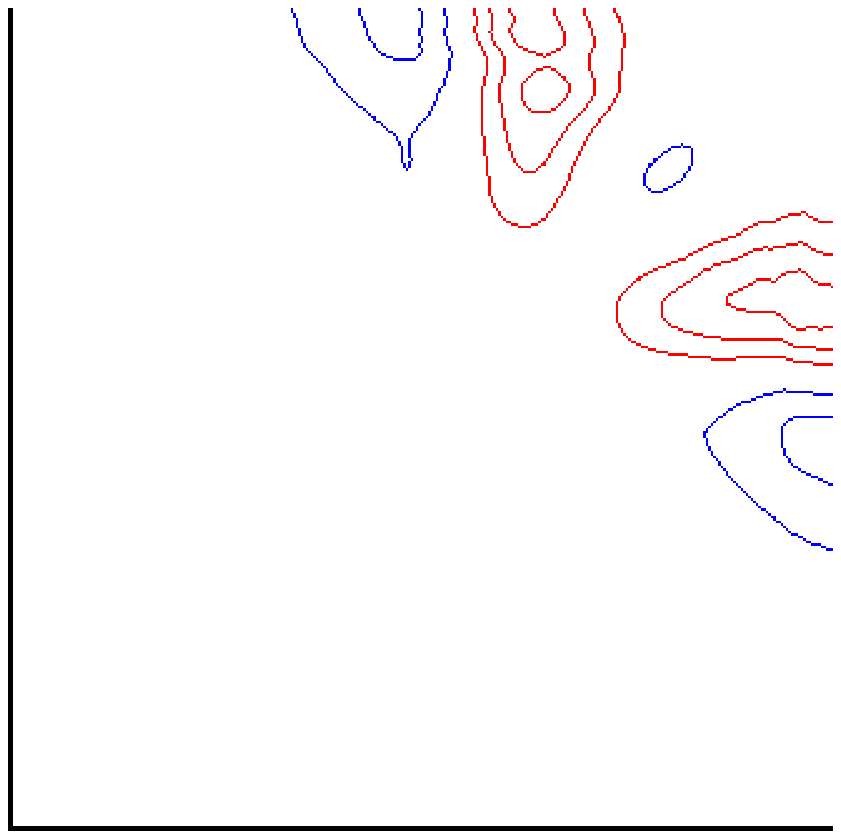}
%\hskip 0.5cm
\includegraphics[clip,width=3.0cm,angle=90]{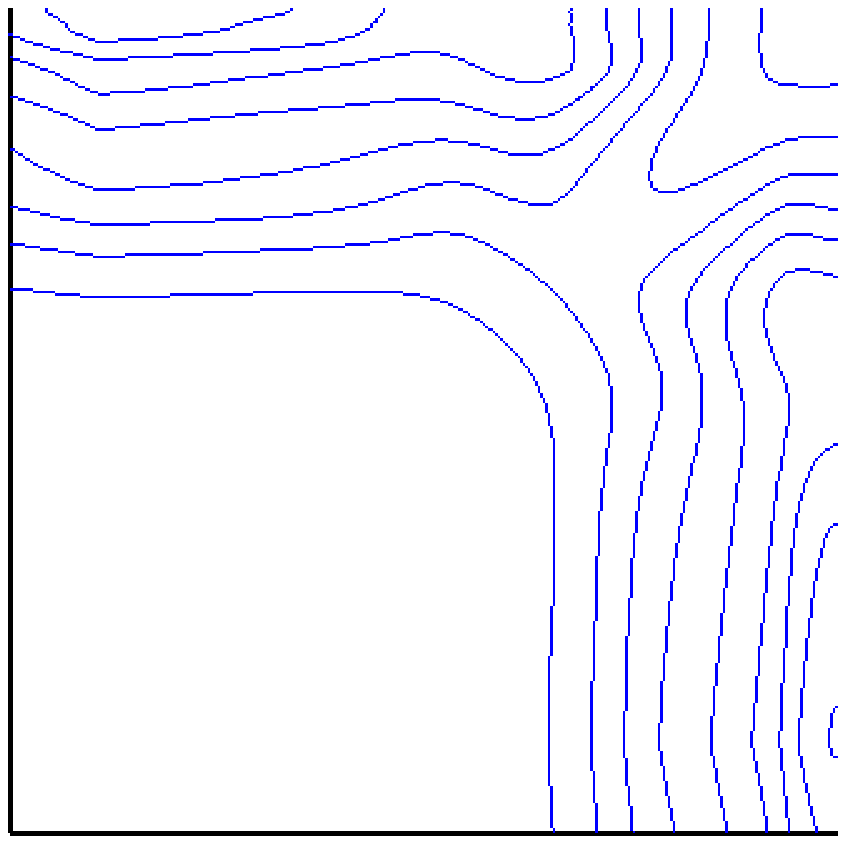}
\vskip 0.20cm
\hskip 1.0cm a) \hskip 3.2cm  b)  \hskip 3.2cm  c)  \hskip 3.2cm  d)
\vskip 0.20cm
\hskip -1.8cm
\includegraphics[clip,width=3.0cm,angle=90]{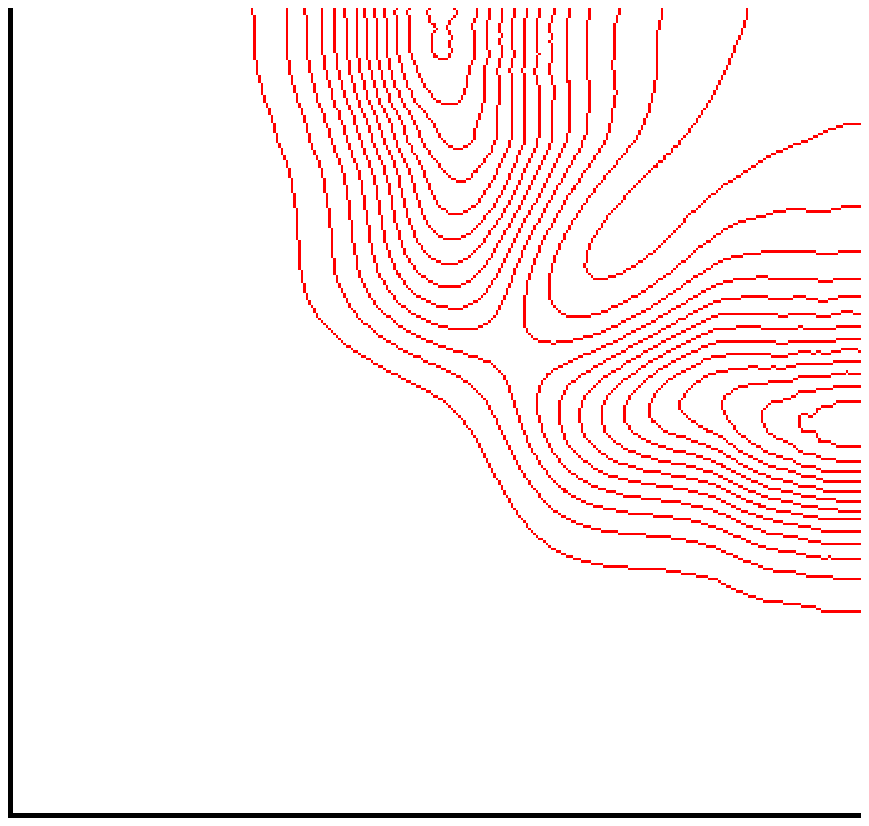}
%\hskip 0.5cm
\includegraphics[clip,width=3.0cm,angle=90]{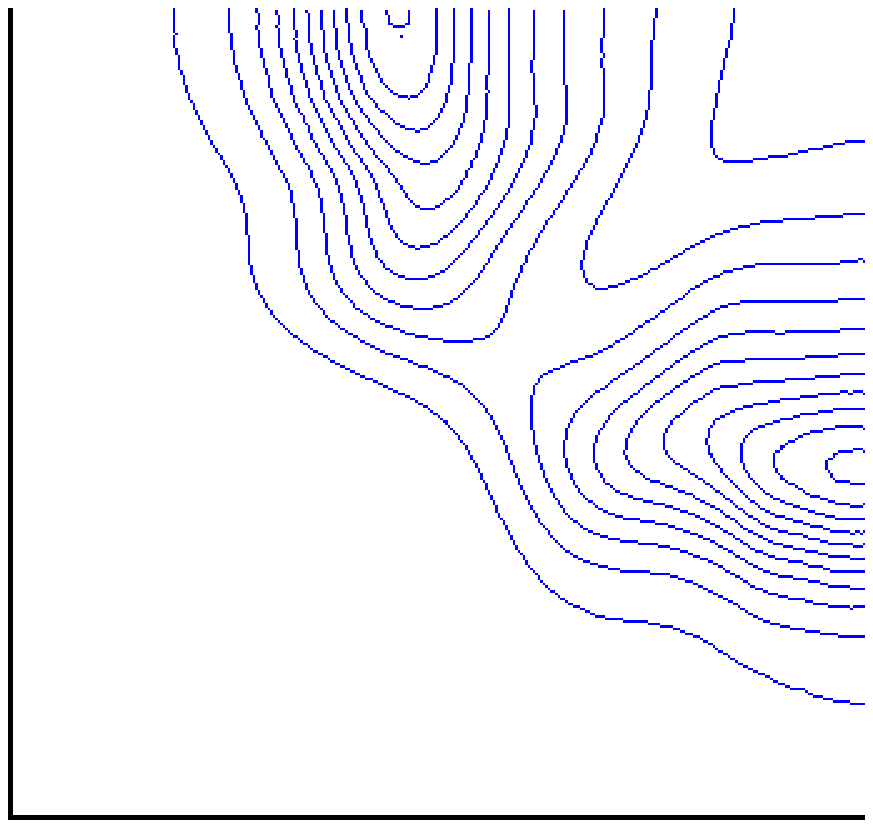}
%\hskip 0.5cm
\includegraphics[clip,width=3.0cm,angle=90]{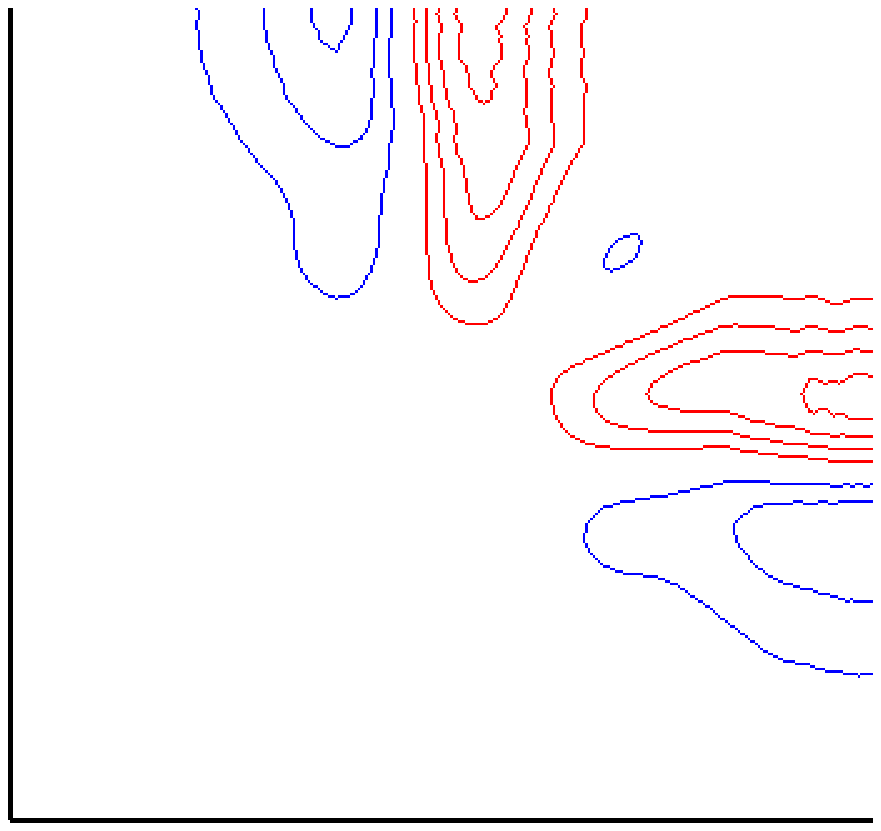}
%\hskip 0.5cm
\includegraphics[clip,width=3.0cm,angle=90]{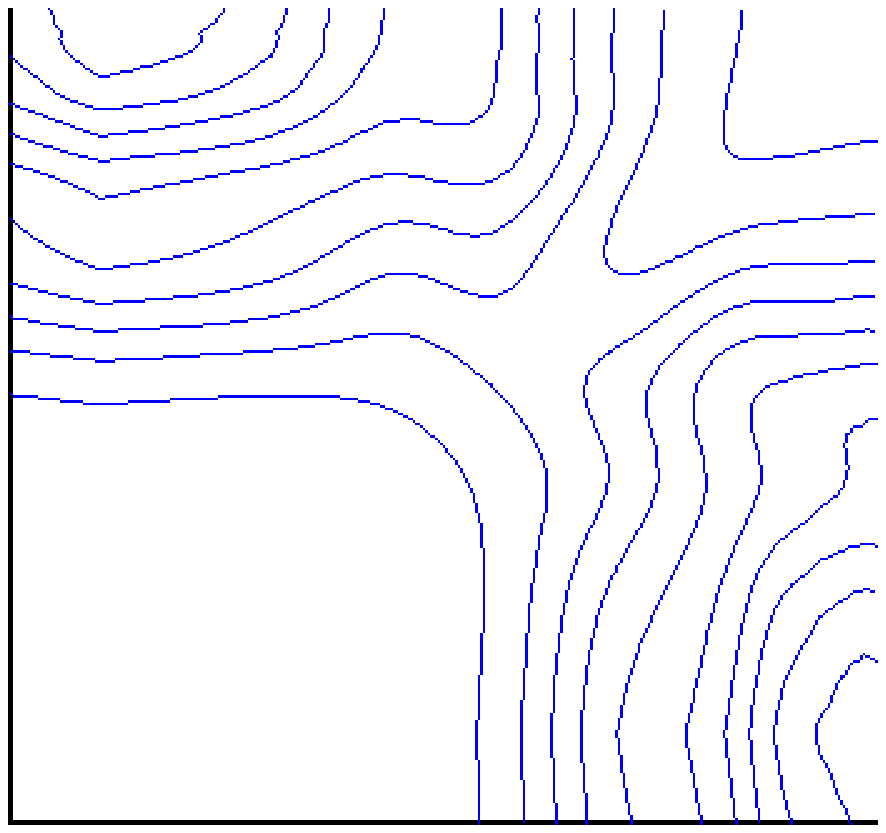}
\vskip 0.20cm
\hskip 1.0cm e) \hskip 3.2cm  f)  \hskip 3.2cm  g)  \hskip 3.2cm  h)
\vskip 0.20cm
\hskip -1.8cm
\includegraphics[clip,width=3.0cm,angle=90]{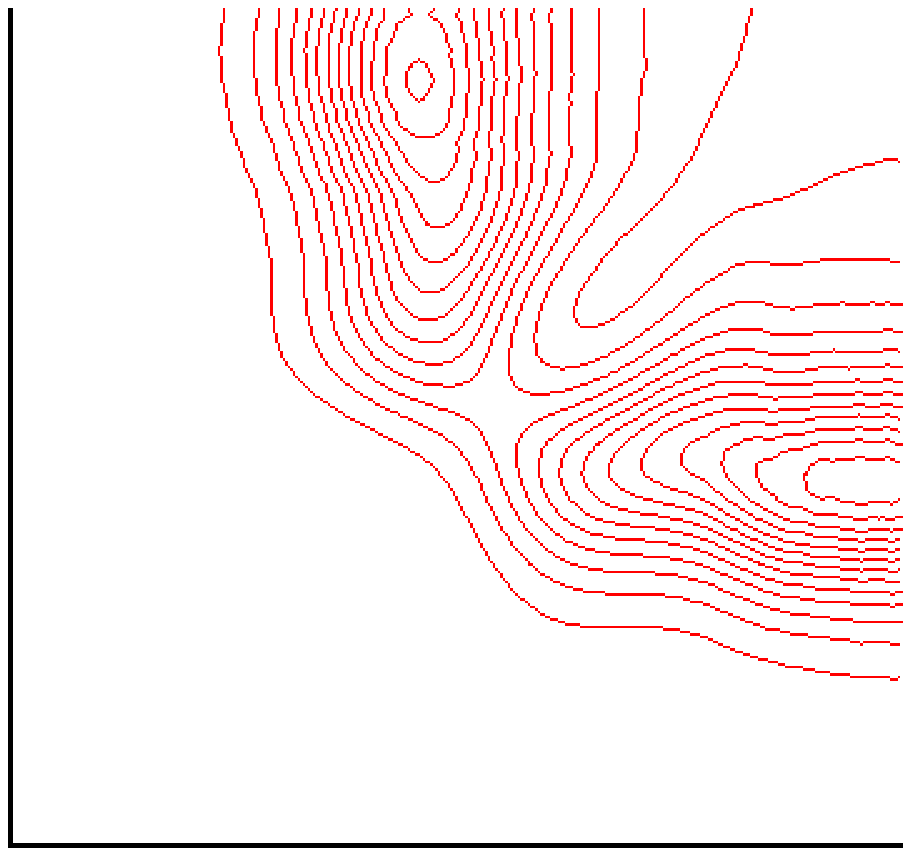}
%\hskip 0.5cm
\includegraphics[clip,width=3.0cm,angle=90]{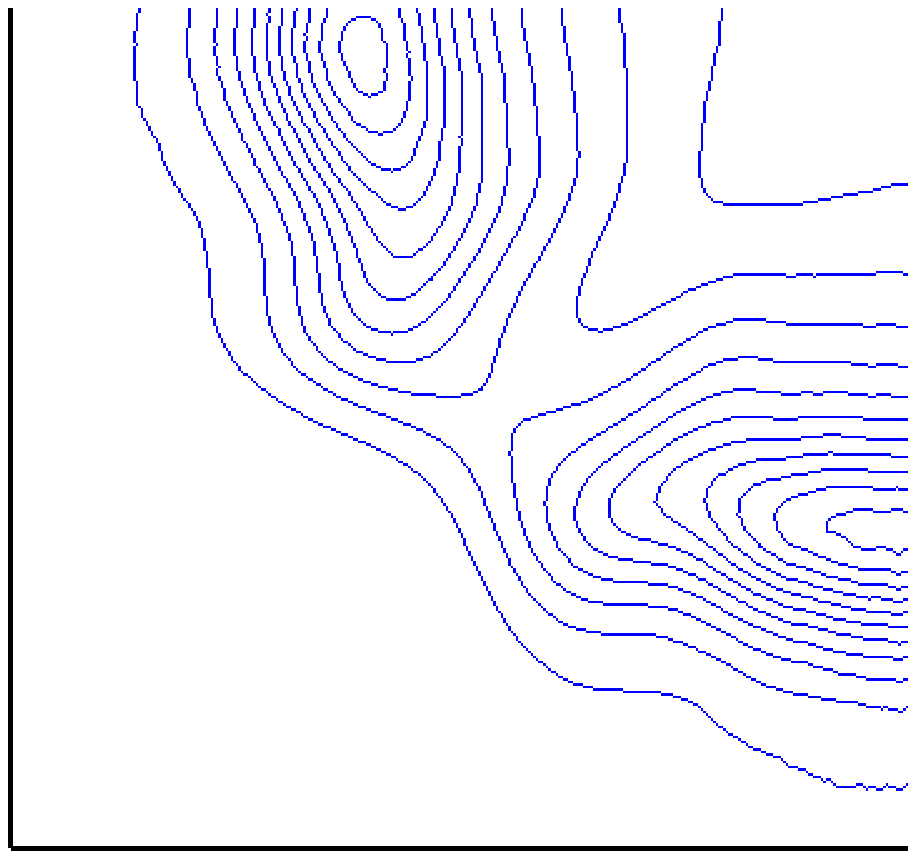}
%\hskip 0.5cm
\includegraphics[clip,width=3.0cm,angle=90]{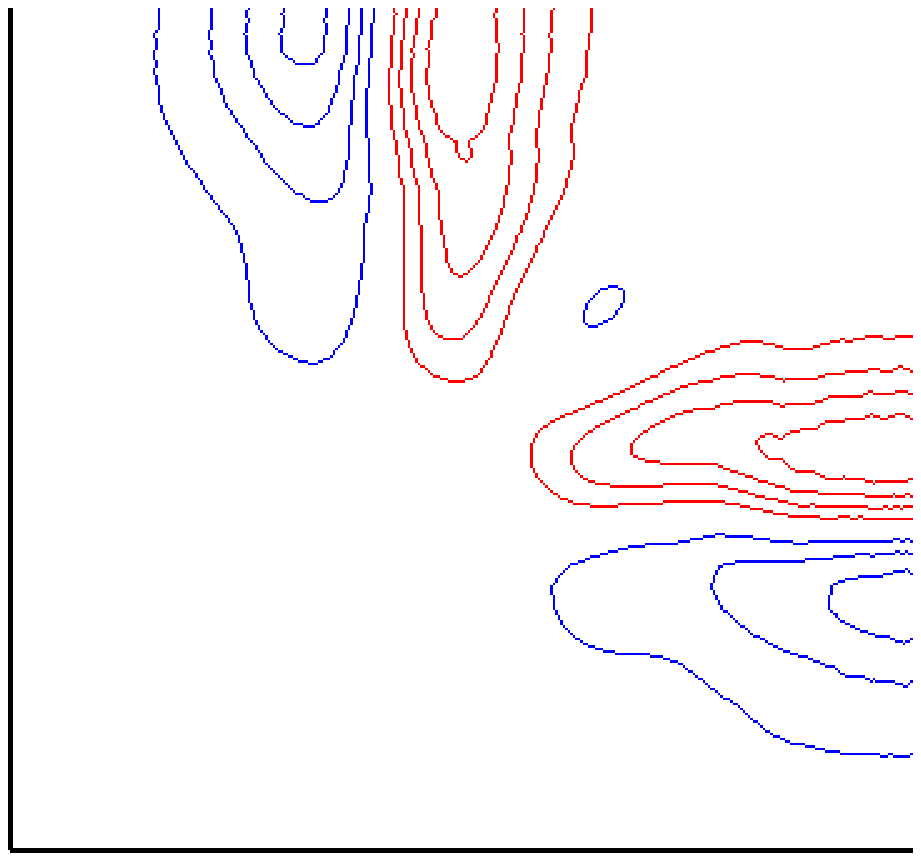}
%\hskip 0.5cm
\includegraphics[clip,width=3.0cm,angle=90]{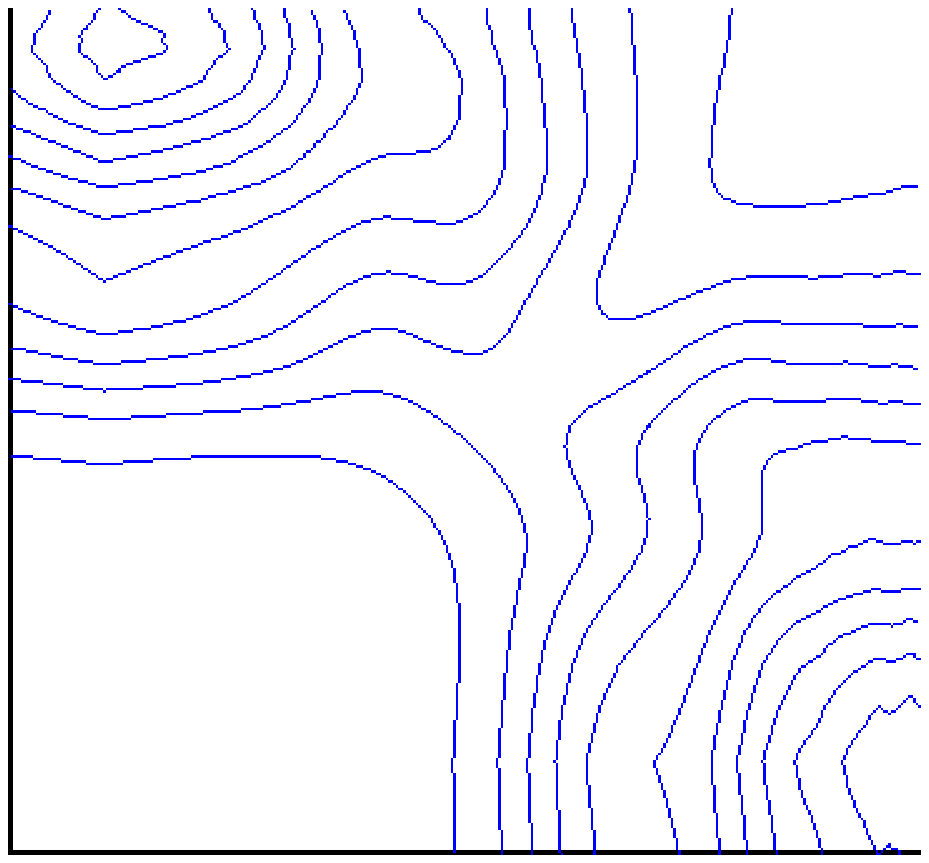}
\vskip 0.20cm
\hskip 1.0cm i) \hskip 3.2cm  j)  \hskip 3.2cm  k)  \hskip 3.2cm  l)
\vskip -0.0cm
\caption{Square duct: contours of the terms in the
turbulent kinetic energy budget (equation~\eqref{eqbudgk}). 
at different $Re$ ((a-d) $Re=2500$,
(e-h) $Re=5000$, (i-l) $Re=7750$.
(a,e,i) production ($P_K^+$);
(b,f,j) total dissipation ($D_K^+$);
(c,g,k) turbulent transfer ($T_K^+$);
(d,h,l) isotropic dissipation ($\epsilon_K^+$).
All quantities are shown in wall units, and 
contours spaced by $\Delta=0.02$.
Logarithmic coordinates are used for the distances from the corner, with 
$d^+ \le 151$ (a-d),
$d^+ \le 284$ (e-h),
$d^+ \le 420$ (i-l).
}
\label{fig9}
\end{figure}

Here, different than usual
the total dissipation $D_K$ is shown 
rather than its decomposition into isotropic dissipation
$\epsilon_K =-\frac{1}{Re}\langle 
(\frac{\partial u_i}{\partial x_k})^2\rangle$,
and viscous diffusion 
$V_K=\frac{1}{Re}\frac{\partial^2 \langle u_iu_i\rangle}{\partial x_k^2}$,
which leads to a single and simpler term to model~\citep{orlandi2018}.
This is corroborated from comparison of
the distributions of $D_K^+$ (figure~\ref{fig9}b,f,j) and 
of $\epsilon_K^+$ (figure~\ref{fig9}d,h,l).
In fact, the former goes 
to zero approaching the walls, whereas the latter has non-zero limit,
being locally balances by viscous diffusion (not shown).
The other interesting output from inspection of
the contours of $P_K^+$ and $D_K^+$ in figure~\ref{fig9} is
that they do not largely differ, having peaks at nearly the same
locations, and distributions with almost the same number of contour levels.
The associated physics then consists of local balance between
production and total dissipation of turbulent kinetic energy.
Hence, it may be inferred that direct
modeling of $D_K$ in RANS turbulence models should be easier than
a separated  closure for $\epsilon_k$ and $V_k$.
The distributions of $T_k^+$ (figure~\ref{fig9}c,g,k),
which accounts for the mean convection (first term in equation~\eqref{eqbudgk}), 
triple velocity correlations (second term), and velocity/pressure gradient correlations (third term)
show that the effect of turbulent transfer is mainly localised in two adjacent positive and negative layers.
Satisfactory universality of the quantities shown in figure~\ref{fig9} 
in wall units is observed, especially at high
Reynolds number. This is further supported by comparison
of the statistics at $Re=7750$ in figure~\ref{fig9},
with those at $Re=1750$, not shown in this paper.

\begin{figure}
\centering
\hskip -1.8cm
\includegraphics[width=6.5cm,angle=0,clip]{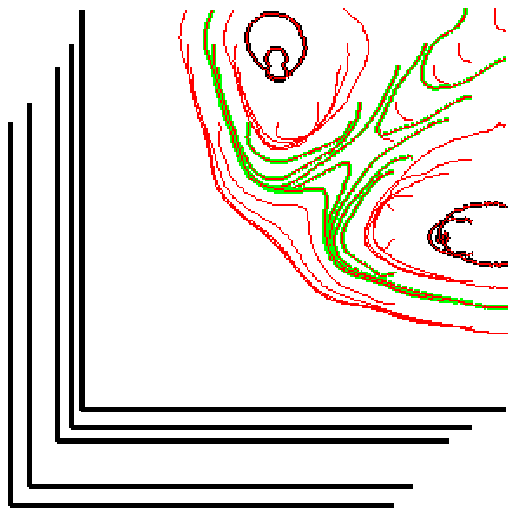}
\includegraphics[width=6.5cm,angle=0,clip]{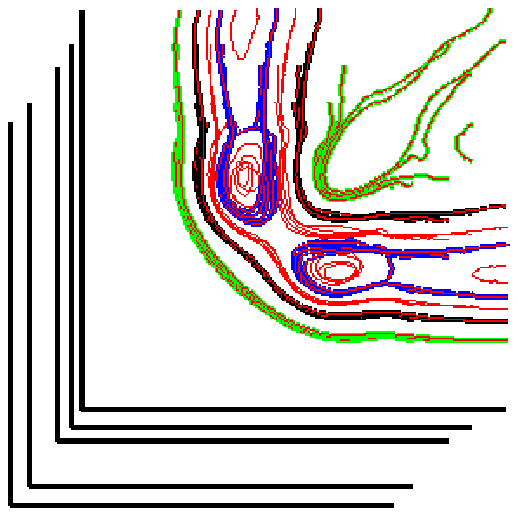}
\vskip -.5cm
\hskip 2.0cm  a)   \hskip 6.0cm  b)
\caption{Square duct: contours of a)
turbulent kinetic energy and b) shear parameter,
in wall units, space coordinates in log scale.
Contours are shown at $Re=1750, 2500, 5000, 7750, 15000$,
superimposed to each other. 
In a) the red contours start from $1$, with increment $\Delta=1$, 
green $q^+=2$, blue $q^+=3$; 
in b) green $S^*=5$, black $S^*=10$, blue $S^*=20$,
and the red contours start from $S^*=5$, with increment $\Delta=5$.
}
\label{fig10}
\end{figure}

To emphasise the Reynolds number dependence of $K^+$,
in figure~\ref{fig10}a the we show iso-contours separated by $\Delta=1$ in red, 
with superposed the $K^+=2$ contour (green) and $K^+=4$ contour (black).
Approximate Reynolds number independence is found, 
both in the peak value and its location.
We point out that in this figure the black straight lines mark the 
the first grid point off the walls, 
at $d^+=0.15$ at $Re=1750$, and at $d^+=0.5$ at $Re=15000$,
which confirms that DNS is fully resolved
near the walls. The maximum of $K^+$ occurs near the location of
maximum turbulent  kinetic energy production, as may
be inferred by comparing figure~\ref{fig10}a with
the distribution of $P_k^+$ in figure~\ref{fig9}.
To have information about the formation
of near-wall structures,
the contours of the shear parameter, $S^*=q^2S/\epsilon_k$ (with $q^2=2K$,
$S=\sqrt{S_{ij}S_{ij}}$) are reported in figure~\ref{fig3}b.
Turbulent structures may form for $S^* \gtrsim 10$~\citep{lee_90},
in the region bounded by the blue line,
which we find not to depend on the Reynolds number.

\subsubsection{Flow visualizations}

\begin{figure}
\centering
\vskip -0.0cm
\hskip -1.8cm
\includegraphics[clip,width=5.0cm]{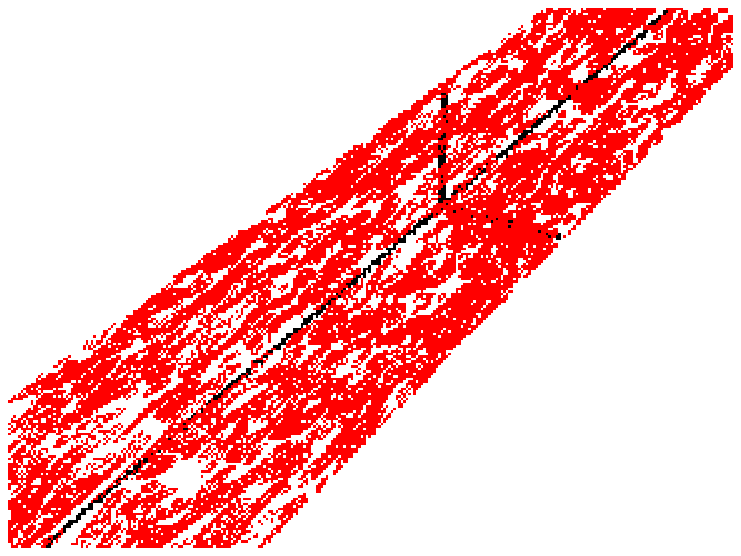}
%\hskip 0.75cm
\includegraphics[clip,width=5.0cm]{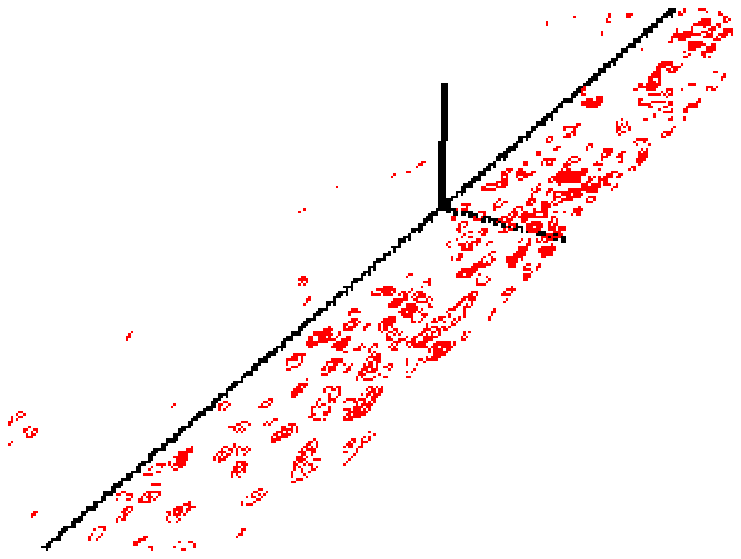}
%\hskip 0.75cm
\includegraphics[clip,width=5.0cm]{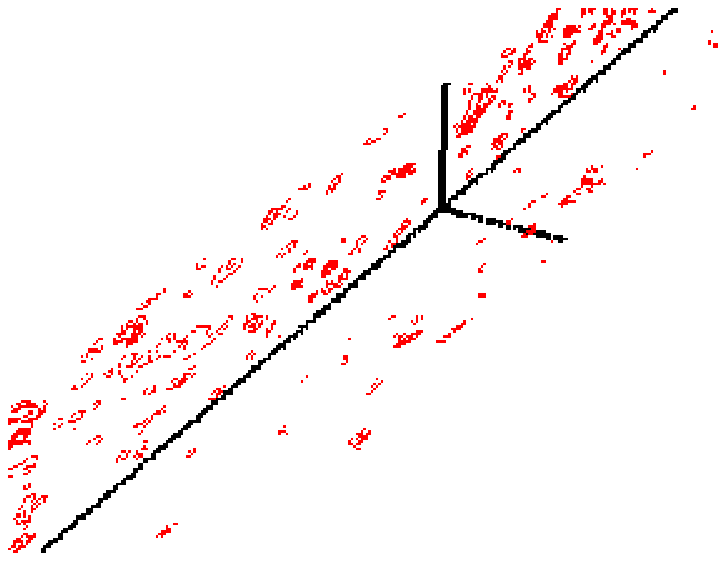}
\vskip -0.0cm
\hskip 1.0cm  a) \hskip 5.0cm  b) \hskip 5.0cm  c)
\vskip -0.0cm
\hskip -1.8cm
\includegraphics[clip,width=5.0cm]{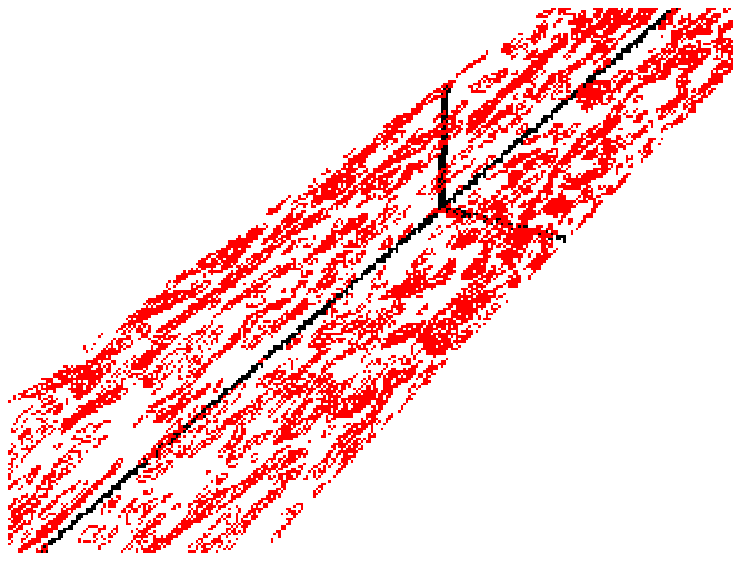}
%\hskip 0.75cm
\includegraphics[clip,width=5.0cm]{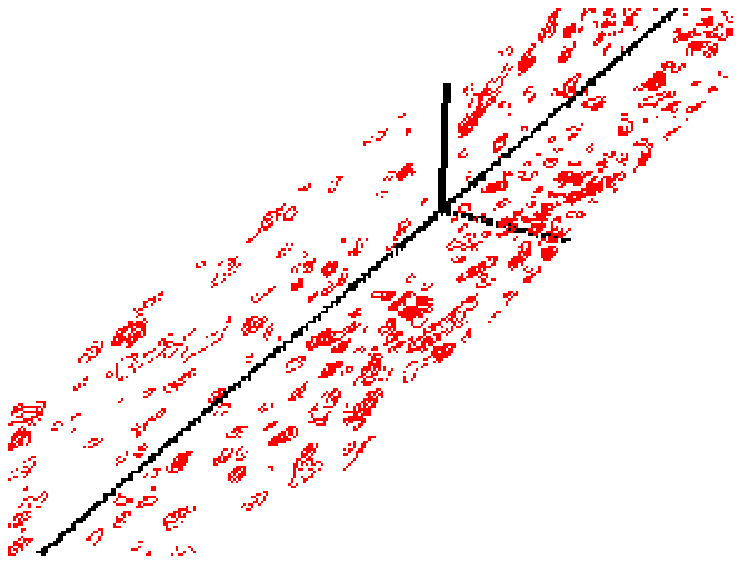}
%\hskip 0.75cm
\includegraphics[clip,width=5.0cm]{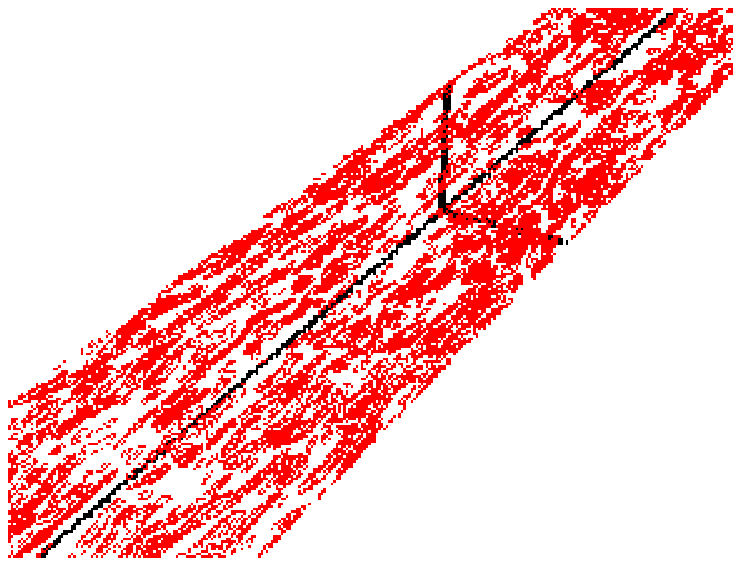}
\vskip -0.0cm
\hskip 1.0cm  d) \hskip 5.0cm  e) \hskip 5.0cm  f)
\vskip -0.0cm
\caption{Square duct: instantaneous visualizations of turbulent normal stresses
in one quarter of the domain at $Re=7750$,
in wall-parallel planes at a distance $d^+\approx 12.5$.
Panels (a-c), stresses in Cartesian basis 
(a) $R_{11}$, b)  $R_{22}$, c)  $R_{33}$);
panels (d-e), stresses in strain eigenvector basis
(d)  $R_{\alpha \alpha}$, e)  $R_{\beta \beta}$, f)  $R_{\gamma \gamma}$).
The contour lines start from $1$, with increment $\Delta=1$.
}
\label{fig11}
\end{figure}

The coherent structures of wall turbulence can be visualised through
several quantities, the most widely used being the fluctuating streamwise
velocity and wall-normal vorticity.
The distributions of $R_{ii}^+$ obtained
by averaging in time and in the streamwise directions were
discussed in figure~\ref{fig7}, from which
the signature of the different types of coherent
structures could be qualitatively deduced. 
Fuller characterization of the coherent structures  
may be obtained through contours of the quantity $r_{ii}=u''_i u''_i(x_1,x_2,x_3)$
in wall-parallel planes.
Here $u''_i=u_i-\tU_i(x_2,x_3)$ indicates velocity fluctuations 
with respect to the instantaneous streamwise-average value, 
denoted with the tilde.
The eigenvalues and the eigenvectors of $\tS_{ij}(x_2,x_3)$
allow to project the velocity fluctuations
to get $r_{\lambda\lambda}=(u''_\lambda u''_\lambda)$.
Comparison between $r_{ii}^+$ and $r_{\lambda\lambda}^+$,
in two planes at distances $d^+=12$ from the walls, are shown in 
figure~\ref{fig11}. The three top panels highlight that 
streamwise elongated structures are only visible in the
the contours of $r_{11}$.  On the other hand,
the other two stresses are localised in short
circular patches, implying the occurrence of intermittent
bursts emanating from the walls,
preferentially located in regions with small $r_{11}$.
In addition, one may appreciate that the distribution of 
$r_{33}$ in the $x_1-x_2$ plane is similar to that of 
$r_{22}$ in the $x_1-x_3$ plane. 
In summary, inspection of the $r_{ii}^+$ contours in the top panels of figure~\ref{fig11}
shows strong anisotropy, which could not be inferred from the
the distribution of $R{ii}$ shown in figure~\ref{fig7}.
The observed complex flow structures dynamics is driving 
complex physics, leading to difficulties in modeling turbulent flows in ducts.
The previously discussed simpler features of turbulent stresses in the 
strain eigenvector basis, and in particular the symmetric distribution of the stresses
around the corner bisector and the increase of isotropy may
instead lead to a simpler dynamics to model.
The larger number of contours in
figure~\ref{fig11}f than those in figure~\ref{fig11}d
is due to the stronger effects of the compressive
than those of the extensional stress. Despite these small 
differences in magnitude, similar  flow
structures are found, which may lead to 
the possibility to improve turbulence closures. The
$r_{\beta\beta}$ structures, localised
in patches of small magnitude are visible
in figure~\ref{fig11}e, which
produce smooth and less intense
distribution of $R_{\beta\beta}$ in figure~\ref{fig7}f.
The contours in the three bottom figures
of figure~\ref{fig11} show good coincidence
of $r_{\alpha\alpha}$ and $r_{\gamma\gamma}$.
On the other hand, $r_{\beta\beta}$ is mainly localised in regions 
with small $r_{\gamma\gamma}$.

\begin{figure}
\centering
\vskip -0.0cm
\hskip -1.8cm
\includegraphics[clip,width=5.0cm]{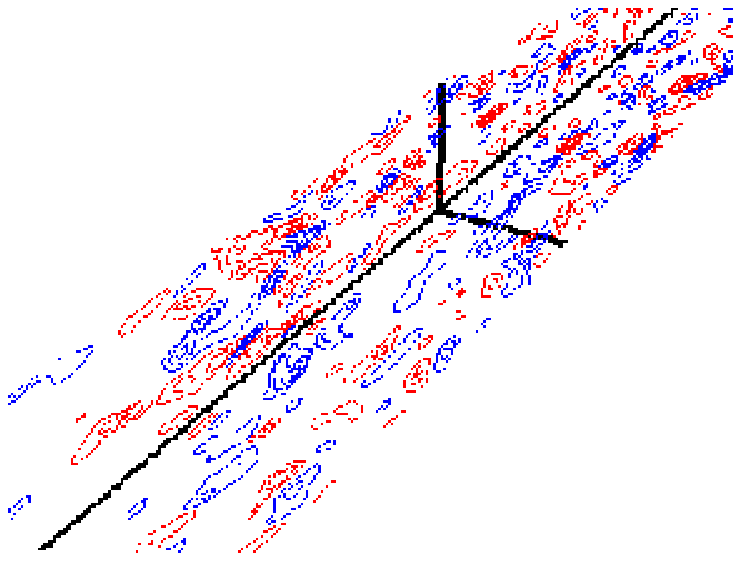}
%\hskip 0.75cm
\includegraphics[clip,width=5.0cm]{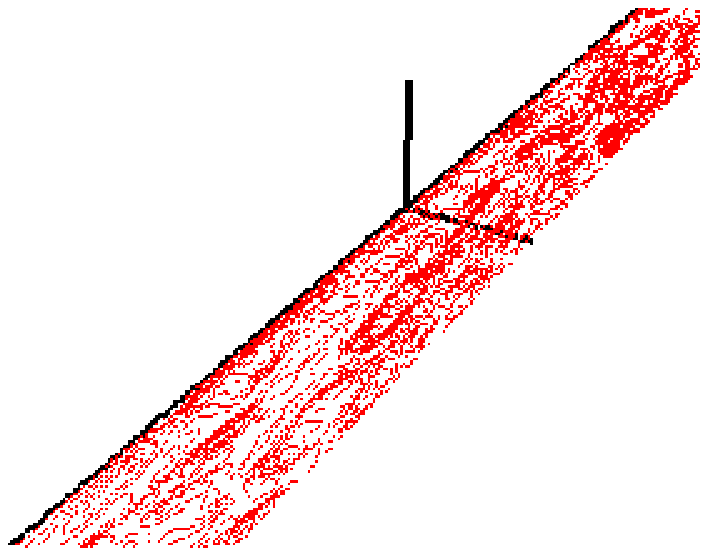}
%\hskip 0.75cm
\includegraphics[clip,width=5.0cm]{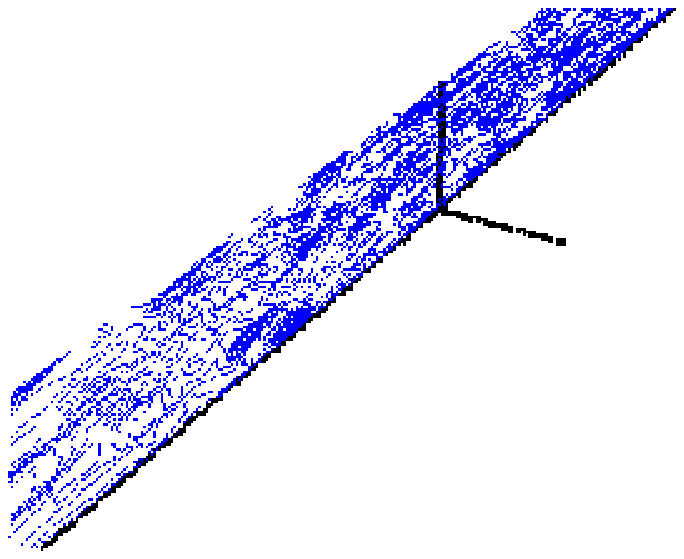}
\vskip -0.0cm
\hskip 1.0cm  a) \hskip 5.0cm  b) \hskip 5.0cm  c)
\vskip -0.0cm
\hskip -1.8cm
\includegraphics[clip,width=5.0cm]{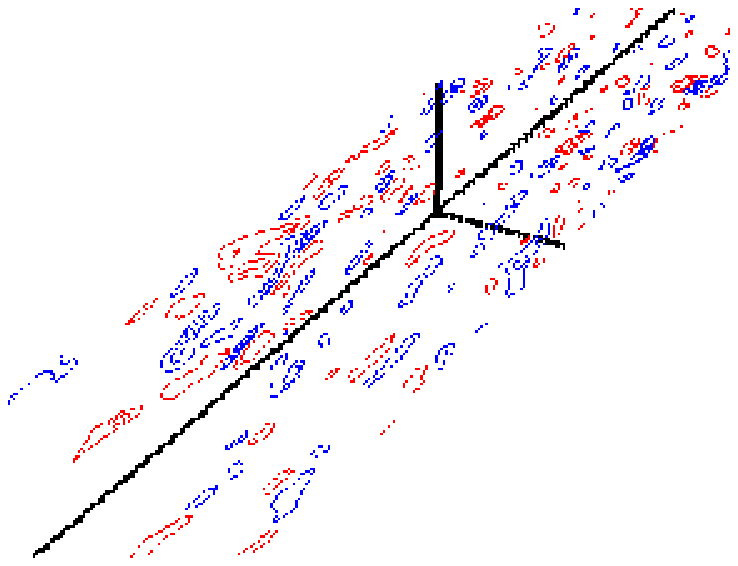}
%\hskip 0.75cm
\includegraphics[clip,width=5.0cm]{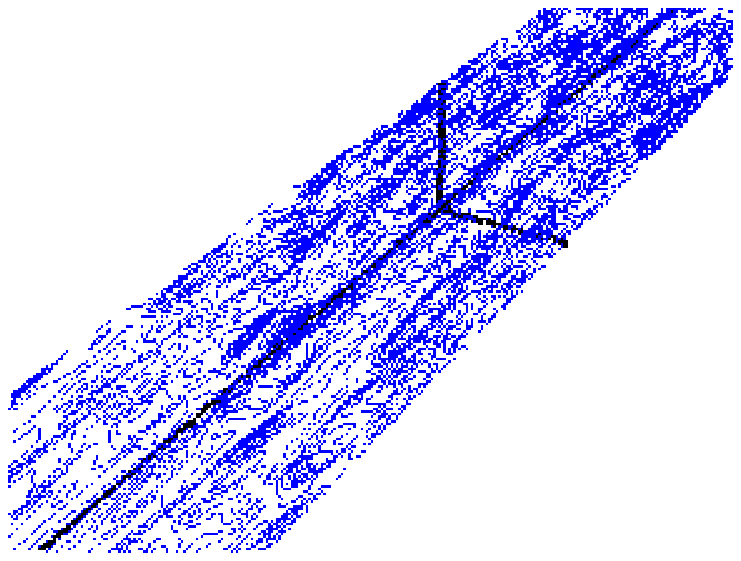}
%\hskip 0.75cm
\includegraphics[clip,width=5.0cm]{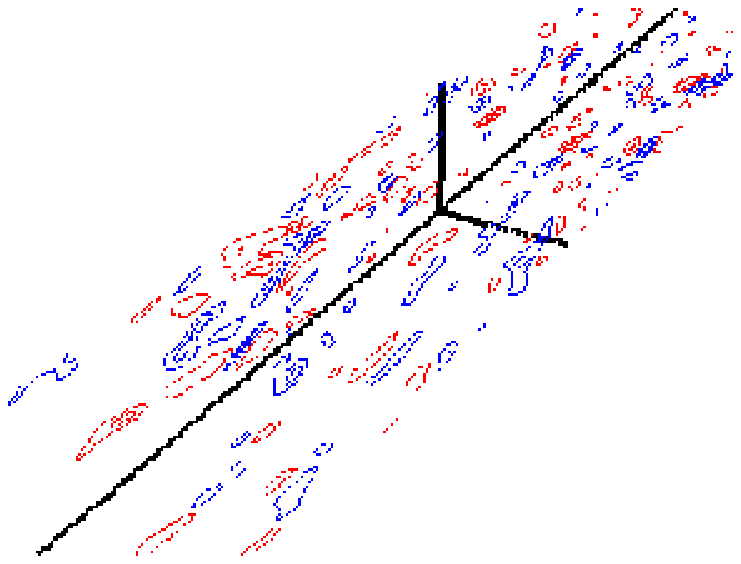}
\vskip -0.0cm
\hskip 1.0cm  d) \hskip 5.0cm  e) \hskip 5.0cm  f)
\vskip -0.0cm
\caption{Square duct: instantaneous visualizations of vorticity components 
in one quarter of the domain at $Re=5000$,
in wall-parallel planes at a distance $d^+\approx 2$.
Panels (a-c), in Cartesian basis,
a)  $\omega_{1}$, b)  $\omega_{2}$, c)  $\omega_{3}$;
panels (d-e), in strain eigenvector basis,
d)  $\omega_{\alpha}$, e)  $\omega_{\beta}$, f)  $\omega_{\gamma}$.
The contour lines start from $\pm 2$ with increments $\Delta=\pm 2$, 
red positive and blue negative.
}
\label{fig12}
\end{figure}

The projection of the flow variables in the 
strain eigenvector basis allows to better understand where
friction is localised. This is obtained through
$\omega_3|$, which is proportional to the wall shear stress
$\tau_w=\frac{1}{Re}\der{u_1}{x_2}$ at the $x_1-x_3$ wall,
and $\omega_2|$, which is proportional to the wall shear stress
$\tau_w=\frac{1}{Re}\der{u_1}{x_3}$ at the $x_1-x_2$ wall.
Hence, visualizations 
of the vorticity components may provide an idea about
shape and length of the near-wall structures. 
In planar channels,
\cite{orlandi2018} observed the prevalence of ribbon-like
with respect to rod-like structures up
to a distance $d^+\approx 2$,
and which account for the spatial distribution of the
wall shear stress. Hence,
in figure~\ref{fig12} we present visualizations 
in wall-parallel planes (at a distance $d^+\approx 2$)
of the vorticity components in the Cartesian basis (top panels),
and in the strain eigenvector basis (bottom panels).
Wide regions without vorticity  appear in visualizations at $Re=5000$ 
producing a picture with 
structures more visible than those at higher Reynolds numbers.
The top panels show that the $\omega_2$ and $\omega_3$
are localised in very long structures, similar to those in
figure~\ref{fig11}a. $\omega_1$ is distributed similarly 
as the other components in figure \ref{fig11}b and figure \ref{fig11}c,
providing evidence for the high intermittency of 
$\omega_1$ and $u_2$. Vorticity is nearly aligned with 
the intermediate $S_\beta$, as  depicted
by the $\omega_\beta$ distributions in figure~\ref{fig12}e,
with greater amplitude than $\omega_\alpha$ and $\omega_\gamma$.
The entire wall friction is proportional to
$\omega_\beta$. It has been also verified
that the sum of the negative and positive patches in
figure~\ref{fig12}d and \ref{fig12}f is zero.

\subsubsection{Budgets of mean momentum equations  }

To understand in greater detail the advantages of the 
strain eigenvector basis, it is worth analysing the distribution 
of the mean velocity components, and the
contribution of each term to their transport equations. At 
steady state the momentum equations are
 
\begin{figure}
\centering
\vskip -0.0cm
\hskip -1.8cm
\includegraphics[clip,width=5.0cm,angle=90]{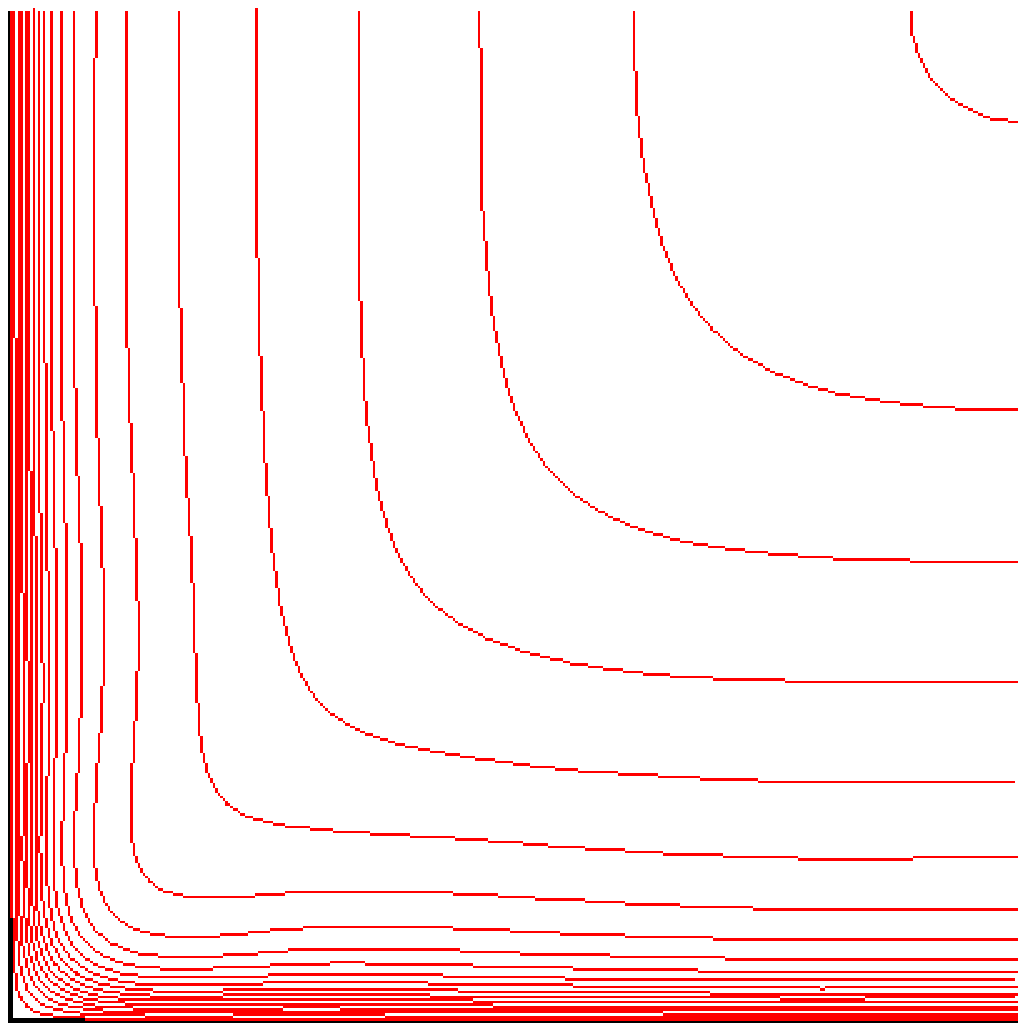}
%\hskip 0.75cm
\includegraphics[clip,width=5.0cm,angle=90]{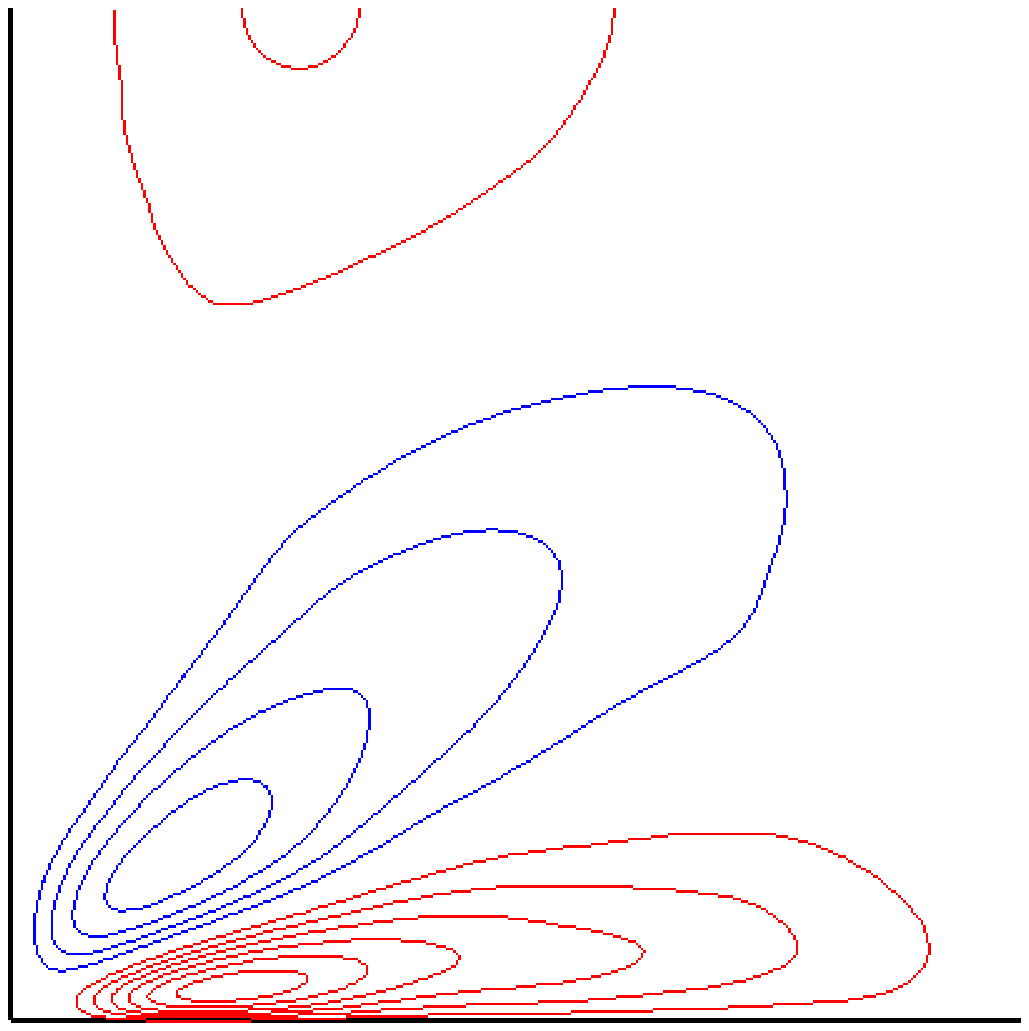}
%\hskip 0.75cm
\includegraphics[clip,width=5.0cm,angle=90]{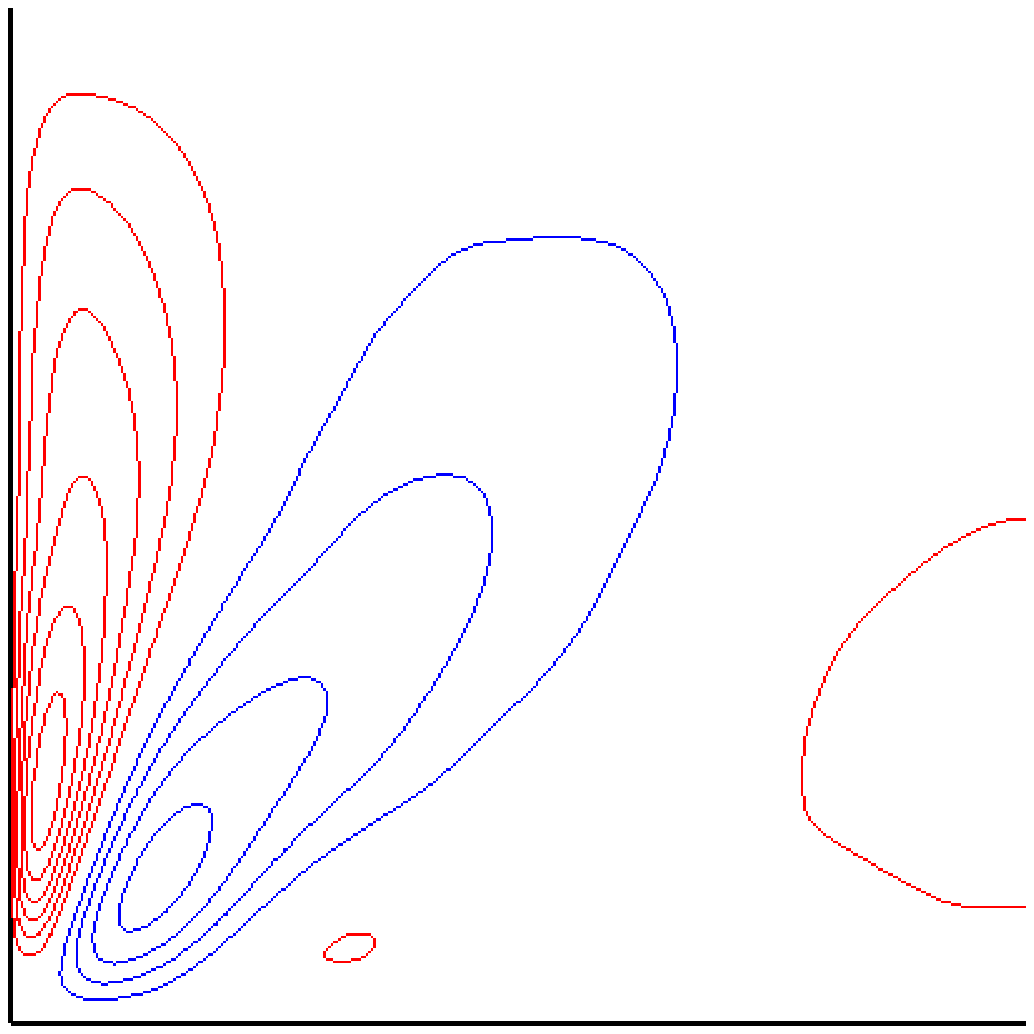}
\vskip -0.0cm
\hskip 1.0cm  a) \hskip 4.0cm  b) \hskip 4.0cm  c)
\vskip 0.0cm
\hskip -1.8cm
\includegraphics[clip,width=3.0cm,angle=90]{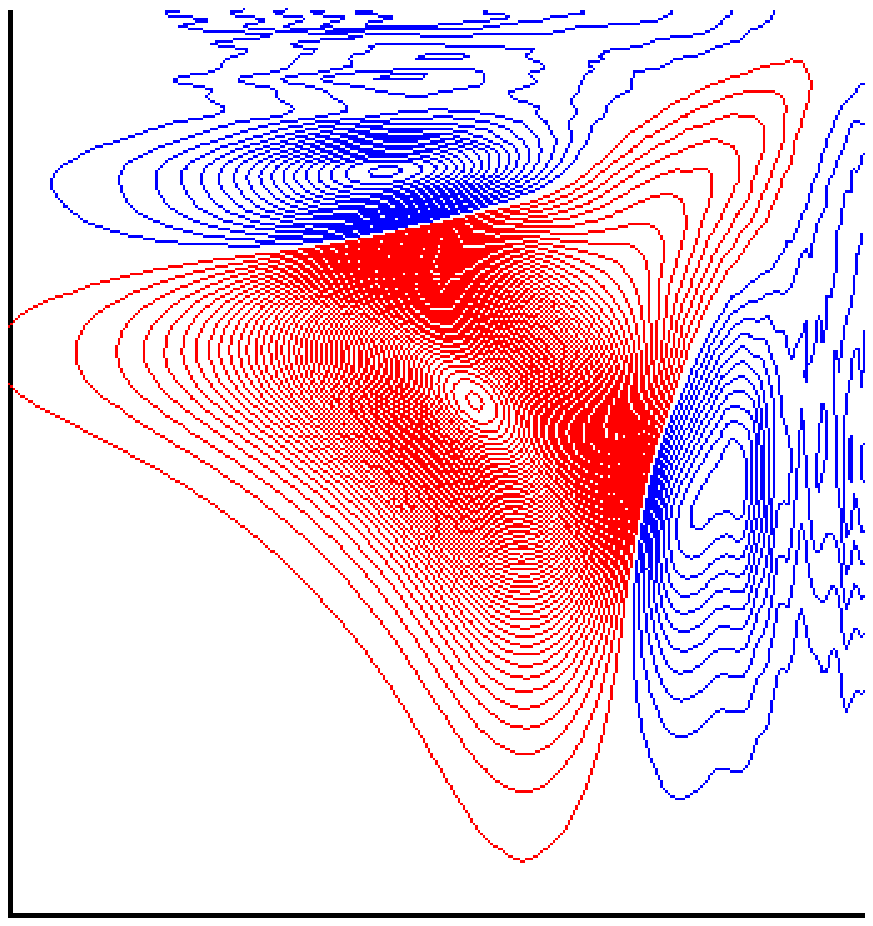}
%\hskip 0.75cm
\includegraphics[clip,width=3.0cm,angle=90]{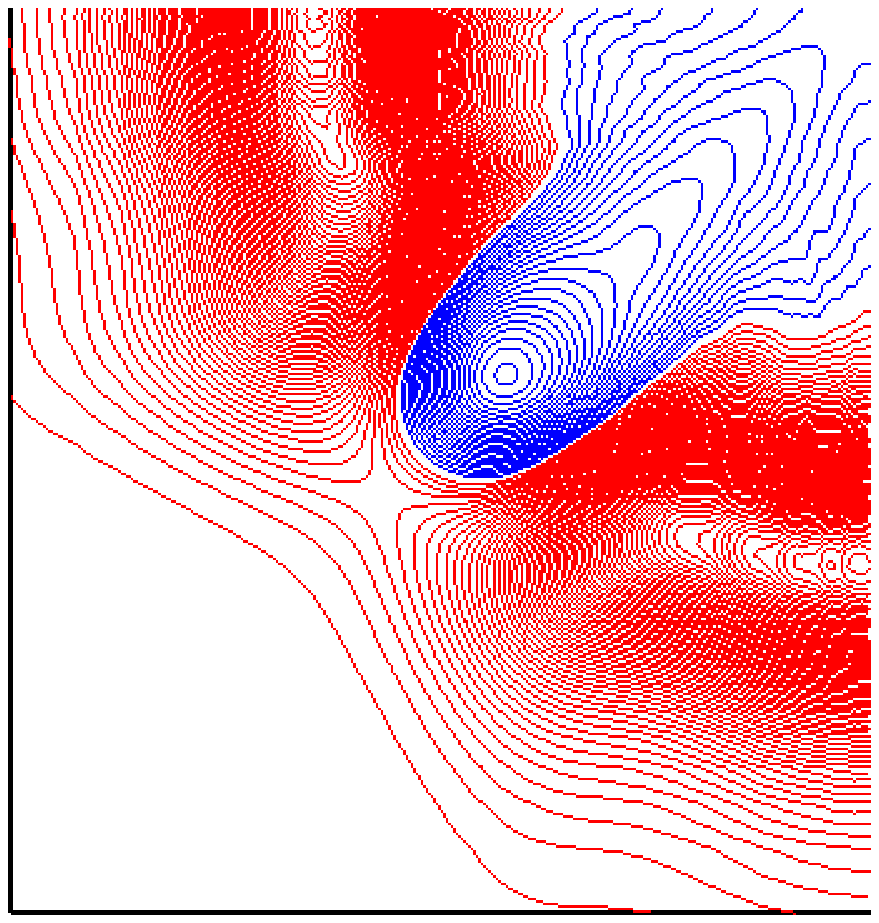}
%\hskip 0.75cm
\includegraphics[clip,width=3.0cm,angle=90]{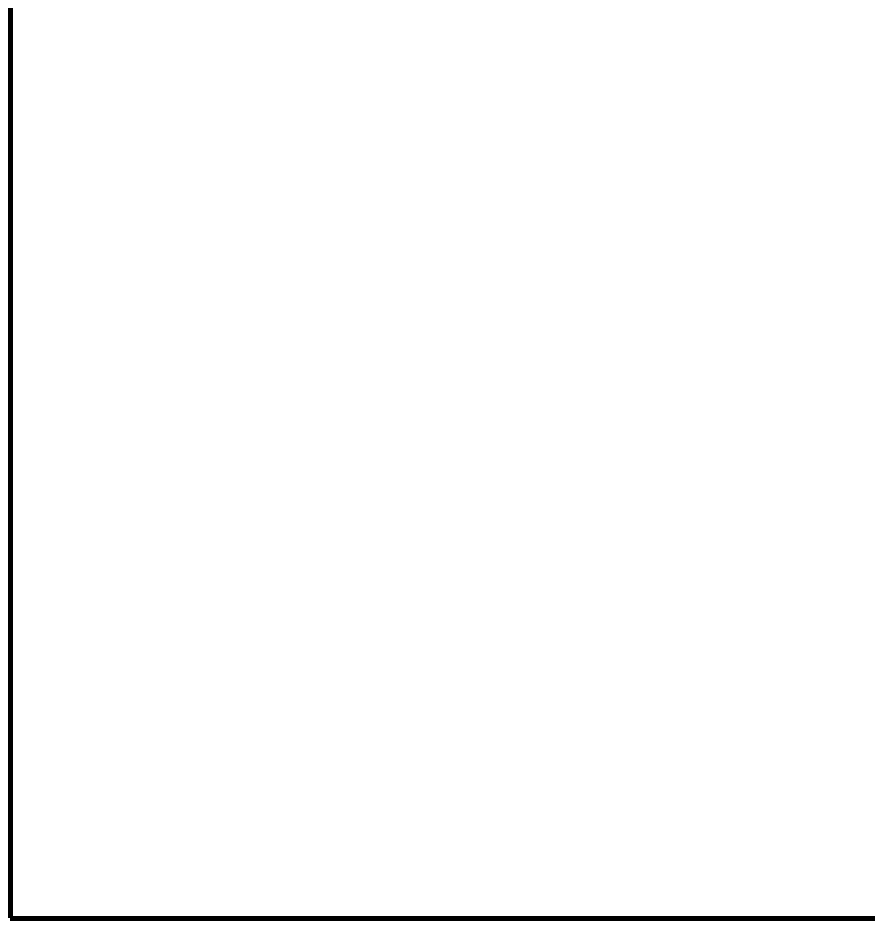}
%\hskip 0.75cm
\includegraphics[clip,width=3.0cm,angle=90]{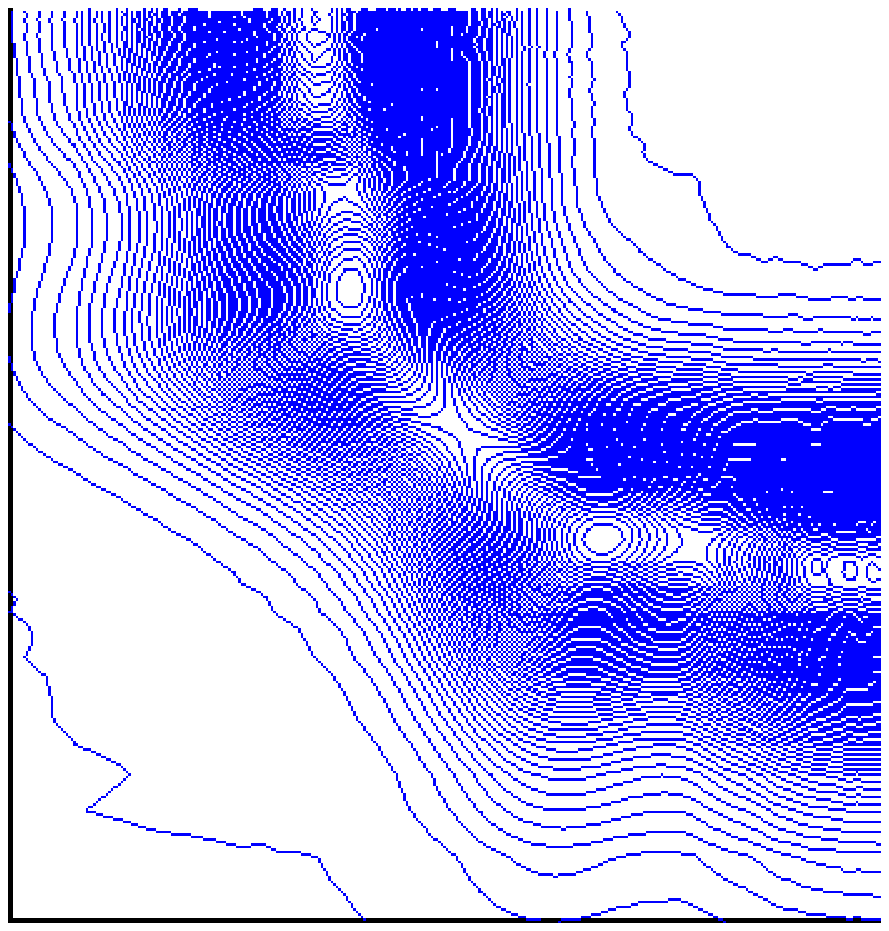}
\vskip -0.0cm
\hskip 1.0cm  d) \hskip 3.0cm  e) \hskip 3.0cm  f) \hskip 3.0cm  g)
\vskip 0.0cm
\hskip -1.8cm
\includegraphics[clip,width=3.0cm,angle=90]{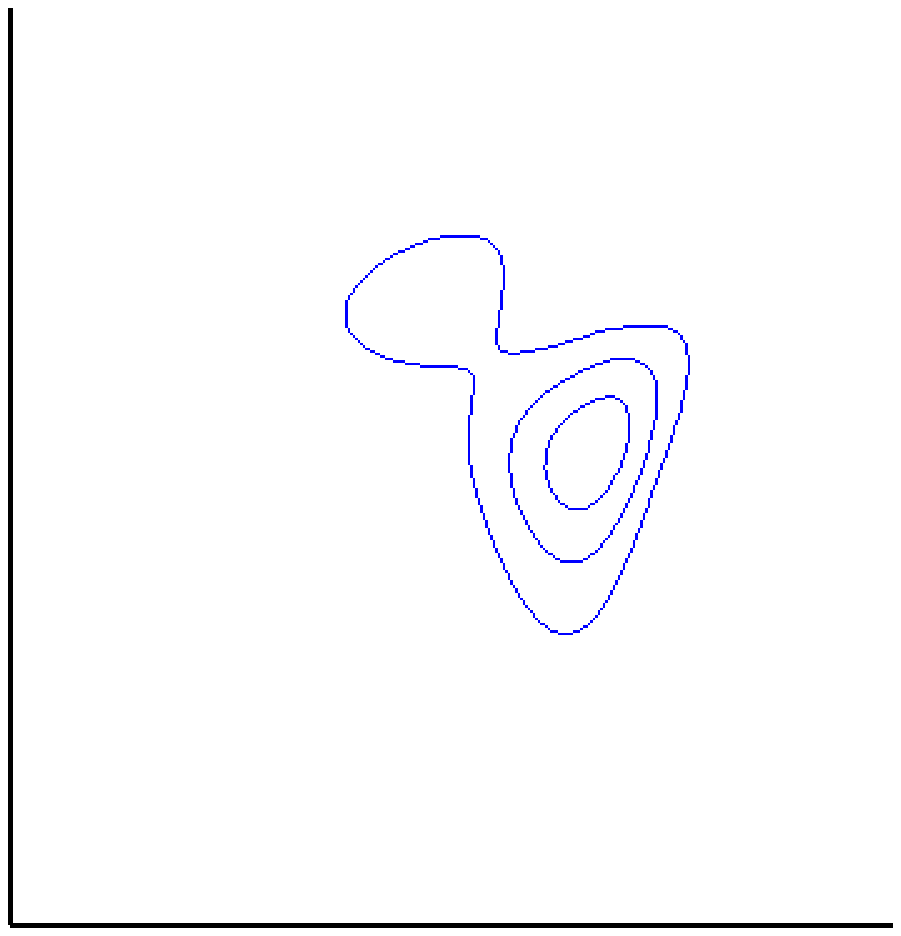}
%\hskip 0.75cm
\includegraphics[clip,width=3.0cm,angle=90]{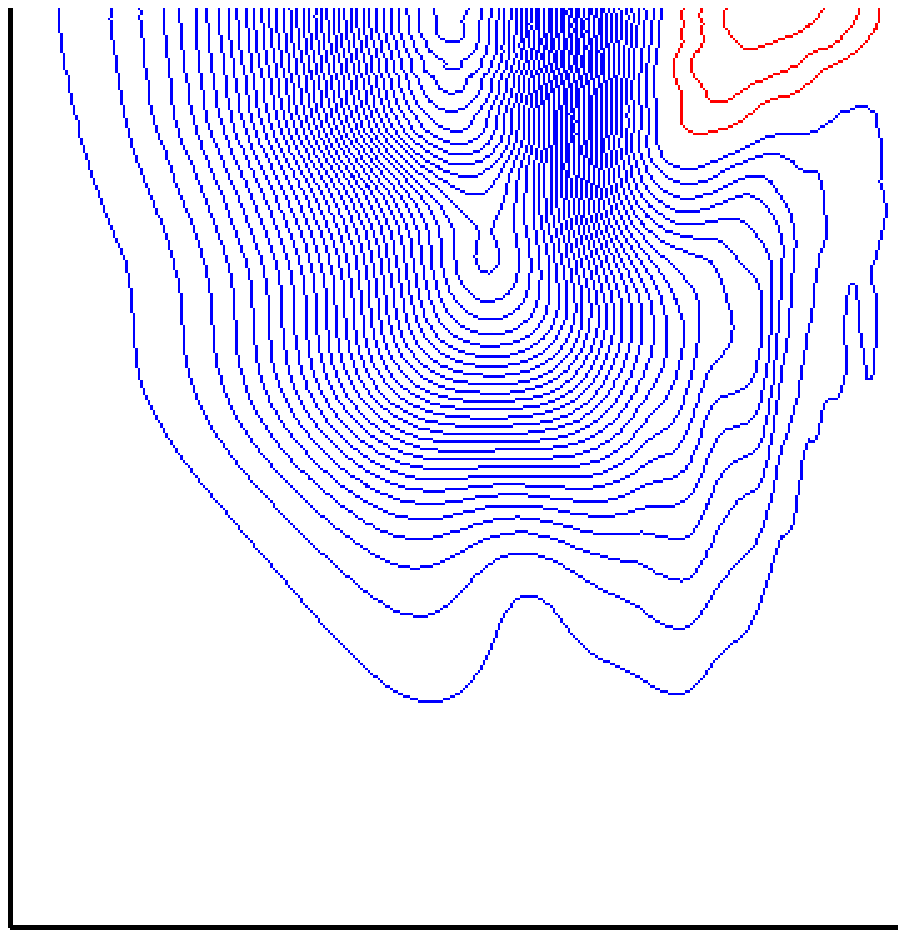}
%\hskip 0.75cm
\includegraphics[clip,width=3.0cm,angle=90]{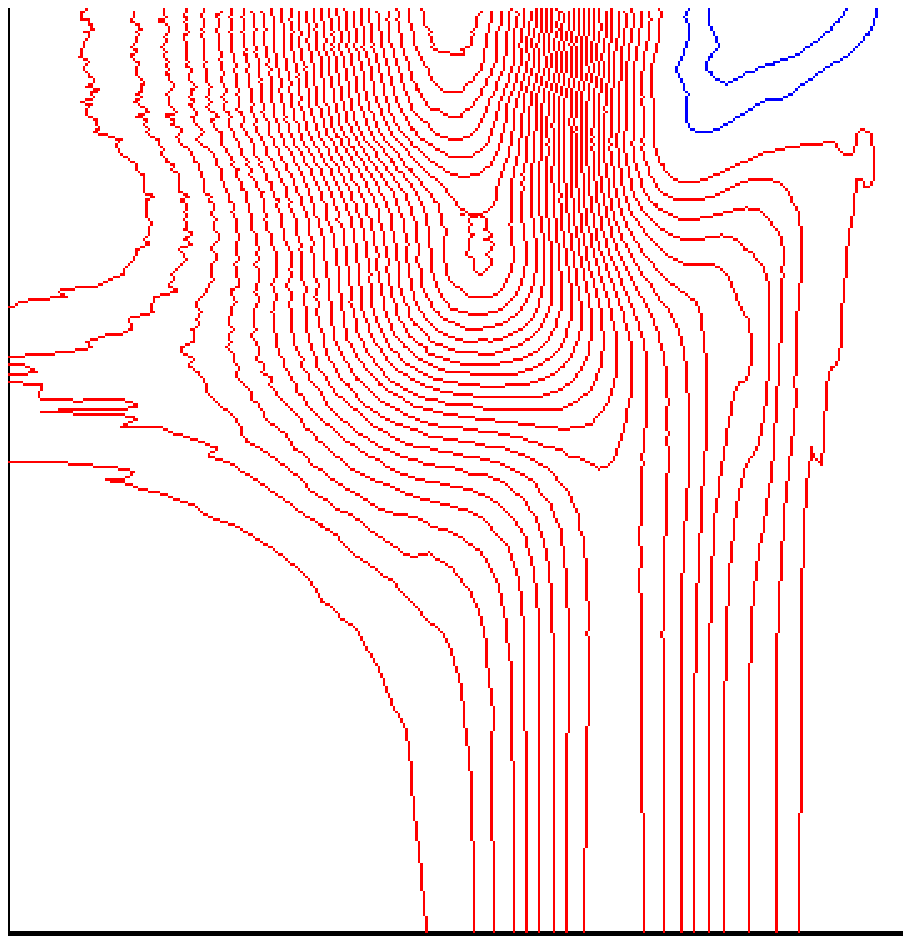}
%\hskip 0.75cm
\includegraphics[clip,width=3.0cm,angle=90]{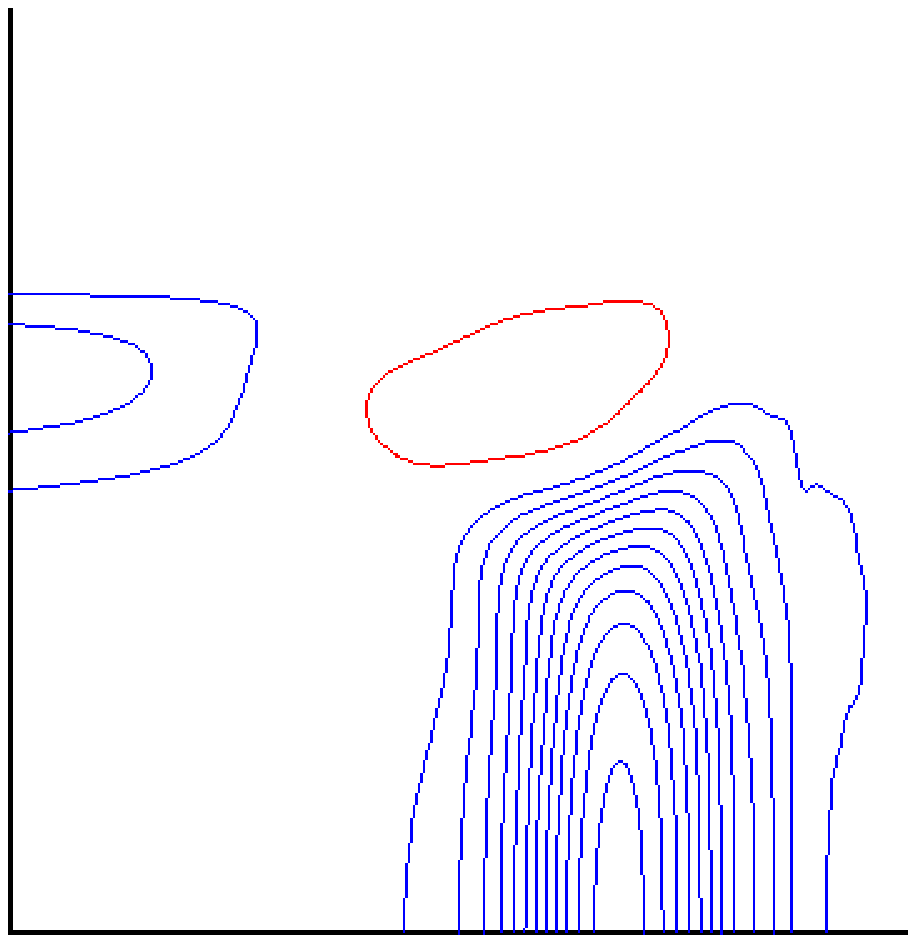}
\vskip -0.0cm
\hskip 1.0cm  h) \hskip 3.0cm  i) \hskip 3.0cm  j) \hskip 3.0cm  k)
\caption{Square duct: contours of mean velocity components at
$Re=5000$, with positive values in red and negative values in blue.
a) $U_1^+$, in intervals $\Delta U^+=1$; b) $U_2^+$; c) $U_3^+$,
both in intervals $\Delta U^+=.05$.
In panels a), b), c) linear coordinates are used  from $0$ to $280$.
In the lower panels we show the contributions to the budget equations~\eqref{eqUk}
for $U_1$ (d-g), and $U_3$ (h-k), namely
$C_{i}^+$ (d,h); $T_{i}^+$ (e,i), $P_{i}^+$ (f,j), $L_{i}^+$ (g,k).
Inner-scaled coordinates are shown in logarithmic scale, 
positive contours are shown in red, negative in blue, in increments $\Delta=0.001$.
}
\label{fig13}
\end{figure}

\begin{equation}
\underbrace{
\frac{\partial   U_iU_k}{\partial x_k} 
}_{C_i}
+
\underbrace{
\frac{\partial  \langle u_iu_k\rangle}{\partial x_k} 
}_{T_i}
+
\underbrace{
\frac{\partial P}{\partial x_i}
}_{P_i}
\underbrace{
- 
\frac{1}{Re} \nabla^2 U_i 
}_{L_i}
=\Pi \delta_{1i} ,
\label{eqUk}
\end{equation}
\noindent 
where $C_i$ denotes the contribution from mean cross-stream flow,
$T_i$ the turbulent stresses, $P_i$ the mean pressure gradient,
$L_i$ the viscous contribution, and $\Pi \delta_{1i}$
acts in the streamwise momentum equation, and stands for
the external force necessary to balance the frictional resistance.
Contours of the velocity components $U_1^+$
and $U_3^+$, and the contribution of each term in their  transport
equations are shown in figure~\ref{fig13}. 
Figures~\ref{fig13}b and \ref{fig13}c 
confirm that the distributions of 
$U_2$ and the $U_3$ are similar by changing the coordinate $x_3$ with $x_2$.
The contributions of each term in the respective balance equations 
are plotted in one duct quadrant
in inner-scaled logarithmic coordinates,
hence they look different than in \citet{pirozzoli_18}. 
In that paper the terms in the streamwise velocity budget were
presented, showing good independence on the
Reynolds number. Compared to that result,
the present plots at $Re=5000$  emphasise 
the large differences at intermediate $Re$ between the terms in the $U_1$
equation and those of the secondary motion, described by the $U_2$ 
and $U_3$ equations. 
To account for the high levels of $U_1^+$, the contours are shown
with increments $\Delta U^+_1=1$,
whereas for the cross-stream velocity components we use
$\Delta U^+_i=0.05$ ($i=2,3$). 
The few contours for $U_2^+$
and $U_3^+$ thus clearly highlight weakness of the secondary motion 
with respect to the mean streamwise motion. The
convective $C_1^+$ (figure~\ref{fig13}d) and the turbulent $T_1^+$ 
(figure~\ref{fig13}e) contributions have close magnitude, 
but act in different regions.
The global effect of $C_1^+$ and $T_1^+$ is positive,
being $P_1^+=0$ (figure~\ref{fig13}f), 
and the sum of the two is balanced by $L_1^+$ (figure~\ref{fig13}g).
A different behavior is obtained for the terms in the
$U_3$ equation.  
$C_3^+$ (figure~\ref{fig13}h) is very small, 
$T_3^+$ (figure~\ref{fig13}i) is large, 
and balanced by $ P_3^+$. The viscous term $L_3^+$ (figure~\ref{fig13}k) 
is largely negative, and it
balances the amount of positive contribution due 
to $P_3^++T_3^+$. Since the transport equations for $U_2$ and $U_3$ 
should also be solved in a RANS closure, it may be concluded
that it is rather difficult to model the various terms by 
$T_1^+$ and $T_3^+$.

\begin{figure}
\centering
\vskip -0.0cm
\hskip -1.8cm
\includegraphics[clip,width=5.0cm,angle=90]{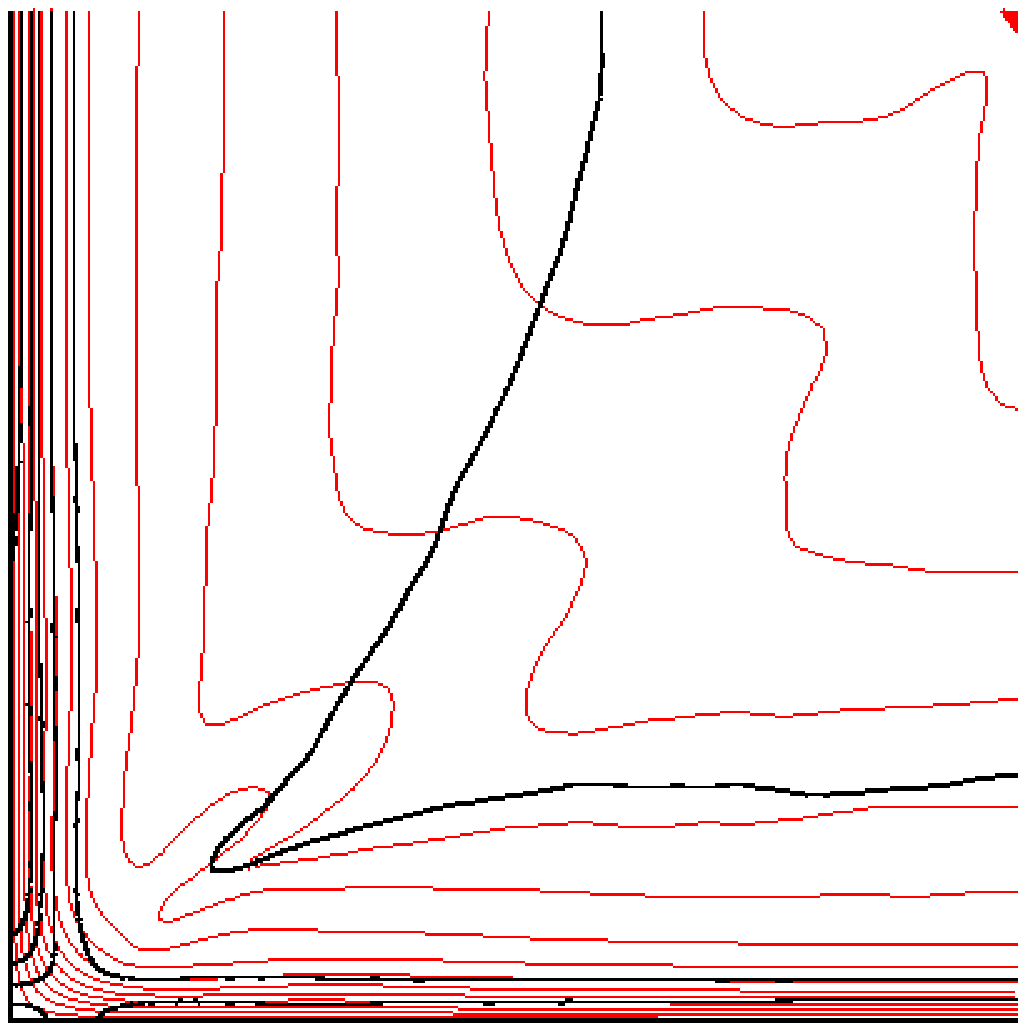}
%\hskip 0.75cm
\includegraphics[clip,width=5.0cm,angle=90]{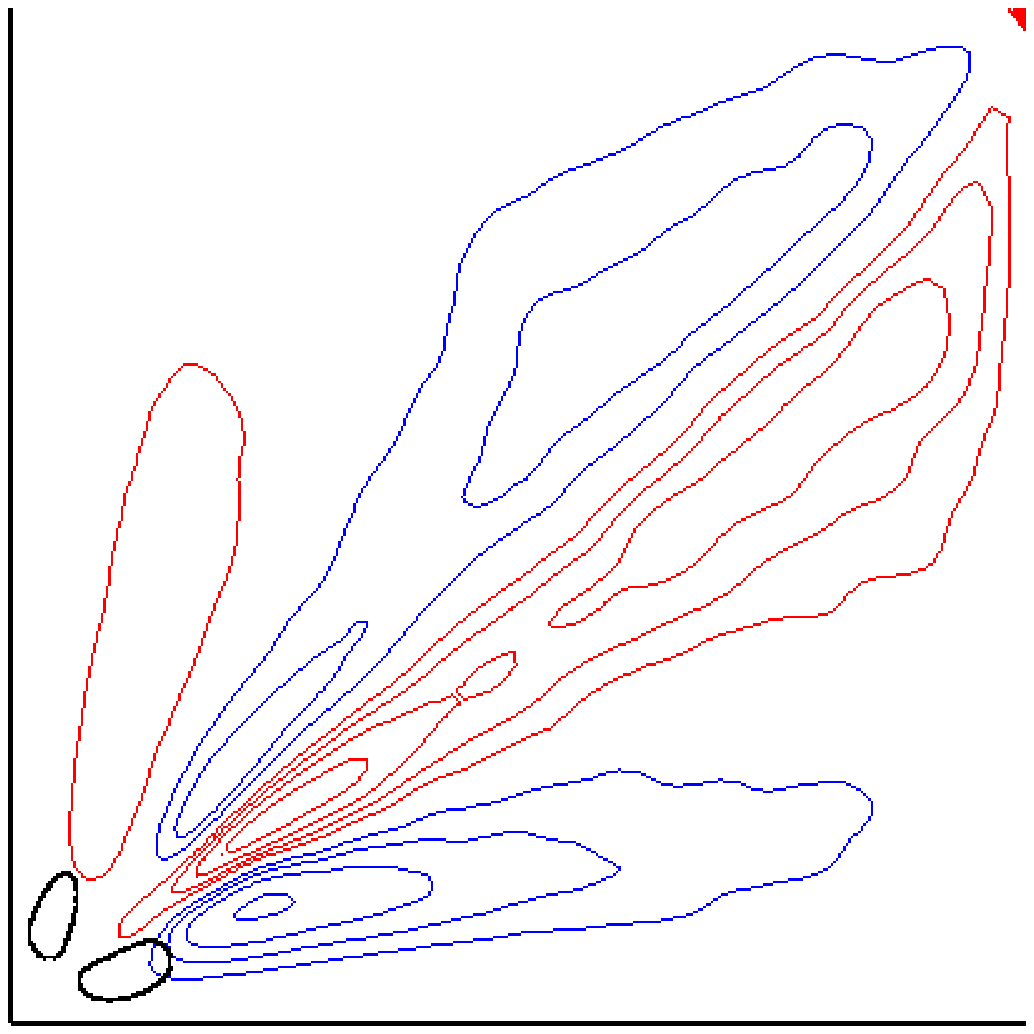}
%\hskip 0.75cm
\includegraphics[clip,width=5.0cm,angle=90]{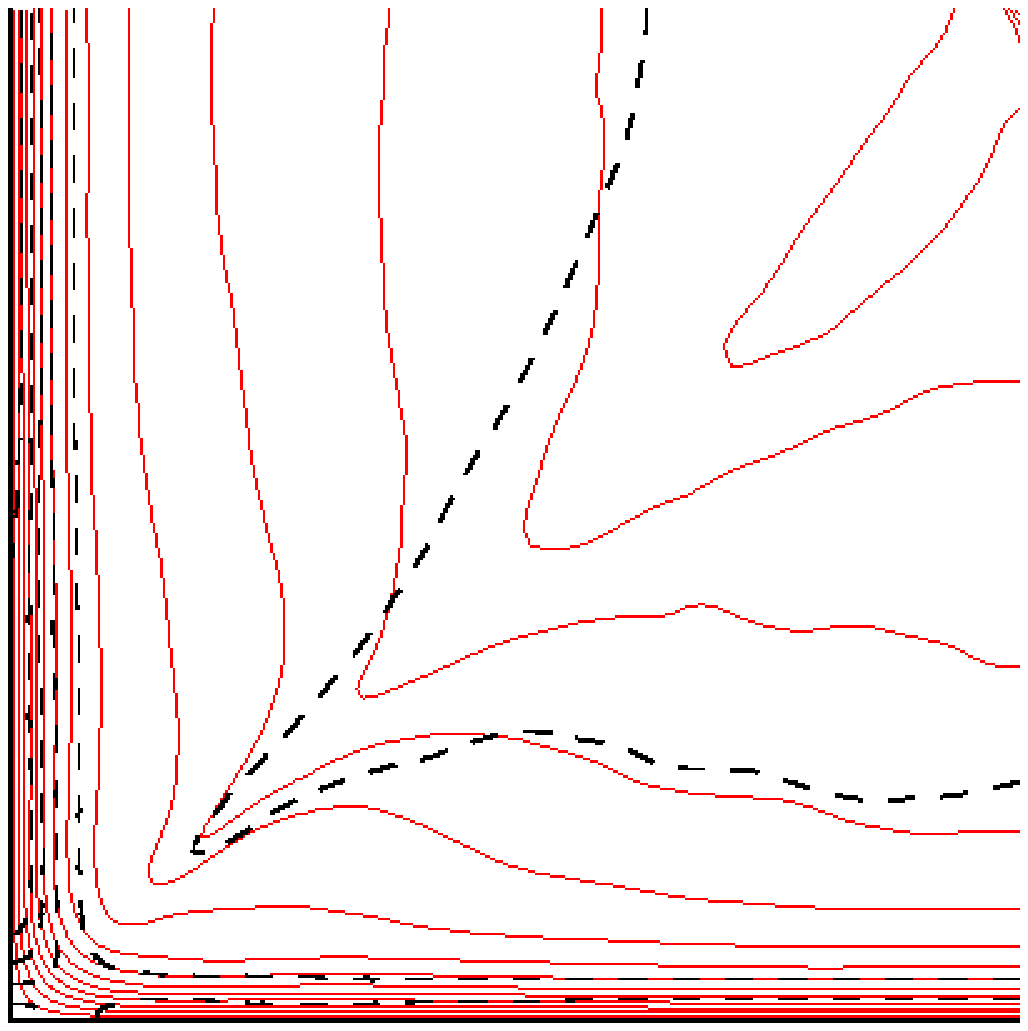}
\vskip -0.0cm
\hskip 1.0cm  a) \hskip 4.0cm  b) \hskip 4.0cm  c)
\vskip 0.0cm
\hskip -1.8cm
\includegraphics[clip,width=3.0cm,angle=90]{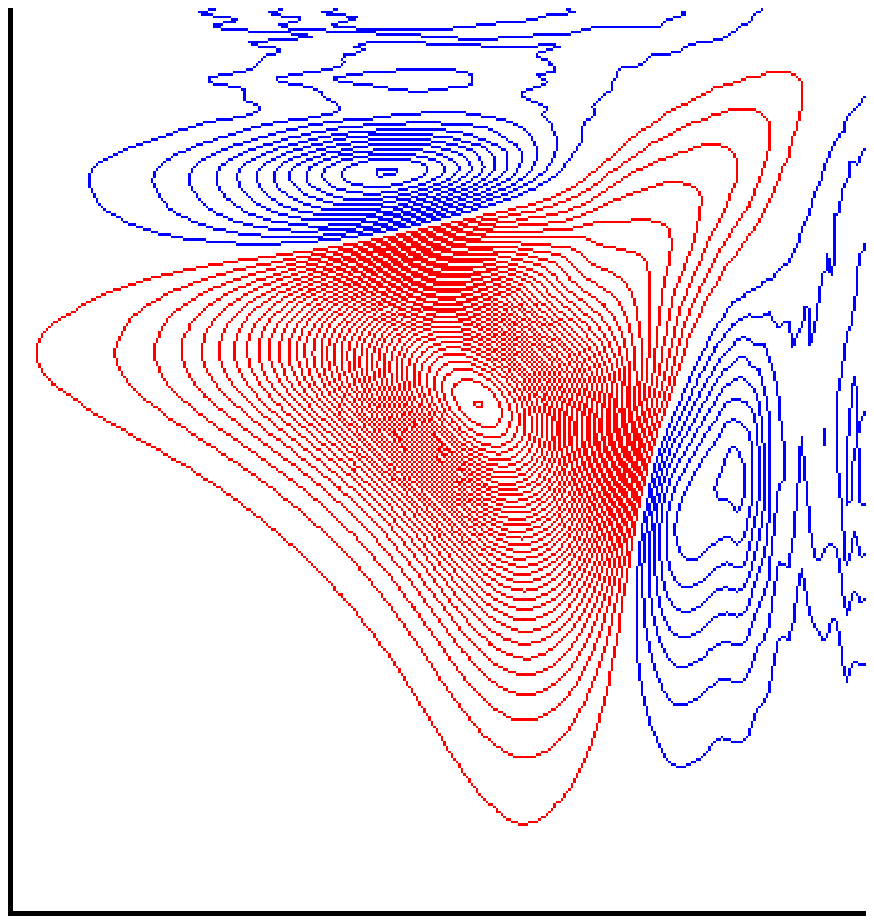}
%\hskip 0.75cm
\includegraphics[clip,width=3.0cm,angle=90]{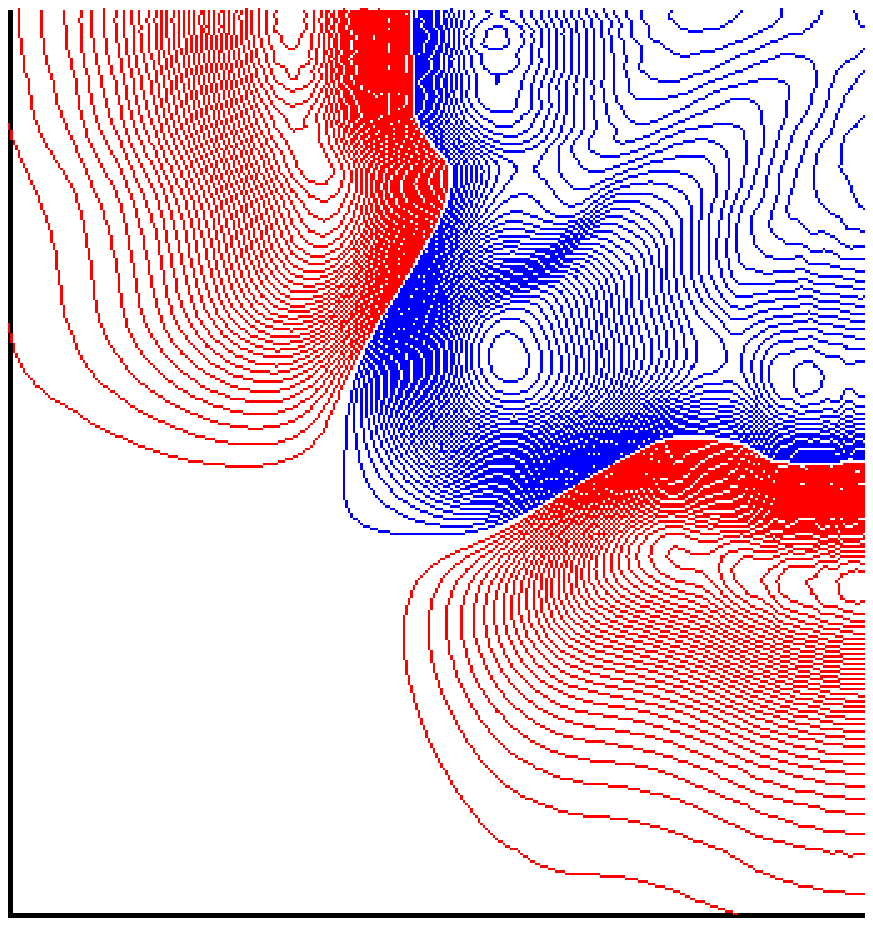}
%\hskip 0.75cm
\includegraphics[clip,width=3.0cm,angle=90]{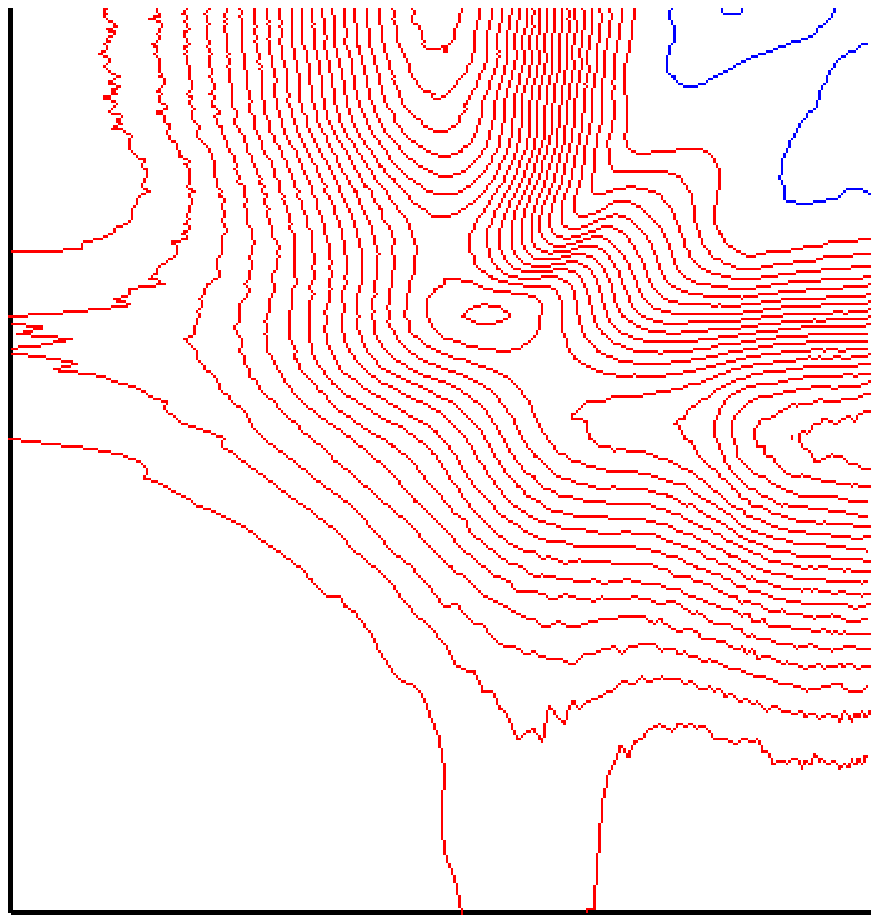}
%\hskip 0.75cm
\includegraphics[clip,width=3.0cm,angle=90]{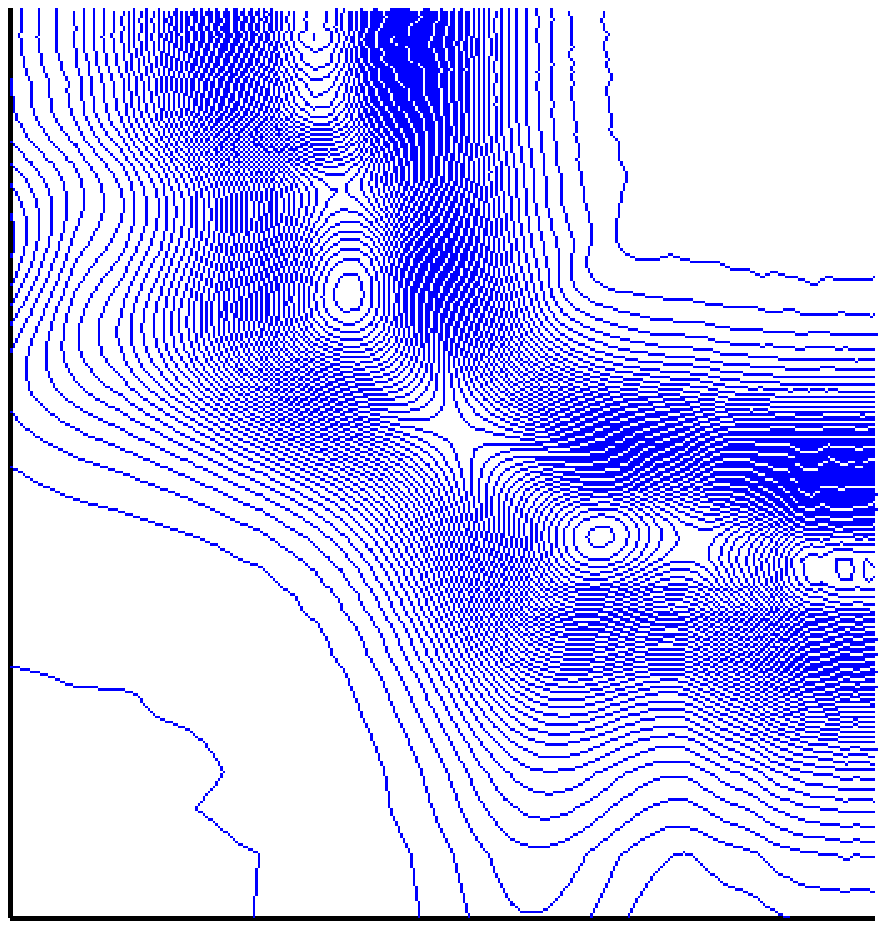}
\vskip -0.0cm
\hskip 1.0cm  d) \hskip 3.0cm  e) \hskip 3.0cm  f) \hskip 3.0cm  g)
\vskip 0.0cm
\hskip -1.8cm
\includegraphics[clip,width=3.0cm,angle=90]{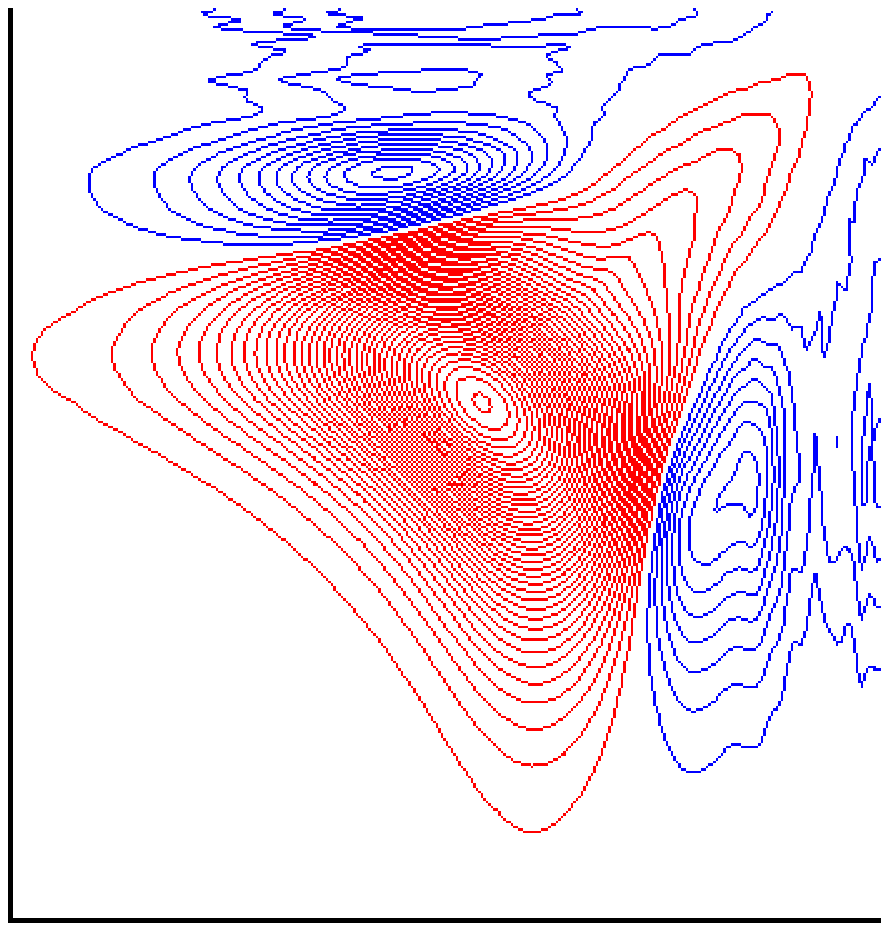}
%\hskip 0.75cm
\includegraphics[clip,width=3.0cm,angle=90]{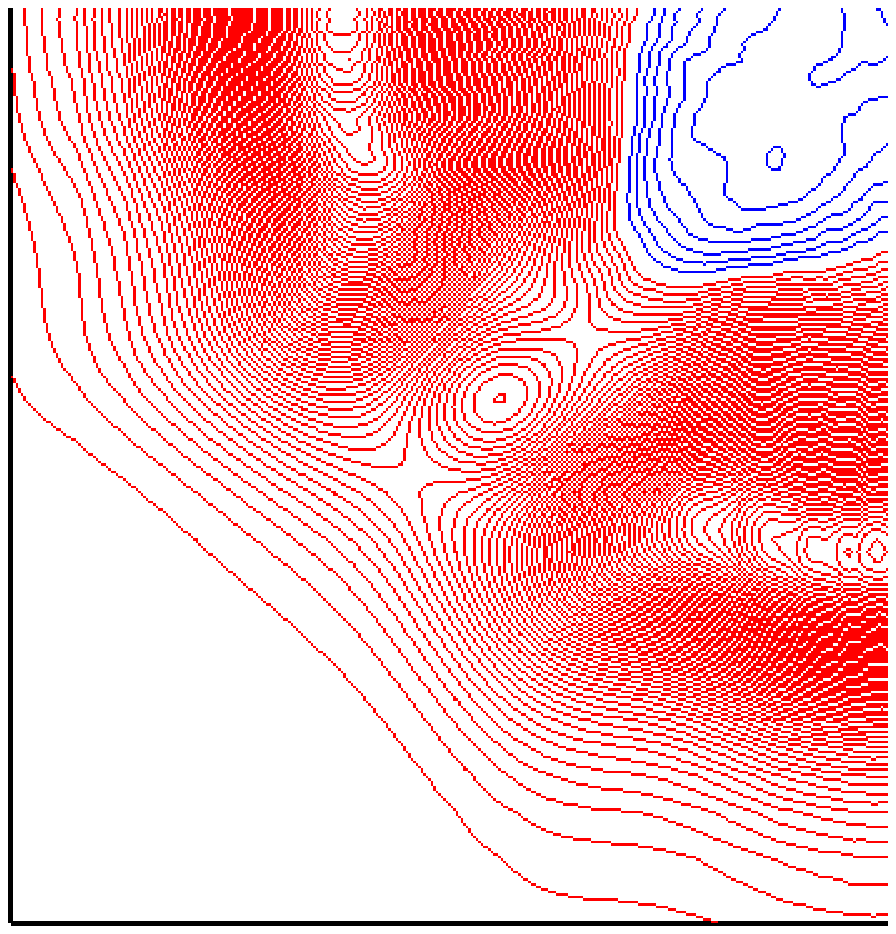}
%\hskip 0.75cm
\includegraphics[clip,width=3.0cm,angle=90]{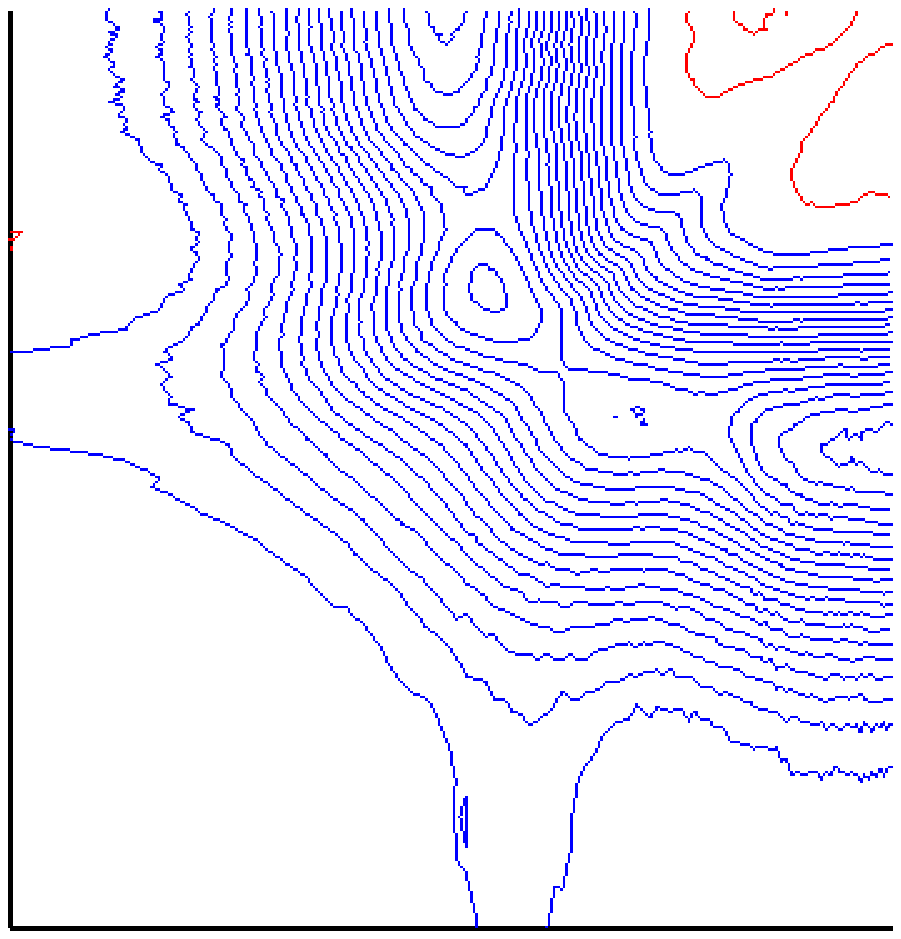}
%\hskip 0.75cm
\includegraphics[clip,width=3.0cm,angle=90]{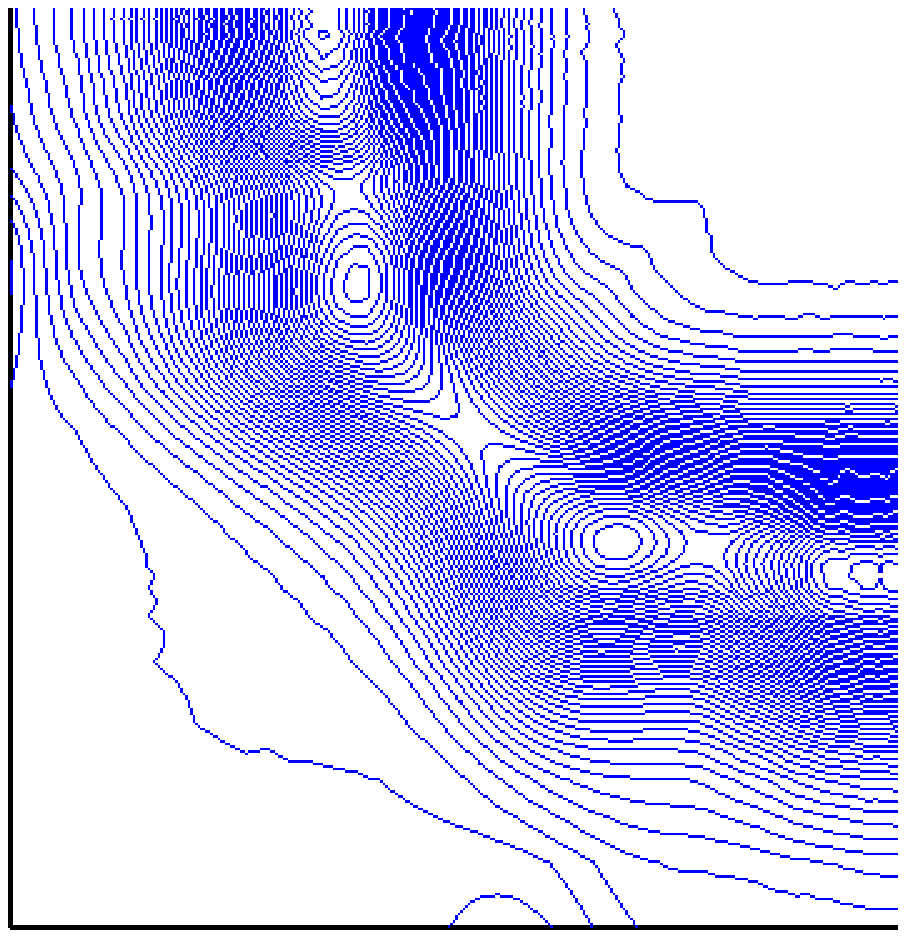}
\vskip -0.0cm
\hskip 1.0cm  h) \hskip 3.0cm  i) \hskip 3.0cm  j) \hskip 3.0cm  k)
\caption{Square duct: contours of mean velocity components at 
$Re=5000$ in strain eigenvector basis, with positive values in red 
and negative values in blue, in intervals $\Delta U^+=1$.
a) $U_\alpha^+$, superimposed to contours of $S_\alpha$;
b) $U_\beta^+$, superimposed to contours of $S_\beta$;
c) $U_\gamma^+$, superimposed to contours of $S_\gamma$,
the black contours starting from $S^+_\lambda=\pm 0.01$, 
in intervals $\Delta S^+=\pm 0.1$ solid positive, dashed negative.
In a), b), c) linear coordinates are used  from $0$ to $280$.
In the lower panels we show the contributions to the budget equations in 
the strain eigenvector basis 
for $U_\alpha$,
d)  $C_\alpha^+$, e)  $T_\alpha^+$, f)  $P_\alpha^+$, g)  $L_\alpha^+$,
and $U_\gamma$ 
h)  $C_\gamma^+$, i)  $T_\gamma^+$, j)  $P_\gamma^+$, k)  $L_\gamma^+$,
Inner-scaled coordinates are shown in logarithmic scale, 
positive contours are shown in red, negative in blue, in increments $\Delta=0.001$.
}
\label{fig14}
\end{figure}

To verify whether a simple budget of the momentum equations emerges
in the strain eigenvector basis,
the various budget terms have been evaluated by projecting
the quantities reported in  figure~\ref{fig13} for $U_1$ and $U_3$,
in addition to to those for $U_2$ 
along the eigenvectors of $S_{ij}$.
The resulting figures 
are shown in figure~\ref{fig14}, 
where the same contour increments are used as 
in figure~\ref{fig13}. 
Before it was observed 
that $|S_\alpha|$ is of the same order than
$|S_\gamma|$, therefore  it follows that $U_\gamma^+$ and $U_\alpha^+$,
are similar, whereas
$U_\beta^+$ is smaller than the
other two components. Differences between $U_\gamma^+$ and $U_\alpha^+$
arise near the corner bisector, with stronger transport of $U_\gamma^+$
towards the corner, due to $S_\gamma$ being larger
than $S_\alpha$. The contours of the contributions
to the momentum equation in the strain eigenvector basis show
that the convective terms ($C_\alpha^+$ in figure~\ref{fig14}d, 
$C_\gamma^+$ in figure~\ref{fig14}h) 
are similar, as well as the viscous terms
($L_\alpha^+$ in figure~\ref{fig14}g, 
$L_\gamma^+$ in figure~\ref{fig14}k),
whereas the pressure gradient contributions
have opposite sign ($P_\alpha^+$ in figure~\ref{fig14}f and 
$P_\gamma^+$ in figure~\ref{fig14}j).
Hence, the different shape of $U_\alpha^+$ and $U_\gamma^+$ 
near the corner bisector is mainly due to the different 
turbulent contributions in this region, which is negative for
$T_\alpha^+$ (figure~\ref{fig14}e), and positive for $T_\gamma^+$
(figure~\ref{fig14}i).

\begin{figure}
\centering
\vskip -0.0cm
\hskip -1.8cm
\psfrag{ylab} {$C_f       $}
\psfrag{xlab}{ $Re$}
\includegraphics[width=14.0cm]{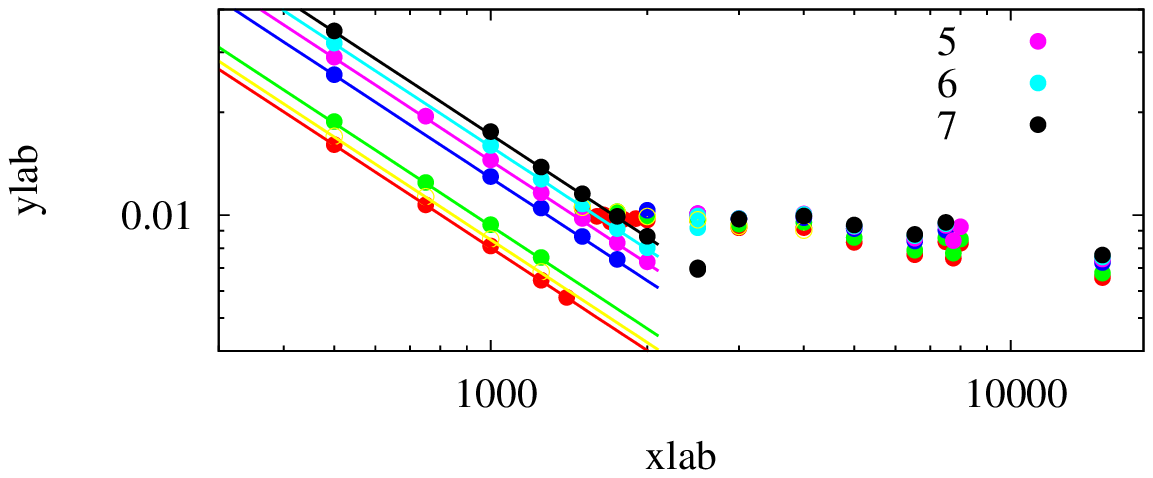}
\hskip 0.5 cm a)
\vskip -0.6cm
\hskip -1.8cm
\psfrag{ylab} {$C_f       $}
\psfrag{xlab}{ $Re_{D}$}
\includegraphics[width=14.0cm]{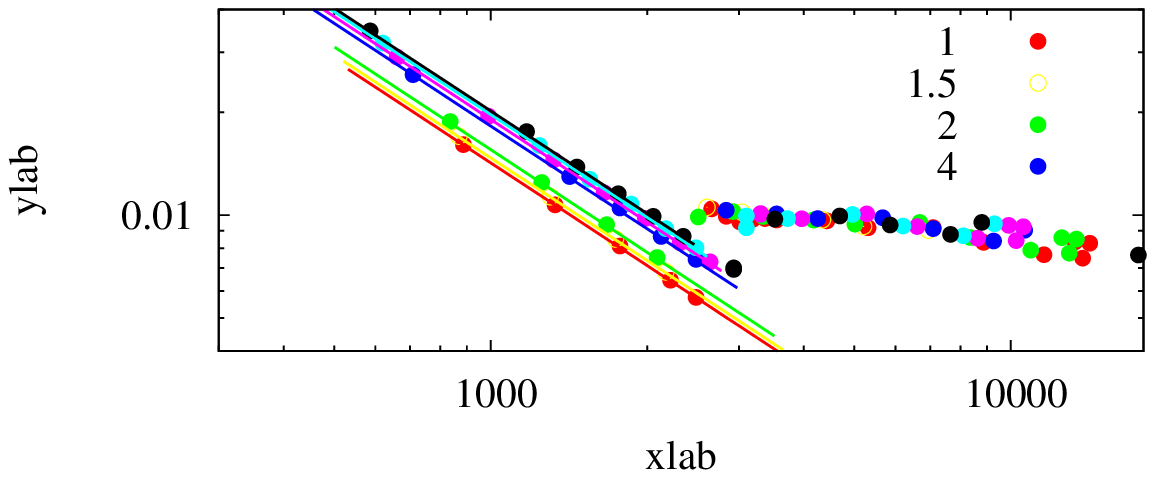}
\hskip 0.5cm b)
\vskip -0.6cm
\hskip -1.8cm
\psfrag{ylab} {$C_f       $}
\psfrag{xlab}{ $Re_3$}
\includegraphics[width=14.0cm]{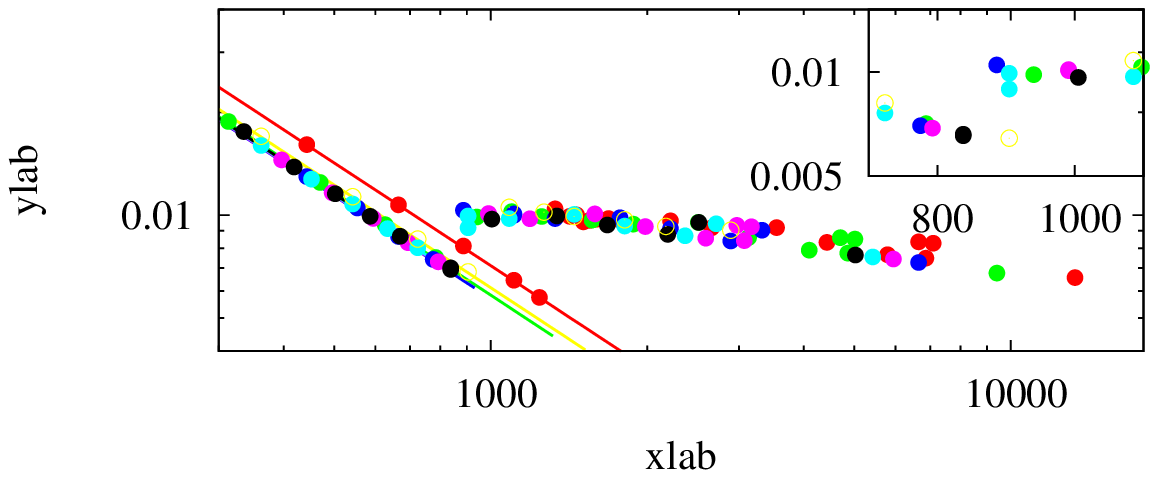}
\hskip 0.5cm c)
\vskip -0.2cm
\caption{
Rectangular ducts: friction coefficient $C_f$ versus 
a) computational Reynolds number $Re$;
b) Reynolds number based on the hydraulic diameter $Re_{D}$;
c) Reynolds number based on the short side half-length. 
The aspect ratio $A_R$ is indicated in the legend of panels a) and b),
and a zoom around the transitional $Re$ is shown in panel (c);
the solid lines correspond to equation (3-48) of \citet{white_74}.
}
\label{fig15}
\end{figure}

\section{Rectangular ducts }

\subsubsection{Friction factor  }

\begin{figure}
\centering
\vskip -0.0cm
\hskip -1.8cm
\includegraphics[clip,width=5.0cm,angle=90]{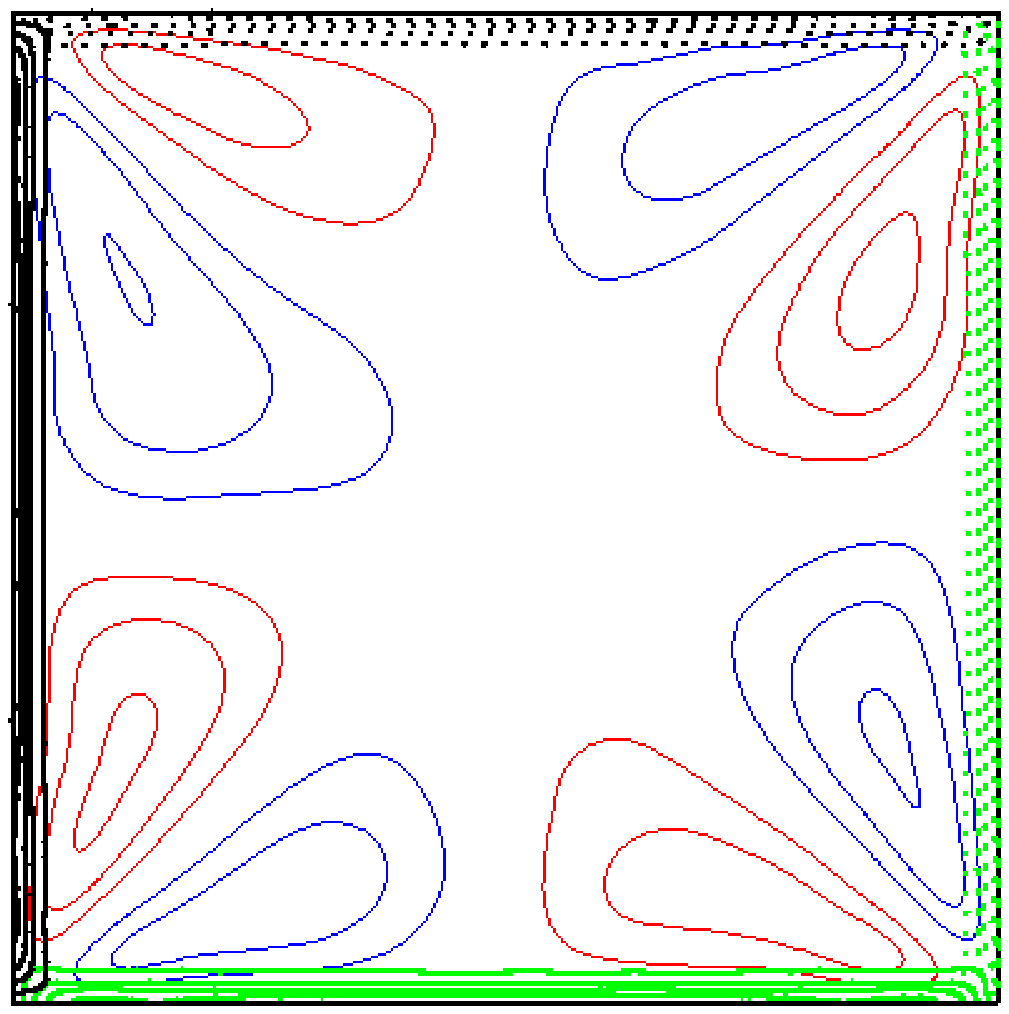}
\hskip +0.0cm
\includegraphics[clip,width=5.0cm,angle=90]{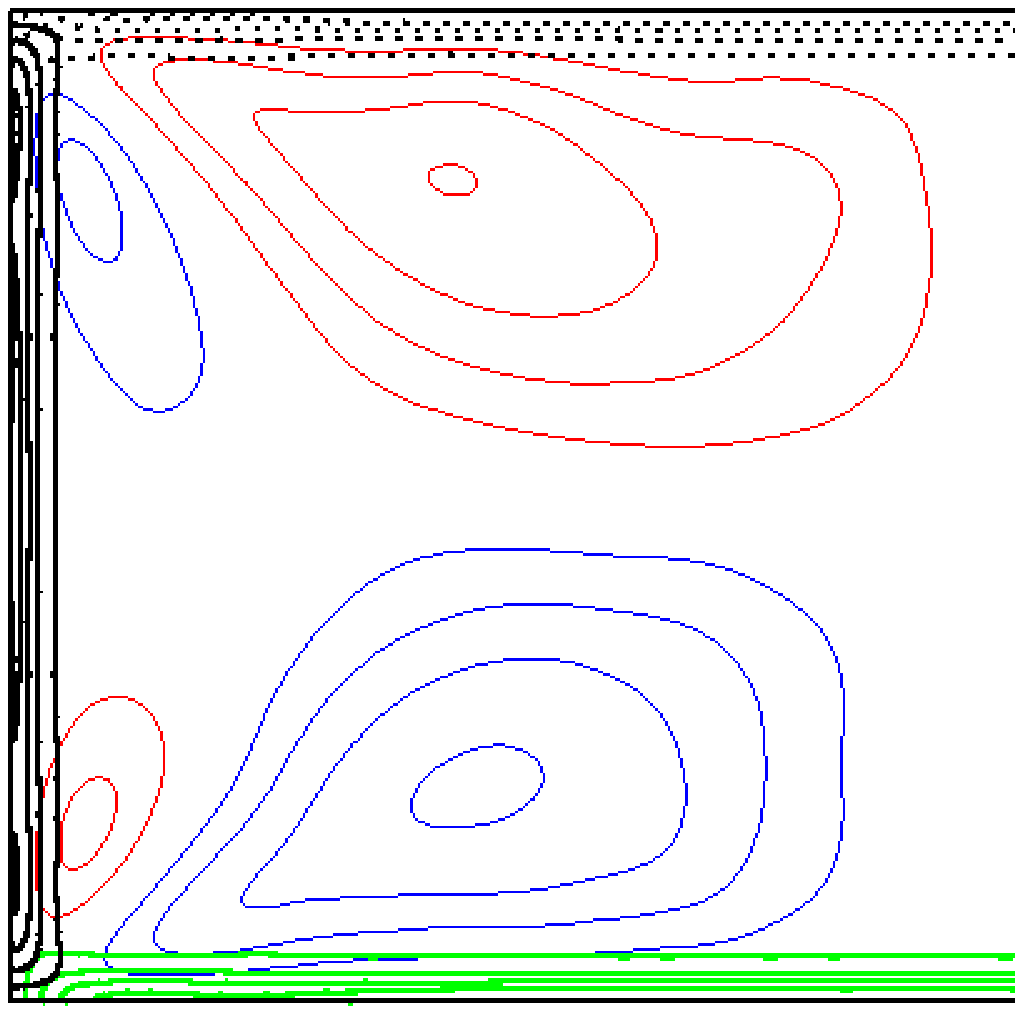}
\hskip +0.0cm
\includegraphics[clip,width=5.0cm,angle=90]{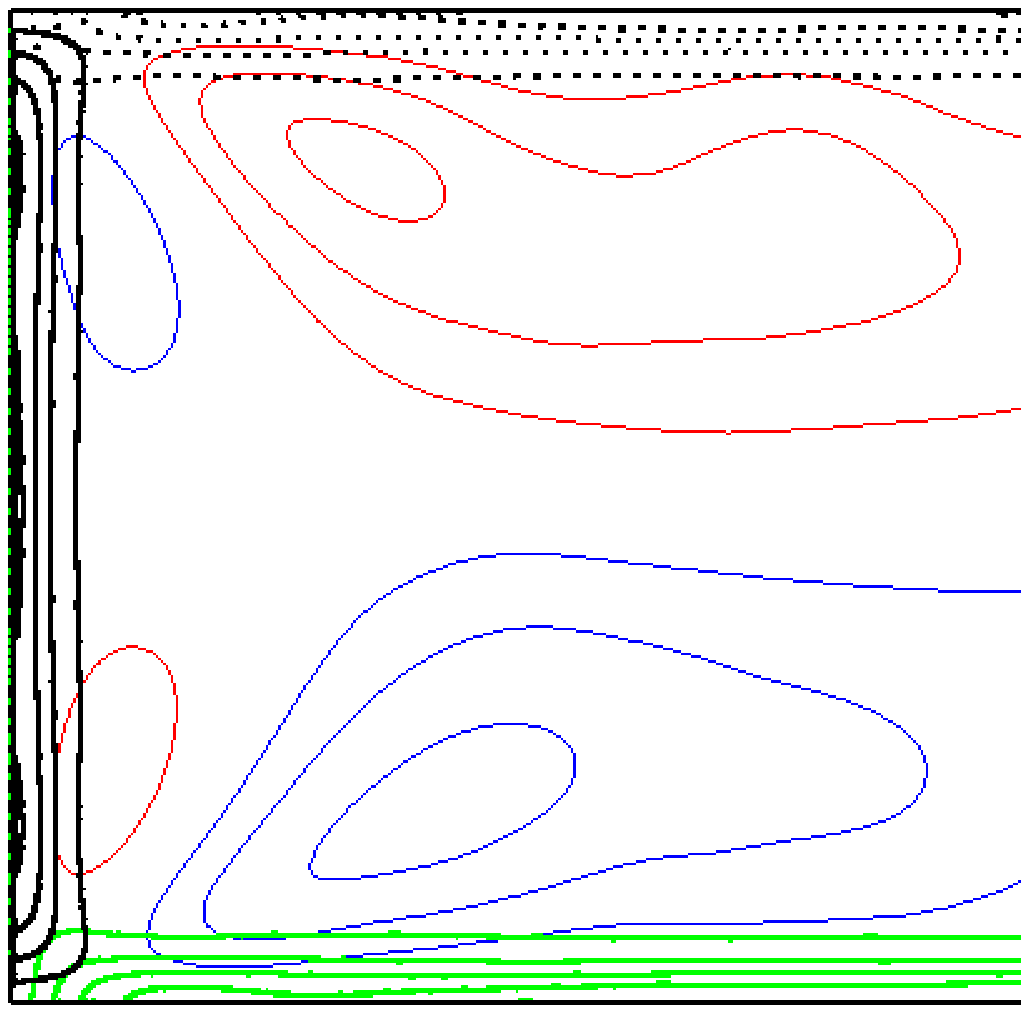}
\vskip -.1cm
\hskip -1.0cm  a)   \hskip 4.0cm  b) \hskip 5.5cm  c)
\vskip 0.0cm
\hskip -1.8cm
\psfrag{ylab} {$\tau_w    $}
\psfrag{xlab}{ $s/L_3$}
\includegraphics[width=5.5cm]{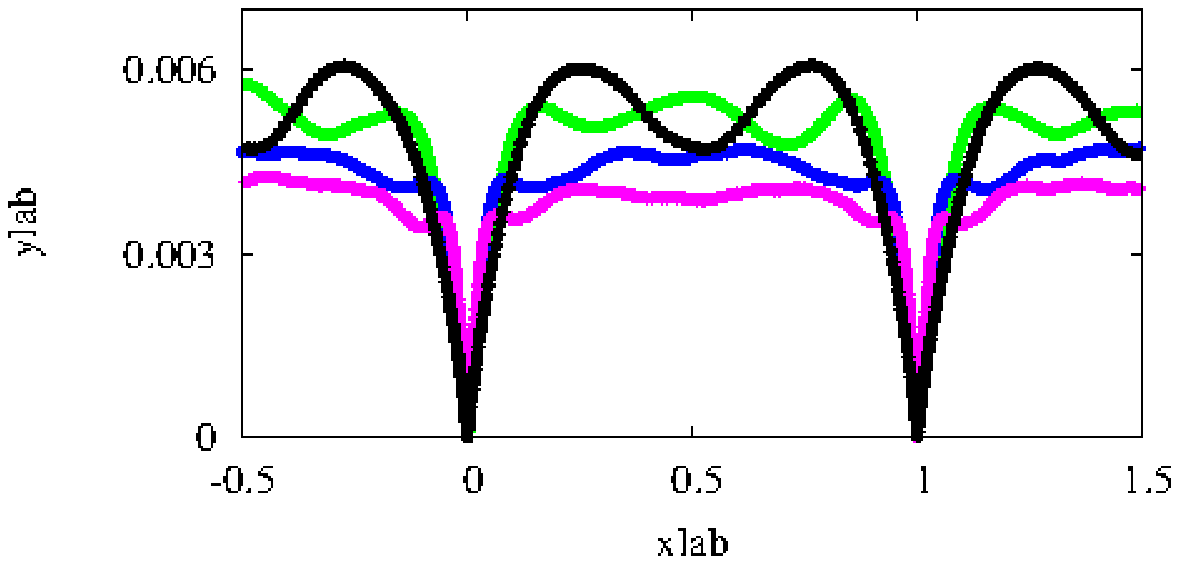}
\hskip -0.0cm
\psfrag{ylab} {$  $}
\includegraphics[width=5.5cm]{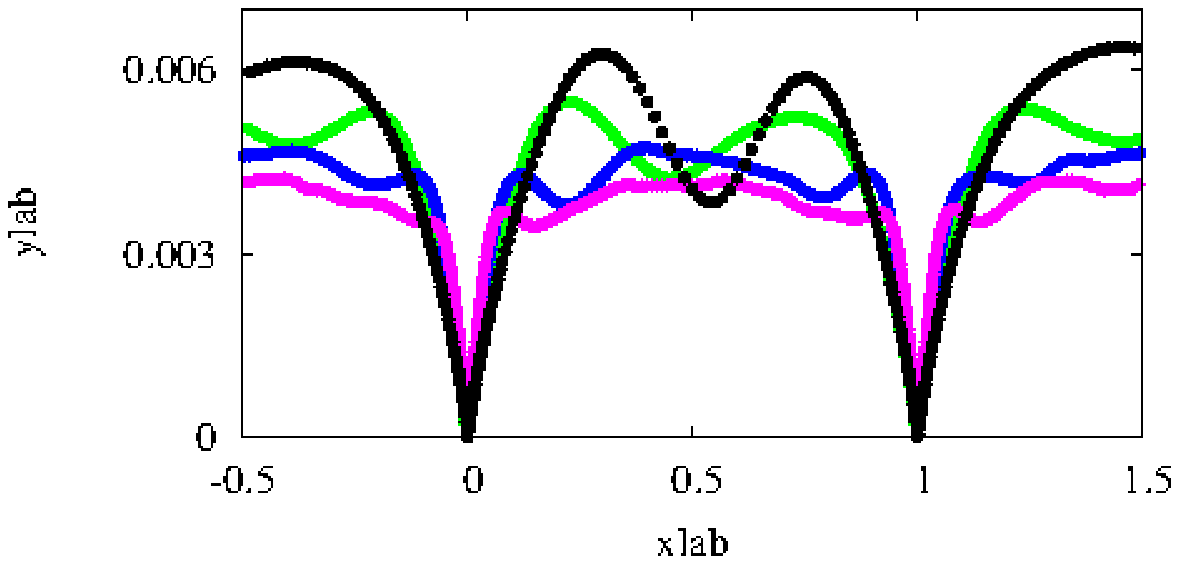}
\hskip -0.0cm
\psfrag{ylab} {$  $}
\includegraphics[width=5.5cm]{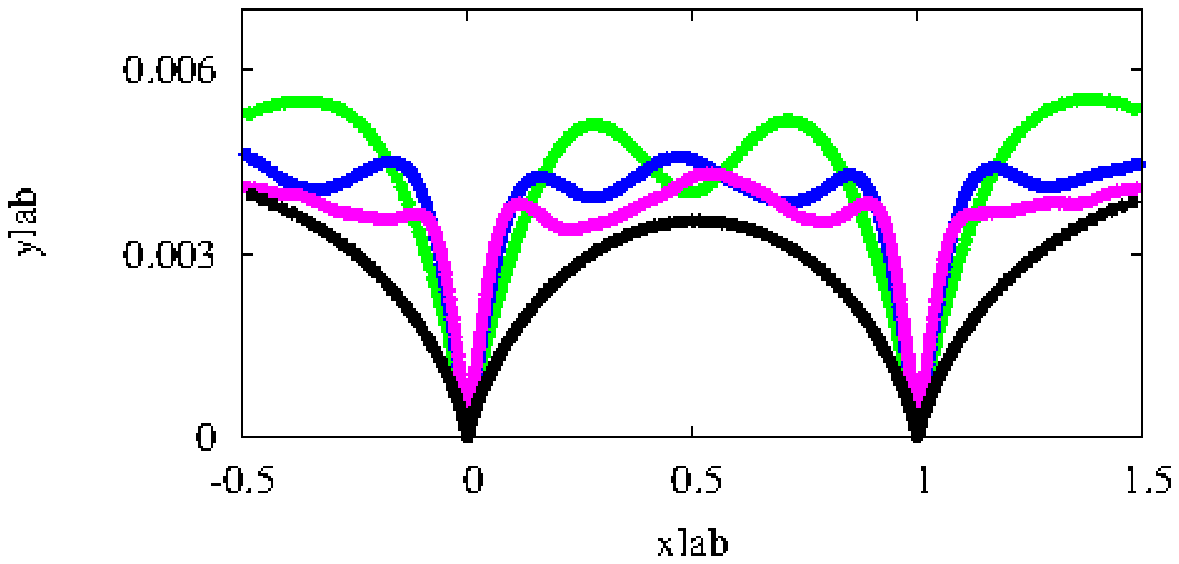}
\vskip 0.5cm
\hskip -1.8cm
\includegraphics[clip,width=5.0cm,angle=90]{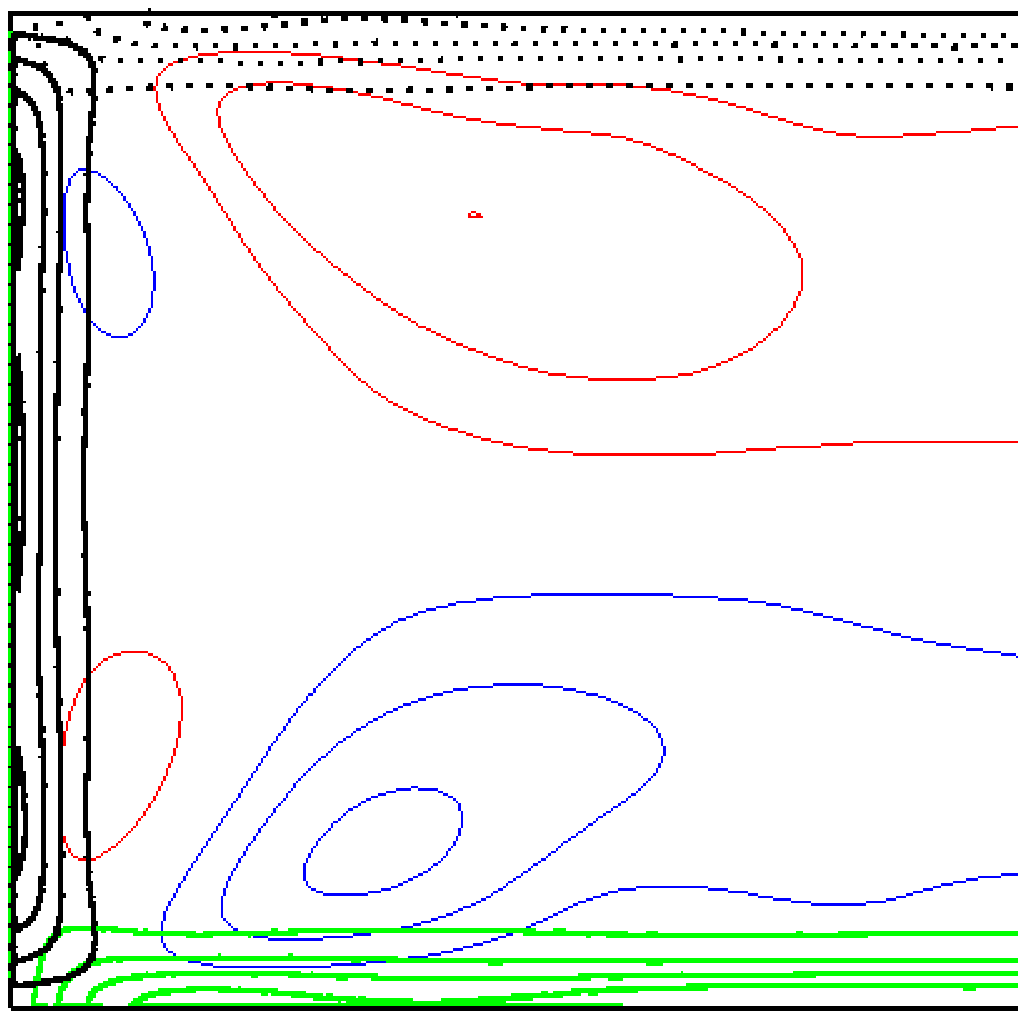}
\includegraphics[clip,width=5.0cm,angle=90]{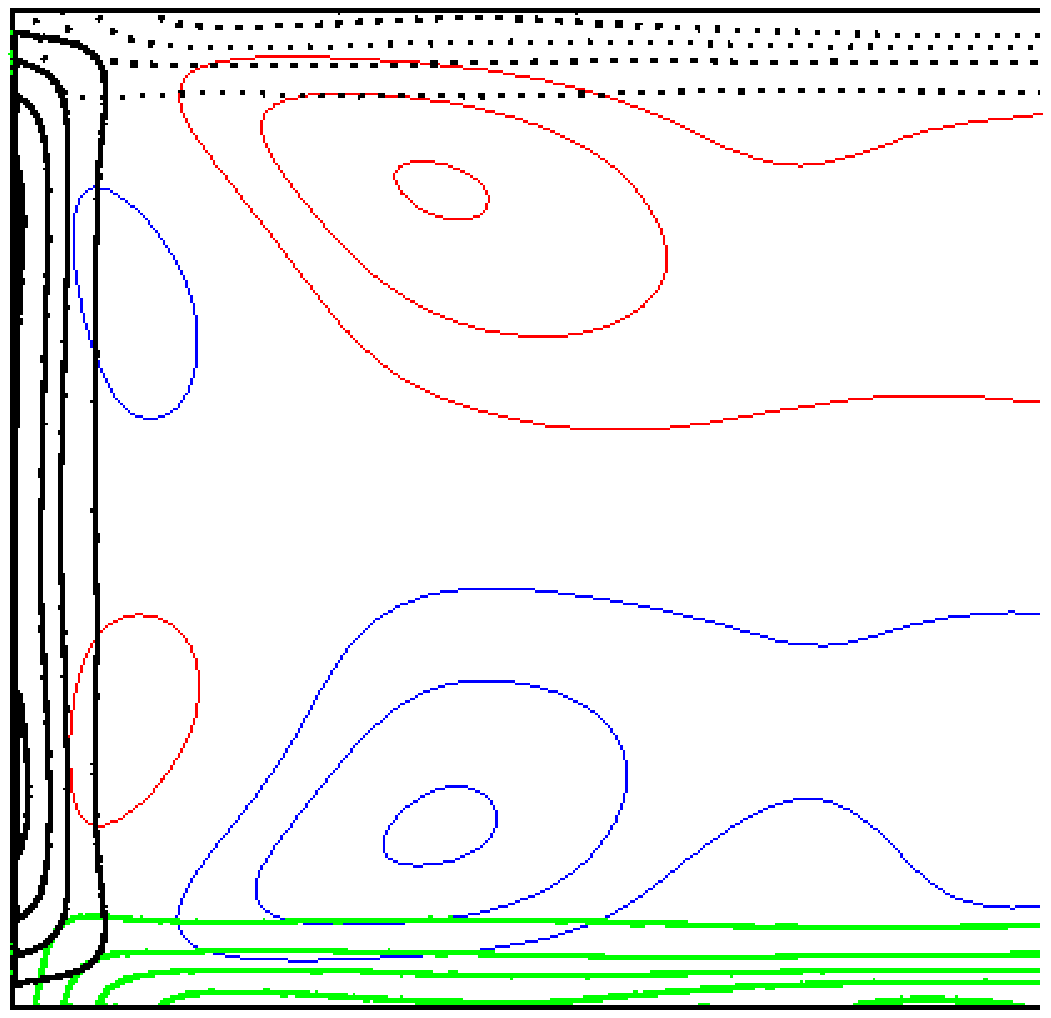}
\includegraphics[clip,width=5.0cm,angle=90]{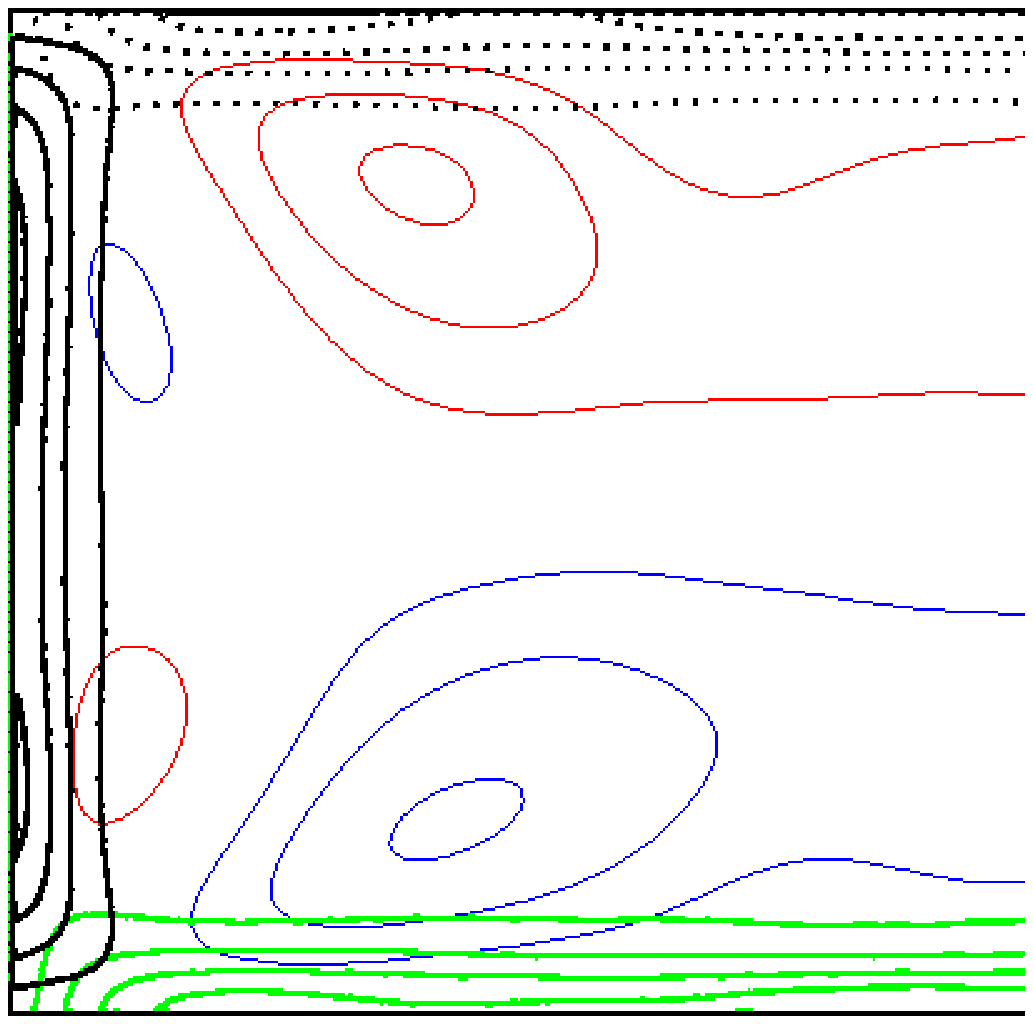}
\vskip -.1cm
\hskip -1.0cm  d)   \hskip 3.5cm  e) \hskip 4.5cm  f)
\vskip 0.0cm
\hskip -1.8cm
\psfrag{ylab} {$\tau_w    $}
\psfrag{xlab}{ $s/L_3 $}
\includegraphics[width=5.5cm]{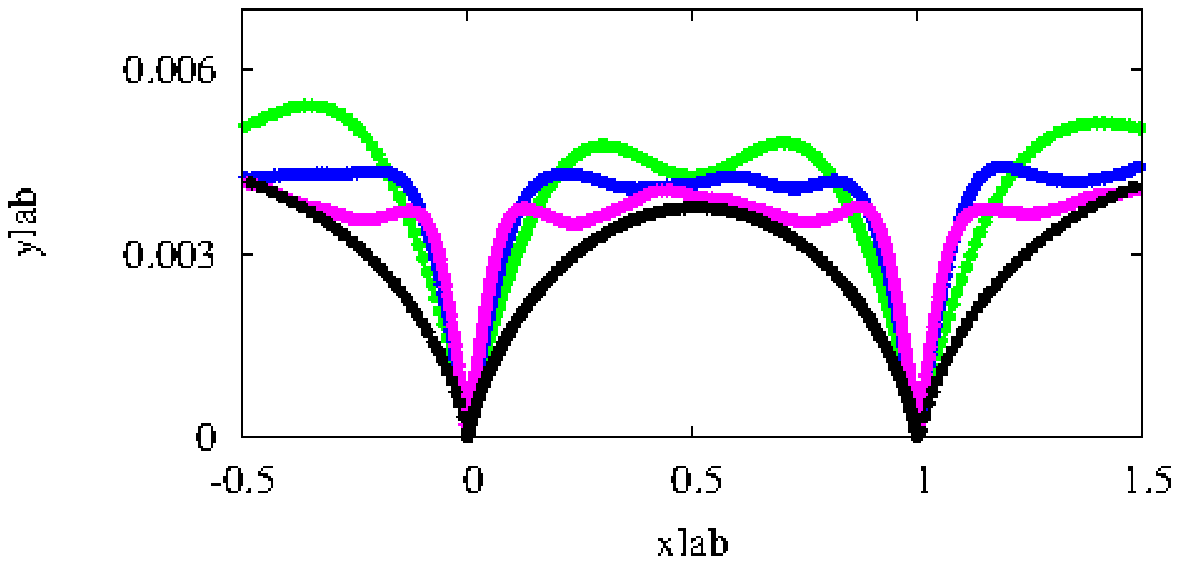}
\hskip -0.0cm
\psfrag{ylab} {$  $}
\includegraphics[width=5.5cm]{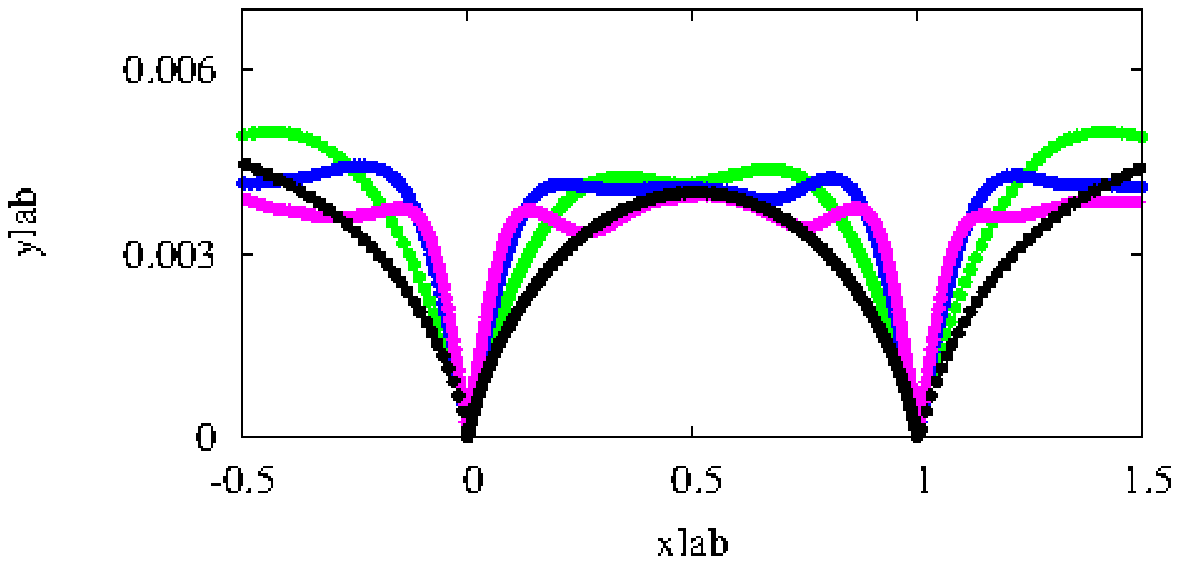}
\hskip -0.0cm
\psfrag{ylab} {$  $}
\includegraphics[width=5.5cm]{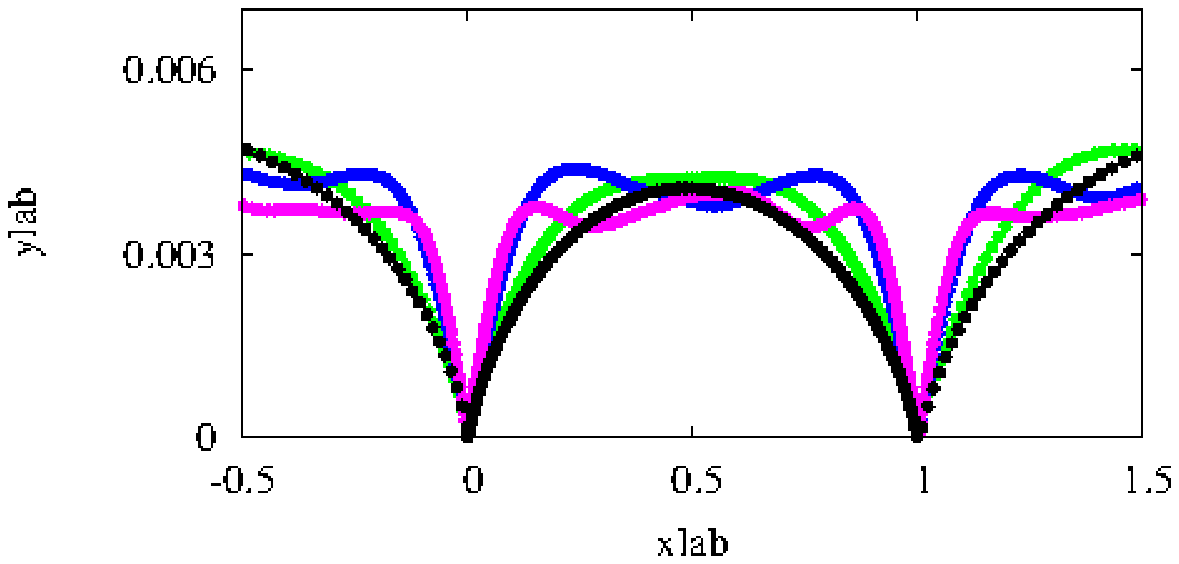}
\vskip -.1cm
\caption{Rectangular ducts: contours of vorticity components $\omega_2$ and $\omega_3$, 
superimposed to the secondary stream-function $\psi$ at $Re=5000$ and various
aspect ratios, a) $AR=1$, b) $AR=2$, c) $AR=4$, d) $AR=5$, e) $AR=6$, f) $AR=7$.
The small panels under each figure show the wall shear stress ($\tau_w$) 
along part of the duct perimeter, to show more clearly the behavior
along the short side. The data are shown at $Re=1750$ (black),
at $Re=2500$ (green), $Re=5000$ (blue), $Re=7750$ (cyan).
}
\label{fig15bis}
\end{figure}
We previously mentioned that in rectangular channels the Reynolds number
based on the short side half length marks the transition from the laminar to the
turbulent regime. To demonstrate that $C_f$ has a nearly universal
behavior with $Re_3$ both in the laminar and in the turbulent
regime, in figure~\ref{fig15} we show the maps of $C_f$ versus
the three Reynolds number indicated in table~\ref{tab1}. The 
common feature is that satisfactory collapse of the data occurs
in the fully turbulent regime,
regardless of the definition of the Reynolds number.
On the other hand, in the laminar regime the poorest scaling 
is obtained with $Re$ (figure~\ref{fig15}a), which may
be understood because the choice of a reference length based on 
the radius of an equivalent pipe does not account for
the shape of the duct. In the three figures, lines with the
same colour as the solid symbols are evaluated from the analytical 
expression (3-48) given at Pg.113
of \citet{white_74}.
Also the classical Reynolds number based
on the hydraulic diameter is not suitable to account for the shape
of the duct cross section, as may be inferred from figure~\ref{fig15}b. 
On the other hand, the choice of the short
side as the reference length yields good collapse both in the laminar and
in the turbulent regime. 
Regardless of the aspect ratio, 
the transitional Reynolds number is found at
$Re_3 \approx 850$, as may be seen in the inset of figure~\ref{fig15}c. 
In this figure it may be noticed that the behavior
for $A_R=1$ (the red solid symbols) is a bit different, which probably implies
that the absence of symmetry about the corner bisector 
in rectangular ducts generates stronger disturbances due to the 
secondary motion, which cause earlier transition.

\subsubsection{Wall friction profiles}

Contours of the secondary stream function and profiles 
of the wall shear stress along
the perimeter of rectangular ducts, similar to those
given in figure~\ref{fig4} for square ducts, 
may provide insight to understand the influence of the aspect ratio.
Simulations at $Re=5000$ are appropriate for this purpose.
In all panels of figure~\ref{fig15bis}, both directions
are normalised by $L_3/2$, hence the vertical coordinate
ranges from $-1$ to $+1$.   
Comparison between figure~\ref{fig15bis}a (for $A_R=1$)
and figure~\ref{fig15bis}b (for $A_R=2$) shows that in a region of equal
size in the vertical and horizontal directions  a
change in the strength of the recirculating regions may
be appreciated. The recirculating region
near the short vertical wall reduces in size and 
strength. On the other hand the recirculating region near
the horizontal wall increases in size and strength.
This asymmetry causes the formation of strong disturbances
propagating from the corner towards the central region, which
explains why the critical Reynolds number in square ducts is higher 
than that in rectangular ducts. The growth of the stronger
recirculating and the location of the maximum is fixed,
which suggests that $L_3$ is the appropriate length
scale at low Reynolds numbers. However, 
transition to the turbulent regime does not occur
without the small recirculating region near the short side.
This was observed in similar plots at $Re=2500$, and is corroborated
by the $C_f$ plots of figure~\ref{fig15}, where the black
dots corresponding to $A_R=7$ at $Re=2500$ are aligned with
the laminar values. Although barely visible
in the profiles of $\tau_w(s)$ for $A_R=7$, at $Re\le 2500$
there is only one peak in the short side, whereas at
$Re=5000$ there are two peaks which are generated by
the secondary recirculating regions near the short side.
At $Re=2500$, $A_R=6$, two peaks with small
undulation are produced, which is enough to have 
the $C_f$ in figure~\ref{fig15} no longer aligned with the
laminar value. At $A_R=5$ the two peaks are visible
in figure~\ref{fig15bis} of $\tau_w(s)$ at $Re=2500$, hence
the corresponding value of $C_f$ in figure \ref{fig15} coincides
with the values of the simulations with smaller aspect ratio.
At $A_R=5$ the profiles of $\tau_w(s)$ only have one peak in
the short side at $Re=1750$, which is found also for $A_R=4$,
and two peaks finally form for $A_R=2$ as confirmed in
figure~\ref{fig15}a, where at $Re=1750$ the value of $C_f$ is
not aligned with the laminar trend.

\subsubsection{ Mean flow  }

\begin{figure}
\centering
\vskip -0.0cm
\hskip -1.8cm
\includegraphics[clip,width=4.5cm]{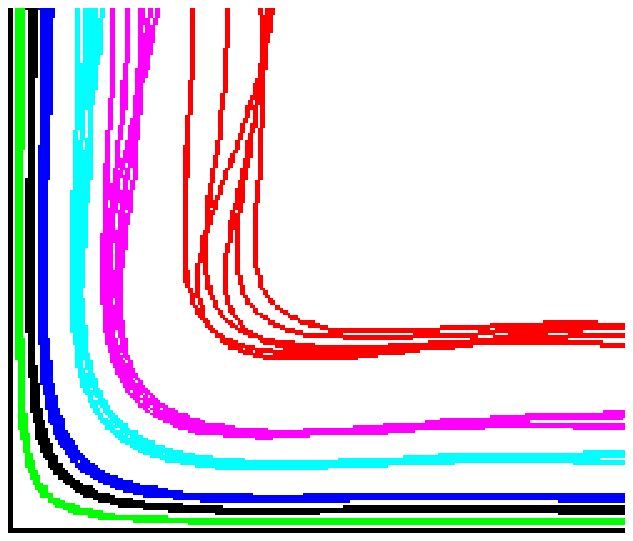}
\hskip -0.5cm
\includegraphics[clip,width=4.5cm]{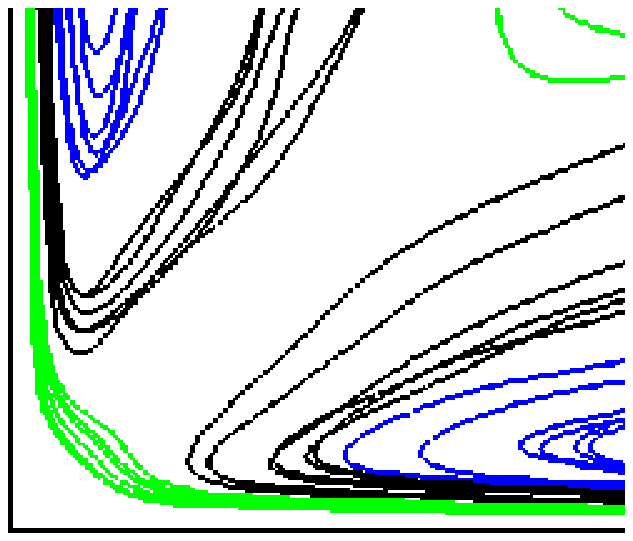}
\hskip -0.5cm
\includegraphics[clip,width=4.5cm]{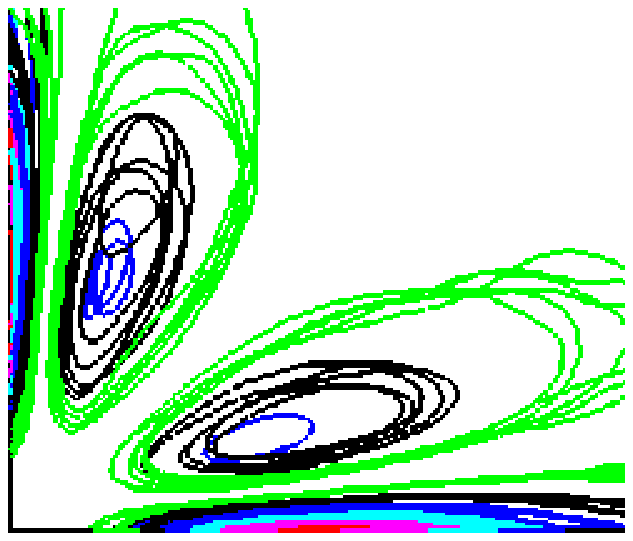}
\hskip -0.5cm
\includegraphics[clip,width=4.5cm]{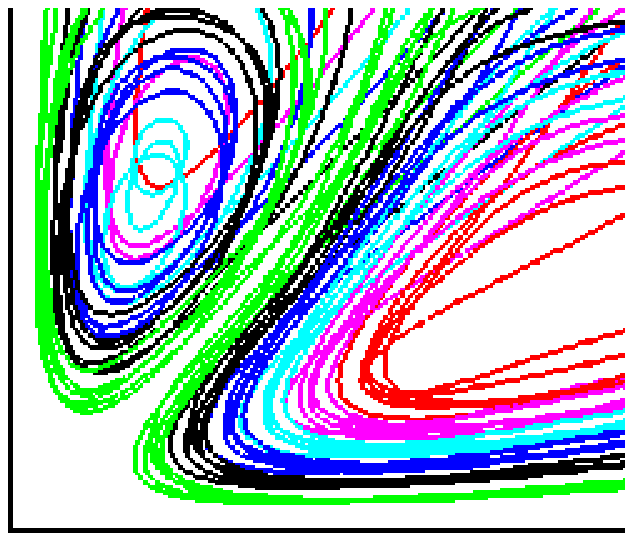}
\vskip -0.2cm
\hskip 1.0cm  a) \hskip 3.5cm  b) \hskip 3.5cm  c) \hskip 3.5cm  d)
\caption{Rectangular duct: contours of a) $U_1^+$, green 2 black 4 blue 6 cyan 10 mag 12 red 14;
b) $q^+$ green 1 black 2 blue 3;
c) $\omega_1^+$ green .01 black .02 blue .03 cyan .04 mag .05 red .06;
b) $\psi^+$ green 1 black 2 blue 3 cyan 4 mag 5.
All data are shown at $Re=5000$ for $A_R=1$, $A_R=2$, $A_R=4$, $A_R=5$, $A_R=6$, $A_R=7$.
Space coordinates are scaled with respect to the averaged friction velocity, $\overline{u_\tau}$.
}
\label{fig16}
\end{figure}

The flow near the corner in rectangular ducts does not change
dramatically from the case of square ducts.
Contours of $U_1^+$,
$q^+$ $\omega_1^+$ and $\psi^+$ in the corner regions are shown in figure \ref{fig16}, 
at fixed $Re=5000$, for different values of the aspect ratio, superimposed to each other to have a
global picture of whether the behavior is drastically affected.
The two space coordinates and all quantities are here scaled
with the averaged friction velocity. The contours of $U_1^+$
(figure~\ref{fig16}a) are superimposed each other, especially near
walls, whereas some difference may be appreciated far from the walls
near the shorter side. Increasing the aspect 
ratio to $A_R=4$ the $U_1^+=14$ iso-line (red) moves parallel towards the central
region. Further increasing $A_R$, the $U_1^+=14$ iso-line
near the bisector moves towards the short side of the duct. 
In the long side region the
contours are flat near the wall, and undulations appear far from the wall.
The turbulent kinetic
energy distributions near the walls are independent of the aspect ratio,
as shown by the green lines ($q^+=1$) in figure~\ref{fig16}b. 
On the other hand, the blue contours ($q^+=3$), in the long side region 
move far from the corner by increasing $A_R$. In this region, also the black 
contours ($q^+=2$) shrink indicating large variations of 
the turbulent kinetic energy distribution. In the short side region 
small variations are see in figure~\ref{fig16}b.
This behavior might be ascribed to the effect of the 
secondary motion, but in fact 
the contours of $\omega_1^+$ in figure~\ref{fig16}c
show only marginal variation with $A_R$. This vorticity component 
is linked to the small scales in the near-wall region.
The strong vorticity layers 
attached to the horizontal and vertical walls are found to scale well with
the averaged friction velocity. The large scale secondary motion
depicted through the stream function in figure~\ref{fig16}d,
consists on two recirculating regions of 
different size, the bigger one along the long side.
It is important to stress that asymmetry only appears only
for $A_R>1$, and in agreement with the previous discussion, disappears
at $A_R=1$.

\subsubsection{ Turbulent kinetic energy budgets }

\begin{figure}
\centering
\vskip -0.0cm
\hskip -1.8cm
\includegraphics[clip,width=4.5cm]{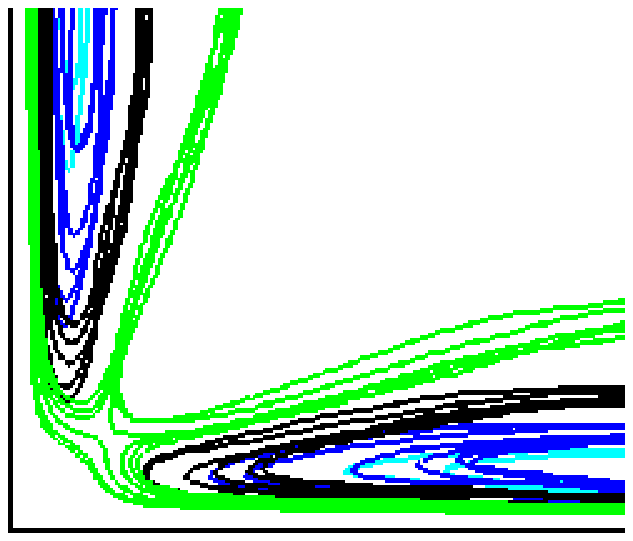}
\hskip -0.5cm
\includegraphics[clip,width=4.5cm]{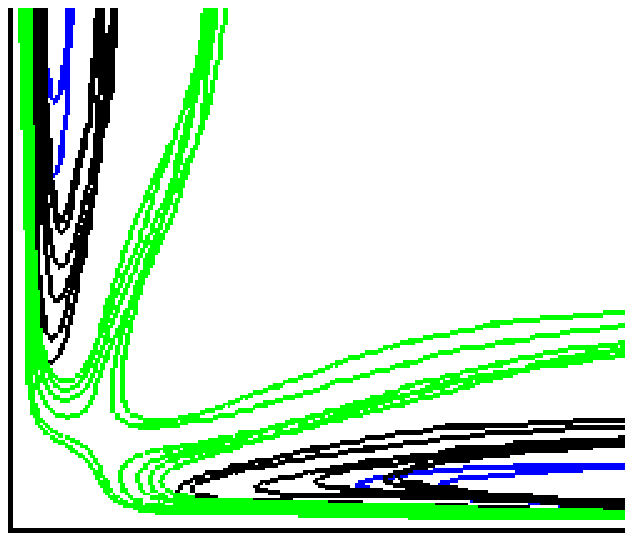}
\hskip -0.5cm
\includegraphics[clip,width=4.5cm]{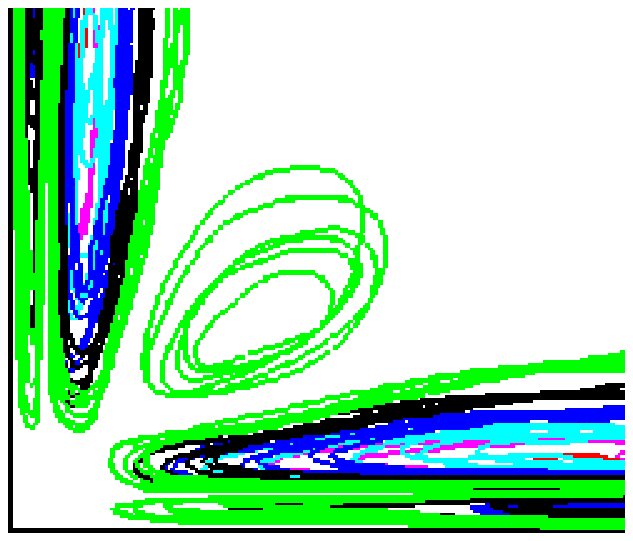}
\hskip -0.5cm
\includegraphics[clip,width=4.5cm]{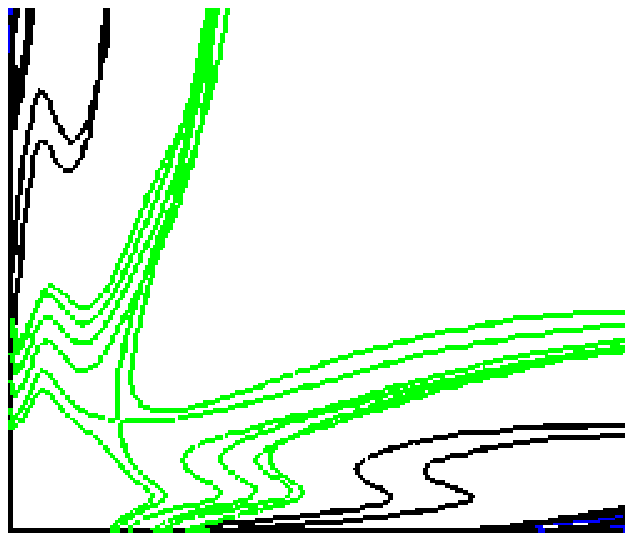}
\vskip -0.2cm
\hskip 1.0cm  a) \hskip 3.5cm  b) \hskip 3.5cm  c) \hskip 3.5cm  d)
\caption{Rectangular duct: contours of 
a ) $P_K^+$; 
b) $D_K^+$ ;
c) $T_K^+$, green $\pm=.0125$, black $\pm=.025$, blue $\pm=.0375$,
cyan $\pm=.05$, magenta $\pm=.0625$, red $\pm=.075$, solid positive and
dotted negative;
d) $\epsilon_K^+$. 
In panels a), b) and d) 
green $\pm=.05$, black $\pm=.1$, blue $\pm=.15$,
cyan $\pm=.2$, magenta $\pm=.25$, red $\pm=.3$, solid positive
dotted negative.
All data are shown at $Re=5000$ for $A_R=1$, $A_R=2$, $A_R=4$, $A_R=5$, $A_R=6$, $A_R=7$.
Space coordinates are scaled with respect to the averaged friction velocity, $\overline{u_\tau}$.
}
\label{fig17}
\end{figure}

The good scaling in wall units of the 
mean motion $U_1^+$, of the secondary motion through $\omega_1^+$,
and of the turbulent kinetic energy $q^+$, and their rather good independence
on the aspect ratio, is also found 
for each term of the simplified turbulent kinetic energy budget,
shown in figure~\ref{fig17}.
The DNS results depict the occurrence of large values
for $D_K^+$ and $P_K^+$ near the walls, with those of $D_K^+$ closer to the walls 
than those of $P_K^+$. The latter has its peak
far from the corner at a distance from the walls
approximately $15$ wall units. In agreement
with the previous discussion for the square duct, 
good equilibrium between production and total dissipation
is corroborated by the small values of $T_K^+$.
Figure~\ref{fig17}c further shows 
alternation of negative and positive layers
near the walls, depending on the relative magnitude of $P_K^+$ and
$D_K^+$. The rather good scaling of the isotropic dissipation rate 
$\epsilon_K^+$ in wall units arises in figure~\ref{fig17}d. Comparison
between this figure and figure~\ref{fig17}b demonstrates that
modeling $D_K^+$ should be easier than modeling $\epsilon_K^+$.

\begin{figure}
\centering
\vskip -0.0cm
\hskip -1.8cm
\includegraphics[clip,width=3.5cm]{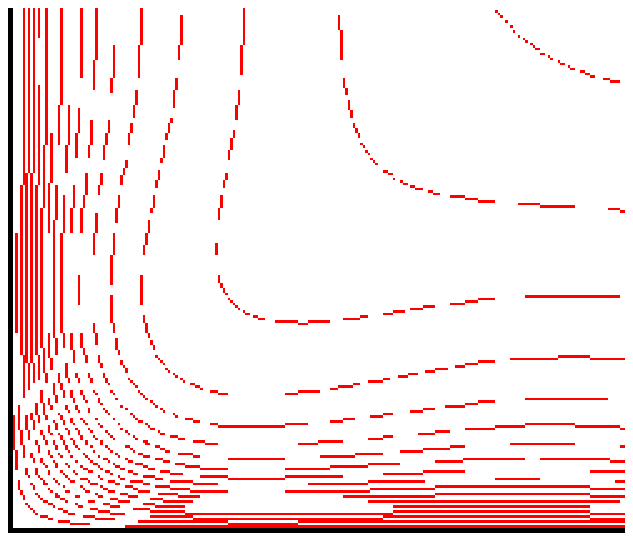}
\hskip -0.5cm
\includegraphics[clip,width=3.5cm]{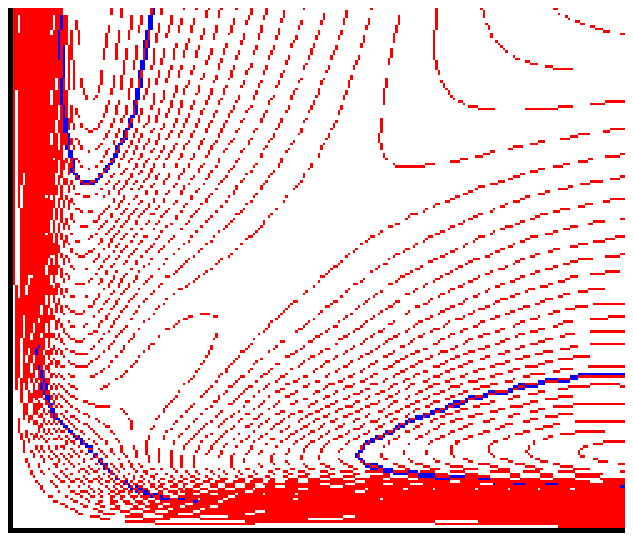}
\hskip -0.5cm
\includegraphics[clip,width=3.5cm]{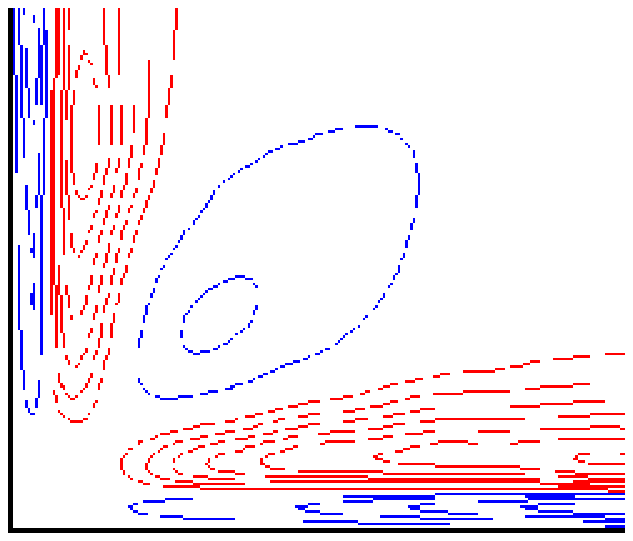}
\hskip -0.5cm
\includegraphics[clip,width=3.5cm]{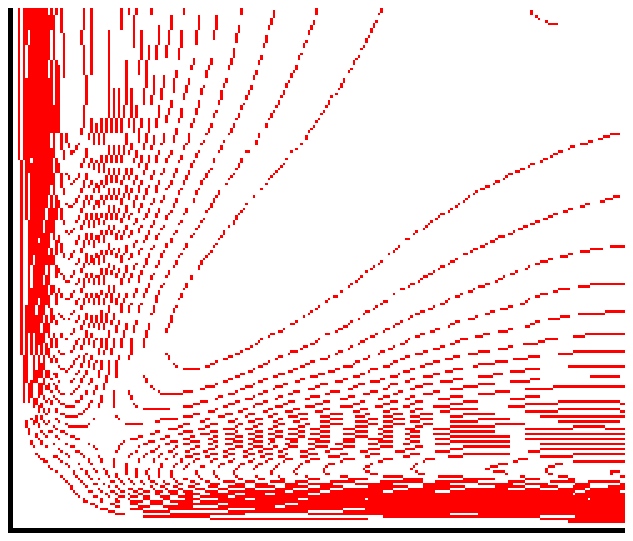}
\hskip -0.5cm
\includegraphics[clip,width=3.5cm]{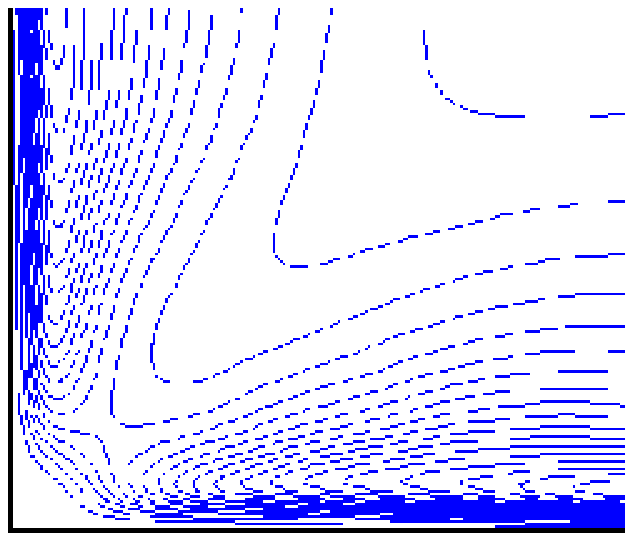}
\vskip -0.0cm
a) \hskip 2.7cm b) \hskip 2.7cm c) \hskip 2.7cm d) \hskip 2.7cm e) \\
%\hskip 1.0cm  aa) \hskip 2.5cm  ab) 
%\hskip 2.5cm  ac) \hskip 2.5cm  ad) \hskip 2.5cm  ae)
\vskip -0.0cm
\hskip -1.8cm
\includegraphics[clip,width=3.5cm]{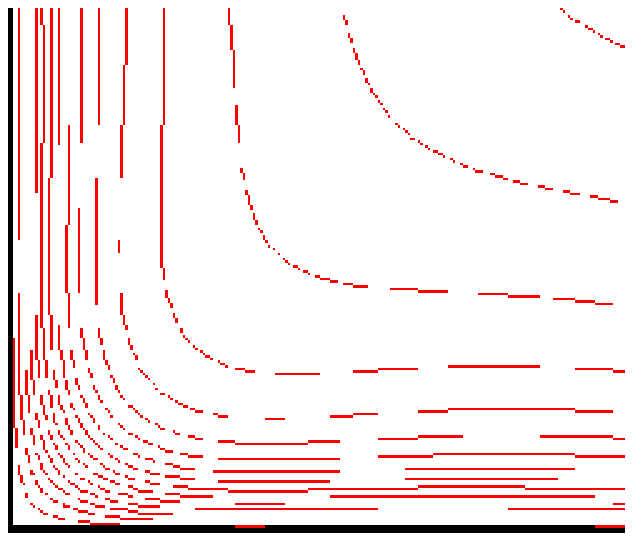}
\hskip -0.5cm
\includegraphics[clip,width=3.5cm]{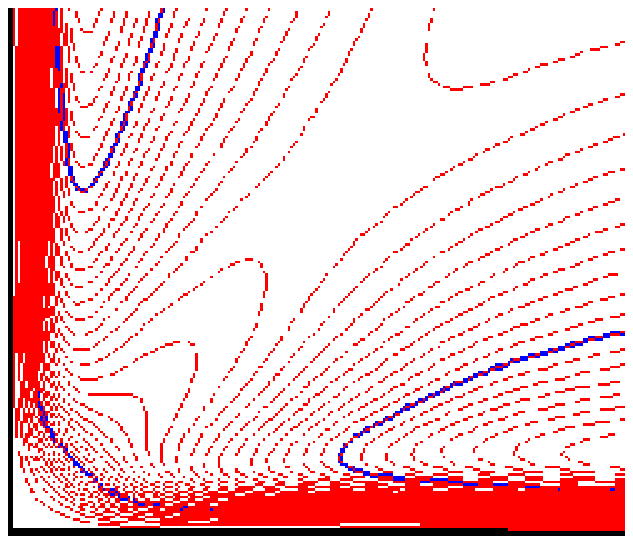}
\hskip -0.5cm
\includegraphics[clip,width=3.5cm]{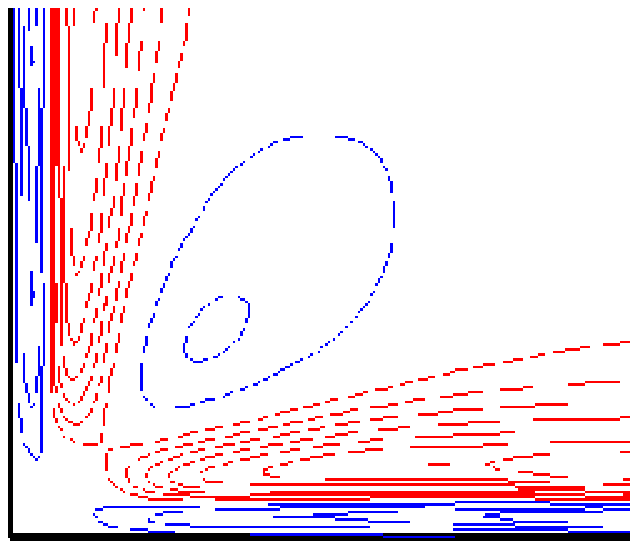}
\hskip -0.5cm
\includegraphics[clip,width=3.5cm]{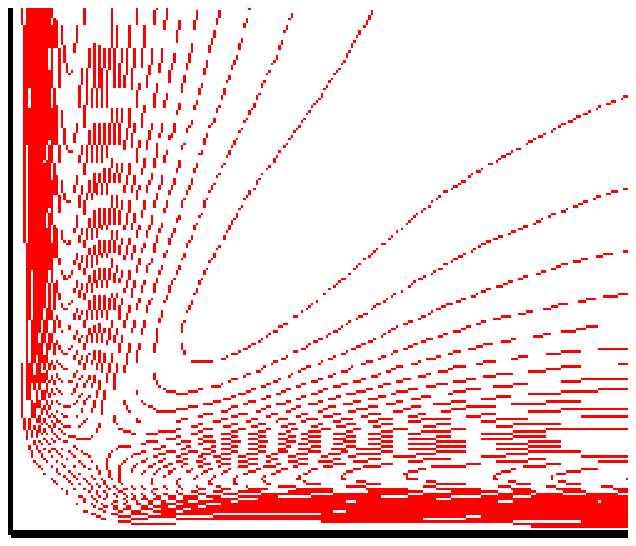}
\hskip -0.5cm
\includegraphics[clip,width=3.5cm]{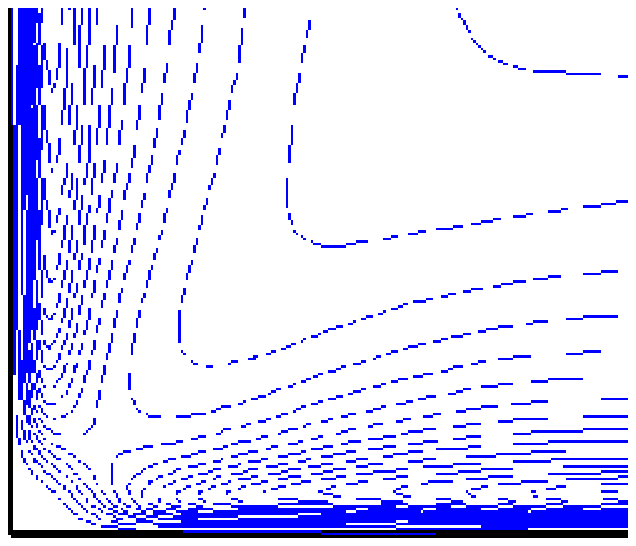}
\vskip -0.0cm
f) \hskip 2.7cm g) \hskip 2.7cm h) \hskip 2.7cm i) \hskip 2.7cm j) \\
%\hskip 1.0cm  ba) \hskip 2.5cm  bb) 
%\hskip 2.5cm  bc) \hskip 2.5cm  bd) \hskip 2.5cm  be)
\vskip -0.0cm
\hskip -1.8cm
\includegraphics[clip,width=3.5cm]{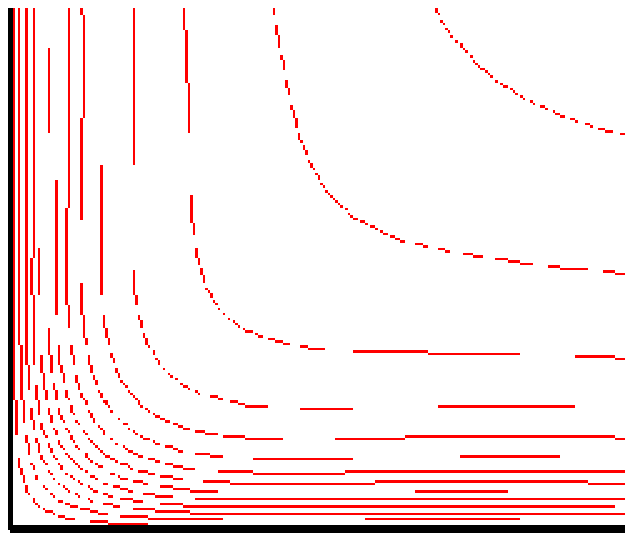}
\hskip -0.5cm
\includegraphics[clip,width=3.5cm]{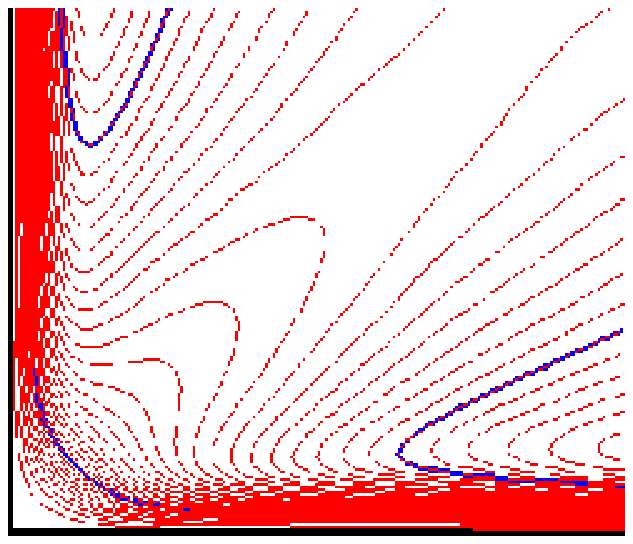}
\hskip -0.5cm
\includegraphics[clip,width=3.5cm]{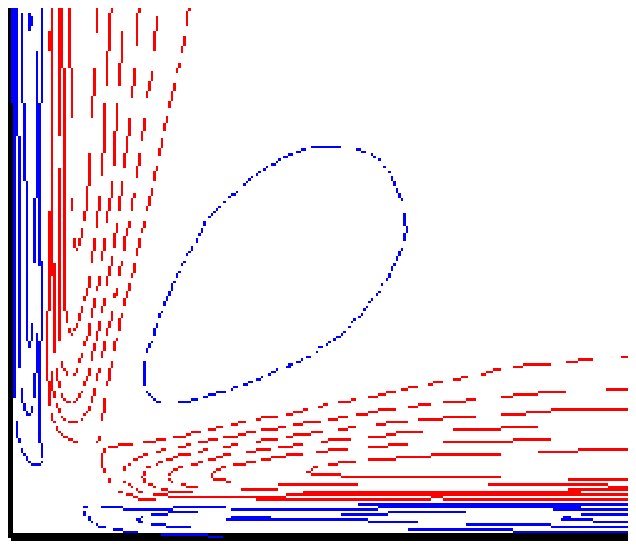}
\hskip -0.5cm
\includegraphics[clip,width=3.5cm]{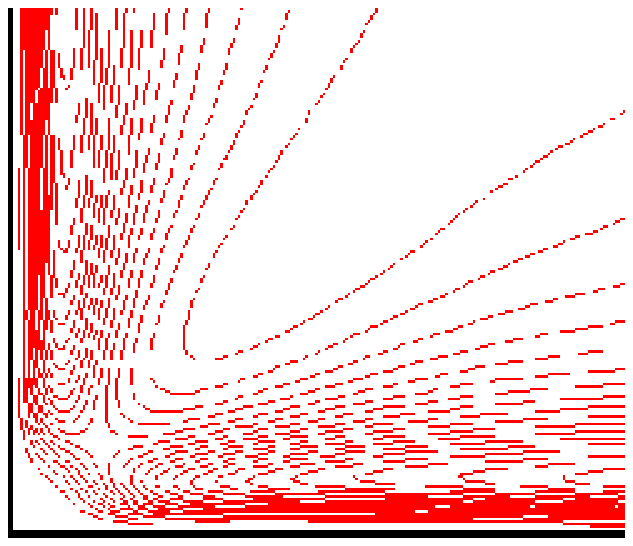}
\hskip -0.5cm
\includegraphics[clip,width=3.5cm]{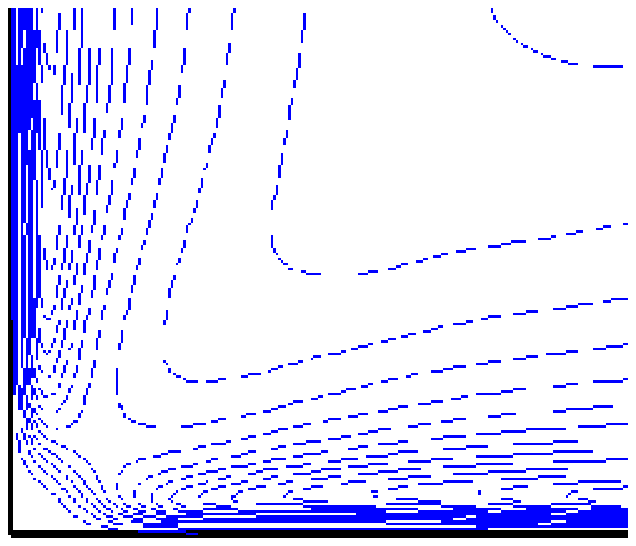}
\vskip -0.0cm
k) \hskip 2.7cm l) \hskip 2.7cm m) \hskip 2.7cm n) \hskip 2.7cm o) \\
%\hskip 1.0cm  ca) \hskip 2.5cm  cb) 
%\hskip 2.5cm  cc) \hskip 2.5cm  cd) \hskip 2.5cm  ce)
\vskip -0.0cm
\vskip -0.0cm
\caption{Rectangular duct ($A_R=1$): contours of
$U_1^+$ (a,f,k); $q^+$ (b,g,l), $C_K^+$ (c,h,m),
$P_K^+$ (d,i,n), $D_K^+$ (e,j,o),
at $Re=2500$ (a-e), $Re=7750$ (f-j),
$Re=15000$ (k-o).
Only a $100^+ \times 100^+$ box is shown near the corner.
Positive contours are shown in red, and negative in blue, 
with increments $\Delta=1$ for $U_1^+$ and $q^+$,
and $\Delta=.001$ for the budgets terms.
The blue line denotes $q^+=1$.
}
\label{fig18}
\end{figure}

To investigate whether the scaling in wall units holds by increasing 
the Reynolds number, in figure~\ref{fig18} we show the contours, 
of mean velocity, turbulent kinetic energy and 
terms of the turbulent kinetic energy budget, for $A_R=1$.
Comparison of the $U_1^+$ contours
demonstrates that the distortion is greater at lower $Re$,
implying that the influence of the secondary motion is stronger,
in large part of the quadrant. At high Reynolds numbers 
($Re=7750$ and $15000$)
the contours are quite similar, and similar variations of $q^+$ 
may be observed at all $Re$. The panels reporting
$P_K^+$ and $D_k^+$, show once more
good balance between turbulent kinetic energy production
and total dissipation. The regions with $D_K^+$ higher than
$P_K^+$ correspond to the thin regions with negative 
$C_K^+$ near the two walls. Far from the corner and
in the region close to the corner bisector $D_K^+$ slightly
overcomes $P_k^+$, however this is a region with weak turbulence
according to the contours of $q^+$. The results
shown in figure~\ref{fig18} thus corroborate the previous
findings about good scaling in wall units
of mean motion, turbulence and budgets in a square duct.
However, in the previous visualizations logarithmic scales for the
coordinates were used to emphasise the behavior near the corner, 
whereas linear scales are used here to also analyse the behavior far from the
corner.

\begin{figure}
\centering
\vskip -0.0cm
\hskip -1.8cm
\includegraphics[clip,width=3.5cm]{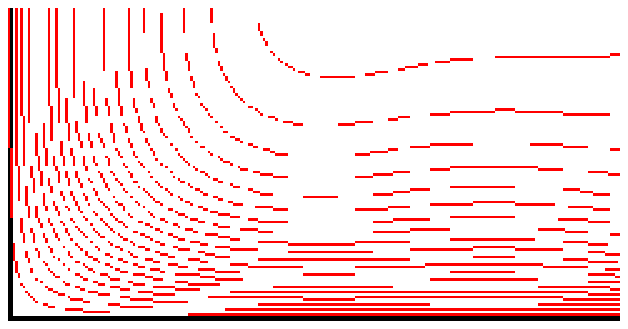}
\hskip -0.5cm
\includegraphics[clip,width=3.5cm]{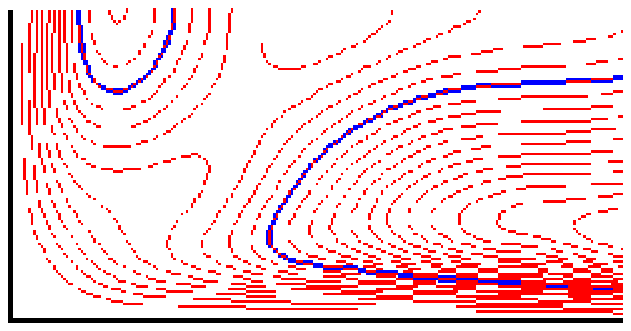}
\hskip -0.5cm
\includegraphics[clip,width=3.5cm]{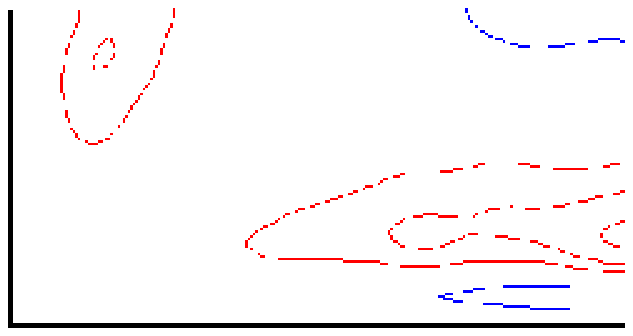}
\hskip -0.5cm
\includegraphics[clip,width=3.5cm]{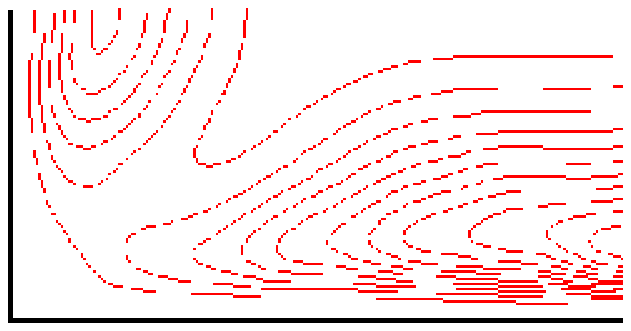}
\hskip -0.5cm
\includegraphics[clip,width=3.5cm]{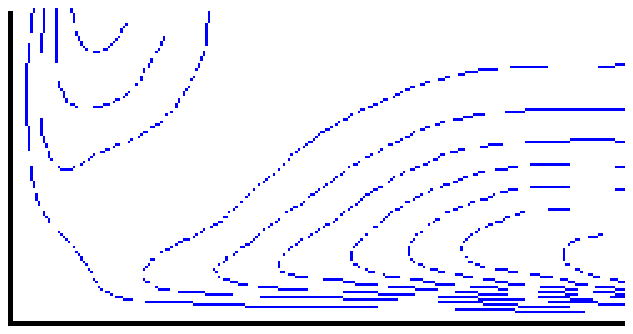}
\vskip -0.0cm
a) \hskip 2.7cm b) \hskip 2.7cm c) \hskip 2.7cm d) \hskip 2.7cm e) \\
%\hskip 1.0cm  aa) \hskip 2.5cm  ab) 
%\hskip 2.5cm  ac) \hskip 2.5cm  ad) \hskip 2.5cm  ae)
\vskip -0.0cm
\hskip -1.8cm
\includegraphics[clip,width=3.5cm]{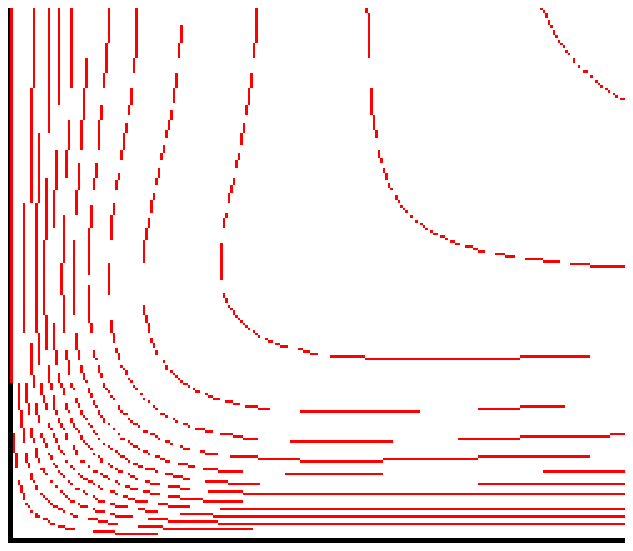}
\hskip -0.5cm
\includegraphics[clip,width=3.5cm]{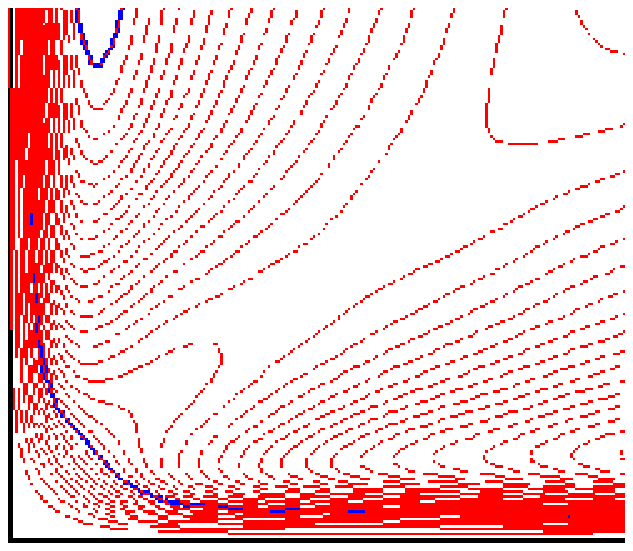}
\hskip -0.5cm
\includegraphics[clip,width=3.5cm]{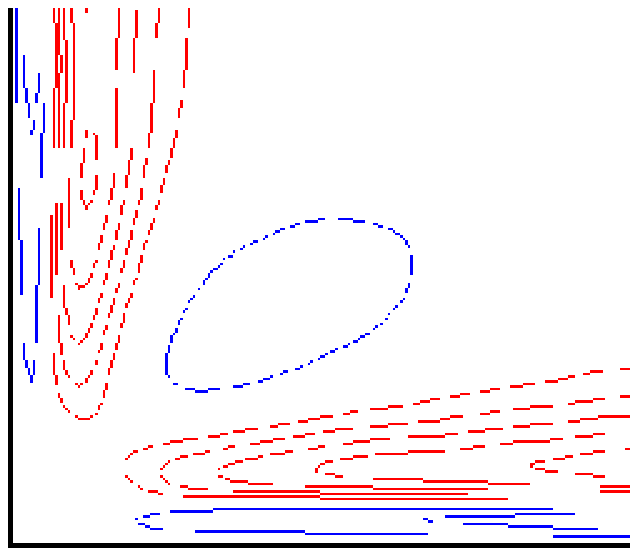}
\hskip -0.5cm
\includegraphics[clip,width=3.5cm]{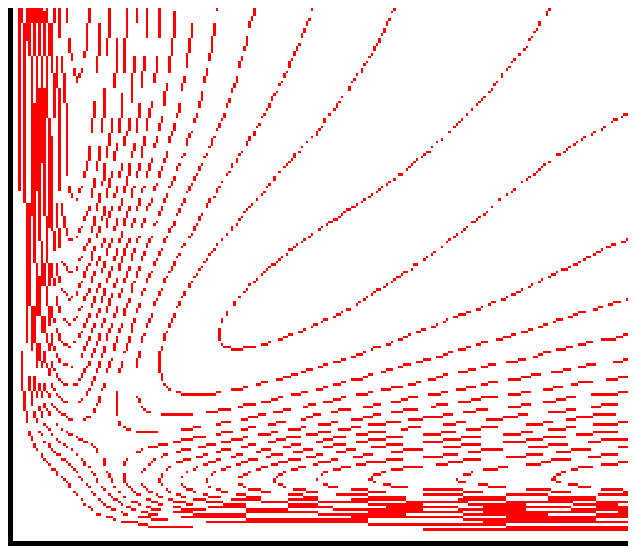}
\hskip -0.5cm
\includegraphics[clip,width=3.5cm]{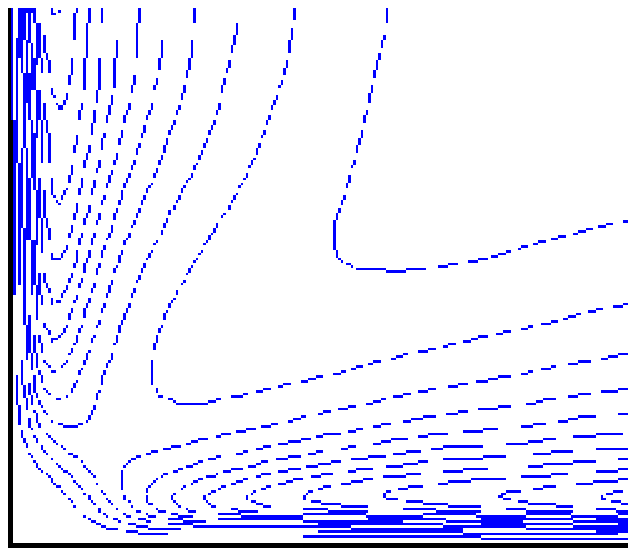}
\vskip -0.0cm
f) \hskip 2.7cm g) \hskip 2.7cm h) \hskip 2.7cm i) \hskip 2.7cm j) \\
%\hskip 1.0cm  ba) \hskip 2.5cm  bb) 
%\hskip 2.5cm  bc) \hskip 2.5cm  bd) \hskip 2.5cm  be)
\vskip -0.0cm
\hskip -1.8cm
\includegraphics[clip,width=3.5cm]{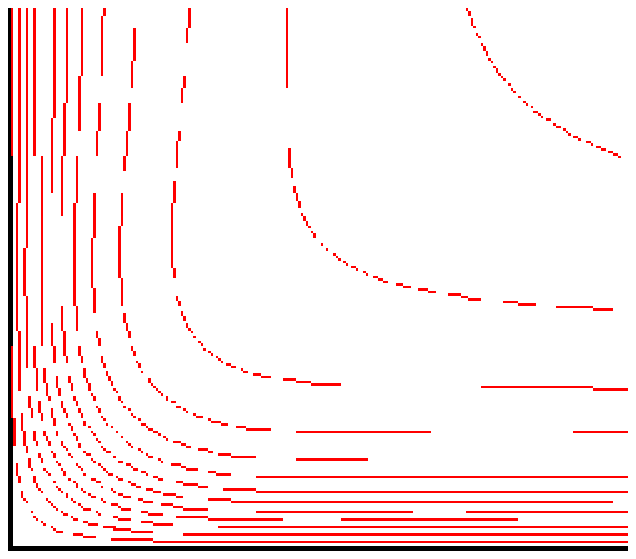}
\hskip -0.5cm
\includegraphics[clip,width=3.5cm]{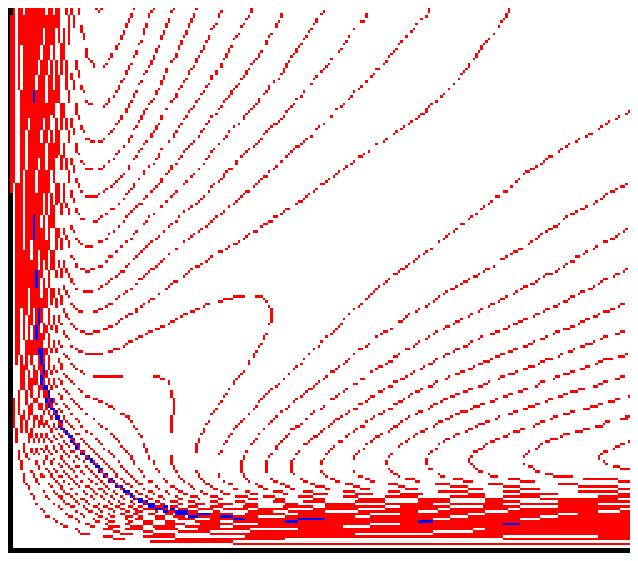}
\hskip -0.5cm
\includegraphics[clip,width=3.5cm]{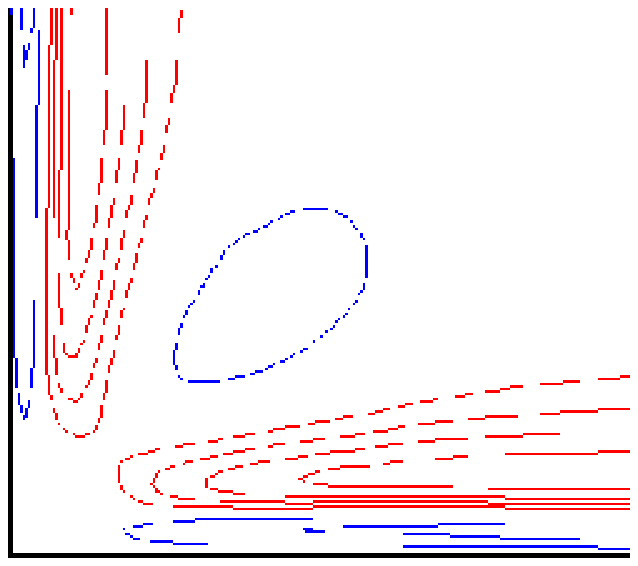}
\hskip -0.5cm
\includegraphics[clip,width=3.5cm]{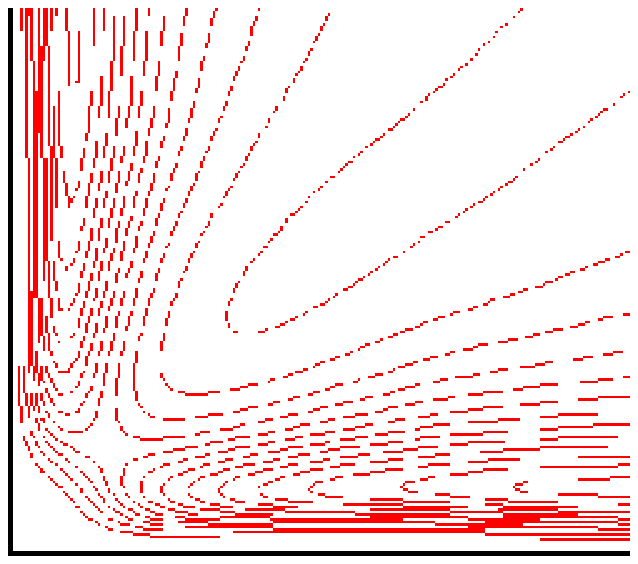}
\hskip -0.5cm
\includegraphics[clip,width=3.5cm]{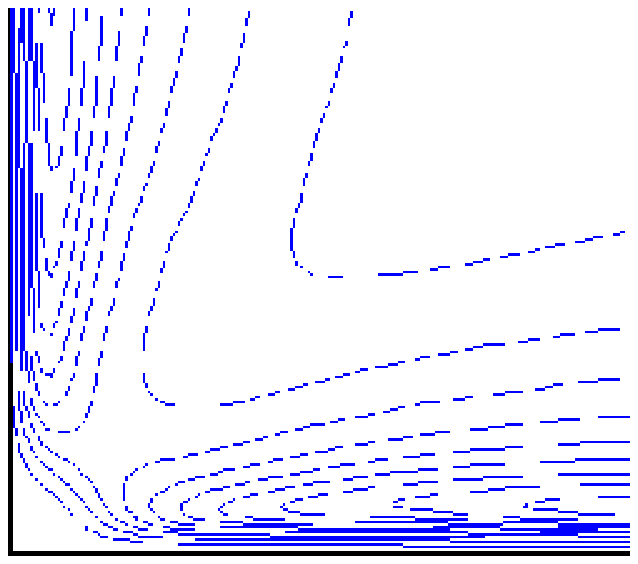}
\vskip -0.0cm
k) \hskip 2.7cm l) \hskip 2.7cm m) \hskip 2.7cm n) \hskip 2.7cm o) \\
%\hskip 1.0cm  ca) \hskip 2.5cm  cb) 
%\hskip 2.5cm  cc) \hskip 2.5cm  cd) \hskip 2.5cm  ce)
\vskip -0.0cm
\hskip 1.0cm  $U^+$ \hskip 2.5cm  $q^+$ \hskip 2.5cm  $C_K^+$
%\hskip 2.5cm  $Ar5$ 
\hskip 2.5cm  $P_K^+$ \hskip 2.5cm  $D_K^+$
\vskip -0.0cm
\caption{Rectangular duct: contours of
$U_1^+$ (a,f,k); $q^+$ (b,g,l), $C_K^+$ (c,h,m),
$P_K^+$ (d,i,n), $D_K^+$ (e,j,o),
at $Re=2500$, $A_R=6$ (a-e), $Re=7750$, $A_R=7$ (f-j),
$Re=15000$, $A_R=7$ (k-o).
Only a $100^+ \times 100^+$ box is shown near the corner.
Positive contours are shown in red, and negative in blue, 
with increments $\Delta=1$ for $U_1^+$ and $q^+$,
and $\Delta=.001$ for the budgets terms.
The blue line denotes $q^+=1$.
}
\label{fig19}
\end{figure}

It is thus worth analysing whether 
mean and turbulent quantities also scale with the mean friction
velocity for $A_R \ne 1$. A rectangular duct with $A_R=7$ has been considered
to compare the distribution of the various quantities 
with those in figure~\ref{fig18}, at $Re=7750$ and $Re=15000$.
However, for $A_R=7$, $Re=2500$, the flow is laminar, as may be argued
from the $C_f$ shown in figure~\ref{fig15}a (black dots).
On the other hand, turbulent flow is found at $A_R=6$, $Re=2500$,
thus in figure~\ref{fig19} we show the behavior of the quantities at $Re=2500$
for $A_R=6$. At this low $Re$ the choice to analyse
a $100^+ \times 100^+$ box around the corner leads to a vertical
size in the top panels of figure~\ref{fig19} which is
shorter than the horizontal one, in fact $L_3^+/2=61$, whereas 
$L_2^+/2= 367$. These differences, in particular in the turbulent kinetic 
energy, lead to different distributions close to the horizontal and
vertical walls. The blue line in figure~\ref{fig19}b,
corresponding to $q^+=1$, near the vertical wall is localised in a 
smaller region than that near the horizontal wall. This is
corroborated by the contours of $P_K^+$ in figure~\ref{fig19}d and 
of $D_K^+$ in figure~\ref{fig19}e. The total dissipation $D_K^+$
near the vertical wall does not overcome $P_k^+$,
whereas this occur near the horizontal wall, and this difference may be 
also inferred from the contours of $C_K^+$ in figure~\ref{fig19}c. 
At $Re=7750$, $L_3^+/2=170$, hence the $100^+ \times 100^+$ box
should not show large differences 
near the walls. Indeed, both the contours of $U_1^+$ and $q^+$
in the middle panels show better symmetry around the corner bisector.
Symmetry is further supported by the distributions of
the three terms in the turbulent kinetic energy budget, which
are very similar to those at the same $Re$ in
figure~\ref{fig18}. Visualizations in a wider region show
that at larger wall distances than $100^+$, all quantities 
do not change, and behave similarly near walls. This explains
why $C_f$ in the fully turbulent regime scales well regardless
of the definition of Reynolds number. At $Re=15000$,
$L_3^+/2=310$, hence in
bottom figures the contours close to the corners show better symmetry
than that at $Re=7750$.
From figure~\ref{fig19} it may be asserted that at high Reynolds 
number, that is in the fully turbulent regime, 
mean and turbulent motion scale well with the averaged friction velocity
also for rectangular ducts.
Under these conditions, the
corrections caused by the corner are concentrated in a small region,
and large part of the duct is occupied by turbulent flow not
different from a canonical planar channel.
Completely different is the behavior near the transitional
Reynolds number, in which the corner bisector symmetry is lost. 
In ducts with high aspect ratios, 
the near-wall structures are constrained near the short side, 
thus turbulence is not sustained, and although the other side is 
long enough, $q^+$ remains nearly zero.

\section{Concluding remarks}

In this paper have been reported the results of DNS
in rectangular ducts with different aspect ratio in
the laminar, transitional and fully turbulent regimes.
The case with $A_R=1$, the square duct, largely investigated
in real and numerical experiments has been considered
to validate the numerical method. The results compared well
with those available in literature. A particular emphasis
has been directed in the transitional regime showing
the formation of a secondary motion consisting in four recirculating 
regions instead that the eight characteristic of the fully 
turbulent regime. In presence of four regions the $C_f$, even
if greater than that should be found without secondary
motion decays linearly with the Reynolds number as in the laminar
regime. Going from four to eight regions the profiles of $\tau_w$
from the corner increase with a behavior rather different and
reaching at high Reynolds numbers a constant profile in a
large part of the duct. In addition it has been found
that the strength of the secondary motion is large at
low and decreases by increasing the Reynolds number.
Therefore at low $Re$ the secondary motion can
be of interest to increase the mixing or the heat
transfer. At high $Re$ the reduction of the strength
and the shrinking of the secondary motion in a
small region near the corner leads to a behavior of
the $C_f$ as well as the profiles of the mean motion and
the turbulent statistics similar to those in the canonical
two-dimensional channel. At high $Re$ there is not
large difference on the definition of the reference
length in plotting the $C_f$ versus the bulk Reynolds
number. On the other hand, in particular in presence
of rectangular ducts with $A_R> 1$ the transitional
Reynolds number is independent of the aspect ratio
by taking as reference length half of the smaller side.
Having observed that for the square duct the critical $Re$
is different from that for $A_R>1$ it has been argued
that for the  ducts with $A_R>1$ the asymmetric disturbances
emanating from the corner act as a trip device. 

Whose interested to apply RANS closures to simulate
flows in practical applications for rectangular ducts
are aware of the difficulties to reproduce the anisotropy
and the asymmetries of the turbulent Reynolds stresses.
In this paper it has been shown that by taking
as reference system that oriented along the
eigenvectors of the strain tensor instead of 
the Cartesian reference frame the anisotropy of
the Reynolds stresses is reduced and the 
normal stresses are symmetric with respect the diagonal.
This can be an useful results in constructing more
reliable RANS closures. The further result that
the behavior of the total dissipation is simpler to model
than the isotropic rate of turbulent kinetic energy could
be useful in RANS closures.

In rectangular ducts at high Reynolds numbers it has been
found that the mean motion, the turbulent kinetic energy
as well as its budget in wall units scale rather well 
with the Reynolds number and with the aspect ratio in
the region near the corner. At low $Re$ the short side
plays an important role, if the friction Reynolds number
evaluated by the short side is low the turbulence can not be
sustained and the flows remains laminar.

\section{Acknowledgements}
We acknowledge that some of the results reported in this paper have been 
achieved using the PRACE Research Infrastructure resource MARCONI
based at CINECA, Casalecchio di Reno, Italy.
\bibliographystyle{jfm}
\bibliography{references}
\end{document}